\documentclass[a4paper,12 pt,twosided,openright,oldfontcommands]{memoir}

\setlrmarginsandblock{3.5cm}{2.5cm}{*}
\setulmarginsandblock{2.5cm}{*}{1}
\checkandfixthelayout 
\usepackage[thinc]{esdiff}
\usepackage{physics}
\usepackage{amsmath,amsthm,amssymb}
\usepackage{tikz}
\usepackage{nameref}
\usepackage{feynmf}
\usepackage{grffile}
\usepackage{url}
\usepackage{graphbox}

\usepackage[backend=bibtex,natbib=true,sorting=none,url=false,style=numeric-comp]{biblatex}
\usepackage{ragged2e}
\usepackage{nomencl}
\defbibheading{subbibliography}{%
  \section*{Chapter \ref*{refsegment:\therefsection\therefsegment} - \nameref*{refsegment:\therefsection\therefsegment}}}
\defbibheading{synopsis}{%
  \section*{\nameref*{refsegment:\therefsection\therefsegment}}}
\defbibheading{appendix}{%
  \section*{Appendix \ref*{refsegment:\therefsection\therefsegment}}}
\usetikzlibrary{shadows}
\usetikzlibrary{shapes}
\usepackage{array}
\usepackage[T1]{fontenc}
\usepackage{multirow,bigdelim}
\usepackage{newtxtext}
\usepackage[cmintegrals]{newtxmath}
\usepackage{hyperref}
\hypersetup{
    pdftitle={Exploring Oscillation Dip and Valley, Non-Standard Interactions, and Earth's Core using Atmospheric Neutrinos at ICAL-INO detector \& Studies on Response Uniformity of RPC},    % title
    pdfauthor={Anil Kumar},     % author
    colorlinks=true,       % false: boxed links; true: colored links
    linkcolor=blue,          % color of internal links (change box color with linkbordercolor)
    citecolor=red,        % color of links to bibliography
    urlcolor=blue           % color of external links
}

\usepackage{graphicx}
\chapterstyle{madsen}
\definecolor{darkgreen}{RGB}{0,54,0}
\definecolor{darkred}{RGB}{87,0,0}
\definecolor{darkindi}{RGB}{54,0,128}
\definecolor{olivegreen}{RGB}{109,113,46}
\definecolor{forestgreen}{RGB}{0,110,51}

\def\epsmutau{\varepsilon_{\mu\tau}}

\usepackage[font=small,format=hang,labelfont={sc}]{caption,subfig}
\setsecnumdepth{subsection}
\makenomenclature
\begin{document}
%\begin{titlingpage}
\begin{adjustwidth}{0 cm}{1 cm}
%\begin{adjustwidth}{1.5 cm}{0 cm}
\pagenumbering{roman}
\begin{center}
\thispagestyle{empty}
\Large{\textbf{Studies on Response Uniformity of RPC and Exploring Oscillation Dip and Valley, Non-Standard Interactions, and Earth's Core using Atmospheric Neutrinos at ICAL-INO detector}}\\
\vspace{38 pt}
\large{\textbf{\emph{By\\
\ \\
Anil Kumar}}}
\ \\
\ \\
PHYS01201604017\\%(Your Enrollment No.)
\ \\
Bhabha Atomic Research Centre\\
\ \\ \vspace{1cm}
\large{\textbf{\emph{A thesis submitted\\
to the Board of Studies in\\
Physical Sciences\\
\ \\
\vspace{1cm}
In partial fulfillment of requirements\\
For the Degree of}}}\\
\ \\
\linespread{1.5}
\vspace{1cm}
{\fontfamily{phv}\selectfont \textbf{DOCTOR OF PHILOSOPHY}}\\
\emph{of}\\
\textbf{HOMI BHABHA NATIONAL INSTITUTE}\\
\ \\ \vspace{1cm}
\includegraphics[width=3 cm]{"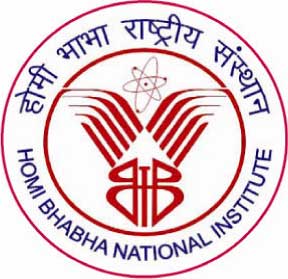"}\\
\textbf{May, 2022}%(example. June, 2011)
\end{center}
\end{adjustwidth}
%\end{titlingpage}
%%%%%%%%%%%%%%%%%%%%%%%%%%%%%%%%%%%%%%%%
\cleartooddpage
\begin{center}
{\Large\textbf{Homi Bhabha National Institute \\}}
{\large\textbf{\ \\Recommendations of the Viva Voce Committee}}
\end{center}
As members of the Viva Voce Committee, we certify that we have read the
dissertation prepared by \textbf{Anil Kumar} entitled \textbf{``Studies on Response Uniformity of RPC and Exploring Oscillation Dip and Valley, Non-Standard Interactions, and Earth's Core using Atmospheric Neutrinos at ICAL-INO detector''} and recommend that it may be accepted as fulfilling the thesis requirement for the award of Degree of Doctor of Philosophy.
\begin{center}
\begin{tabular}{p{0.74\linewidth}p{0.19\linewidth}}
\hline
& \\
Chairman - Prof. Subhasis Chattopadhyay & \textbf{Date:} \\
& \\
 & \\ \hline
  & \\
Guide / Convener - Prof. Sanjib Kumar Agarwalla & \textbf{Date:}\\
 & \\ 
 & \\ \hline
 & \\
 Co-guide - Prof. Supratik Mukhopadhyay & \textbf{Date:}\\
 & \\ 
 & \\ \hline
  & \\

 Examiner - Prof. Subhendu Rakshit & \textbf{Date:}\\ %( comment out this row when submitting for evaluation, You may keep this row while submitting the final copy to HBNI)
 & \\ 
  & \\ \hline
  & \\
Member 1 - Prof. Nayana Majumdar & \textbf{Date:}\\
 & \\ 
 & \\ \hline
 & \\
Member 2 - Prof. Aruna Kumar Nayak & \textbf{Date:}\\
 & \\ 
 & \\ \hline
 & \\
Member 3 - Prof. Nita Sinha & \textbf{Date:}\\ %(If any)
 & \\
\end{tabular}
\end{center}
\rule{\linewidth}{2.5 pt}\\

Final approval and acceptance of this thesis is contingent upon the candidate's
submission of the final copies of the thesis to HBNI.

We hereby certify that we have read this thesis prepared under our direction and recommend that it may be accepted as fulfilling the thesis requirement.\\
\begin{center}
%\begin{tabular}{p{0.185\linewidth}p{0.4\linewidth}>{\centering\arraybackslash}p{0.40\linewidth}}
\begin{tabular}{p{0.185\linewidth}>{\centering\arraybackslash}p{0.4\linewidth}>{\centering\arraybackslash}p{0.40\linewidth}}
\textbf{Date:}& & \\[10 pt]
\textbf{Place:} & \textbf{Prof. Supratik Mukhopadhyay} (Co-guide)& \textbf{Prof. Sanjib Kumar Agarwalla} (Guide)\hfill
\end{tabular}
\end{center}
%%%%%%%%%%%%%%%%%%%%%%%%%%%%%%%%%%%%%%%
%%%%%%%%%%%%%%%%%%%%%%%%%%%%%%%%%%%%%%%
\cleartooddpage
\thispagestyle{empty}
\DoubleSpacing
\begin{center}
{\large\textbf{\textsc{Statement by the Author}}}
\end{center}
This dissertation has been submitted in partial fulfillment of requirements for an advanced degree at Homi Bhabha National Institute (HBNI) and is deposited in the library to be made available to borrowers under rules of the HBNI.

Brief quotations from this dissertation are allowable without special permission, provided that accurate acknowledgement of source is made. Requests for permission for extended quotation from or reproduction of this manuscript in whole or in part may be granted by the Competent Authority of HBNI when in his or her judgment the
proposed use of the material is in the interests of scholarship. In all other instances, however, permission must be obtained from the author.
\vspace{24 pt}\\
\begin{center}
	\begin{tabular}{p{0.17\linewidth}p{0.55\linewidth}>{\centering\arraybackslash}p{0.3\linewidth}}
		{May 2022}& & \\
		{Bhubaneswar} &  & \emph{Anil Kumar}\hfill
	\end{tabular}
\end{center}
\cleartooddpage
\begin{center}
{\large\textbf{\textsc{Declaration}}}
\end{center}
I hereby declare that the investigation presented in the thesis has been carried out by me. The work is original and has not been submitted earlier as a whole or in part for a degree/diploma at this or any other Institution/University.
\vspace{24 pt}\\
\begin{center}
	\begin{tabular}{p{0.17\linewidth}p{0.55\linewidth}>{\centering\arraybackslash}p{0.3\linewidth}}
		{May 2022}& & \\
		{Bhubaneswar} &  & \emph{Anil Kumar}\hfill
	\end{tabular}
\end{center}
\cleartooddpage
\thispagestyle{empty} 
\begin{center}
	{\LARGE\textbf{List of Publications Arising from the Thesis}}
\end{center}

\section*{Publications in Refereed Journal}
\subsection*{Published}
\begin{enumerate}
\item \textbf{Validating the Earth's Core using Atmospheric Neutrinos with ICAL at INO} \\
\underline{Anil Kumar}, Sanjib Kumar Agarwalla \\
\href{https://doi.org/10.1007/JHEP08(2021)139}{Journal of High Energy Physics, 08 (2021) 139} \\
e-Print arXiv: \href{https://arxiv.org/abs/2104.11740}{2104.11740 [hep-ph]} 

\item \textbf{A New Approach to Probe Non-Standard Interactions in Atmospheric Neutrino Experiments} \\
\underline{Anil Kumar}, Amina Khatun, Sanjib Kumar Agarwalla, Amol Dighe \\
\href{https://doi.org/10.1007/JHEP04(2021)159}{Journal of High Energy Physics, 04 (2021) 159} \\
e-Print arXiv: \href{https://arxiv.org/abs/2101.02607}{2101.02607 [hep-ph]}

\item \textbf{From oscillation dip to oscillation valley in atmospheric neutrino experiments} \\
\underline{Anil Kumar}, Amina Khatun, Sanjib Kumar Agarwalla, Amol Dighe \\
\href{https://doi.org/10.1140/epjc/s10052-021-08946-8}{The European Physical Journal C, volume 81 (2021) 2, 190} \\
e-Print arXiv: \href{https://arxiv.org/abs/2006.14529}{2006.14529 [hep-ph]}
\end{enumerate}
\section*{Symposium and Conference Proceedings}
\begin{enumerate}
\item \textbf{Exploring NSI using oscillation dip and valley in atmospheric neutrino experiments} \\
\underline{Anil Kumar}, Amina Khatun, Sanjib Kumar Agarwalla, Amol Dighe \\
Proceedings of The 17th International Conference on Topics in Astroparticle and Underground Physics (TAUP2021), Online Conference, 26 August - 30 September, 2021\\
\href{https://doi.org/10.1088/1742-6596/2156/1/012119}{Journal of Physics: Conference Series 2156 (2021) 012119}

\item \textbf{Probing the Earth's Core using Atmospheric Neutrinos at INO
} \\
\underline{Anil Kumar}, Sanjib Kumar Agarwalla \\
Proceedings of The European Physical Society Conference on High Energy Physics (EPS-HEP2021), Online Conference, 26-30 July, 2021 \\
\href{https://pos.sissa.it/398/257/pdf}{PoS(EPS-HEP2021)257}, e-Print arXiv: \href{https://arxiv.org/abs/2110.08333}{2110.08333 [hep-ph]}

\item \textbf{Probing NSI in Atmospheric Neutrino Experiments using Oscillation Dip and Valley} \\
\underline{Anil Kumar}, Amina Khatun, Sanjib Kumar Agarwalla, Amol Dighe \\
Proceedings of the XXIV DAE-BRNS High Energy Physics Symposium 2020, NISER, Bhubaneswar, India, 14-18 December, 2020 \\
\href{https://doi.org/10.1007/978-981-19-2354-8_96}{Springer Proc.Phys. 277 (2022) 525-529}, e-Print arXiv: \href{https://arxiv.org/abs/2104.06955}{2104.06955 [hep-ph]} 

\item \textbf{Effect of Variation of Surface Resistivity of Graphite Layer in RPC} \\
\underline{Anil Kumar}, V. Kumar, S. Mukhopadhyay, S. Sarkar, and N. Majumdar \\
Proceedings of the XXIII DAE-BRNS High Energy Physics Symposium 2018, IIT, Madras, India, 10-14 December, 2018\\
\href{https://doi.org/10.1007/978-981-33-4408-2_100}{Springer Proc.Phys. 261 (2021) 725-730}
 
\end{enumerate}

\section*{Symposium and Conference Attended}
\subsection*{Oral Presentations}
\begin{enumerate}
	\item \textbf{Validating the Earth's Core using Atmospheric Neutrinos with ICAL at INO}, National Science Day, Tezpur University, Mar 1, 2022
	\item \textbf{A New Approach to Probe Non-Standard Interactions in Atmospheric Neutrino Experiments}
	\begin{itemize}
		\item The 22nd International Workshop on Neutrinos from Accelerators (NuFact 2021, online), Sep 6 to 11, 2021
		\item The 17th International Conference on Topics in Astroparticle and Underground Physics, (TAUP 2021, online), Aug 26 - Sep 3, 2021
		\item APS April Meeting 2021 - online, Apr 17-20, 2021
		\item XIX International Workshop on Neutrino Telescopes (Neutel 21), Padova (Italy) - online, Feb 18-26, 2021
		\item XXIV DAE-BRNS High Energy Physics (HEP) Symposium (held online), NISER, Bhubaneswar, India, Dec 14 - 18, 2020.
	\end{itemize} 

	\item \textbf{From Oscillation Dip to Oscillation Valley in Atmospheric Neutrino Experiments}, Virtual Neutrino Theory mini-workshop, Sep 21-23, 2020
\end{enumerate}

\subsection*{Poster Presentations}
\begin{enumerate}
	\item \textbf{A New Approach to Probe Non-Standard Interactions in Atmospheric Neutrino Experiments}
	\begin{itemize}
		\item Lepton Photon 2021, online, Jun 10 - 14, 2022
		\item The 28th International Workshop on Weak Interactions and Neutrinos (WIN 2021, online), Jun 7 - 12, 2021
		\item Invisibles21 virtual Workshop, May 31 to Jun 4, 2021
	\end{itemize}
	\item \textbf{Validating the Earth's Core using Atmospheric Neutrinos with ICAL at INO}, The 22nd International Workshop on Neutrinos from Accelerators (NuFact 2021, online), Sep 6 to 11, 2021

	\item \textbf{Neutrino tomography of Earth using ICAL@INO}, XXIV DAE-BRNS High Energy Physics (HEP) Symposium (held online), NISER, Bhubaneswar, India, Dec 14 - 18, 2020.
	\item \textbf{From oscillation dip to oscillation valley in atmospheric neutrino experiments}, Neutrino 2020 conference (held online), Jun 22 - Jul 2, 2020
	\item \textbf{Effect of Variation of Surface Resistivity of Graphite layer in RPC},
	XXIII DAE-BRNS High Energy Physics (HEP) Symposium 2018 held at IIT Madras, India, Dec 10-14, 2018
	\item \textbf{Exploring Neutrino properties using Atmospheric Neutrinos at ICAL},
	International Neutrino Summer School 2018 at Schloss Waldthausen near Mainz, Germany, May 21 to Jun 1, 2018
\end{enumerate}

\vspace{24 pt}
\begin{center}
	\begin{tabular}{p{0.17\linewidth}p{0.55\linewidth}>{\centering\arraybackslash}p{0.3\linewidth}}
		{May 2022}& & \\
		{Bhubaneswar} &  & \emph{Anil Kumar}\hfill
	\end{tabular}
\end{center}
%%%%%%%%%%%%%%%%%%%%%%%%%%%%%%%%%%%%%%%
\cleartooddpage
\thispagestyle{empty}
\begin{center}
{\Large\textsc{Dedicated to\\}}
\ \\
\emph{My Parents}
\end{center}
\cleartooddpage 
\begin{center}
\thispagestyle{empty}
{\Large\textbf{\textsc{Acknowledgements\\}}} 
\end{center}
The long journey to Ph.D. would not have been possible without the support and encouragement of many people. Here, I would like to take the opportunity to thank them. 

First of all, I would like to thank my supervisor, Prof. Sanjib Kumar Agarwalla, and co-supervisor, Prof. Supratik Mukhopadhyay, for invaluable guidance and support throughout the Ph.D. journey. The continuous motivation and encouragement from Prof. Agarwalla helped me to keep on progressing despite the difficult times of the COVID-19 pandemic. His continuous push to keep working hard has helped me to achieve all this success. The encouragement from Prof. Supratik Mukhopadhyay to try new experimental ways taught me how to be an independent researcher. 

Next, I would like to thank my collaborators Prof. Amol Dighe, Dr. Amina Khatun, and Prof. Nayana Majumdar. I learned from Prof. Dighe how to do research passionately and see the deep physics insights. I found myself privileged to have worked with Prof. Dighe. In the starting phase of my research, the help from Dr. Amina Khatun regarding programming, tools, and physics was invaluable, and I am really thankful to have a supportive senior and collaborator like her.  

I acknowledge my doctoral committee members, Prof. Subhasis Chattopadhyay, Prof. Sanjib Kumar Agarwalla, Prof. Supratik Mukhopadhyay, Prof. Nayana Majumdar, Prof. Aruna Kumar Nayak, and Prof. Nita Sinha for their useful suggestions and comments. I would like to acknowledge the Department of Atomic Energy (DAE), Govt. of India, for their financial support. During my Ph.D., I worked at the Insitute of Physics (IOP), Saha Insitute of Nuclear Physics (SINP), and Tata Insitute of Fundamental Research (TIFR). I would like to thank these institutes for providing me with the research infrastructure,  academic support, and administrative support. I acknowledge the India-based Neutrino Observatory (INO) for providing me a platform to pursue research in neutrino physics. I would like to thank the members of the INO collaboration for their useful suggestions, comments, and reviews. I acknowledge Prof. Amol Dighe, Prof. S. Uma Sankar, Prof. M. V. N. Murthy, Prof. Vivek Datar, Prof. Srubabati Goswami, Prof. D. Indumathi, Prof. N. K. Mondal, Prof. G. Rajasekaran, Prof. P. Roy, Prof. S. P. Behera, Dr. Satyanarayana, Prof. Gobinda Majumder, Prof. Sandip Sarkar, Prof. P. Denton, Prof. S. Petcov, Prof. J. Salvado, Prof. A. Smirnov, Prof. F. Halzen, Prof. P. Coyle, Prof. E. Lisi, Prof. S. Palomares-Ruiz for their helpful suggestions and comments on our work.

I am thankful to my fellow research scholars, Sadashiv, Anuj, Avnish, Dr. Ashish, Masoom, Sudipta, Pragyan, Ritam, Soumya, Vishal, Dr. Sridhar, Dr. Jaydeep, Promita, Ram, Dr. Prasant, Tanay, Subhendu, Dr. Neha, Dr. Suryanarayan, Dr. Aparajita, Dr. Dhruv, Dr. Pethuraj, Hariom, Mamta, Jim, Deepak, Roni, Dr. Abhijit, Dr. Nizam, and Dr. Apoorva. I acknowledge the technical support from Sh. Nagaraj, Sh. Ravindra R Shinde, Sh. Mandar, Sh. Yuvraj, Sh. Pataleshwar, Sh. Vishal, Sh. Saibal and Sh. Makrand.

Finally, I would like to thank my mother Rajkali, father Shri Ram, and brother Anuj Kumar for their support and encouragement in fulfilling this dream of doing a Ph.D..\\

\vspace{24 pt}
\begin{center}
	\begin{tabular}{p{0.17\linewidth}p{0.55\linewidth}>{\centering\arraybackslash}p{0.3\linewidth}}
		{May 2022}& & \\
		{Bhubaneswar} &  & \emph{Anil Kumar}\hfill
	\end{tabular}
\end{center}
\cleartooddpage
%\DoubleSpacing
\tableofcontents*
\cleartooddpage
\cleartooddpage
\addcontentsline{toc}{chapter}{Summary}
\chapter*{Summary}
\begin{refsegment}
	
The discovery of neutrino oscillations using atmospheric neutrinos by the Super-Kamiokande experiment opened a new field of research to understand the properties of neutrinos. Neutrino oscillations demand that the neutrinos should be massive which is strong experimental evidence of the physics beyond the Standard Model (BSM) of particle physics. Over the past two decades, most of the neutrino oscillation parameters have been measured with great precision. The Dirac CP phase $\delta_{\rm CP}$ and atmospheric mixing angle $\theta_{23}$ are the least precisely measured oscillation parameters. The other important issues in neutrino oscillation physics are the determination of the neutrino mass ordering and octant of $\theta_{23}$. 

While traveling through the Earth, the upward-going atmospheric neutrinos undergo charged-current weak interactions with the ambient electrons, which give rise to density-dependent matter effects in neutrino oscillations. The matter effects can contribute to the determination of neutrino mass ordering and octant of $\theta_{23}$. The matter effects can also be used as a tool to probe the interior of Earth, which is termed as neutrino tomography of Earth. The proposed 50 kiloton Iron Calorimeter (ICAL) detector at the India-based Neutrino Observatory (INO) would detect atmospheric neutrinos and antineutrinos separately in the multi-GeV range of energies over a wide range of baselines. In this thesis, we present experimental work related to the detector and the physics simulation studies exploring neutrino oscillations, BSM scenarios, and neutrino tomography of Earth.

ICAL consists of stacks of iron layers as passive detector elements, and the Resistive Plate Chamber (RPCs) sandwiched between them as active detector elements. In the experimental work, we study the effects of the non-uniform resistivity of the graphite layer on the response of RPC. A ROOT-based framework is developed to simulate charge transport in the graphite layer. An experimental setup is also developed to measure the non-uniform surface resistivity that is given as an input to the simulation. The simulation of charging of graphite layer predicts that the potential distribution is uniform and independent of non-uniformity in surface resistivity, whereas the time-constant is found to be affected by the non-uniformity in surface resistivity. The experimental measurements are performed, which are found to be in good agreement with simulated results.

In the physics simulation studies, we demonstrate for the first time that the oscillation dip and valley can be observed at ICAL using the up/down ratio for reconstructed $\mu^-$ and $\mu^+$ events separately. The position of the dip and the alignment of the valley are used to measure the oscillation parameter $\Delta m^2_{32}$ in two separate ways using the multiple sets of simulated data for 10 years at ICAL. In our approach, we incorporate statistical fluctuations, systematics errors, and uncertainties in neutrino oscillation parameters.  

We further propose a new approach to probe neutral-current Non-standard Interactions (NSIs) of neutrinos using oscillation dip and valley. Due to the presence of non-zero NSI parameter $\epsilon_{\mu\tau}$, the oscillation dip shifts in the opposite directions for $\mu^-$ and $\mu^+$, whereas the oscillation valley bends in the opposite directions for $\mu^-$ and $\mu^+$. We use the difference in dip locations and contrast in curvatures of oscillation valleys for $\mu^-$ and $\mu^+$ to constrain $\epsilon_{\mu\tau}$ in two separate ways. 

The information about the interior of Earth has been obtained using gravitational measurements and seismic studies. Neutrino absorption and oscillations can be used as a complementary tool to probe the internal structure of Earth, paving the way for multi-messenger tomography of Earth. We show that ICAL can detect core-passing neutrinos and antineutrinos with good directional resolution. We demonstrate that the presence of Earth's core can be validated using atmospheric neutrinos at ICAL by ruling out the two-layered density profile of mantle-crust with respect to the three-layered density profile of core-mantle-crust.  

We believe that the analyses performed in this thesis have contributed to the understanding of the properties of neutrinos and have enriched the physics potential of ICAL.
\end{refsegment}

%%%%%%%%%%%%%%%%%%%%%%%%%%%%%%%%%%%%%%%
\cleartooddpage
\listoffigures
\cleartooddpage
\listoftables
%\cleartooddpage
%\include{abrev} % The list of abbreviations can be stored in abrev.tex. For that you have to first type the abbreviations and then run the command
%%%%%%%%%%%%%%%%%%%%% Issue command %%%%%%%%%%%%%%%%%%%%%%%%%
%                                                           %
%                                                           %
%                                                           %
%     makeindex thesis.nlo -s nomencl.ist -o thesis.nls     %
%                                                           %
%                                                           %
%                                                           %
%%%%%%%%%%%%%%%%%%%%%%%%%%%%%%%%%%%%%%%%%%%%%%%%%%%%%%%%%%%%%
%\nocite{*}
%%%%%%%%%%%%%%%%%%%%%%%%%%%%%%%%%%%%%%%
\cleartooddpage
\pagenumbering{arabic}
%%%%%% If you want to divide your thesis into parts %%%%%%%
%% \part{Name of the part}
%%%%%%%%%%%%%%%%%%%%%%%%%%%%%%%%%%%%%%%%%%%%%%%%%%%%%%%%%%%
\chapter{Introduction}
\label{chap:intro}
\begin{refsegment}

Neutrinos were invented by Wolfgang Pauli in 1930 to explain the problem of energy conservation in beta decay~\cite{Pauli_1930}. In the process of beta decay, a neutron converts into a proton with the emission of an electron. Energy conservation implies that the available energy before the reaction should be shared by the product particles. Since the mass of the daughter nucleus was much larger than that of the electron, the nucleus was almost at rest, and all the available energy was supposed to be carried by the electron. Further, the energy and momentum conservation demands that the energies of emitted electrons should be constant. The experimentally observed energy spectrum of the emitted electron was found to be continuous. Some of the electrons were emitted with the maximum energy, which was nearly equal to the available initial energy, but in most cases, the energies of the electrons were less than the available initial energy. This missing energy could not be attributed to anything and puzzled the scientist so much that Neils Bohr was ready to even abandon the law of conservation of energy.  

To solve this problem of beta decay, Wolfgang Pauli proposed that there is another particle emitted in the beta decay, which should be charge-less and have a tiny mass. Since neutron, proton, and electron are spin half particles, this particle was also expected to be spin half to conserve the total spin in beta decay.
Pauli further told that this particle may not be detected experimentally due to extremely small cross section. This particle was later named as ``neutrino'' by Edoardo Amaldi which means ``little neutral one'' in Italian. The complete beta decay looks like 
\begin{align}
n \rightarrow p^+ + e^- + \bar{\nu}_e\,,
\end{align}
where, $\bar{\nu}_e$ is the electron-type antineutrino emitted during the decay of neutron.

In Sec.~\ref{sec:neutrino_discovery}, we describe the path-breaking discovery of neutrinos. In Sec.~\ref{sec:neutrino_sources}, we present a brief description of various sources of neutrinos. Section~\ref{sec:neutrino_anomalies} discusses the discrepancy in expected and observed neutrino fluxes for solar and atmospheric neutrinos giving rise to neutrino anomalies which were solved by the discovery of neutrino oscillations. In the end, we present the outline of this thesis in Sec.~\ref{sec:thesis_layout}.

%=====================================
\section{Discovery of Neutrinos}
\label{sec:neutrino_discovery}
%=====================================

The detection of neutrino was believed to be extremely difficult because of the tiny interaction cross section~\cite{Cowan:1956rrn,Reines:1956rs}. Since neutrino has no charge, it cannot interact electromagnetically. There was a possibility that antineutrino can be detected via the process of inverse beta decay
\begin{align}
\bar{\nu}_e + p^+  \rightarrow n + e^+\,,
\end{align}
where, an antineutrino interacts with a proton to produce a neutron and a positron. The small interaction cross section can be compensated by using an intense source of antineutrino and a bigger detector. A few antineutrino events can be detected if we wait for a sufficiently long time. 

In the 1950s, Frederick Reines and Clyde Cowan decided to use the inverse beta decay process to detect the antineutrinos emitted from the nuclear reactor at Savannah River Plant, which was the brightest source of antineutrinos~\cite{Cowan:1956rrn,Reines:1956rs}. The inverse beta decay process was suitable for the detection of antineutrino because the resulting positron and neutron have characteristic signals, which helps in identifying inverse beta decay events with high confidence. The positron annihilates with the ambient electron and emits two gamma rays (prompt signal) having energies of about 511 keV. The neutron undergoes scattering and, after a few microseconds, gets captured with the emission of gamma rays (delayed signal). The coincidence of prompt and delayed signals indicates the event corresponding to inverse beta decay. 

Reines and Cowan used water as a target which provided a large number of protons for the interaction of reactor antineutrinos~\cite{Cowan:1956rrn,Reines:1956rs}. Cadmium chloride was dissolved in water to enhance the cross section for neutron capture. The water target was sandwiched between the liquid scintillators, where 5-inch photo-multiplier tubes (PMTs) were used to observe the scintillation light. The gamma rays emitted during positron annihilation and neutron capture were detected in these scintillation detectors. The whole setup was located underground with an encasing of paraffin and lead shield to achieve an excellent shielding from reactor neutrons, reactor gamma rays, and cosmic muons. 

In 1956, the antineutrinos were detected via the process of inverse beta decay in this experiment in Savannah River Plant~\cite{Cowan:1956rrn,Reines:1956rs}. A reactor-power-dependent signal of 2.88 $\pm$ 0.22 counts per hour was observed which was in agreement with the predicted cross section of about $6 \times 10^{-44}$ cm\textsuperscript{2}. The ratio of signal to reactor-dependent background was about 20 to 1, whereas the ratio of signal to reactor-independent background was about 3 to 1. The total experiment runtime was about 1371 hours which included reactor up as well as down periods. 

Reines and Cowan discovered the antineutrino of electron flavor. Another type of neutrino associated with muon was discovered at Alternating Gradient Synchrotron (AGS) located at the Brookhaven National Laboratory~\cite{Danby:1962nd}. A high-energy proton beam was bombarded on a beryllium target to produce secondary particles like pions and kaons, which further decay into neutrinos. In the detector, these neutrinos produced only muons but no electrons, which was a strong hint that these neutrinos are different than the ones produced in the beta decay.  

The discovery of the third generation lepton called tau raised an expectation that a third type of neutrino corresponding to tau lepton should also exist. The tau neutrino was detected in DONuT (direct observation of nu-tau) experiment in nuclear emulsion target~\cite{DONUT:2000fbd}. A proton beam was directed to the tungsten beam dump resulting in charmed mesons, which decayed to provide a beam of tau neutrinos. In a set of 203 neutrino events, four events meet the criteria for $\tau$ decays without any other lepton from the primary vertex. These events provided evidence that the charged-current (CC) interactions of $\tau$ neutrinos were observed. Now, we discuss various sources of neutrinos. 

%=====================================
\section{Sources of Neutrinos}
\label{sec:neutrino_sources}
%=====================================

Neutrinos are produced by various sources, which can be natural or artificial. The neutrinos produced from these sources can have different flavors and cover a wide range of energies. Now, we describe the neutrino from different sources and their properties.  

%=====================================
\subsection{Neutrinos from Natural Sources}
%=====================================

%=====================================
\subsubsection{Solar Neutrinos}
\label{sec:solar_neutrinos}
%=====================================

The Sun shines due to the energy released in the exothermic thermonuclear reaction as described by the Standard Solar Model (SSM). Inside the core of the Sun, the hydrogen nuclei fused together to form helium nuclei with the emission of positrons and electron neutrinos. The effective nuclear reaction looks like following
\begin{align}
4p \rightarrow He + 2e^+ + 2\nu_e + 26.7 \text{~MeV}.
\end{align}
Inside the core of the Sun, the electron neutrinos are mainly produced in two types of fusion reactions which are the proton-proton (\textit{pp}) cycle and the carbon-nitrogen-oxygen (CNO) cycle. About 98\% solar neutrinos are contributed by the \textit{pp} chain. Inside the Sun, the neutrinos are also produced by the electron-capture decay of \textsuperscript{7}Be, beta decay of \textsuperscript{8}B, and proton-electron-proton (\textit{pep}) fusion. The energy of solar neutrinos lies in the range of 0.8 MeV to 15 MeV. 

The solar neutrinos were detected for the first time by the Homestake experiment performed by R. Davis and his group in 1968~\cite{Davis:1968cp}. The detector consisted of 520 tons of chlorine (liquid tetrachloroethylene, C\textsubscript{2}Cl\textsubscript{2}) located 4850 ft underground in the Homestake gold mine in South Dakota. The detection was performed by the neutrino capture on \textsuperscript{37}Cl which converts into \textsuperscript{37}Ar with the emission of an electron. The neutrino events are identified by the decay of \textsuperscript{37}Ar, which is observed by a small proportional counter. The weak-interaction nature of the neutrinos allows them to escape the Sun immediately and bring information about the core of the Sun. Solar neutrinos can be used to probe the solar metallicity, the temperature of the core, the evolution of the Sun, and the density profile of the Sun, etc. Kamiokande experiment reconstructed the image of the core of the Sun using the real-time observation of solar neutrinos~\cite{Kamiokande-II:1989hkh}.

%=====================================
\subsubsection{Supernova Neutrinos}
\label{sec:supernova_neutrinos}
%=====================================

The stars shine by combining the lighter elements (mostly hydrogen) to form heavier elements in thermonuclear reactions. When a star exhausts the nuclear fuel, the core of the star collapses, and the star undergoes a supernova (SN) explosion~\cite{Giunti:2007ry}. During the supernova, the star releases a tremendous amount of energy, which is of the same order as the energy produced during the whole life of the star. During the supernova, protons merge with electrons to form neutrons with the emission of a large number of neutrinos ($\sim10^{58}$). The supernova neutrinos consist of all flavors and have energies of the order of 10 MeV. 

The density of matter is high inside the collapsing core, but neutrinos are able to come out immediately due to their weak-interaction nature. These neutrinos are able to reach the Earth a few hours before the light from the supernova can be seen. In 1987, neutrino detectors on Earth observed the neutrinos from a supernova (SN1987A) in the nearby galaxy called Large Magellanic Cloud. At that time, Kamiokande II~\cite{Kamiokande-II:1987idp}, IMB~\cite{Bionta:1987qt}, and BAKSAN~\cite{Alekseev:1988gp} were the active neutrino detectors, and a total of 24 neutrinos were observed in about 12 seconds from this supernova on 23rd February 1987. The visible light from SN1987A was observed three hours after the detection of neutrinos.

%=====================================
\subsubsection{Atmospheric Neutrinos}
\label{sec:atm_neutrinos}
%=====================================

Atmospheric neutrinos are produced during the interactions of primary cosmic rays coming from outer space with the nuclei of gases in the Earth's atmosphere~\cite{Giunti:2007ry,Vitagliano:2019yzm}. The primary cosmic rays are isotropic in nature and can have energies from about a few MeV to as high as $10^{20}$ eV. The isotropic nature of cosmic rays indicates that their origin is outside the solar system. The cosmic rays consist of mostly protons and helium nuclei. 

The interactions of primary cosmic rays with the nuclei of gases present in the atmosphere produce secondary particles like pions and kaons. These are unstable particles and decay further to produce leptons and neutrinos. The decay of pions mostly result in the production of muons and muon type of neutrinos
\begin{align}\label{eq:pion_decay}
	\pi^- &\rightarrow \mu^- + \bar{\nu}_\mu\,,\\
	\pi^+ &\rightarrow \mu^+ + \nu_\mu.
\end{align}
These muons can decay further to produce electrons along with muon and electron types of neutrinos
\begin{align}\label{eq:muon_decay}
	\mu^- & \rightarrow e^- + \nu_\mu + \bar{\nu}_e\,, \\
	\mu^+ & \rightarrow e^+ + \bar{\nu}_\mu + \nu_e\,. 
\end{align}
These neutrinos and antineutrinos produced during interaction of cosmic rays are called atmospheric neutrinos. At low energies of around 1 GeV, the decay chain of pions and muons give rise to the ratio of muon to electron flavor of about 2, i.e.
\begin{align}\label{eq:flavor_ratio}
	\frac{\nu_\mu + \bar{\nu}_\mu}{\nu_e + \bar{\nu}_e} \sim 2. 
\end{align}
On the other hand, at high energies, the muon may reach the surface of Earth without decaying which increases this ratio
\begin{align}
	\frac{\nu_\mu + \bar{\nu}_\mu}{\nu_e + \bar{\nu}_e} > 2.
\end{align}
The decays of pions and kaons mainly produce neutrinos of muon and electron types. Note that a small number of tau neutrinos may also get produced during the decay of charmed mesons but at energies above the TeV range. 

The atmospheric neutrinos possess a wide range of energies, starting from a few MeV to about TeV. The atmospheric neutrinos are produced at an average height of about 15 km above the surface of Earth, which is also the lowest path traveled by the downward-going atmospheric neutrinos. On the other hand, the upward-going neutrinos travel all the way through the bulk of Earth and can have pathlengths\footnote{The net distance $L_\nu$ traveled by a neutrino (its ``pathlength'' or ``baseline'') is related to its zenith angle via
\begin{equation}\label{eq:zen-bl-rel}
	L_\nu = \sqrt{(R+h)^2 - (R-d)^2\sin^2\theta_\nu} \,-\, (R-d)\cos\theta_\nu \, ,
\end{equation}
where $R$, $h$, and $d$ denote the radius of Earth, the average height from the surface at which neutrinos are produced, and the depth of the detector under the surface of Earth, respectively. In the analyses of this thesis, we take $R = 6371$ km, $h = 15$ km, and $d=0$ km. Note that the zenith angle $\cos\theta_\nu=1$ for downward-going neutrinos and $\cos\theta_\nu=-1$ for upward-going neutrinos.} 
as high as about 12750 km. Since the cosmic rays are isotropic in nature, the atmospheric neutrinos going in upward and downward directions are expected to be equal. At low energies, the geomagnetic field of Earth disturbs the isotropy of cosmic rays, which translates to a small up-down asymmetry in atmospheric neutrinos. But at high energies, the atmospheric neutrinos are expected to be up-down symmetric. 

The atmospheric neutrinos were observed first time in 1965 by detectors in the Kolar Gold Field mine in India~\cite{Achar:1965cha,Achar:1965ova}. The detectors based on scintillators and neon flash tubes were located at a depth of about 7600 ft (7500 meters of water equivalent) below the ground to get shielding from cosmic muons. The residual cosmic muons at such depth are peaked along the downward-going vertical direction. The observation of muon tracks in the horizontal direction indicated that the observed events were unlikely to be cosmic muons and were most probably due to the interaction of atmospheric neutrinos, which were expected to be isotropic and could give rise to the horizontal events. During the same time, another experiment in the East Rand Proprietary Gold Mine in South Africa also detected the horizontal events corresponding to atmospheric neutrinos~\cite{Reines:1965qk}.

%=====================================
\subsubsection{Ultra High-energy Astrophysical Neutrinos}
\label{sec:astrophysical_neutrinos}
%=====================================

The ultra-high-energy neutrinos are produced in the cosmic sources like active galactic nuclei (AGN) and gamma-ray bursts. The energy of these astrophysical neutrinos lies in the range of TeV to PeV. In these sources, the charged particles like protons and nuclei are accelerated to high energies. The interactions of these high-energy charged particles with gases and photons result in the production of unstable particles like pions and kaons, which decay further to produce neutrinos. Unlike photons, the astrophysical neutrinos pass through the galactic medium undeflected, and the deflections due to galactic or extragalactic magnetic fields are also negligible. Therefore, these neutrinos are potential probes to understand the high-energy phenomena happening in astrophysical sources.

The astrophysical neutrinos with energy in the range of TeV to PeV were detected first time by the IceCube neutrino observatory at the South pole in 2013~\cite{IceCube:2013low,IceCube:2014stg}. IceCube detected about 37 astrophysical neutrinos in three years and excluded the hypothesis of atmospheric origin with a 5.7$\sigma$ confidence level. In the recently updated data, about 102 events corresponding to astrophysical neutrinos were observed at IceCube in 7.5 years~\cite{IceCube:2020wum}. 

%=====================================
\subsubsection{Geoneutrinos}
\label{sec:Geoneutrinos}
%=====================================

The Earth contains radioactive isotopes like \textsuperscript{238}U, \textsuperscript{232}Th, and \textsuperscript{40}K having half-lives comparable to or more than the age of Earth~\cite{Bellini:2021sow}. These long-lived isotopes undergo beta decay to emit electron antineutrinos that are known as geoneutrinos. These radioactive decays also result in the production of energy along with antineutrinos. This energy is called radiogenic heat, which is roughly half the amount of heat dissipated by Earth. The geoneutrinos have energies of the order of MeV. 

The geoneutrinos have been detected by KamLAND~\cite{Araki:2005qa} and Borexino~\cite{Borexino:2015ucj}, experiments using the process of inverse beta decay where the electron antineutrinos interact with the protons to produce positrons and neutrons. The positron gives a prompt signal in the form of 511 keV annihilation photons, whereas the neutron thermalizes and gets captured after some time to produce a delayed signal. The precise measurement of geoneutrinos can reveal the distribution of radioactive isotopes inside Earth, which will help us to understand the chemical composition and dynamics of heat production inside the Earth.

%=====================================
\subsubsection{Relic Neutrinos}
\label{sec:relic_neutrinos}
%=====================================

Relic neutrinos are by-products of Big-bang~\cite{Giunti:2007ry}. In the early stages of the universe, neutrinos were in thermal equilibrium with the hot plasma through weak interactions. As the universe expanded, the interaction rate of neutrino decreased. Eventually, the neutrinos got decoupled when the expansion rate became more than the interaction rate. Relic neutrinos are the second most abundant particles in the universe after photons. The present temperature of relic neutrinos is very small (1.95 K), which corresponds to the energy of the order of $10^{-4}$ eV\textsuperscript{2}. Due to this small temperature, the interaction cross section of relic neutrinos is extremely low, which makes the direct detection of relic neutrinos very difficult.  

%=====================================
\subsection{Neutrino from Artificial Sources}
%=====================================

%=====================================
\subsubsection{Reactor Neutrinos}
\label{sec:reactor_neutrinos}
%=====================================

Nuclear reactors are an intense source of electron antineutrino flux. The energy required for power generation in nuclear reactor is the product of the beta-decay reactions of isotopes like \textsuperscript{235}U, \textsuperscript{238}U, \textsuperscript{239}Pu, and \textsuperscript{241}Pu. The reactor antineutrinos have energy in the range of 0.1 to 10 MeV with mean energy around 4 MeV. In the Uranium fission reaction, six electron antineutrinos are produced with a total energy of about 200 MeV. Therefore, a nuclear reactor with 1 GW power will produce about $2\times10^{20}$ electron antineutrinos per second. Reactor antineutrinos were detected first time by Reines and Cowan using the process of inverse beta-decay as discussed in Sec.~\ref{sec:neutrino_discovery}.

%=====================================
\subsubsection{Accelerator Neutrinos}
\label{sec:accelerator_neutrinos}
%=====================================

The particle accelerators are used to produce a beam of muon-type neutrinos and antineutrinos. A high-energy proton beam from the accelerator is bombarded on a target where unstable particles like pions and kaons are produced. These secondary particles are focused using a magnetic horn such that a beam of $\pi^-$ or $\pi^+$ is obtained. The focused beam of pions is passed through a decay tunnel where $\pi^-$ ($\pi^+$) decays to produce $\mu^-$ ($\mu^+$) and $\bar{\nu}_\mu$ ($\nu_\mu$). At the end of the tunnel, the muons and undecayed mesons are absorbed in the beam dump, and a beam consisting of muon-type neutrinos or antineutrinos is obtained. An advantage of the accelerator neutrinos is that we can produce the neutrino beam with a peak at the desired energy. 

The accelerator neutrinos are used to understand the fundamental properties of neutrinos. In the long-baseline experiments, a neutrino beam is used as a source, and a far detector is placed at a fixed distance. For example, the K2K experiment~\cite{K2K:2004iot} in Japan was the first long-baseline experiment that used the muon neutrinos beam produced at the KEK accelerator, which was detected about 250 km away by the water-Cherenkov-based Super-Kamiokande (Super-K) detector~\cite{Super-Kamiokande:2017yvm}. 

%=====================================
\section{Neutrino Anomalies}
\label{sec:neutrino_anomalies}
%=====================================

Till now, we learned that neutrinos originate from different sources, and most of them have been detected by various experiments. However, it was found that the observed solar neutrino flux was much less than the expected flux. This deficit could not be explained by any error in experiments or solar models. Similarly, the atmospheric neutrinos were also found to be less than the expectation. Now, we describe these anomalous behaviors and their solution.  

%=====================================
\subsection{Solar Neutrino Anomaly}
\label{sec:solar_neutrino_anomaly}
%=====================================

The solar neutrinos were first detected in the Homestake experiment as described in Sec.~\ref{sec:solar_neutrinos}. The Homestake experiment observed the capture rate for the solar neutrino to be $2.56 \pm 0.25$ SNU\footnote{1 Solar Neutrino Unit (SNU) = $10^{-36}$ captures per second per target atom.}~\cite{Cleveland:1998nv} which was less than the predictions of $9.3 \pm 1.2$ SNU~\cite{Bahcall:1995bt} by the Standard Solar Model. This deficit in the flux of electron neutrinos from the Sun was also confirmed by the Galium-based experiments like SAGE~\cite{SAGE:1999nng} and GALLEX~\cite{GALLEX:1998kcz} where the observed capture rate was about half the expected rate~\cite{Bahcall:1995bt}. Further, the Kamiokande~\cite{Kamiokande-II:1989hkh} and the Super-K~\cite{Super-Kamiokande:2005wtt,Super-Kamiokande:2008ecj,Super-Kamiokande:2010tar} experiments based on water Cherenkov principle observed the solar neutrinos in real-time with information on energy and direction of events. The Kamiokande and the Super-K experiment also observed the deficit in solar neutrinos flux. This mystery of missing electron neutrinos from the Sun was known as the ``solar neutrino anomaly''.  

Since various neutrino experiments based on different principles observed the electron neutrino flux to be lower than the expectation, there was nothing wrong with the neutrino experiments. On the other hand, the Standard Solar Model was also robust because of the agreement with helioseismic measurements. Another possibility was that there was something happening to electron neutrinos while traveling from the Sun to the Earth. Since the solar neutrinos have energies below about 30 GeV, the charged-current interactions cannot be used to detect muon or tau neutrinos where the masses of the resulting leptons are more than the available energy. Therefore, if electron neutrinos change their flavor to muon or tau neutrinos while traveling from the Sun to the Earth, then the solar neutrino flux observed using charged-current interactions will be lower than the expectations. The only way to observe neutrinos of all flavors was to use neutral-current interactions. 

The SNO experiment~\cite{SNO:1999crp} was designed to observe solar neutrinos using charged-current as well as neutral-current interactions. The SNO experiment consisted of a target of heavy water containing deuterium, which has proton as well as neutron. Since deuterium has binding energy as low as 2 MeV, solar neutrino of any flavor with energy up to 30 MeV can easily break it into protons and neutrons. The SNO experiment was sensitive to the total flux corresponding to all flavors ($\nu_e$, $\nu_\mu$, $\nu_\tau$) via neutral-current interactions
\begin{align}
\nu + e^- &\rightarrow \nu + e^-\,, \\
\nu + d &\rightarrow n + p + \nu.
\end{align}
Additionally, the charged-current interaction was also possible for electron neutrinos
\begin{align}
\nu_e + d \rightarrow p + p + e^-\,,
\end{align} 
which gave rise to the sensitivity towards solar neutrino flux corresponding to electron flavor only. In 2002, the observed flux of electron neutrinos through charged-current interactions at SNO was found to be one-third of the total flux corresponding to all flavors through neutral-current interactions~\cite{SNO:2002tuh}. This implied that the electron neutrinos produced inside the core of the Sun got converted into muon and tau neutrinos during propagation. The total observed flux of solar neutrinos corresponding to all three flavors was also found to be in good agreement with the predictions of the Standard Solar Model. Therefore, the solar neutrino anomaly was explained using the phenomenon of flavor conversion during propagation which was known as ``neutrino oscillations''. Though the deficit in the solar neutrino flux was observed first, the neutrino oscillations were discovered first time using atmospheric neutrinos, which we will discuss next. 

%=====================================
\subsection{Atmospheric Neutrino Anomaly}
%=====================================

Proton is a stable particle in the Standard Model (SM) of particle physics, but the Grand Unified Theories predict that the proton may decay~\cite{Langacker:1980js}. In the early 1980s, the experiments like Kamiokande~\cite{Kamiokande-II:1988sxn}, IMB~\cite{Casper:1990ac}, and Soudan-2~\cite{Allison:1996yb} were built to look for proton decay where atmospheric neutrino events were acting as a background. Therefore, these experiments started measuring the properties of atmospheric neutrinos for a better understanding of the background. In Sec.~\ref{sec:atm_neutrinos}, we learned that the atmospheric neutrinos consist of muon and electron flavors such that the ratio of muon to electron flavor is expected to be 2 (see Eq.~\ref{eq:flavor_ratio}). Though the predictions in atmospheric neutrino fluxes had large uncertainties, the predicted ratio of muon to electron neutrino fluxes had small uncertainties. Hence, these experiments measured the double ratio, which is defined as the ratio of data and Monte Carlo (MC) for flavor ratio
\begin{align}
R = \frac{(N_\mu/N_e)_\text{data}}{(N_\mu/N_e)_\text{MC}}\,,
\end{align} 
where, $N_\mu$ ($N_e$) is the sum of neutrinos and antineutrinos with muon (electron) flavor. If the observed data matched the predictions, then the value of $R$ was expected to be 1, but it was found to be significantly less than 1. Therefore, either the events with muon flavor were less, or the events with electron flavor were more than the expectations. This was known as the ``atmospheric neutrino anomaly''. 

The Super-K experiment, which was the successor of Kamiokande, also observed the value of $R$ lower than 1, but Super-K also had the capability to measure the direction and energy of neutrino events~\cite{Super-Kamiokande:1998wen}. The atmospheric neutrinos were expected to be up-down symmetric due to the isotropic nature of cosmic rays, as discussed in Sec.~\ref{sec:atm_neutrinos}. On the other hand, Super-K observed that the upward-going multi-GeV ``muon-like'' events showed a deficit compared to predictions, whereas the downward-going events matched the predictions~\cite{Super-Kamiokande:1998tou}. As far as the ``electron-like'' events were concerned, those were consistent with the predictions in the whole range of zenith angle. There was neither deficit nor excess for multi-GeV electron-like events. These observations can be explained by the idea that the muon neutrinos going in the upward direction have traveled long pathlengths that are large enough for muon neutrinos to oscillate into other flavors. Since the observed electron neutrino flux was almost the same as predicted, they may not have significant participation in the oscillations, and the muon neutrinos must have mainly oscillated to tau neutrinos. No deficit in muon neutrino flux was observed in the downward direction because pathlengths were not large enough for oscillations to develop. This discovery of neutrino oscillations in 1998 solved the atmospheric neutrino anomaly~\cite{Super-Kamiokande:1998kpq}. We will discuss more about neutrino oscillations in chapter~\ref{chap:neutrino_oscillations}.

\section{Layout of the Thesis}
\label{sec:thesis_layout}
Neutrinos were postulated to explain the process of beta decay. Neutrinos were believed to be massless particles. However, the discovery of neutrino oscillations by atmospheric neutrinos provided evidence of the non-zero mass of neutrinos. The massive neutrinos are a strong hint towards the theories beyond the Standard Model (BSM) of particle physics. Neutrino oscillations provide a unique opportunity to explore many BSM physics scenarios. Over the past two decades, most of the neutrino oscillation parameters have been measured with good precision, where atmospheric neutrinos have played an important role. This thesis focuses on the experimental aspect of the upcoming Iron Calorimeter (ICAL) detector at the India-based Neutrino Observatory (INO) and the physics simulations to study three-flavor neutrino oscillations, non-standard interactions (NSI), and the tomography of Earth using atmospheric neutrinos at ICAL.

In chapter~\ref{chap:neutrino_oscillations}, we describe neutrino oscillations in detail and briefly discuss the current and upcoming neutrino oscillation experiments. Chapter~\ref{chap:ICAL} describes the working of the ICAL detector and the details of event simulation, which is used to perform the analyses of this thesis. In chapter~\ref{chap:RPC_response}, we present the study on the response uniformity of Resistive Plate Chamber where the effect of non-uniform surface resistivity is shown. In chapter~\ref{chap:dip_valley}, we demonstrate that the oscillation dip and valley can be reconstructed at ICAL, and the value of atmospheric oscillation parameters can be measured using the oscillation dip and valley. Chapter~\ref{chap:NSI} presents a new approach to probe neutral-current non-standard interactions of neutrinos using oscillation dip and valley. In chapter~\ref{chap:tomography}, we present the sensitivity with which the presence of Earth's core can be validated using the atmospheric neutrinos at ICAL. Finally, we provide a summary and concluding remarks in chapter~\ref{chap:summary}.

\end{refsegment}
 %(foldername/filename without extension)
\cleartooddpage
\chapter{Neutrino Oscillations}
\label{chap:neutrino_oscillations}
\begin{refsegment}
	
The idea of neutrino oscillation was first proposed by Bruno Pontecorvo  in the form of transition between neutrinos and antineutrinos~\cite{Pontecorvo:1957qd}. The theory of neutrino flavor mixing was given by Ziro Maki, Masami Nakagawa, and Shoichi Sakata~\cite{Maki:1962mu} in 1962 after the discovery of muon neutrinos. The neutrino oscillation is a quantum mechanical phenomenon where the flavor of neutrino changes while traveling. Neutrinos are produced and detected via weak interactions in flavor eigenstates. On the other hand, neutrinos travel in mass eigenstates in vacuum. Neutrino does not exist in flavor and mass eigenstates simultaneously. Therefore, while traveling from source to detector, the neutrino flavor states start evolving and mixing with each other. Neutrino oscillations also demand that neutrinos should have non-degenerate masses, which is strong evidence for theories beyond the Standard Model of particle physics. 

In this chapter, we discuss the theory of neutrino oscillations, neutrino oscillation experiments, and the status of neutrino oscillation parameters. In Sec.~\ref{sec:vacuum_osc}, we describe neutrino oscillations in vacuum. While passing through Earth, neutrinos undergo charged-current coherent forward elastic scattering with ambient electrons modifying the neutrino oscillation patterns. We describe this phenomenon of matter effect in Sec.~\ref{sec:matter_osc}. In Sec.~\ref{sec:osc_exp}, we discuss neutrino oscillation experiments that have contributed to the measurement of neutrino oscillation parameters. The present status and unsolved issues of neutrino oscillation parameters are described in sections~\ref{sec:status_osc_par} and \ref{sec:unsolved_issues}, respectively. Section~\ref{sec:future_exp} discusses the upcoming neutrino oscillation experiments that aim to solve the issues in neutrino oscillations. Finally, we summarize in Sec.~\ref{sec:prob_conclusion}.

\section{Neutrino Oscillations in Vacuum}
\label{sec:vacuum_osc}

The weak-interaction Hamiltonian possesses the flavor eigenstates   $\ket*{\nu_\alpha}$ where $\alpha$ can be $e$, $\mu$, or $\tau$. On the other hand, the propagation Hamiltonian has mass eigenstates $\ket*{\nu_j}$ where $j = 1, 2$, or 3. Since the weak-interaction Hamiltonian and propagation Hamiltonian don't commute with each other, the neutrino cannot exist in these two type of eigenstates simultaneously. The mixing between flavor eigenstates and mass eigenstates results in the neutrino oscillations where the flavor eigenstates  can be  expressed as a superposition of mass eigenstates 
\begin{equation}
\ket*{\nu_\alpha} = \sum_{j} U_{\alpha j}^* \ket*{\nu_j}\,,
\end{equation}
where, $U$ is a unitary matrix i.e., $U^\dagger = U^{-1}$. Since $\ket*{\nu_j}$ is the eigenstate of the free Hamiltonian during propagation with an eigenvalue $E_j$, the time evolution of stationary state can be expressed as
\begin{align}
\operatorname{H}\ket*{\nu_j(t)} &= E_j \ket*{\nu_j(t)}\,,\\
i\frac{\partial{\ket*{\nu_j(t)}}}{\partial t} &= E_j \ket*{\nu_j(t)}\,\\
\implies \ket*{\nu_j (t)} &= e^{-iE_j t}\ket*{\nu_j (t=0)}\,.
\end{align}
Let us consider that the flavor state $\ket*{\nu_\alpha}$ is produced at time $t = 0$ and after time $t$, the state evolves to
\begin{equation}
\ket*{\nu (t)} = \sum_{j} U_{\alpha j}^* e^{-iE_j t} \ket*{\nu_j(t=0)}.
\end{equation}
The probability amplitude for evolution of initial state $\ket*{\nu_\alpha}$ to the final state $\ket*{\nu_\beta}$ at a later time $t$ is
\begin{align}
\braket*{\nu_\beta}{\nu (t)} &= \sum_{j} U_{\alpha j}^* e^{-iE_j t} \braket*{\nu_\beta}{\nu_j}\,, \nonumber \\
&= \sum_{j} \sum_{i} U_{\beta i} U_{\alpha j}^* e^{-iE_j t} \braket*{\nu_i}{\nu_j}\,, \nonumber \\
&= \sum_{j} U_{\beta j} U_{\alpha j}^* e^{-iE_j t}
\end{align}
where, we use the orthogonality condition $\braket*{\nu_i}{\nu_j} = \delta_{ij}$. Therefore, the probability of evolution of eigenstate $\ket*{\nu_\alpha}$ to another eigenstate $\ket*{\nu_\beta}$ turns out to be
\begin{align}
P(\nu_\alpha \rightarrow \nu_\beta) &= \left|\braket*{\nu_\beta}{\nu (t)}\right|^2 = \left|\sum_{j} U_{\beta j} U_{\alpha j}^* e^{-iE_j t}\right|^2 \label{eq:prob_amp_rel} \\
&= \sum_{j} \sum_{k} U_{\alpha j}^* U_{\beta j} U_{\alpha k} U_{\beta k}^*  e^{-i(E_j - E_k) t}. \label{eq:prob_amp_rel_expand}
\end{align}
Using the identity 
\begin{equation}\label{eq:complex_identity}
|z_1 + z_2 + z_3 + \ldots|^2 = \sum_j |z_j|^2 + 2  \mathfrak{Re} \sum_{j>k} z_j z_k^*\,,
\end{equation}
we can write Eq.~\ref{eq:prob_amp_rel} as
\begin{align}\label{eq:prob_real}
P(\nu_\alpha \rightarrow \nu_\beta) = \sum_j \left|U_{\alpha j}\right|^2 \left|U_{\beta j}\right|^2 + 2  \mathfrak{Re} \sum_{j>k}  U_{\alpha j}^* U_{\beta j} U_{\alpha k} U_{\beta k}^*  e^{-i(E_j - E_k) t}.
\end{align}
After taking square of unitary relation, 
\begin{equation}
\sum_{k} U_{\alpha k} U_{\beta k}^* = \delta_{\alpha\beta}\,,
\end{equation}
we obtain, 
\begin{equation}
\left|\sum_{j} U_{\alpha j} U_{\beta j}^*\right|^2 = \delta_{\alpha\beta}\,.
\end{equation}
Now, using Eq.~\ref{eq:complex_identity}, we get
\begin{equation}\label{eq:unitary_expand}
\sum_{j} \left|U_{\alpha j}\right|^2 \left|U_{\beta j}\right|^2 = \delta_{\alpha\beta} -  2  \mathfrak{Re} \sum_{j>k}  U_{\alpha j}^* U_{\beta j} U_{\alpha k} U_{\beta k}^*\,,
\end{equation}
Using Eq.~\ref{eq:prob_real} and \ref{eq:unitary_expand}, we obtain
\begin{align}
P(\nu_\alpha \rightarrow \nu_\beta) &= \delta_{\alpha\beta} -  2  \mathfrak{Re} \sum_{j>k}  U_{\alpha j}^* U_{\beta j} U_{\alpha k} U_{\beta k}^* + 2  \mathfrak{Re} \sum_{j>k}  U_{\alpha j}^* U_{\beta j} U_{\alpha k} U_{\beta k}^*  e^{-i(E_j - E_k) t}\,, \nonumber\\
&= \delta_{\alpha\beta} -  2  \mathfrak{Re} \sum_{j>k}  U_{\alpha j}^* U_{\beta j} U_{\alpha k} U_{\beta k}^* \nonumber\\
&\quad + 2  \mathfrak{Re} \sum_{j>k}  U_{\alpha j}^* U_{\beta j} U_{\alpha k} U_{\beta k}^*  [\cos{((E_j - E_k) t)} - i \sin{((E_j - E_k) t)}]\,, \nonumber \\
&= \delta_{\alpha\beta} -  4   \sum_{j>k}  \mathfrak{Re} \left( U_{\alpha j}^* U_{\beta j} U_{\alpha k} U_{\beta k}^* \right) \sin^2 \left(\frac{(E_j - E_k) t}{2}\right) \nonumber\\
&\quad +2    \sum_{j>k} \mathfrak{Im} \left( U_{\alpha j}^* U_{\beta j} U_{\alpha k} U_{\beta k}^*\right) \sin \left((E_j - E_k) t\right)\,. \label{eq:prob_real_image}
\end{align}

The energy of neutrino mass eigenstate $E_j$  with mass $m_j$ can be expressed as 
\begin{align}\label{eq:energy_approx}
E_j = \sqrt{p^2 + m^2_j} \approx p + \frac{m_j^2}{2p} \approx p + \frac{m_j^2}{2E}\,,
\end{align}
where, we keep only linear terms in $m_j/E$ under relativistic approximation, i.e.,  $m_j \ll E$. Since the neutrino masses are less than 1 eV with their energies of the order of MeV or more for the relevant experiments, this is always a valid approximation where $p \approx E$. Under relativistic approximation, neutrino travels with almost of the speed of light, hence, we take the path-length of neutrino $L = t$ in natural units with $\hbar = c = 1$. Since the interference require the neutrinos to have same $p$, we consider coherent neutrino with momentum $p$. Note that the momentum $p$ give rise to a common phase $e^{-ipt}$ which is irrelevant for oscillation probability in Eq.~\ref{eq:prob_amp_rel}. Using Eqs.~\ref{eq:prob_real_image} and \ref{eq:energy_approx}, we get
\begin{align}
P(\nu_\alpha \rightarrow \nu_\beta) &= \delta_{\alpha\beta} -  4  \mathfrak{Re} \sum_{j>k}  U_{\alpha j}^* U_{\beta j} U_{\alpha k} U_{\beta k}^* \sin^2 \left(\frac{\Delta m^2_{jk} L}{4E}\right) \nonumber\\
&\quad +2  \mathfrak{Im} \sum_{j>k}  U_{\alpha j}^* U_{\beta j} U_{\alpha k} U_{\beta k}^* \sin \left(\frac{\Delta m^2_{jk} L}{2E}\right)\,, \label{eq:prob_real_image_dmsq}
\end{align}
where, $\Delta m^2_{jk} = m_j^2 - m_k^2$. We can see that the oscillation probability depends on the difference of mass square instead of absolute mass and $L/E$, where $E$ and $L$ are the energy and the path-length traveled by neutrino from source to detector. Now, we describe the neutrino oscillations in two-flavor framework. 

\subsection{Two-flavor Oscillations in Vacuum}
\label{sec:2f_vac_osc}

In two-flavor neutrino oscillation framework, the flavor eigenstates of $\nu_\alpha$ and $\nu_\beta$ are expressed as a superposition of mass eigenstates of $\nu_1$ and $\nu_2$
\begin{align}
\begin{pmatrix}
\nu_\alpha \\
\nu_\beta 
\end{pmatrix} 
= \begin{pmatrix}
\cos\theta & \sin\theta \\
-\sin\theta & \cos\theta 
\end{pmatrix}
\begin{pmatrix}
\nu_1 \\
\nu_2
\end{pmatrix},
\end{align}
here, mixing matrix depends on only one mixing angle $\theta$. This superposition can also be written as
\begin{align}
\ket*{\nu_\alpha} &= \cos\theta \ket*{\nu_1} + \sin\theta \ket*{\nu_2}\,\label{eq:initial_fl_state}\\
\ket*{\nu_\beta} &= -\sin\theta \ket*{\nu_1} + \cos\theta \ket*{\nu_2}
\end{align}
At time $t = 0$, we start with a flavor state $\ket*{\nu_\alpha}$ as given in Eq.~\ref{eq:initial_fl_state} and after time $t$, it  evolves to
\begin{align}
\ket*{\nu_\alpha(t)} &= \cos\theta ~e^{-\frac{im_1^2 t}{2E}} \ket*{\nu_1} + \sin\theta ~e^{-\frac{im_2^2 t}{2E}} \ket*{\nu_2}.
\end{align}
The probability amplitude for transition  of flavor state $\ket*{\nu_\alpha}$ to $\ket*{\nu_\beta}$ is 
\begin{align}
\braket*{\nu_\beta}{\nu_\alpha(t)} &= -\sin\theta\cos\theta ~e^{-\frac{im_1^2 t}{2E}}+ \sin\theta\cos\theta ~e^{-\frac{im_2^2 t}{2E}}. 
\end{align}
The transition probability for $\ket*{\nu_\alpha}$ can be given as
\begin{align}
P(\nu_\alpha \rightarrow \nu_\beta) &= \left|\braket*{\nu_\beta}{\nu_\alpha(t)}\right|^2= \sin^2 2\theta \sin^2 \left(\frac{\Delta m^2 L}{4E} \right)\,,
\end{align}
where, $\Delta m^2 = m_2^2 - m_1^2$. The transition probability $P(\nu_\alpha \rightarrow \nu_\beta)$ is non-zero only if $\theta \neq 0$ and masses are non-degenerate, i.e., $m_1 \neq m_2$. The survival probability $P(\nu_\alpha \rightarrow \nu_\alpha)$ is given as 
\begin{align}
P(\nu_\alpha \rightarrow \nu_\alpha) = 1 - \sin^2 2\theta \sin^2 \left(\frac{\Delta m^2 L}{4E} \right)\,. \label{eq:survival_2f}
\end{align}
Now if we express $\Delta m^2$ in eV\textsuperscript{2}, $L$ in km (m), and $E$ in GeV (MeV), then Eq.~\ref{eq:survival_2f} can be written as
\begin{align}
P(\nu_\alpha \rightarrow \nu_\alpha) = 1 - \sin^2 2\theta \sin^2 \left(1.27\,\frac{\Delta m^2 [\text{eV}^2] \,L[\text{km}]}{E[\text{GeV}]} \right)\,, \label{eq:survival_2f_GeV}
\end{align}
or
\begin{align}
P(\nu_\alpha \rightarrow \nu_\alpha) = 1 - \sin^2 2\theta \sin^2 \left(1.27\,\frac{\Delta m^2 [\text{eV}^2] \, L[\text{m}]}{E[\text{MeV}]} \right)\,. \label{eq:survival_2f_MeV}
\end{align}

Now, we describe neutrino oscillations in vacuum in three-flavor framework.

\subsection{Three-flavor Oscillations in Vacuum}
\label{sec:3f_vac_osc}

In three-flavor paradigm, the flavor state $\nu_e$, $\nu_\mu$, and $\nu_\tau$ are expressed in terms of mass eigen states $\nu_1$, $\nu_2$, and $\nu_3$ as following 
\begin{align}
\begin{pmatrix}
\nu_e \\
\nu_\mu \\
\nu_\tau 
\end{pmatrix} 
&= \begin{pmatrix}
1 & 0 & 0 \\
0 & c_{23} & s_{23} \\
0 & -s_{23} & c_{23} \\
\end{pmatrix}	
\begin{pmatrix}
c_{13} & 0 & s_{13}e^{-i\delta_\text{CP}} \\
0 & 1 & 0 \\
-s_{13}e^{i\delta_\text{CP}} & 0 & c_{13} \\
\end{pmatrix}
\begin{pmatrix}
c_{12} & s_{12} & 0 \\
-s_{12} &  c_{12} &0 \\
0 & 0 & 1 \\
\end{pmatrix}
\begin{pmatrix}
\nu_1 \\
\nu_2 \\
\nu_3
\end{pmatrix},\nonumber \\
&= \begin{pmatrix}
c_{13}c_{12}&  c_{13}s_{12} & s_{13}e^{-i\delta_{CP}}\\ 
- s_{12}c_{23}-c_{12}s_{23}s_{13}e^{i\delta_{CP}}   &c_{23}c_{12} -s_{12}s_{23}s_{13}e^{i\delta_{CP}}& s_{23}c_{13} \\
s_{23}s_{12}-c_{23}s_{13}c_{12}e^{i\delta_{CP}}& -s_{23}c_{12}-c_{23}s_{13}s_{12}e^{i\delta_{CP}} &c_{23}c_{13}  
\end{pmatrix} 
\begin{pmatrix}
\nu_1 \\
\nu_2 \\
\nu_3
\end{pmatrix}.
\end{align}
where, $s_{ij}$ and $c_{ij}$ represent $\sin\theta_{ij}$ and $\cos\theta_{ij}$, respectively. This mixing matrix is called the Pontecorvo-Maki-Nakagawa-Sakata (PMNS) matrix and it consists of three mixing angles $\theta_{12}$, $\theta_{13}$, and $\theta_{23}$ with one Dirac CP phase $\delta_\text{CP}$. Apart from these parameters of PMNS matrix, the neutrino oscillations in three-flavor framework also depend upon two independent mass-squared differences which are $\Delta m^2_{32}$ and $\Delta m^2_{21}$. The experimental observations have indicated that $ |\Delta m^2_{32}| \gg \Delta m^2_{21}$.

In atmospheric neutrinos, the flavor mixing occurs mainly between $\nu_\mu$ and $\nu_\tau$, which are governed by the mixing angle $\theta_{23}$ and mass-squared difference $\Delta m^2_{32}$. Hence, these are known as atmospheric oscillation parameters. The oscillations in the solar neutrino sector are controlled by $\theta_{12}$ and $\Delta m^2_{21}$, which are known as solar oscillation parameters. As far as $\theta_{13}$ is concerned, it is known as the reactor mixing angle, and it connects atmospheric and solar oscillations. Now, we will discuss matter effects in neutrino oscillations. 

\section{Matter Effects}
\label{sec:matter_osc}

\begin{figure}
	\centering
	\includegraphics[align=c, height=0.49\linewidth]{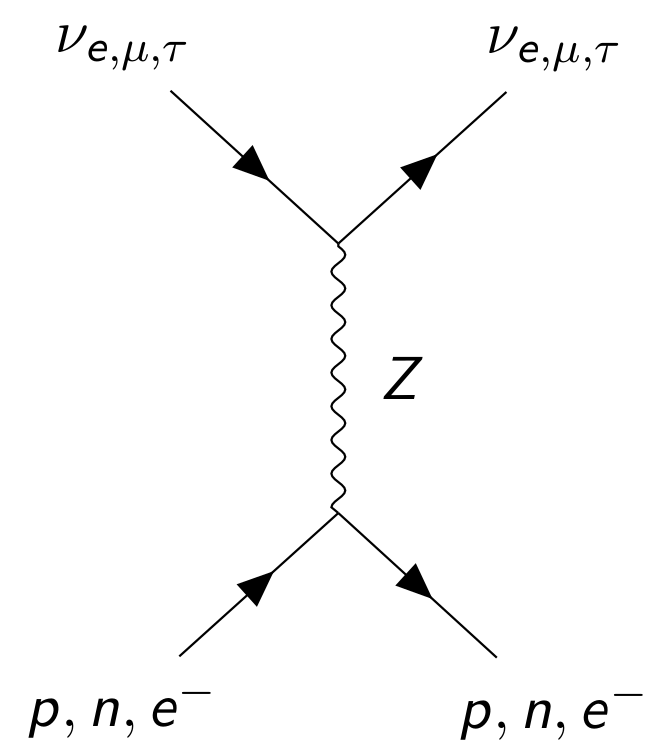}   \hspace*{.2in}
	\includegraphics[align=c, height=0.49\linewidth]{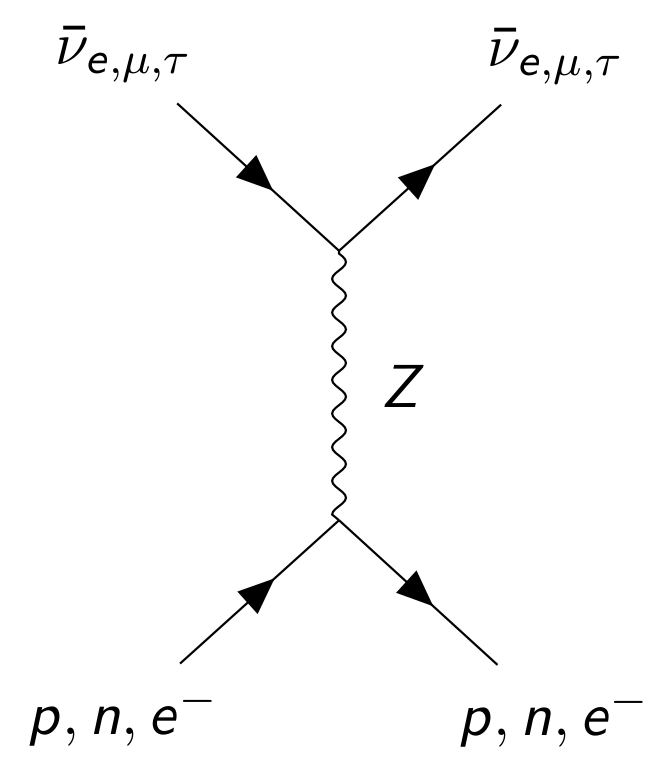}
	\caption{Feynman diagrams showing neutral-current interactions of neutrinos ($\nu_e$, $\nu_\mu$, $\nu_\tau$ ) and antineutrinos ($\bar{\nu}_e$, $\bar{\nu}_\mu$, $\bar{\nu}_\tau$) with $p$, $n$ and $e^-$ inside Earth in the left and right panels, respectively.}
	\label{fig:NC_matter_effect}
\end{figure}

So far, we have discussed the neutrino oscillations during propagation through vacuum. Now, we shall talk about what happens when neutrinos pass through matter regions such as the interiors of the Earth. During propagation through Earth, neutrinos and antineutrinos undergo weak interactions with the matter fermions such as electrons ($e^-$), protons ($p$), and neutrons ($n$). Neutrinos and antineutrinos of all three flavors ($e$, $\mu$, $\tau$) can interact with these matter fermions via $Z$-mediated coherent neutral-current (NC) interactions as shown by Feynman diagrams in Fig.~\ref{fig:NC_matter_effect}. This forward elastic NC scattering results in an effective potential as given by
\begin{align}
V_\text{NC} = -\frac{G_F N_n}{\sqrt{2}},
\end{align}
where $G_F$ is the Fermi coupling constant, and $N_n$ is the neutron number density inside the medium. Since neutral-current potential $V_\text{NC}$ is independent of neutrino flavor, it does not affect the neutrino oscillation probabilities. For antineutrino, the neutral-current potential $V_\text{NC} \rightarrow - V_\text{NC}$.

\begin{figure}
	\centering
	\includegraphics[align=c, height=0.49\linewidth]{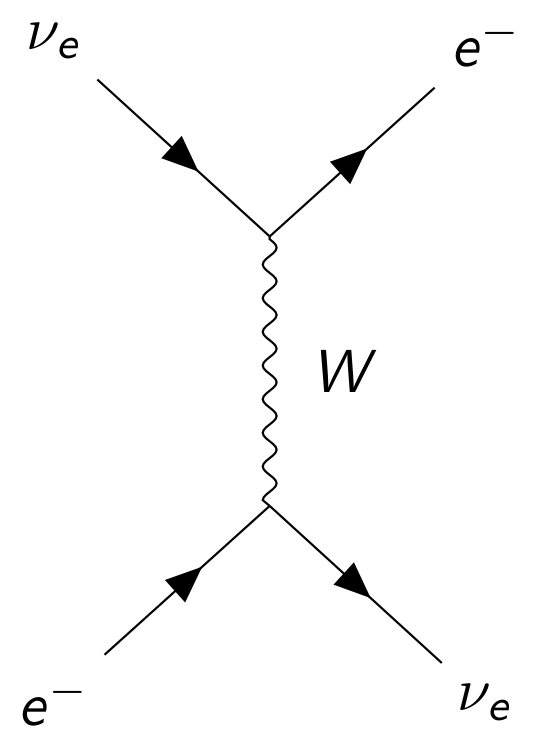}   \hspace*{.2in}
	\includegraphics[align=c, width=0.49\linewidth]{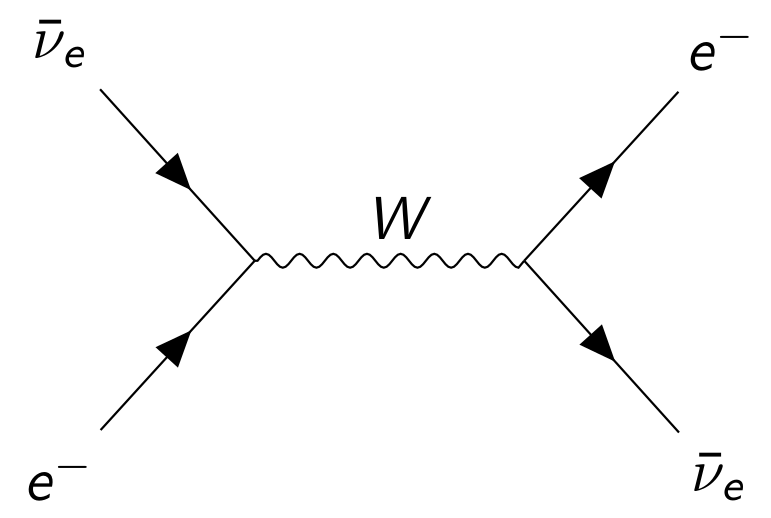}
	\caption{Feynman diagrams showing charged-current interactions of neutrinos ($\nu_e$) and antineutrinos ($\bar{\nu}_e$) with electrons inside Earth in the left and right panels, respectively.}
	\label{fig:CC_matter_effect}
\end{figure}

We know that there are electrons inside Earth but no muons and tau leptons. Hence, the $W$-mediated charged-current (CC) interactions with ambient electrons in the matter are only possible for electron neutrinos and antineutrinos as shown by Feynman diagrams in Fig.~\ref{fig:CC_matter_effect}. The coherent forward elastic charged-current scattering results in an effective potential as given by
\begin{align}
V_\text{CC} = \sqrt{2} G_F N_e,
\end{align} 
where, $N_e$ represent the electron number density inside matter. Since this charged-current potential $V_\text{CC}$ is not felt by muon and tau neutrinos,  it is capable of modifying the neutrino oscillation patterns. For antineutrino, the charged-current potential $V_\text{CC} \rightarrow - V_\text{CC}$.  By inserting the value of $G_F$ and expressing $N_e$ in terms of matter density $\rho$, the charged-current potential $V_\text{CC}$ for $\nu_e$ turn out to be  
\begin{equation}\label{eq:matter_pot}
V_{CC} \approx 7.6 \times Y_e \times 10^{-14} \left[\frac{\rho}{\text{g/cm}^3}\right]~\text{eV}\, ,
\end{equation}
where, $Y_e = N_e/(N_p + N_n)$ corresponds to the relative electron number density inside the matter with respect the density of protons $N_p$ and neutrons $N_n$. As far as the propagation of neutrinos through Earth is concerned, $\rho$ denotes the matter density of various layers inside the Earth for a given density profile. In all the analyses presented in this thesis, we assume the Earth to be electrically neutral and isoscalar where $N_n \approx N_p = N_e$ which results in $Y_e = 0.5$.

The effective Hamiltonian in the presence of matter considering three flavors of neutrino can be expressed as
\begin{align}\label{eq:eff_H_mass}
H_\text{eff} = \frac{1}{2E} U \begin{pmatrix}
m_1^2 & 0 & 0 \\
0 & m^2_{2} & 0 \\
0 &  0 & m^2_{3}
\end{pmatrix} U^\dagger + 
\begin{pmatrix}
V_\text{NC} + V_\text{CC} & 0 & 0 \\
0 & V_\text{NC} & 0 \\
0 & 0 & V_\text{NC} 
\end{pmatrix}
\end{align}
After absorbing the terms proportional to identity matrix as phases, Eq.~\ref{eq:eff_H_mass} boils down to  
\begin{align}
H_\text{eff} = \frac{1}{2E} U \begin{pmatrix}
0 & 0 & 0 \\
0 & \Delta m^2_{21} & 0 \\
0 &  0 & \Delta m^2_{31}
\end{pmatrix} U^\dagger + 
\begin{pmatrix}
V_\text{CC} & 0 & 0 \\
0 & 0 & 0 \\
0 & 0 & 0 
\end{pmatrix}. 
\end{align}
For antineutrino, $V_\text{CC} \rightarrow - V_\text{CC}$ and $U \rightarrow U^*$. 

Now, we describe the matter effect in a simple two-flavor framework where analytic expressions are available. The effective Hamiltonian for two neutrino flavors can be expressed as 
\begin{align}
H_\text{eff} &= \frac{1}{2E}
\begin{pmatrix}
\cos\theta & \sin\theta \\
-\sin\theta & \cos\theta 
\end{pmatrix}
\begin{pmatrix}
0 & 0  \\
0 & \Delta m^2
\end{pmatrix} 
\begin{pmatrix}
\cos\theta & -\sin\theta \\
\sin\theta & \cos\theta 
\end{pmatrix} + 
\begin{pmatrix}
V_\text{CC} & 0 \\
0 & 0
\end{pmatrix}\\
&= \frac{1}{2E}\begin{pmatrix}
\Delta m^2 \sin^2\theta + A & \Delta m^2\sin\theta \cos\theta \\
\Delta m^2 \sin\theta \cos\theta & \Delta m^2 \cos^2\theta
\end{pmatrix}, 
\end{align}
where, $\Delta m^2 = m_2^2 - m_1^2$ and $A = 2E V_\text{CC}$. Absorbing the term proportional to identity matrix leads to
\begin{align}\label{eq:Hamiltonion_matter_2f}
H_\text{eff} = \frac{1}{4E}\begin{pmatrix}
-\Delta m^2 \cos 2\theta + 2A & \Delta m^2 \sin 2\theta\\
\Delta m^2 \sin 2\theta & \Delta m^2 \cos 2\theta
\end{pmatrix}. 
\end{align}
The eigenvalues of Hamiltonian matrix (Eq.~\ref{eq:Hamiltonion_matter_2f}) are
\begin{align}
\frac{m_{1m}^2}{2E} &= \frac{1}{2E}\left( \frac{A}{2} - \frac{1}{2}\sqrt{(\Delta m^2 \cos 2\theta - A)^2 + (\Delta m^2 \sin 2\theta)^2}\right), \\
\frac{m_{2m}^2}{2E} &= \frac{1}{2E}\left( \frac{A}{2} + \frac{1}{2}\sqrt{(\Delta m^2 \cos 2\theta - A)^2 + (\Delta m^2 \sin 2\theta)^2}\right).
\end{align}
The effective mass-squared difference in matter becomes
\begin{align}\label{eq:eff_dmsq}
\Delta m^2_m \equiv  m_{2m}^2 - m_{1m}^2 = \sqrt{(\Delta m^2 \cos 2\theta - A)^2 + (\Delta m^2 \sin 2\theta)^2}.
\end{align}
The effective Hamiltonian $H_\text{eff}$ can be diagonalized using a rotation matrix $O$ 
\begin{align}
O = \begin{pmatrix}
\cos\theta_m & \sin\theta_m \\
-\sin\theta_m & \cos\theta_m 
\end{pmatrix},
\end{align}
such that 
\begin{align}
O^T H_\text{eff} O = 
\begin{pmatrix}
m^2_{1m} & 0 \\
0 & m^2_{2m}
\end{pmatrix},
\end{align}
where, the mixing angle $\theta_m$ in matter is given as
\begin{align}\label{eq:eff_mixing_angle}
\tan 2\theta_m = \frac{\Delta m^2 \sin 2\theta}{\Delta m^2 \cos 2\theta - A}.
\end{align}

Note that unlike the case of vacuum oscillations, the effective mixing angle $\theta_m$ and effective mass-squared difference $\Delta m^2_m$ in matter depend on the energy ($E$) and  density of matter $\rho$ (hence, path-length ($L$) of neutrino). Even if the vacuum mixing angle $\theta$ is small, the effective mixing angle $\theta_m$ in the presence of matter can still have large value of $45^\circ$ corresponding to maximal mixing (see Eq.~\ref{eq:eff_mixing_angle}), if
\begin{align}\label{eq:MSW_condition}
\Delta m^2 \cos 2\theta = A,
\end{align}
which is known as Mikheyev-Smirnov-Wolfenstein (MSW) resonance. At the same time, the effective mass-squared difference $\Delta m^2_m$ becomes minimum at MSW resonance (see Eq.~\ref{eq:eff_dmsq}). For the case of neutrino $A$ is positive, hence the condition given in Eq.~\ref{eq:MSW_condition} is fulfilled if $\Delta m^2 > 0$. On the other hand, this MSW resonance for antineutrino ($A < 0$) occurs if $\Delta m^2 < 0$. Note that if $A \ll \Delta m^2$, the matter effect is not significant and on the other hand if $A \gg \Delta m^2$, the neutrinos oscillations are suppressed ($\theta_m = \pi/2$).

Note that though the mixing angle and mass-squared difference get modified in the presence of matter, the dynamics of neutrino mixing remain the same.  The analytic expressions for oscillation probabilities in the presence of constant-density matter can be obtained by replacing the vacuum parameters ($\theta$ and $\Delta m^2$) with the effective parameters ($\theta_m$ and $\Delta m^2_m$),
\begin{align}
P(\nu_\alpha \rightarrow \nu_\beta) &= \sin^2 2\theta_m \sin^2 \left(\frac{\Delta m^2_m L}{4E} \right)\,, \\
P(\nu_\alpha \rightarrow \nu_\alpha) &= 1 - \sin^2 2\theta_m \sin^2 \left(\frac{\Delta m^2_m L}{4E} \right)\,. 
\end{align}

Now, we will discuss neutrino oscillation experiments that have contributed to the measurement of neutrino oscillation parameters.

\section{Neutrino Oscillation Experiments}
\label{sec:osc_exp}

Neutrino oscillation parameters have been measured by various kinds of experiments using neutrinos from the atmosphere, Sun, accelerators, and reactors. Now, we describe some of these neutrino oscillation experiments. 

\begin{itemize}
	
	\item \textbf{GALLEX:} GALLEX (Gallium Experiment)~\cite{GALLEX:1992gcp} was a solar neutrino experiment at the Gran Sasso Underground Laboratory from 1992 to 1997. The detector had a rock coverage of about 3300 m water equivalent. GALLEX used about 100-ton gallium chloride solution ($\sim$30 ton Galium) as target to detect solar neutrinos using the inverse beta decay reaction $^{71}\text{Ga}(\nu_e,e)^{71}\text{Ge}$. The energy threshold of this reaction for detecting solar neutrino was about 233 keV. The solar neutrino flux was measured by extracting the neutrino induced $^{71}\text{Ge}$ and counting its decay events. The observed rate of interaction for solar neutrinos was about [77.5 $\pm$ 7] SNU during phase I-IV~\cite{GALLEX:1998kcz} while the expected rate was around 130 SNU. 
		
	\item \textbf{SAGE:} The Russian-American Gallium Experiment (SAGE)~\cite{SAGE:1999nng} was a solar neutrino experiment at the Baksan Neutrino Observatory (BNO) in the northern Caucasus Mountains, Russia. The underground observatory had an rock shielding of about 4700 m of water equivalent. SAGE used about 50 ton of liquid Gallium metal to detect solar neutrinos using inverse beta decay reaction $^{71}\text{Ga}(\nu_e,e)^{71}\text{Ge}$ with an energy threshold of 233 keV similar to the GALLEX experiment. During 1990 to 2007, SAGE observed the interaction rate of solar neutrinos to be $65^{+3.1}_{-3.0} \text{ (stat) }^{+2.6}_{-2.8}\text{ (syst) }$ SNU~\cite{SAGE:2009eeu} which was only half of the predicted rate of about 130 SNU. Note that the GALLEX and SAGE experiments had not measured the energy, direction, and timing of solar neutrino events. 
	
	\item \textbf{Kamiokande:} The Kamioka Nucleon Decay Experiment (Kamiokande)~\cite{Kamiokande-II:1987idp,Suzuki:2022ldt} was built to search for proton decay.  Since atmospheric neutrinos acted as a background for the signals corresponding to proton decays, Kamiokande had to perform the precision measurement of atmospheric neutrino flux.  Kamiokande consisted of 3000 tons of water in a cylindrical tank with a diameter of 15.5 m and a height of about 16 m.  The fiducial volume of the detector was about 2140 tons.  Kamiokande worked on the principle of Cherenkov radiation.  Neutrino interactions inside water produce energetically charged particles that emit Cherenkov photons when they travel faster than the velocity of light in the medium.  These photons were detected using 1000 20-inch PMTs mounted on the wall, bottom, and top surfaces.  The Kamiokande experiment can distinguish electron and muon types of neutrinos.  The real-time measurement of solar neutrinos was also performed by Kamiokande experiment~\cite{Kamiokande-II:1989hkh}. In 1987, the Kamiokande experiment also observed neutrinos from the Supernova SN1987A~\cite{Kamiokande-II:1987idp}. 
	
	\item \textbf{Super-K:} The Super-Kamiokande (Super-K) experiment~\cite{Super-Kamiokande:2017yvm} is a successor of Kamiokande experiment. Super-K consists of 50 kt of water with a fiducial mass of 22.5 kt. The Cherenkov photons emitted by secondary particles during neutrino interactions are detected using about 11000 large-area PMTs fixed on the walls, top and bottom surfaces of the cylindrical detector. Super-K experiment is capable of detecting solar, supernova, atmospheric, and accelerator neutrinos. Unlike GALLEX and SAGE, Kamiokande and Super-K experiment was able to perform real-time observation of solar neutrinos by measuring timing, energy, and direction of events~\cite{Super-Kamiokande:2005wtt,Super-Kamiokande:2008ecj,Super-Kamiokande:2010tar}. The super-K experiment was the first experiment to discover the evidence of neutrino oscillations using atmospheric neutrino data in 1998~\cite{Super-Kamiokande:1998kpq}. Super-K experiment has contributed significantly in measuring atmospheric neutrino oscillation parameters~\cite{Super-Kamiokande:2017yvm}. 
	
	\item \textbf{SNO:} The Sudbury Neutrino Observatory (SNO) experiment~\cite{SNO:1999crp} was a solar neutrino experiment located at the INCO, Ltd., Creighton mine in Sudbury, Ontario, Canada. The detector was placed underground at a depth of about 2000 m with a rock overburden of about 6000 m water equivalent. The detector consisted of 1000 tonnes of heavy water in a transparent acrylic sphere with a diameter of 12 m. About 9400 inward-looking PMTs were mounted on the sphere to detect Cherenkov light emitted by the secondary particles produced in the neutrino interactions. SNO was capable of detecting charged-current as well as flavor-independent neutral-current interactions of solar neutrinos as described in Sec.~\ref{sec:solar_neutrino_anomaly}. The SNO experiment resolved the solar neutrino problem by detecting the neutral-current neutrino interactions corresponding to all flavors, which matched the expected solar neutrino flux~\cite{SNO:2002tuh,SNO:2011hxd}. 
	
	\item \textbf{Borexino:} The Borexino experiment~\cite{Borexino:2008gab} is a liquid scintilator detector located deep underground at the Gran Sasso laboratory in Italy with a rock overburden of about 3800 m water equivalent. The aim of Borexino experiment is to perform real-time measurement of low energy solar neutrinos which include monochromatic solar neutrinos emitted during electron capture decay of $^7$Be and other possible contributions also such as CNO, \textit{pep}, \textit{pp}, and $^8$B neutrinos. The low-energy neutrinos of all flavors are detected via elastic scattering of electrons whereas electron antineutrinos can be detected through the process of inverse beta decay. 
	
	The liquid scintillation technique have advantage over Cherenkov and radiochemical detectors in term of detection of low energy neutrinos but liquid scintillators require low background environment. The scintillation light produced during neutrino interaction is detected by PMTs mounted on the spherical walls of the detector and the energy of the neutrino event is reconstructed. Note that the directional information of neutrino is lost in the liquid scintillator due to an isotropic emission of scintillation photons. Borexino experiment has already observed \textsuperscript{7}Be neutrinos~\cite{Borexino:2007kvk,Borexino:2008dzn,Bellini:2011rx}, \textit{pep} neutrinos \cite{Borexino:2011ufb}, \textsuperscript{8}B neutrinos~\cite{Borexino:2008fkj}, \textit{pp} neutrinos~\cite{BOREXINO:2014pcl,BOREXINO:2018ohr}, CNO neutrinos~\cite{BOREXINO:2020aww} from the Sun. The observations of neutrinos from these different reactions helps us to understand the important issues such as solar metallicity, temperature of the core, evolution of the sun, and density profile of the sun etc. 
	
	\item \textbf{IceCube:} The IceCube Neutrino Observatory~\cite{IceCube:2016zyt} is a kilometer-cube detector built deep inside the ice at the South pole. The primary aim of IceCube is to observe astrophysical neutrino at PeV-scale. However, IceCube can also detect atmospheric neutrinos at energies above hundreds of GeV. IceCube work on the principle of Cherenkov radiation emitted by secondary particles, which are produced during the neutrino interactions inside the ice. The Cherenkov photons are detected by Digital Optical Modules (DOMs), which consist of the large-area PMTs. About 5160 DOMs are deployed on 86 vertical strings between 1450 m and 2450 m below the surface of the ice. The neutrino interactions in IceCube can result in a hadronic shower at the vertex and an outgoing muon with a long track. 
	
	DeepCore~\cite{IceCube:2011ucd} is a sub-array in the IceCube detector having densely spaced DOMs deployed at a depth lower than 1750 m from the surface of the ice. DeepCore has eight new specialized strings along with the seven standard IceCube strings. Apart from dense packing, DeepCore also consists of DOMs using PMTs with higher quantum efficiency than the standard modules. This densely packed geometry and increased efficiency enable DeepCore to achieve a lower energy threshold of about 10 GeV compared to about 100 GeV for IceCube. IceCube and DeepCore have already performed oscillation studies using atmospheric neutrino~\cite{IceCube:2014flw,IceCube:2017lak}.

	\item \textbf{K2K:} The KEK to Kamioka (K2K)~\cite{K2K:2004iot} was a long-baseline neutrino oscillation experiment to observe neutrino oscillations in the form of the disappearance of muon neutrinos. In the K2K experiment, the neutrino beam originated at the KEK facility and was directed towards the Super-K detector located at a distance of about 250 km. The muon neutrino beam had mean energy of about 1.3 GeV. The K2K experiment also consisted of near detectors at a distance of about 300 m to minimize the systematics. The observation of the energy-dependent disappearance of muon neutrinos at K2K from 1999 to 2004 independently confirmed the neutrino oscillation~\cite{K2K:2004iot} which was discovered earlier at Super-K using atmospheric neutrino data.
	
	\item \textbf{MINOS:} The Main Injector Neutrino Oscillation Search (MINOS)~\cite{MINOS:1998kez} was also a long-baseline experiment to study neutrino oscillations. The muon neutrino (or antineutrino) beam for MINOS was produced at the Fermilab and detected by a 5.4 kt far detector placed at a distance of about 735 km in the Soudan mine in Minnesota. MINOS also consisted of a 1 kt near detector at a distance of about 1 km. The near and far detectors were designed by sandwiching 2.54 cm thick iron plates and 1 cm thick scintillators such that the detectors can act as tracking calorimeters. A magnetic field of 1.3 T enabled MINOS to distinguish neutrinos ($\nu_\mu$) and antineutrinos ($\bar{\nu}_\mu$).  MINOS measured the atmospheric neutrino oscillation parameter with high precision~\cite{MINOS:2014rjg}. 
	
	\item \textbf{OPERA:} The OPERA experiment~\cite{Shibuya:1997qr,OPERA:2010pne,OPERA:2021xtu} was a long-baseline experiment to observe neutrino oscillations via appearance of $\nu_\tau$ in $\nu_\mu \rightarrow \nu_\tau$ channel. In OPERA, a muon neutrino beam was directed from CERN to the Gran Sasso Underground Laboratory at a distance of about 730 km. The average neutrino energy was about 17 GeV. In OPERA, the neutrino interactions took place inside a 1.25 kt target of lead plates (passive element) sandwiched with nuclear emulsion films (active element), which acted as a high-accuracy tracking device. This type of detector is historically known as Emulsion Cloud Chamber and is suitable for detection of decay of $\tau$ lepton, which is produced during charged-current interaction of $\nu_\tau$. Using the data taken during 2008 to 2012, OPERA discovered the $\nu_\mu \rightarrow \nu_\tau$ oscillations at $6.1\sigma$ level by observing 10 $\nu_\tau$ charged-current interactions candidates~\cite{OPERA:2018nar}. 
	
	\item \textbf{T2K:} The Tokai-to-Kamioka (T2K)~\cite{T2K:2011qtm} is a long-baseline experiment to study neutrino oscillations. T2K consists of the J-PARC facility at Tokai as a source of muon neutrino beam and Super-K as a far detector located 295 km away. Super-K is present at $2.5^\circ$ off-axis with respect to the muon beam such that the narrow spectrum of energy is obtained with a peak at 0.6 GeV. T2K also uses on-axis and off-axis near detectors at a distance of about 280 m from the source to minimize the systematic uncertainties. One of the main aims of the T2K experiment is to look for CP-violation in the lepton sector~\cite{T2K:2014xyt}. T2K has also contributed to the precision measurement of neutrino oscillation parameters~\cite{T2K:2020}.

	\item \textbf{NOvA:} The NuMI Off-axis $\nu_e$ Appearance (NOvA)~\cite{NOvA:2004blv,Talaga:2016rlq} experiment is another presently running accelerator-based long-baseline neutrino oscillation experiment. NOvA consists of a muon neutrino ($\nu_\mu$ or $\bar{\nu}_\mu$) beam produced at the NuMI beamline at Fermilab and a 14 kt far detector located at a distance of about 810 km in Ash River, Minnesota. NOvA also has a 0.3 kt near detector at a distance of about 1 km to measure unosillated neutrino flux and minimize systematics. The near and far detectors are tracking calorimeters based on the technology of liquid Scintillation. About 37\% mass of the detector is present in the form of PVC (polyvinyl chloride) and 63\% in the liquid scintillator. The detectors are located at 14.6 mrad off-axis with respect to the neutrino beam such that the neutrino energy peaks around 2 GeV with a narrow energy range of about 1 to 3 GeV. NOvA has access to both $\nu_e$ appearance and $\nu_\mu$ disappearance channel, which give rise to the sensitivity towards the precision measurement of atmospheric oscillation parameter, octant of $\theta_{23}$, neutrino mass ordering, and CP violation~\cite{NOvA:2020}. 

	\item \textbf{KamLAND:} The Kamioka Liquid scintillator Anti-Neutrino Detector (KamLAND) \cite{KamLAND:2002uet,KamLAND:2004mhv} was a reactor-based neutrino oscillation experiment located in Japan. KamLAND detected antineutrino from about 53 nuclear reactors having an average baseline of about 180 km. The KamLAND detector was placed at the same site where Kamiokande was located earlier, with an average rock overburden of about 2700 m water equivalent. The target for antineutrino interaction was a 1 kt of ultra-pure liquid scintillator in a spherical nylon balloon having a diameter of 13 m. The volume between the balloon and an outer spherical stainless steel vessel with a diameter of 18 m was filled with purified mineral oil to reduce the external radioactive background. The antineutrinos were detected using the process of inverse beta decay. The scintillation photons produced during interaction were detected using 1879 PMTs mounted on the inner surface of stainless steel vessel. A 3.2 kt water-Cherenkov-based outer detector with 225 PMTs was used to absorb gamma rays and neutrons coming from the surrounding rocks and veto the comic muons. KamLAND observed the disappearance of $\bar{\nu}_e$ which helped in measuring the solar neutrino oscillation parameters~\cite{KamLAND:2002uet,KamLAND:2004mhv,KamLAND:2013rgu}.
		
	\item \textbf{Double Chooz:} Double Chooz~\cite{DoubleChooz:2012gmf} was a reactor antineutrino experiment located in France. It detected the electron antineutrinos from the two reactor cores of the Chooz nuclear power plant using the process of inverse beta decay. Double Chooz consisted of two identical liquid scintillator detectors, which were located at distances of about 400 m and 1050 m from the nuclear core. Both detectors were filled with 8 tons of gadolinium-loaded liquid scintillators as a target for electron antineutrinos. Double Chooz exploited disappearance of electron antineutrinos with the help of near and far detectors to measure the value of reactor mixing angle $\theta_{13}$ as $\sin^2 2\theta_{13} = 0.123 \pm 0.023$ (stat.+syst.)~\cite{Meregaglia:2017spw}.
	
	\item \textbf{Daya Bay:} The Daya Bay experiment~\cite{DayaBay:2007fgu,DayaBay:2015ayh} is also a reactor antineutrino experiment located in China. Daya Bay consists of eight identical antineutrino detectors which detect antineutrinos emitted by the six reactor cores. Each antineutrino detector has three nested cylindrical vessels. The inner vessel made up of acrylic contains 20 tonnes of gadolinium-doped (0.1\%) liquid scintillator, which acts as a primary target for electron antineutrinos. This target is surrounded by an acrylic vessel containing 20 tonnes of undoped liquid scintillator, which increases the efficiency of detecting gamma photons produced in the target volume. The outer-most vessel is made up of stainless steel, which contains 37 tonnes of mineral oil to reduce the radioactive background. The reactor antineutrinos are detected using the process of inverse beta decay. The scintillation photons emitted during antineutrinos interactions are detected by 192 PMTs mounted uniformly on the inner surface of the stainless steel cylinder. Daya Bay has performed the most precise measurement of the reactor mixing $\theta_{13}$ as $\sin^2 2\theta_{13} = 0.0856 \pm 0.0029$~\cite{DayaBay:2018yms}.
	
	\item \textbf{RENO:} The Reactor Experiment for Neutrino Oscillation (RENO)~\cite{RENO:2010vlj} detects antineutrinos from the Hanbit Nuclear Power Plant in South Korea. RENO consists of two identical near and far antineutrino detectors that are placed at 294 and 1383 m, respectively, from the six reactor cores. Each detector consists of an inner detector and an outer veto detector. The inner detector contains a cylindrical stainless steel vessel that has two nested acrylic vessels. The innermost acrylic vessel contains 16.5 tons of gadolinium-doped liquid scintillator that acts as a target for neutrino interactions. This is surrounded by an undoped liquid scintillator to catch gamma rays emitted from the neutrino interactions in the target. The $\gamma$-catcher is followed by mineral oil that is used to reduce the radioactive background. The neutrinos are detected by the process of inverse beta decay. The neutrino interactions result in the scintillation light, which is detected by 354 low background 10-inch PMTs mounted on the inner wall of the stainless steel container. The outer detector contains highly purified water and 67 PMTs mounted on the surface of the concrete vessel. During 2011 to 2020, the far (near) detector observed 120383 (989736) antineutrino candidate events and the reactor mixing angle $\theta_{13}$ was measured as $\sin^2 2\theta_{13} = 0.0892 \pm 0.0044 (\text{stat.}) \pm 0.0045 (\text{syst})$~\cite{RENO:2020,Shin:2020mue}.
\end{itemize}

\section{Present status of Neutrino Oscillation Parameters}
\label{sec:status_osc_par}
The neutrino oscillations were discovered more than two decades ago by the Super-K experiment using atmospheric neutrino data~\cite{Super-Kamiokande:1998kpq}. Since then, a large number of neutrino oscillation experiments have contributed to the measurement of the values of neutrino oscillation parameters. Table~\ref{tab:global_fit} summarizes the best-fit values of neutrino oscillation parameters with $1\sigma$ and $3\sigma$ uncertainties as obtained from current global fit~\cite{Esteban:2020cvm,NuFIT:2021} of data from various neutrino oscillation experiments. The best-fit results from other global fit analyses~\cite{deSalas:2020pgw,Capozzi:2021fjo} are also consistent with these results. 

%%%%%%%%%%%%%%%%%%%%%%%%%%%%%%%%%%%%%%%%%%%%%%%%%%%%%%%%%%%%%%%%
\begin{table}[h]
	\centering
	\renewcommand{\arraystretch}{1.6}% for the vertical padding
	\begin{tabular}{|c|cc|}
		\hline
		& best fit point $\pm 1\sigma$ & $3\sigma$ range\\ \hline
		$\sin^2\theta_{12}$ & $0.304^{+0.013}_{-0.012}$ & $0.269 \rightarrow 0.343$ \\
		$\theta_{12} /^\circ$ & $33.45^{+0.78}_{-0.75}$ & $31.27 \rightarrow 35.87$ \\
		$\frac{\Delta m^2_{21}}{10^{-5}\text{ eV}^2}$ & $7.42^{0.021}_{-0.20}$ & $6.82 \rightarrow 8.04$\\
		$\sin^2 \theta_{23}$ (NO) & $0.450^{+0.019}_{-0.016}$ & $0.408 \rightarrow 0.603$ \\
		$\theta_{23} /^\circ$ (NO) & $42.1^{+1.1}_{-0.9}$ & $39.7 \rightarrow 50.9$ \\
		$\sin^2 \theta_{23}$ (IO) & $0.507^{+0.016}_{-0.022}$ & $0.410 \rightarrow 0.613$ \\
		$\theta_{23} /^\circ$ (IO) & $49.0^{+0.9}_{-1.3}$ & $39.8 \rightarrow 51.6$ \\
		$\sin^2\theta_{13}$ (NO) & $0.02246^{+0.00062}_{-0.00062}$ & $0.02060 \rightarrow 0.02435$\\
		$\theta_{13} /^\circ$ (NO) & $8.62^{+0.12}_{-0.12}$ & $8.25 \rightarrow 8.98$\\
		$\sin^2\theta_{13}$ (IO) & $0.02241^{+0.00074}_{-0.00062}$ & $0.02055 \rightarrow 0.02457$\\
		$\theta_{13} /^\circ$ (IO) & $8.61^{+0.14}_{-0.12}$ & $8.24 \rightarrow 9.02$\\
		$\delta_\text{CP}/^\circ$ (NO) & $230^{+36}_{-25}$ & $144 \rightarrow 350$ \\
		$\delta_\text{CP}/^\circ$ (IO) & $278^{+22}_{-30}$ & $194 \rightarrow 345$ \\
		$\frac{\Delta m^2_{31}}{10^{-3} \text{ eV}^2}$ (NO) & $+2.510^{+0.027}_{-0.027}$ & $+2.430 \rightarrow +2.593$ \\
		$\frac{\Delta m^2_{32}}{10^{-3} \text{ eV}^2}$ (IO) & $-2.490^{+0.026}_{-0.028}$ & $-2.574 \rightarrow -2.410$ \\
		\hline 
	\end{tabular}
	\caption{The best-fit values of three-flavor neutrino oscillation parameters with $3\sigma$ range as obtained in~\cite{Esteban:2020cvm,NuFIT:2021} from global fit of data from various neutrino oscillation experiments. The other global fit studies~\cite{deSalas:2020pgw,Capozzi:2021fjo} have also obtained almost similar results.}
	\label{tab:global_fit}
\end{table}
%%%%%%%%%%%%%%%%%%%%%%%%%%%%%%%%%%%%%%%%%%%%%%%%%%%%%%%%%%%%%%%%

The parameters $\theta_{12}$ and $\Delta m^2_{21}$ have been measured precisely using KamLAND~\cite{KamLAND:2013rgu} and solar neutrino experiments~\cite{Cleveland:1998nv,SAGE:2009eeu,Kaether:2010ag,Super-Kamiokande:2005wtt,Super-Kamiokande:2008ecj,Super-Kamiokande:2010tar,SNO:2011hxd}. The mixing angle $\theta_{13}$ is measured with high precision using disappearance of electron antineutrinos at reactor experiments like Daya Bay~\cite{DayaBay:2018yms}, Double Chooz~\cite{Meregaglia:2017spw}, and RENO~\cite{RENO:2020,Shin:2020mue}. The precision measurement of atmospheric mass-squared difference $\Delta m^2_{31}$ is contributed by the long-baseline accelerator~\cite{MINOS:2014rjg,T2K:2020,NOvA:2020}, reactor~\cite{DayaBay:2018yms,RENO:2020,Shin:2020mue} and atmospheric~\cite{Super-Kamiokande:2017yvm,IceCube:2014flw,IceCube:2017lak} neutrino experiments. As far as the sign of $\Delta m^2_{31}$ is concerned, it is still not known. The accelerator~\cite{MINOS:2014rjg,T2K:2020,NOvA:2020} and atmospheric~\cite{Super-Kamiokande:2017yvm,IceCube:2014flw,IceCube:2017lak} neutrino experiments measured the disappearance of muon neutrinos and antineutrinos which give rise to the sensitivity towards $\sin^2\theta_{23}$ but it still has large uncertainties. The CP-phase $\delta_\text{CP}$ is least precisely measured oscillation parameter where some bounds are obtained using accelerator experiments like T2K~\cite{T2K:2020} and NOvA~\cite{NOvA:2020}. 

The analyses in this thesis have been performed using the benchmark values of neutrino oscillation parameters mentioned in Table~\ref{tab:osc-param-value}. We use the effective atmospheric mass-squared difference\footnote{The effective atmospheric mass-squared difference is related to $\Delta m^2_{31}$ and $\Delta m^2_{21}$ as follows~\cite{deGouvea:2005hk,Nunokawa:2005nx}
\begin{equation}
\Delta m^2_\text{eff} = \Delta m^2_{31} - \Delta m^2_{21} (\cos^2\theta_{12} - \cos \delta_\text{CP} \sin\theta_{13}\sin2\theta_{12}\tan\theta_{23}).
\end{equation}} $\Delta m^2_\text{eff}$ to consider mass ordering (MO), the positive and negative value of $\Delta m^2_\text{eff}$ corresponds to normal ordering (NO) with $m_1 < m_2 < m_3$ and inverted ordering (IO) with $m_3 < m_1 < m_2$, respectively.

%%%%%%%%%%%%%%%%%%%%%%%%%%%%%%%%%%%%%%%%%%%%%%%%%%%%%%%%%%%%%%%%
\begin{table}[h]
	\centering
	\begin{tabular}{|c|c|c|c|c|c|c|}
		\hline
		$\sin^2 2\theta_{12}$ & $\sin^2\theta_{23}$ & $\sin^2 2\theta_{13}$ &
		$|\Delta m^2_{32}|$ (eV$^2$) & $\Delta m^2_{21}$ (eV$^2$) & $\delta_\text{CP}$
		& Mass Ordering\\
		\hline
		0.855 & 0.5 & 0.0875 & $2.46\times 10^{-3}$ & $7.4\times10^{-5}$
		& 0 & Normal (NO)\\
		\hline 
	\end{tabular}
	\caption{The benchmark values of neutrino oscillation parameters that we use in the analyses in this thesis~\cite{Kumar:2020wgz}.}
	\label{tab:osc-param-value}
\end{table}
%%%%%%%%%%%%%%%%%%%%%%%%%%%%%%%%%%%%%%%%%%%%%%%%%%%%%%%%%%%%%%%%

\section{Unsolved issues in Neutrino Physics}
\label{sec:unsolved_issues}

The precisions of neutrino oscillation parameters have improved significantly over the past two decades, but there are still some unsolved issues related to neutrino oscillations and properties of neutrinos which are as follows
\begin{itemize}
	\item \textbf{Neutrino mass ordering:} The magnitude of the atmospheric mass-squared splitting $\Delta m^2_{31}$ has been measured using acceleration, reactor, and atmospheric neutrino experiments, but the sign of $\Delta m^2_{31}$ is not known. The case of positive value of $\Delta m^2_{31}$ corresponding to $m_3 > m_2 > m_1$ is called normal ordering (NO) or normal hierarchy (NH) whereas the case of negative value of $\Delta m^2_{31}$ corresponding to $m_2 > m_1 > m_3$ is called inverted ordering (IO) or inverted hierarchy (IH). The determination of the correct mass ordering scenario is one of the most important aims of future neutrino oscillation experiments. 
	
	\item \textbf{Octant of $\theta_{23}$:} The disappearance of muon neutrinos and antineutrino in accelerator and atmospheric neutrino experiments have been used to measure $\sin^2 2\theta_{23}$ but we do not know whether $\theta_{23} < 45^\circ$ (lower octant) or $\theta_{23} > 45^\circ$ (higher octant). The determination of the octant of $\theta_{23}$ is the aim of next-generation neutrino oscillation experiments.
	
	\item \textbf{CP violation in lepton sector:} The future long-baseline experiments are focusing on the measurement of the value of CP phase $\delta_\text{CP}$ and the determination of a possibility of CP violation in the lepton sector. A value of $\delta_\text{CP}$ other than $0$ and $\pi$ would represent the CP violation, whereas $\pi/2$ and $3\pi/2$ would correspond to the maximum CP violation. The CP violation in the neutrino sector can help us to probe the possible contribution of neutrinos in producing the matter-antimatter asymmetry of the universe. 
	
	\item \textbf{Precision measurement of oscillation parameters:} In the past two decades, the understanding of neutrino oscillation physics has improved significantly. Now, the present and future experiments are focusing on the precision measurement of neutrino oscillation parameters. Solar oscillation parameters ($\theta_{12}$, $\Delta m^2_{21}$) and reactor mixing angle ($\theta_{13}$) are measured with good precision. The least precisely measured parameters are $\delta_{\rm CP}$ and $\theta_{23}$. The huge amount of data in the next generation experiments are going to improve the precision measurement of oscillation parameters at an unprecedented level. 
	
	\item \textbf{Sterile neutrinos:} The measurement at ALEPH detector on Large Electron-Positron collider showed that there are three active neutrino flavor states~\cite{ALEPH:1989kcj} which take part in the Standard Model interactions. Neutrino oscillation is a tool to probe if there exists more states known as sterile neutrinos, which do not interact via Standard Model interactions. The Liquid Scintillator Neutrino Detector (LSND) observed an excess of events in $\bar{\nu}_\mu \rightarrow \bar{\nu}_e$ appearance data that could not be explained by the three-flavor neutrino oscillations and could only be accounted by the presence of a sterile neutrino with mass-squared splitting $\Delta m^2 \sim 0.2 - 10$ eV$^2$~\cite{LSND:2001aii}. A recent measurement at the MiniBooNE experiment~\cite{MiniBooNE:2018esg} is also consistent with the LSND observation. Therefore, probing the possibility of sterile neutrinos is also the aim of future neutrino experiments. 
	
	\item \textbf{Non-standard interactions:} The discovery of neutrino oscillations has provided evidence that neutrinos are massive. The presence of massive neutrinos is a strong hint for the theories beyond the Standard Model (BSM) of particle physics. Non-standard interactions (NSI) of neutrinos are one of the possibilities in BSM scenarios. The neutrino oscillation patterns get modified due to the presence of neutral-current NSI during the propagation of neutrino. Therefore, neutrino oscillations can be used to probe the presence of neutral-current NSI. 

	\item \textbf{Neutrino tomography of Earth:} The upcoming atmospheric neutrino experiments like Hyper-K, ORCA, IceCube Upgrade, DUNE atmospheric, and ICAL will gather a huge amount of atmospheric neutrino data, which can play an important role in the neutrino tomography of Earth. The upward-going neutrinos experience the matter effects due to weak interactions with the ambient electrons. This matter effect depends on the density of the regions through which neutrino has traversed. Therefore, the matter effects in neutrino oscillations can be an effective tool to probe the internal structure of Earth. 

\end{itemize}

Though the work presented in this thesis focuses mainly on neutrino oscillations, there are other unsolved issues related to the properties of neutrinos which are actively being pursued in various experiments as discussed below

\begin{itemize}
	
	\item \textbf{Dirac or Majorana:} Another unsolved issue is the nature of neutrinos. It is not known whether the neutrino is its own antiparticle, in which case, it will be a Majorana-type particle. On the other hand, the neutrino will correspond to the Dirac-type if neutrino and antineutrino turn out to be different particles. The process of neutrinoless double-beta decay ($0\nu\beta\beta$)~\cite{Furry:1939qr,Schechter:1981bd,DellOro:2016tmg,Dolinski:2019nrj,Jones:2021cga,Cirigliano:2022oqy} involves conversion to two neutrons into two protons and two electrons without any neutrino, i.e., $n + n \rightarrow p^+ + p^+ + e^- + e^-$. The lepton number conservation gets violated by two units in $0\nu\beta\beta$, and this process is possible only if neutrinos and antineutrinos are the same particles. Therefore, experiments all over the world are looking for $0\nu\beta\beta$, but this phenomenon is not observed yet~\cite{DellOro:2016tmg,Dolinski:2019nrj,Cirigliano:2022oqy}.
	
	\item \textbf{Absolute mass of neutrino:} Neutrino oscillations have helped us to measure the mass-squared difference, but the absolute scale of neutrino mass is not yet known. The information about the absolute neutrino mass can be obtained by performing the precise measurement of the energy spectrum of the emitted electrons in beta decay. Using the tritium beta decay, the Karlsruhe Tritium Neutrino (KATRIN) experiment has provided an upper bound on the effective neutrino mass\footnote{The effective neutrino mass for electron flavor ($m_\beta$) is defined by the following relation 
		\begin{equation}
		m_\beta^2 = \sum_{j=1}^{3} |U_{ej}|^2 m^2_j\,,
		\end{equation}
		where $m_j$ are the mass eigenvalues. 
	} for electron flavor to be $m_\beta < 0.8$ eV~\cite{KATRIN:2021uub}. The upper limits on the absolute mass of neutrino can also be obtained using cosmological observations and neutrinoless double-beta decay~\cite{Lisi:2022nka}. Measuring the absolute mass of the neutrino is one of the important goals of future experiments. 
\end{itemize}

\section{Proposed Neutrino Oscillation Experiments}
\label{sec:future_exp}
Now, we describe the next-generation neutrino oscillation experiments that have been proposed to resolve the above-mentioned issues in neutrino oscillation physics. 

\begin{itemize}
	\item \textbf{DUNE:} The Deep Underground Neutrino Experiment (DUNE)~\cite{DUNE:2020lwj,DUNE:2020ypp} is an upcoming accelerator-based long-baseline experiment to study neutrino oscillations. DUNE will have a muon neutrino (or antineutrino) beam originating at Fermilab and detected at a distance of about 1300 km at the far detector in Stanford. The 40 kt far detector will be located underground at a depth of about one mile. The far detector consists of ultra-pure cryogenic liquid Argon, which acts as both target as well as a tracker for neutrino interactions using the advanced technology of the Time Projection Chamber (TPC). DUNE will also consist of near detectors to measure unoscillated neutrino flux and minimize the systematics. The $\nu_\mu$ and $\bar{\nu}_\mu$ beam will peak at the neutrino energy of about 2.5 GeV. The efficient detection of electron neutrinos provides access to the appearance channel $\nu_\mu \rightarrow \nu_e$, which opens up a unique avenue to probe CP violation in the neutrino sector. The matter effect developed for the baseline of 1300 km through Earth's crust is sufficient enough to measure neutrino mass ordering and octant of $\theta_{23}$ at DUNE. The atmospheric neutrinos can also be observed at DUNE, which will enhance the capability of DUNE to precisely measure the neutrino oscillation parameters. 
	
	\item \textbf{Hyper-K:} The Hyper-Kamiokande (Hyper-K)~\cite{Hyper-K:2021} is a proposed atmospheric neutrino experiment and the successor of Super-K based on the water-Cherenkov technology. With a fiducial mass of about 190 kt, Hyper-K is about 8 times larger in size compared to Super-K. Hyper-K will have a cylindrical water tank with a height of about 71 m and a diameter of about 68 m. Along with its larger size, Hyper-K will also be instrumented with PMTs having higher quantum efficiency compared to that in Super-K. Hyper-K is capable of detecting neutrinos from J-PARK beam, proton decays, solar neutrinos, atmospheric neutrinos, and astronomical neutrinos. The muon and electron neutrinos can be separately detected at Hyper-K by identifying clean and fuzzy rings, respectively. Though it is not possible to distinguish neutrinos and antineutrinos on an event-by-event basis at Hyper-K, large statistics and flavor identification capability can help Hyper-K to observe Earth's matter effects that may give rise to the sensitivity towards neutrino mass ordering and octant of $\theta_{23}$. The large statistics is also going to improve the precision measurement of neutrino oscillation parameters. 
	
	\item \textbf{T2HK:} The Tokai-to-Hyper-K (T2HK) is an upcoming long-baseline experiment that will consist of the J-PARC facility located at Tokai as the source of a neutrino beam and Hyper-K as a far detector at a distance of about 295 km. For T2HK, the J-PARC neutrino beam will be upgraded from 0.5 to 1.3 Mega watt. The higher intensity of the neutrino beam, along with the larger fiducial mass of the detector, would increase the accelerator neutrino event rate at Hyper-K by about 20 times compared to that in the case of T2K. In T2HK, the far detector will also be located at $2.5^\circ$ off-axis with respect to the neutrino beam such that the energy of the neutrino beam peaks at 0.6 GeV. T2HK would also consist of near detectors to minimize the systematic uncertainties. The large statistics of $\nu_e$ appearance events would enable T2HK to measure the leptonic CP phase with the best precision for all possible values of $\delta_{\rm CP}$. 
	
	\item \textbf{ORCA:} The Oscillation Research with Cosmics in the Abyss (ORCA)~\cite{KM3Net:2016zxf} is an upcoming atmospheric neutrino experiment under KM3NeT Consortium. ORCA will be based on the principle of Cherenkov radiation emitted by secondary particles produced in the neutrino interactions inside seawater. ORCA is going to have two detectors modules in the Mediterranean sea at a depth of about 2450 m. ORCA will consist of 115 strings covering a total volume of about 6 MegaTonnes. Each string will consist of 18 optical modules, where each optical module will be made up of 31 PMTs. ORCA will be sensitive to the atmospheric neutrinos in the energy range of about 1 GeV to 100 GeV. The aim of ORCA is to determine neutrino mass ordering, precision measurement of neutrino oscillation parameters, and neutrino astronomy in the GeV range of energy. In a prototype of ORCA, six strings have already been deployed with a horizontal spacing of about 20 m~\cite{ORCA:2021}. This prototype has started observing strong oscillation signals. 
	
	\item \textbf{IceCube Upgrade:} The IceCube Upgrade~\cite{Ishihara:2019aao} is the next stage of the IceCube neutrino observatory for lower energy neutrino events. The IceCube Upgrade will have seven new strings with about 700 optical sensors around the bottom center of the IceCube detector at depths between 2150 m and 2425 m. The horizontal and vertical spacing will be about 20 m and 3 m, respectively. There will be two types of optical modules which are the Multi-PMT Digital Optical Module (mDOM) and the Dual optical sensors in an Ellipsoid Glass (D-Egg). These designs are optimized to improve photon detection efficiency and the calibration capability of the whole IceCube detector. This improved efficiency will help IceCube Upgrade to achieve the energy threshold of about 1 GeV. Upgrade will also give rise to the sensitivity towards the tau neutrinos from $\nu_\mu \rightarrow \nu_\tau$ appearance channel with high precision. With the Upgrade, the neutrino oscillation analysis is going to get improved significantly, and oscillation parameters will be measured more precisely~\cite{Ishihara:2019aao}. In the light of better calibration, the archived data collected over 10 years can be reanalyzed to enhance IceCube's sensitivity towards high-energy cosmic neutrinos.
	
	\item \textbf{ICAL}: The 50 kt Iron Calorimeter (ICAL) detector at the proposed India-based Neutrino Observatory (INO)~\cite{ICAL:2015stm} would consist of stacks of iron layers having a magnetic field of about 1.5 T. The detector will be placed under the mountain with a rock coverage of at least 1 km from all directions. The magnetic field enables ICAL to detect atmospheric muon neutrinos and antineutrinos separately in the multi-GeV energy range over a wide range of baselines. ICAL aims to measure neutrino mass ordering by observing the difference in Earth's matter effects experienced by neutrinos and antineutrinos. ICAL would also contribute to the precision measurement of atmospheric neutrino oscillation parameters $\Delta m^2_{32}$ and $\theta_{23}$. We describe the ICAL detector in detail in chapter~\ref{chap:ICAL}.
\end{itemize}

\section{Summary}
\label{sec:prob_conclusion}
Neutrino oscillations were discovered by the Super-Kamiokande experiment using atmospheric neutrino data and later confirmed by solar, acceleration, and reactor neutrino experiments also. Neutrino oscillations are explained by the phenomena of quantum interference, where neutrino flavor eigenstates are taken as a superposition of mass eigenstates. As a consequence of this, neutrinos change their flavor during propagation. The probabilities of neutrino oscillations during propagation in vacuum are parameterized in terms of three mixing angles ($\theta_{12}$, $\theta_{23}$, and $\theta_{13}$), CP-phase ($\delta_{CP}$), and two mass-squared differences ($\Delta m^2_{32}$ and $\Delta m^2_{21}$). Over the past two decades, most of the neutrino oscillation parameters have been measured precisely by different neutrino oscillation experiments (see Sec.~\ref{sec:osc_exp}). However, the CP-phase $\delta_{CP}$ and the mixing angle $\theta_{23}$ are still uncertain by large amounts. We have summarized the current status of neutrino oscillation parameters in Table~\ref{tab:global_fit}.

The neutrino oscillation probabilities in vacuum depend on the energy ($E_\nu$) and path-length ($L_\nu$) of neutrinos in the form of $L_\nu/E_\nu$. However, neutrinos passing through Earth undergo charged-current coherent forward elastic scattering with the ambient electrons giving rise to a matter potential that modifies the neutrino oscillation patterns. These density-dependent matter effects introduce additional dependence of oscillation probabilities on baseline $L_\nu$, which is not in the form of $L/E$. Matter effects are potential tools to solve important issues such as the determination of neutrino mass ordering and identification of correct octact of $\theta_{23}$. The next-generation neutrino oscillation experiments like DUNE, Hyper-K, ORCA, IceCube Upgrade, and ICAL are going to accumulate a huge amount of data that would be used to solve the present issues in neutrino oscillation parameters and would pave the way for practical applications such as tomography of Earth using neutrinos.

\end{refsegment}

\cleartooddpage
\chapter{ICAL Detector at INO}
\label{chap:ICAL}
\begin{refsegment}

The proposed India-based Neutrino Observatory (INO)~\cite{ICAL:2015stm} is a mega-science project being pursued by a collaboration of many institutes and universities with the aim of building a world-class underground observatory. INO is planned to be built at Pottipuram in Bodi West hills of Theni District of Tamil Nadu, India. INO would host three experiments which are Iron Calorimeter detector (ICAL), Neutrino-less double decay (NDBD)~\cite{NDBD}, and Dark-matter@INO (DINO)~\cite{DINO}. INO will consist of a large underground cavern of size 132 m $\times$ 26 m $\times$ 20 m along with many other small caverns under approximately 1200 m high mountain such that the observatory will have a rock coverage of at least 1 km from all direction. This rock coverage will filter out the unwanted particles like cosmic muons by 1 million times to provide a low background environment for the experiments of INO. The underground cavern can be accessed by a 2100 m long and 7.5 m wide horizontal tunnel. 

In this thesis, we focus on the ICAL detector, which we discuss in the coming sections. In Sec.~\ref{sec:ICAL_geometery}, we talk about ICAL geometry, Resistive Plate Chambers, and magnetic field. The interactions of neutrinos with iron layers at ICAL are described in Sec.~\ref{sec:nu_interactions}. In Sec.~\ref{sec:event_reco}, we discuss event reconstruction at ICAL, where we talk about the detector response for muons and hadrons. The details about the simulation of reconstructed events at ICAL are given in Sec.~\ref{sec:event_simulation}. Next, we present the impact of atmospheric oscillation parameters on reconstructed events at ICAL in Sec.~\ref{sec:impact_osc_par_events}. Finally, we summarize in Sec.~\ref{sec:ICAL_conclusion}

\section{Iron Calorimeter Detector}
\label{sec:ICAL_geometery}
The 50 kt magnetized ICAL detector at INO~\cite{ICAL:2015stm} aims to detect atmospheric muon neutrinos and antineutrinos separately in the multi-GeV energy range over a wide range of baselines starting from about 15 km to 12750 km. The magnetic field of about 1.5 T~\cite{Behera:2014zca} enables ICAL to identify the charge of muons which are produced during the interactions of neutrinos inside the detector. This charge identification  (CID) capability of ICAL is used to identify neutrinos and antineutrinos separately in mulit-GeV energy range. 

The upward-going neutrinos pass through the Earth and experience matter effects in neutrino oscillations. The matter effects modify the oscillation patterns differently for neutrinos and antineutrinos. The CID capability helps in preserving this information and enables ICAL to determine the neutrino mass ordering which is the main goal of ICAL~\cite{Ghosh:2012px,Devi:2014yaa}. ICAL would also help in improving the precision of atmospheric neutrino oscillation parameters $\Delta m^2_{32}$ and $\sin^2\theta_{23}$~\cite{Thakore:2013xqa,Devi:2014yaa}. The ICAL collaboration has also studied how ICAL can play an important role in probing various BSM physics scenarios~\cite{Khatun:2019tad,Kumar:2021lrn,Thakore:2018lgn,Dash:2014fba,Behera:2016kwr,Khatun:2017adx,Choubey:2017eyg,Khatun:2018lzs,Choubey:2017vpr,Tiwari:2018gxz,Sahoo:2021dit,Upadhyay:2021kzf,Sahoo:2022rns}. The ICAL detector can also be used to probe the internal stucture of Earth as we describe in chapter~\ref{chap:tomography} of this thesis. 

\begin{figure}
	\centering
	\hspace{1.5cm}\includegraphics[width=0.7\linewidth]{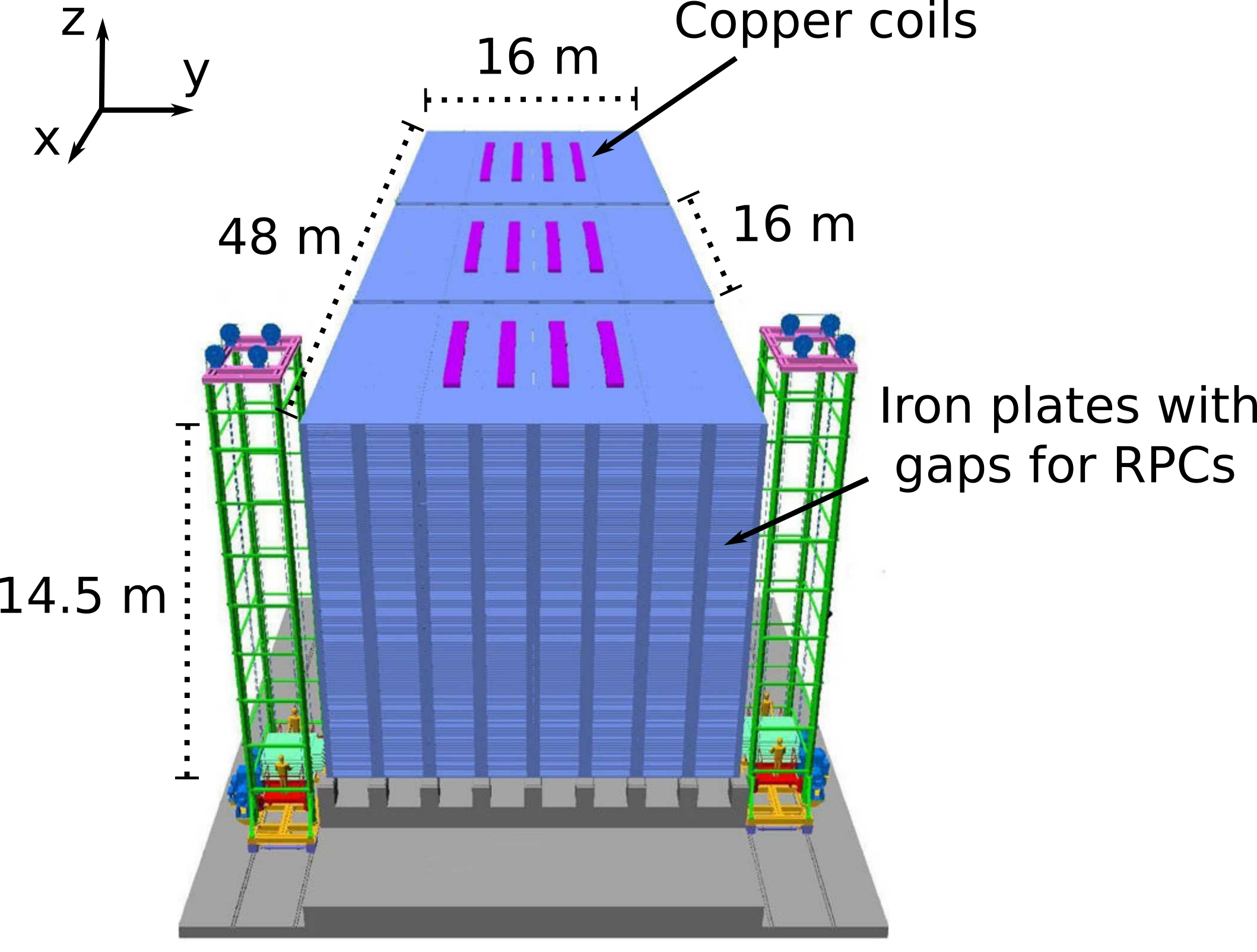}
	\caption{A schematic diagram of the ICAL detector at INO~\cite{INO}.}
	\label{fig:icalann}
\end{figure}

ICAL would consist of three modules, each of which has a size of 16 m $\times$ 16 m $\times$ 14.5 m as shown in Fig.~\ref{fig:icalann}. Each module contains a stack of 151 iron layers having a thickness of 5.6 cm~\cite{ICAL:2015stm}. The iron acts as a passive detector element that provides a target for neutrino interactions. The Resistive Plate Chambers (RPCs) of dimensions 2 m $\times$ 2 m are active detector elements that are sandwiched between the consecutive iron layers with a gap of about 4 cm. Whenever a charged particle passes through an RPC, it produces a signal in the form of hits with (x,y) coordinates. By looking at the pattern of hits, we can figure out the type of events, whether it is ``track-like'' to ``shower-like''. Now, we describe the components of ICAL in detail.

\subsection{Resistive Plate Chamber as Active Detector Element}
\label{sec:RPC}

\begin{figure}
	\centering
	\includegraphics[width=0.7\linewidth]{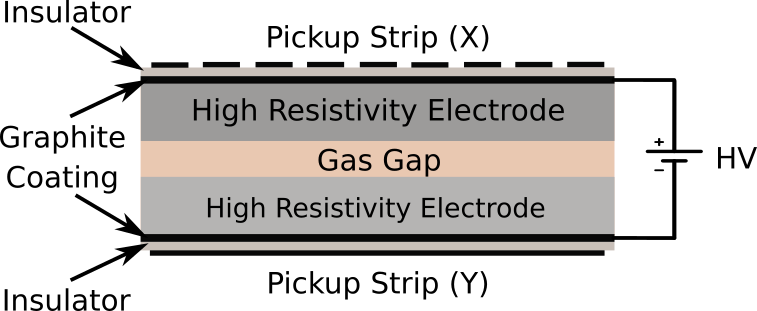}
	\caption{The schematic of Resistive Plate Chamber (RPC)~\cite{Kumar:2021dka}.}
	\label{fig:rpc_schematic}
\end{figure}

RPC~\cite{Santonico:1981,Santonico:1988qi,Knoll,Bheesette:2009yrp,Bhuyan:2012zzc}  is a gaseous detector used in high energy physics to detect charged particles like muons. Figure~\ref{fig:rpc_schematic} illustrates the schematic diagram of an RPC. Inside the RPC, a gas mixture is confined between two parallel plates of high resistivity (typically $10^9$ and $10^{13}$ $\Omega\,\text{cm}$~\cite{Knoll}) material like glass or bakelite. In RPC design for the ICAL detector, the gap between the parallel glass plate of 3 mm thickness is maintained at 2 mm with the help of button spacers. The walls of the chamber are closed with side spacers and path for gas flow is provided in the corner spacers.  Since the high voltage cannot be applied on the resistive plate of glass or bakelite directly, a coating of conductive layer like graphite is applied on the outer surface of the resistive plate. A positive potential of about 5 kV is applied on one surface of the graphite layer and negative potential of about - 5 kV on another surface. The voltage on the graphite layer spreads uniformly due to moderate resistivity of around 1 M$\Omega$ and a uniform electric field of about 50 kV/cm (i.e., 10 kV potential in 0.2 cm gap) is established inside the chamber. The surface resistivitiy of graphite layer should be low enough to allow the induction of signal on the pickup strip.

%=====================================
%=====================================

A charged particle interacts with the gaseous medium inside the RPC and ionizes gas molecules into electron-ion pairs. The electrons and ions experience an electric force in the opposite direction due to the applied electric field. The presence of the electric field inside the gas gap prevents the recombination of electrons and ions as well as separates them by accelerating in the opposite direction. The electrons are accelerated towards the positive anode whereas ions move towards the negative cathode. The electrons are accelerated to such an energy that they can further ionize other gas molecules and result in secondary electrons. These secondary electrons also gain energy from the electric field and ionize more molecules. This process of increase in electron number due to ionization is called avalanche. The positive ions have larger mass due to which they cannot be accelerated to high velocities and hence, they move slowly. 

A readout consists of a pickup panel made up of copper strips that are used to collect signals from RPC by the principle of induction. The pickup panel consists of 64 strips with 2.8 cm width and a gap of 0.2 cm between two neighboring strips such that the pitch is 3 cm~\cite{Singh:2016tzf}. The readout is isolated from graphite layer using a sheet of insulating material like Mylar.  The motion of electrons inside the gas gap induces a potential on the pickup strip following Shockley-Ramo's theorem~\cite{Shockley:1938,Ramo:1939} which is given as input to read-out electronics. Two pickup panels are placed orthogonal to each other such that their strips are aligned in the X and Y directions, respectively. While passing through the RPC, a charged particle leaves a hit in the form of (X,Y) coordinates. The time resolution of RPC is about 1 ns~\cite{Dash:2014ifa,Bhatt:2016rek,Gaur:2017uaf} which enables ICAL to efficiently distinguish between the upward-going and downward-going reconstructed muon events. When the electrons reach the inner surface of the resistive plate, the induced signal drops to zero. The charge pileup results in a decrease in the electric field in the region of avalanche which remains dead until the charge gets neutralized. This induced charge remain localized and the other regions of the detector function normally during this time. The accumulated charge on the inward surface of RPC discharges through the bakelite and graphite. The dead time depends on the resistivities of bakelite and graphite.

%=====================================
\subsubsection{Composition of Gas Mixture in RPCs at ICAL}
\label{sec:gas_mixture}
%=====================================

The composition of the gas mixture depends on several factors such as high gain, low working voltage, good proportionality, and high rate capability. Consider $n_0$ to be the number of primary electrons produced when a charged particle passes through the detector. We denote the distance between the location of production of primary electrons and the anode by $x$. Then, the number of electrons arriving at the anode can be given as follows~\cite{Leo}
\begin{equation}
n = n_0 \exp{( (\alpha - \beta) x)},
\end{equation}
where $\alpha$ is known as the first Townsend coefficient, which is decided by the number of ionization per unit length. On the other hand, $\beta$ is the attachment coefficient which is decided by the number of electrons captured per unit length.

Noble gases are suitable for the low working voltage because of low ionization potential, which in turn decides the first Townsend coefficient. At ICAL, we plan to use Freon (R134a, C\textsubscript{2}F\textsubscript{4}H\textsubscript{2}) to achieve low ionization potential. The primary electrons are produced as electron-ion pairs having a tendency to recombine with the emission of ultraviolet photons. These energetic photons are capable of creating more electron-ion pairs at other locations inside the detector, which can result in numerous avalanches through the chamber, destroying the spatial resolution. These photons can be absorbed by a polyatomic gas, which is usually hydrocarbon gas. In the case of ICAL, we plan to use Isobutane (C\textsubscript{4}H\textsubscript{10}). Since the photon-absorbing gas localizes the avalanche, it is also known as quenching gas.

The excess electrons can be absorbed by an electronegative gas having a high electron affinity. At ICAL, we use SF\textsubscript{6} for absorbing excess electrons. The gas mixture in RPCs at ICAL contains Freon, Isobutane, and SF\textsubscript{6} gasses in the ratio of 95.2 : 4.5 : 0.3, respectively~\cite{Bhuyan:2012zzc,Bhuyan:2014ena,Gaur:2017osj}.

\subsubsection{Mode of Operation of RPCs}
The RPCs can be operated in avalanche mode or streamer mode depending upon the gas mixture and the applied voltage.
\begin{itemize}
	\item \textbf{Avalanche mode:} The RPC operates in avalanche mode at relatively lower voltage and the size of avalanche is limited to a small region. The produced signals have smaller amplitudes (order of mV) and requires pre-amplifier to process the signals. The smaller size ($\sim$1 pC) pulses decrease the electric field in a small area of RPC and thus, have better rate capability. This mode results in a longer lifetime of RPC. At ICAL, we operate RPCs in avalanche mode which can enable the RPCs to operate for decades with less maintenance. 
	
	\item \textbf{Streamer mode:} The RPC operates in streamer mode on application of higher voltage and produce signal with larger amplitude and hence, don't need any pre-amplifiers. Due to larger signals ($\sim$100 pC) the electric field drops for large area and the rate capability is lower (few hundred Hz/cm\textsuperscript{2}) in this mode. Also, the lifetime of RPC decreases in streamer mode.  
\end{itemize}

\subsection{Magnetized Iron Calorimeter Detector as Passive Detector Element}
\label{sec:ICAL_magnet}
The 50 kt magnetized ICAL detector will consists of three modules such that the x-direction is defined as the direction along which the three modules are placed whereas the direction perpendicular to this in horizontal plane is labeled as the y-direction. The z-axis points in the upward direction as shown in Fig.~\ref{fig:icalann}. The center of central module is considered as the origin. Since it is difficult to handle a single iron plate of size 16 m $\times$ 16 m $\times$ $5.6$ cm, plates of smaller sizes of 2 m $\times$ 4 m $\times$ 5.6 cm with a gap of 2 mm will be used to achieve the desired size. In order to identify the charge of muon, ICAL will be magnetized to a magnetic field of about 1.5 T in the y-direction.

ICAL collaboration has performed the detailed simulation study for magnetized ICAL detector in Ref.~\cite{Behera:2014zca}. In this work, various aspects of ICAL magnet are explored and its design was optimized. It was shown that a configuration with continuous slots for current carrying coils provides a more uniform magnetic field than those with discrete slots.  In this configuration, each module contains the two parallel slots at $x = \pm 4$ m and spanning in the y-direction with $-4 \text{ m} \leq y \leq 4$ m. There will be four sets of coils, each of which carries a current of about 5 kA such that a uniform magnetic field higher than 1 T is achieved in y-direction over an area of about 75\%. 

It was also observed that the magnetic field increases with the thickness of iron plate and gets saturated for thickness more than 4 cm, which justifies the choice of 5.6 cm. The higher plate thickness results in the increase in multiple scattering of muons which deteriorates the energy resolution. The gap between two layers placed side by side reduces the magnetic field due to leakage, thus, this gap should be minimum and a gap of 2 mm may be practical. More details about the dependence of magnetic field on various design aspects of ICAL are described in Ref.~\cite{Behera:2014zca}. Now, we describe the interactions of neutrino with the iron nuclei. 

%=====================================
\section{Neutrino Interactions at ICAL}
\label{sec:nu_interactions}
%=====================================

The ICAL detector is sensitive to neutrinos in the energy range of 1 to 25 GeV. In this multi-GeV energy range, the charged-current (CC) neutrino interactions occur via quasi-elastic scattering (QES), resonance scattering, and deep-inelastic scattering (DIS). Apart from this, neutrinos can also interact via neutral-current (NC) interactions.  

The quasi-elastic scattering dominates at neutrinos energies below 2 GeV. In charged-current QES, neutrinos interact with the entire nucleon and produce charged leptons. In the case of QES for neutrino, neutron converts into a proton
\begin{equation}
\nu_\mu + n \rightarrow \mu^- + p.
\end{equation}
On the other hand, QES for antineutrino causes conversion of proton into neutron
\begin{equation}
\bar{\nu}_\mu + p \rightarrow \mu^+ + n.
\end{equation}
In the elastic scattering, a neutrino can scatter from the entire nucleon via neutral-current interaction where neutrino itself is a final state particle. 

The resonance scattering is an inelastic scattering where a neutrino with sufficient energy interacts with nucleon and excites it to an excited state called baryon resonance. The decay of baryon resonance mostly produces a nucleon and a single pion in the final state. In resonance scattering, the production of multiple pions is also possible. The resonance scattering can occur via both CC as well as NC interactions. 

As the energy increases, neutrino starts probing the internal structure of the nucleon. In deep-inelastic scattering, a high-energy neutrino interacts with an individual quark producing a lepton and hadrons in the final state. In DIS, hadrons carry away a significant fraction of incoming neutrino energy. Both CC and NC processes are possible in DIS. Now, we describe how these interactions can be observed at ICAL. 

%=====================================
\section{Event Reconstruction at ICAL}
\label{sec:event_reco}
%=====================================

In the previous section, we discussed that the CC interactions of neutrinos with nucleons in iron result in the production of charged muons. While passing through RPC, the resulting muon deposits energy in the gaseous region with the induction of signals in the perpendicular pickup strips in X and Y directions that provide (x,\;y) coordinate of the hit. The z-coordinate of the hit is obtained from the layer number of RPC. Since the multi-GeV muon is a minimum ionizing particle, it passes through several layers and leaves hits in those layers in the form of a long track. These muon events with long tracks are termed ``track-like'' events. In the presence of the magnetic field, the muon tracks curve in the opposite directions for $\mu^-$ and $\mu^+$. Hence, the charge of muon can be identified by the direction of bending of track in the magnetic field, which results in the ability of ICAL to distinguish between atmospheric neutrinos and antineutrinos in the multi-GeV energy range. 

In Sec.~\ref{sec:nu_interactions}, we describe that the neutrino interactions are also contributed by resonance scattering and deep-inelastic scattering at multi-GeV energies resulting in the production of hadrons. At ICAL, hadrons leave multiple hits in a plane which result in the ``shower-like'' events. At ICAL, the hit pattern is almost the same for different hadrons. Hence, the particle identification of hadron is difficult. 

To understand the detector response for muons and hadrons, the ICAL collaboration has performed a rigorous detector simulation study using the widely used GEANT4 package~\cite{GEANT4:2002zbu}. The details of these simulation studies performed by the ICAL collaboration are given in Refs.~\cite{Chatterjee:2014vta,Devi:2013wxa}. Now, we describe the detector response for muons and hadrons in the next two sections. 

%=====================================
\subsection{Detector Response for Muons}
\label{sec:muon_response}
%===================================== 

The authors of Ref.~\cite{Chatterjee:2014vta} discuss in detail how various response functions for muons have been obtained by performing a rigorous GEANT4-based simulation study by the ICAL collaboration. To simulate the detector response, a huge number of muons are passed through the ICAL detector. The muon leaves hits in various RPC layers in the form of a track. The reconstruction algorithm fits the track using the Kalman filter technique and calculates the vertex, direction ($\cos\theta_\mu^\text{rec}$), energy ($E_\mu^\text{rec}$), and charge of the muon. Note that in Ref.~\cite{Chatterjee:2014vta}, $\cos\theta = 1$ corresponds to the upward-going events. We also use this convention in the present section to describe the muon response. In the remaining part of this thesis,  $\cos\theta = 1$ represents downward-going events. The ICAL reconstruction algorithm requires about a minimum of 8 to 10 hits to reconstruct the muon track, which translates to the energy threshold\footnote{The ICAL detector consists of a stack of iron layers with a thickness of 5.6 cm each having a gap of 4 cm between two successive iron layers to insert active RPCs. In the multi-GeV energy range, muon is a minimum ionizing particle and it deposits energy inside a medium at the rate of about $(1/\rho)\cdot (dE/dx) \sim 2 \text{ MeV\,g}^{-1}\text{\,cm}^2$ as described in the PDG~\cite{ParticleDataGroup:2020ssz}. For the case of iron ($\rho \sim 7.9$ g/cm\textsuperscript{3}), the muon in the GeV energy range will deposit energy of about 16 MeV/cm, and this will lead to an energy loss of about 100 MeV in each layer of iron (thickness of 5.6 cm) in ICAL. We need about a minimum of 8 to 10 hits to reconstruct a muon track at ICAL, which results in an energy threshold of about 1 GeV.} of about 1 GeV. The outcomes of migration matrices are nicely summarized as a function of input muon momentum for various input zenith angles in Ref.~\cite{Chatterjee:2014vta}. The authors in Ref.~\cite{Chatterjee:2014vta} show the reconstruction efficiency in Fig.~13, charge identification (CID) efficiency in Fig.~14, muon energy resolution in Fig.~11, and muon angular resolution in Fig.~6. 

The reconstruction efficiency increases sharply with the input muon energy up to 2 GeV, and then it saturates to around  80\% to 90\% in the muon energy range of 2 to 20 GeV for a wide range of zenith angle starting from $\cos\theta_\mu = 0.35$ to 0.85 as shown in Fig.~13 of  Ref.~\cite{Chatterjee:2014vta}. In ICAL, the number of events are less in the horizontal direction because the reconstruction efficiency is poor in this case due to the horizontally placed layers of RPC where only a few RPC layers receive hits in case of horizontal events. As far as the charge identification is concerned, the ICAL detector is expected to perform quite well in the muon energy range of 1 to 20 GeV since it plans to have the magnetic field of around 1.5 T which will be sufficient enough to get the curvature of the muon track to identify the charge of the muon. The charge identification efficiency at ICAL is about 98\% in the muon energy range of 2 to 20 GeV for various zenith angles in the range of $\cos\theta_\mu = 0.35$ to 0.85 as shown in Fig.~14 of Ref.~\cite{Chatterjee:2014vta}. 

The reconstructed muon energies and directions are fitted with Gaussian distribution to calculate means and standard deviations. The standard deviation represents the detector resolution of the reconstructed parameter. Figure~11 in Ref.~\cite{Chatterjee:2014vta} portrays that the muon energy resolution of the ICAL detector in the muon energy range of 2 to 20 GeV for zenith angles in the range of $\cos\theta_\mu = 0.35$ to 0.85 is approximately 10 to 15\%, which is sufficient enough to capture the information about neutrino oscillation parameters and Earth's matter effects in the multi-GeV energy range for a wide range of baselines. The ICAL detector has an excellent angular resolution of less than $1^\circ$ for a large muon energy range and a wide range of zenith angles, as shown in Fig.~6 of Ref.~\cite{Chatterjee:2014vta}. These numbers tell us that the ICAL detector performs quite well as far as the reconstruction of the four-momenta of muon is concerned, which is important to have the sensitivity of ICAL towards neutrino oscillations and Earth's matter effect.

%=====================================
\subsection{Detector Response for Hadrons}
\label{sec:had_response}
%=====================================

Now, let us elaborate on how the ICAL collaboration obtains the hadron energy response inside the ICAL detector. In the multi-GeV range of energies, the resonance scatterings and deep-inelastic scatterings of neutrinos produce hadrons along with muons. Unlike muons, hadrons produce multiple hits in a single layer of RPC, and this leads to shower-like events. These hadrons take away a significant fraction of the incoming neutrino energy, and the hadron energy deposited in the detector is defined using a variable ${E'}_\text{had} = E_\nu - E_\mu$ in Ref~\cite{Devi:2013wxa}. This Reference discusses in detail how the hadron energy resolution has been obtained by performing a rigorous GEANT4-based simulation study by the ICAL collaboration. The hadron energy response is simulated by passing a huge number of hadrons through ICAL geometry. The distribution of the total number of hits for these hadrons is fitted with the Vavilov distribution function. The mean number of hits and the square root of the variance obtained after fitting is related to the energy and the energy resolution of hadron, respectively. Figure~8 in Ref.~\cite{Devi:2013wxa} shows that the hadron energy resolution is about 40 to 60\% in the energy range of 2 to 8 GeV and about 40\% for energies above 8 GeV. Though the hadron energy resolution is not as good as muon, it is sufficient to capture the possible correlation between four-momenta of muon ($E_\mu^\text{rec}$, $\cos\theta_\mu^\text{rec}$) and hadron energy (${E'}_\text{had}^\text{rec}$) which we treat as independent variables.

%=====================================
\section{Event Simulation at ICAL}
\label{sec:event_simulation}
%=====================================

Now, we describe how we simulate the neutrino events at the ICAL detector.

%=====================================
\subsection{NUANCE Simulation with Unoscillated Atmospheric Neutrino Flux }
\label{sec:neutrino_flux}
%=====================================

In this work, the neutrino interactions in the ICAL detector are simulated using Monte Carlo (MC) neutrino event generator NUANCE~\cite{Casper:2002sd}. The ICAL geometry containing the material information is given as a target for NUANCE. We use the Honda 3D flux~\cite{SajjadAthar:2012dji,Honda:2015fha} of atmospheric neutrinos at the proposed INO site at Theni district of Tamil Nadu, India. In this context, it is worthwhile to mention that the earlier studies of atmospheric neutrino oscillations~\cite{ICAL:2015stm} for the ICAL detector have been performed using the neutrino fluxes calculated for the Kamioka site. The neutrino fluxes vary from place to place due to the different geomagnetic fields of the Earth, and there are important differences in the fluxes at Kamioka and at the proposed site of INO at Theni district of Tamil Nadu, India. Theni is close to the region having the largest horizontal component ($\approx 40 ~\mu T$) of the Earth's magnetic field. In comparison to this, the horizontal component of the geomagnetic field at Kamioka is $\approx 30 ~ \mu T$. As a result, the neutrino flux at the INO site is smaller by approximately a factor of 3 at low energies. However, for neutrino energies around 10 GeV, the fluxes at INO and Kamioka sites are almost the same~\cite{Honda:2015fha}. In this study, we use the neutrino fluxes given in~\cite{SajjadAthar:2012dji,Honda:2015fha}, taking into account the 1 km rock coverage of the mountain. The rock coverage of about 1 km (3800 m water equivalent) is expected to reduce the downward-going cosmic muon background by $\sim 10^6$~\cite{Dash:2015blu}, most of the remaining events will be further vetoed by employing the fiducial volume cut. Also, to take into account the effect of solar modulation on the neutrino fluxes, we use the fluxes with high solar activity (solar maximum) for half the exposure and the fluxes with low solar activity (solar minimum) for the remaining half. 

\begin{figure}
	\centering
	\includegraphics[width=0.49\linewidth]{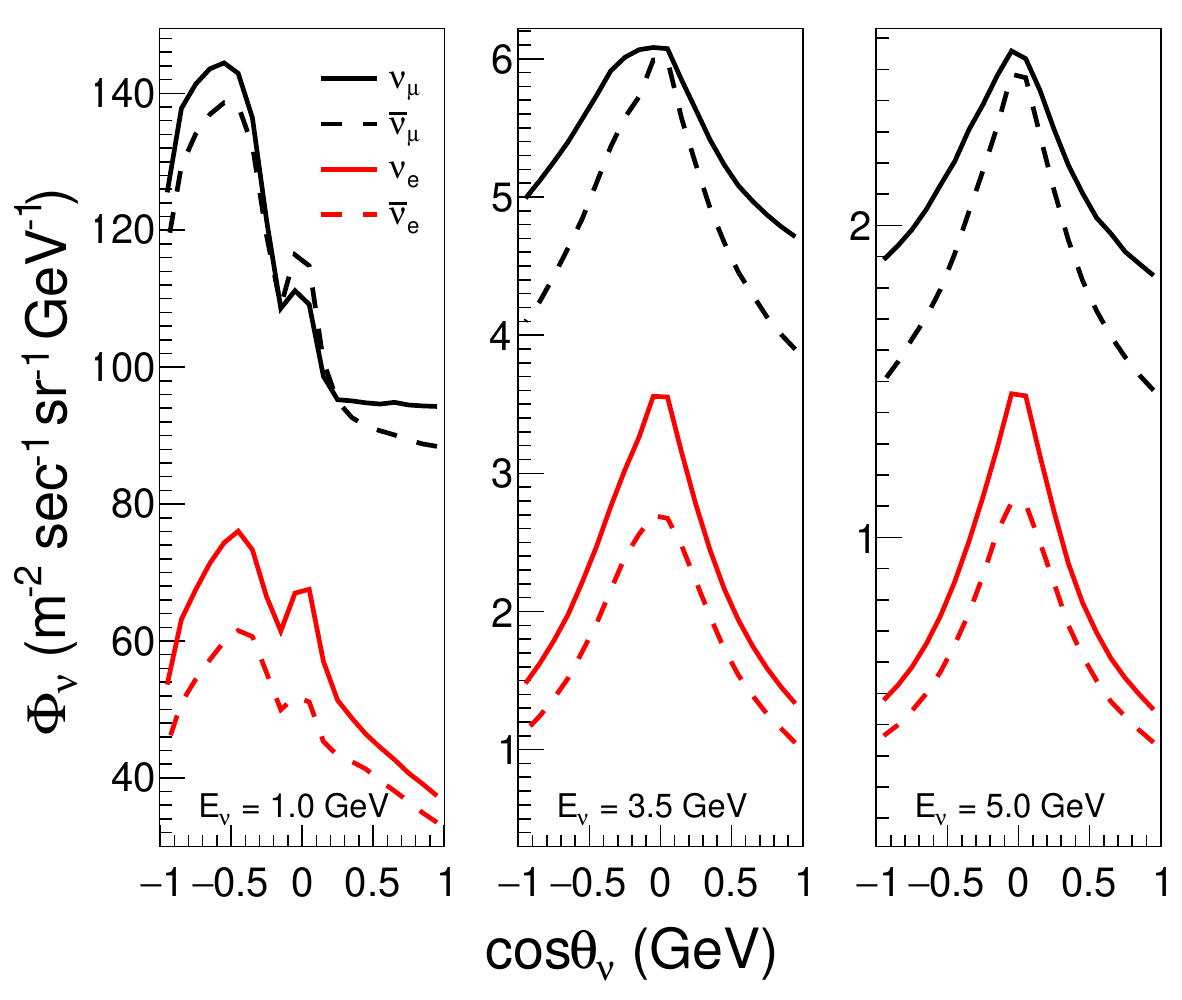}
	\includegraphics[width=0.49\linewidth]{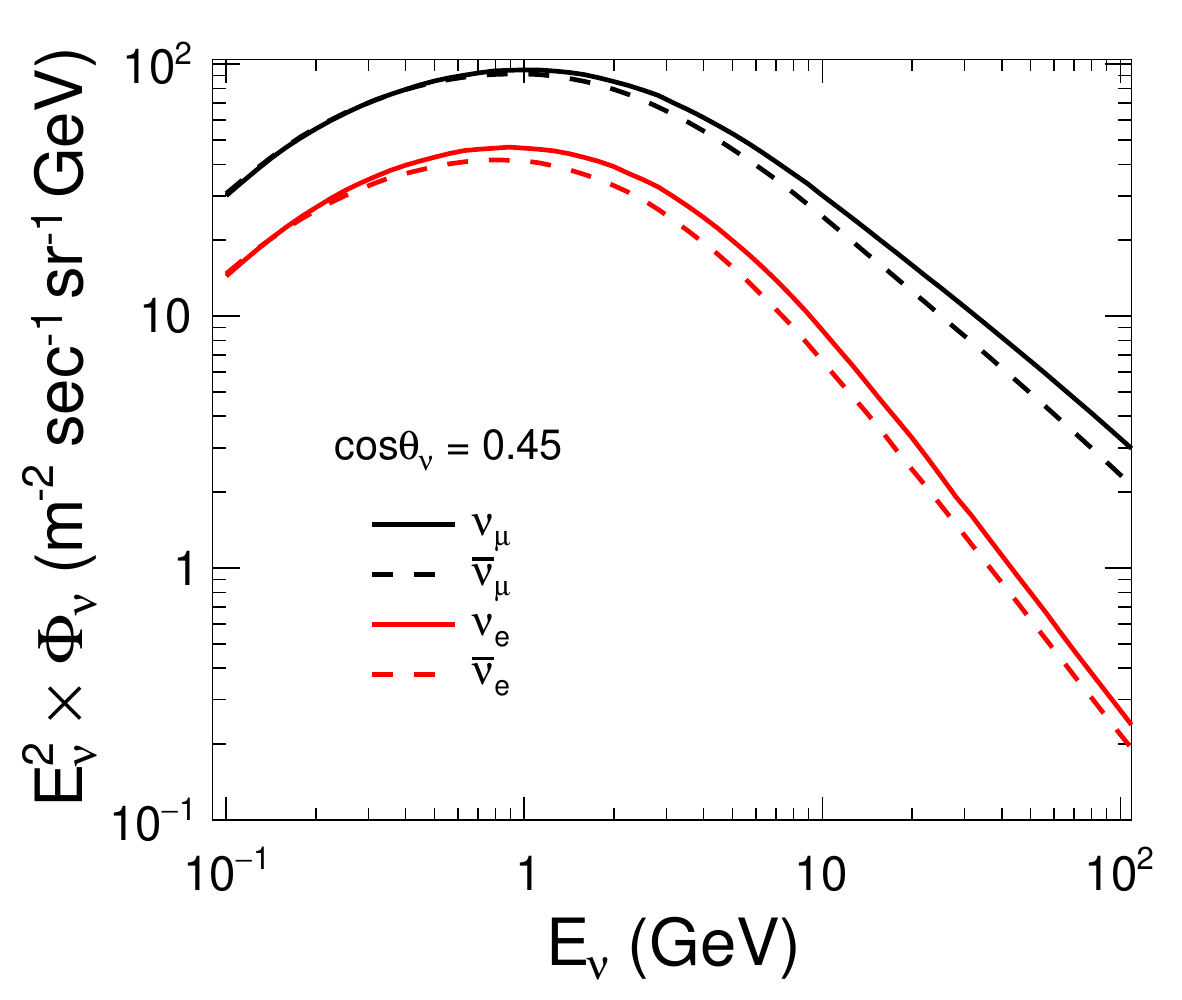}
	\caption{The unoscillated neutrino flux at INO site as a function of $\cos\theta_\nu$ for three different energies of 1 GeV, 3.5 GeV, and 5 GeV in the left panel. Neutrino flux (multiplied with $E_\nu^2$) as function of neutrino energy $E_\nu$ in the right panel. The black and red solid (dashed) curves stand for $\nu_\mu$ ($\bar{\nu}_\mu$) and $\nu_e$ ($\bar{\nu}_e$), respectively. Here, we show the Honda 3D flux~\cite{SajjadAthar:2012dji,Honda:2015fha} for INO site with mountain for solar minima. }
	\label{fig:ino_flux}
\end{figure}

In Fig.~\ref{fig:ino_flux}, we show the neutrino flux at INO site as a function of neutrino direction $\cos\theta_\nu$ (left panel) and energy $E_\nu$ (right panel). In both the panels, black and red solid (dashed) curves denote $\nu_\mu$ ($\bar{\nu}_\mu$) and $\nu_e$ ($\bar{\nu}_e$), respectively. In the left panel, neutrino fluxes are shown for three different energies 1 GeV, 3.5 GeV, and 5 GeV. Note that the scale of the y-axis is different for these energies because the neutrino flux decreases with energy following a power law where neutrino flux depends on energy in the form of $E_\nu^{-2.7}$ in the multi-GeV range of energy. We can observe that neutrino fluxes are maximum for horizontal direction ($\cos\theta_\nu = 0$) because the probability of decay of cosmic muon is more for longer path-lengths available along the horizontal direction. For lower neutrino energy of around 1 GeV and 3.5 GeV, we observe that flux is not symmetric for upward ($\cos\theta_\nu < 0$) and downward ($\cos\theta_\nu > 0$) directions, this happens because of the geomagnetic field of Earth. At higher energies, this up-down asymmetry is not present, which can be seen for neutrino flux at 5 GeV in the left panel of Fig.~\ref{fig:ino_flux}. The flux for $\nu_\mu$ ($\bar{\nu}_\mu$) is larger than that for $\nu_e$ ($\bar{\nu}_e$) because muon neutrinos can be produced during decays of pions as well as muons whereas electron neutrinos are mainly produced during the decay of muons. Also, the flux for neutrino is more than that for antineutrino. The right panels in Fig.~\ref{fig:ino_flux} clearly show that the neutrino flux drops sharply with the increase in energies. Note that we have multiplied the flux by $E_\nu^2$ in the right panel of Fig.~\ref{fig:ino_flux}. Now, we describe how we incorporate the neutrino oscillations in our simulation. 

%=====================================
\subsection{Incorporating Neutrino Oscillations}
\label{sec:reweighting_algorithm}
%=====================================

In this work, we focus on the CC interactions of muon neutrinos and antineutrinos at the ICAL detector. The muon events at ICAL are contributed by both survival ($\nu_\mu \rightarrow \nu_\mu$) as well as appearance ($\nu_e \rightarrow \nu_\mu$) channels. For the case of NO, about 98\% of $\mu^-$ and 99\% of $\mu^+$ events at ICAL come from the survival channel, whereas the remaining contribution is from the appearance channel. The tau neutrino, $\nu_\tau$ may appear due to neutrino oscillation and produce tau lepton during interaction inside the detector. The muon events produced in the tau decay in the ICAL detector are small (around 2\% of total upward-going muons from $\nu_\mu$ interactions~\cite{Pal:2014tre}), and these muon events are mostly softer in energy and below the 1 GeV energy threshold of ICAL. In this work, we have not considered this small contribution. 

Since simulating the neutrino interactions using NUANCE is a computationally intensive task, simulation of the neutrino events for different oscillation parameters is practically unrealistic. Hence, we generate unoscillated neutrino events at ICAL, and the three-flavor neutrino oscillations in the presence of matter are taken into account using a reweighting algorithm as described in Refs.~\cite{Ghosh:2012px,Thakore:2013xqa}. In the first two analyses described in chapters~\ref{chap:dip_valley} and \ref{chap:NSI}, we used the Preliminary Reference Earth Model (PREM) profile~\cite{Dziewonski:1981xy} while considering Earth matter effect in neutrino oscillations. However, while probing the structure of Earth in chapter~\ref{chap:tomography}, we use different density profiles. To minimize the statistical fluctuations of NUANCE, we generate 1000-year MC unoscillated neutrino events at ICAL, which is scaled to 10-year MC after incorporating the neutrino oscillations. Now, we describe how we take into account the detector responses in our simulation.

%=====================================
\subsection{Folding with Detector Responses}
\label{sec:detector_response}
%=====================================

NUANCE gives output in term of true energy ($E_\mu$) and directions ($\cos\theta_\mu$) of resulting muons. NUANCE also provides information about the type and momenta of various hadrons produced during the neutrino interactions. Since the ICAL detector is unable to distinguish the type of hadrons, we do not differentiate between the types of hadrons in our analysis; instead, take the difference between energies of neutrino ($E_\nu$) and muon ($E_\mu$) as the energy of hadron shower (${E}'_\text{had} \equiv E_\nu - E_\mu$). The events are folded with the detector responses for muons~\cite{Chatterjee:2014vta} and hadrons~\cite{Devi:2013wxa} following the procedure mentioned in Refs.~\cite{Ghosh:2012px,Thakore:2013xqa,Devi:2014yaa}. 
The detector responses for muons and hadrons are provided by the ICAL collaboration in terms of the migration matrices or lookup tables which are obtained by performing detailed GEANT4 simulation studies for the ICAL detector as described in sections~\ref{sec:muon_response} and \ref{sec:had_response}. The reconstructed muon energy ($E_\mu^\text{rec}$), muon direction ($\cos\theta_\mu^\text{rec}$), and hadron energy (${E'}_\text{had}^\text{rec}$) are considered to be the observable quantities. Note that we do not reconstruct neutrino energy or its direction. Since the hadron energy response is poor, the addition of energies of muon and hadron results in the deterioration of ICAL sensitivities, and we are not able to take advantage of excellent muon energy response. Hence, we treat hadron energy ${E'}_\text{had}^\text{rec}$ as a separate observable which has been observed to improve the ICAL sensitivities as demonstrated in Refs.~\cite{Devi:2014yaa,Khatun:2019tad}. Now, we present the reconstructed muon events expected at the ICAL detector. 

%=====================================
\subsection{Reconstructed Events Expected at ICAL}
\label{sec:reconstructed_events}
%=====================================

%--------------------------------------------------
\begin{table}[]
	\begin{center}
		\begin{tabular}{|l|c|c|}
			\hline
			& $\mu^-$ events & $\mu^+$ events \cr
			\hline
			U & 1654 & 740 \cr
			\hline
			D & 2960 & 1313 \cr
			\hline
			Total & 4614 & 2053\cr
			\hline
		\end{tabular}
		\caption{The total number of upward-going (U) and downward-going (D) reconstructed $\mu^-$ and $\mu^+$ events expected at the 50 kt ICAL detector in 10 years (total exposure of 500 kt$\cdot$yr)~\cite{Kumar:2020wgz}. We use the benchmark values of oscillation parameters given in Table~\ref{tab:osc-param-value}.}
		\label{tab:total_events}
	\end{center}
\end{table}
%------------------------------------------------------------------

Table~\ref{tab:total_events} presents the total number of reconstructed $\mu^-$ and $\mu^+$ events at 50 kt ICAL detector in 10 years which is equivalent to 500 kt$\cdot$yr exposure. Note that these events are scaled from 1000-yr MC data. For the case of NO, ICAL would detect about 4614 reconstructed $\mu^-$ and 2053 reconstructed $\mu^+$ events in 10 years using three-flavor neutrino oscillations with matter effects considering 25-layered PREM profile of Earth. We can observe that the number of $\mu^+$ events is less than that of the $\mu^-$ because, at these multi-GeV energies, the interaction cross section for antineutrino is lower than that of the neutrino by almost a factor of 2. 

The ns timing resolution of RPCs~\cite{Dash:2014ifa,Bhatt:2016rek,Gaur:2017uaf} enables ICAL to distinguish between upward-going ($\cos\theta_\mu^\text{rec} < 0$) and downward-going ($\cos\theta_\mu^\text{rec} > 0$) reconstructed muon events as denoted by U and D, respectively. ICAL is expected to detect about 1654 upward-going and 2960 downward-going $\mu^-$ events in 10 years, whereas $\mu^+$ events in upward and downward direction will be around 740 and 1313, respectively, in 10 years. Note that the upward-going muon events are nearly half of the downward-going muon events because the longer baselines available to upward-going neutrinos cause them to oscillate to other flavors. On the other hand, the neutrino oscillations are not developed significantly for smaller baselines available to the downward-going neutrinos. 

%=====================================
\section{Impact of Atmospheric Oscillation Parameters on Reconstructed Events}
\label{sec:impact_osc_par_events}
%=====================================

Now, we describe the impact of oscillation parameters on reconstructed muon events at ICAL. We consider full three-flavor oscillation parameters in the presence of matter with PREM profile~\cite{Dziewonski:1981xy}. ICAL is sensitive to atmospheric neutrinos and antineutrinos in the multi-GeV energy range and they are dominantly affected by atmospheric neutrino oscillation parameters $\theta_{23}$ and  $\Delta m^2_\text{eff}$. On the other hand, the solar oscillation parameters $\theta_{12}$ and $\Delta m^2_{21}$ have negligible impact on events at ICAL. As far as the reactor mixing angle $\theta_{13}$ is concerned, it has already been measured with high precision. We take CP-violating parameter $\delta_\text{CP} = 0$. Therefore, we discuss the impact of $\theta_{23}$ and $\Delta m^2_\text{eff}$ in the next sections.

%=====================================
\subsection{Impact of $\theta_{23}$ on Reconstructed Events}
\label{sec:impact_th23_events}
%=====================================

\begin{figure}
	\centering
	\includegraphics[width=0.49\linewidth]{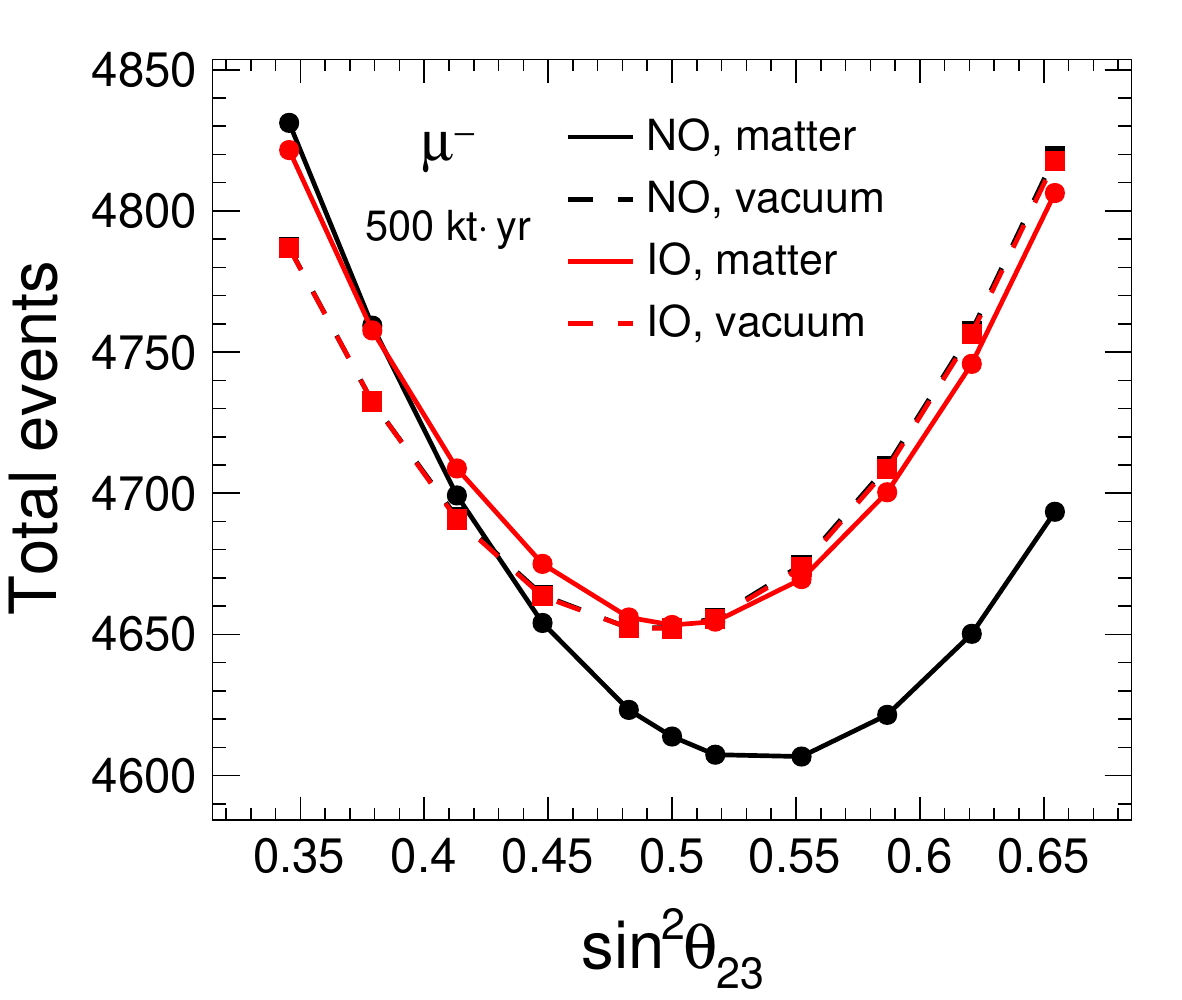}
	\includegraphics[width=0.49\linewidth]{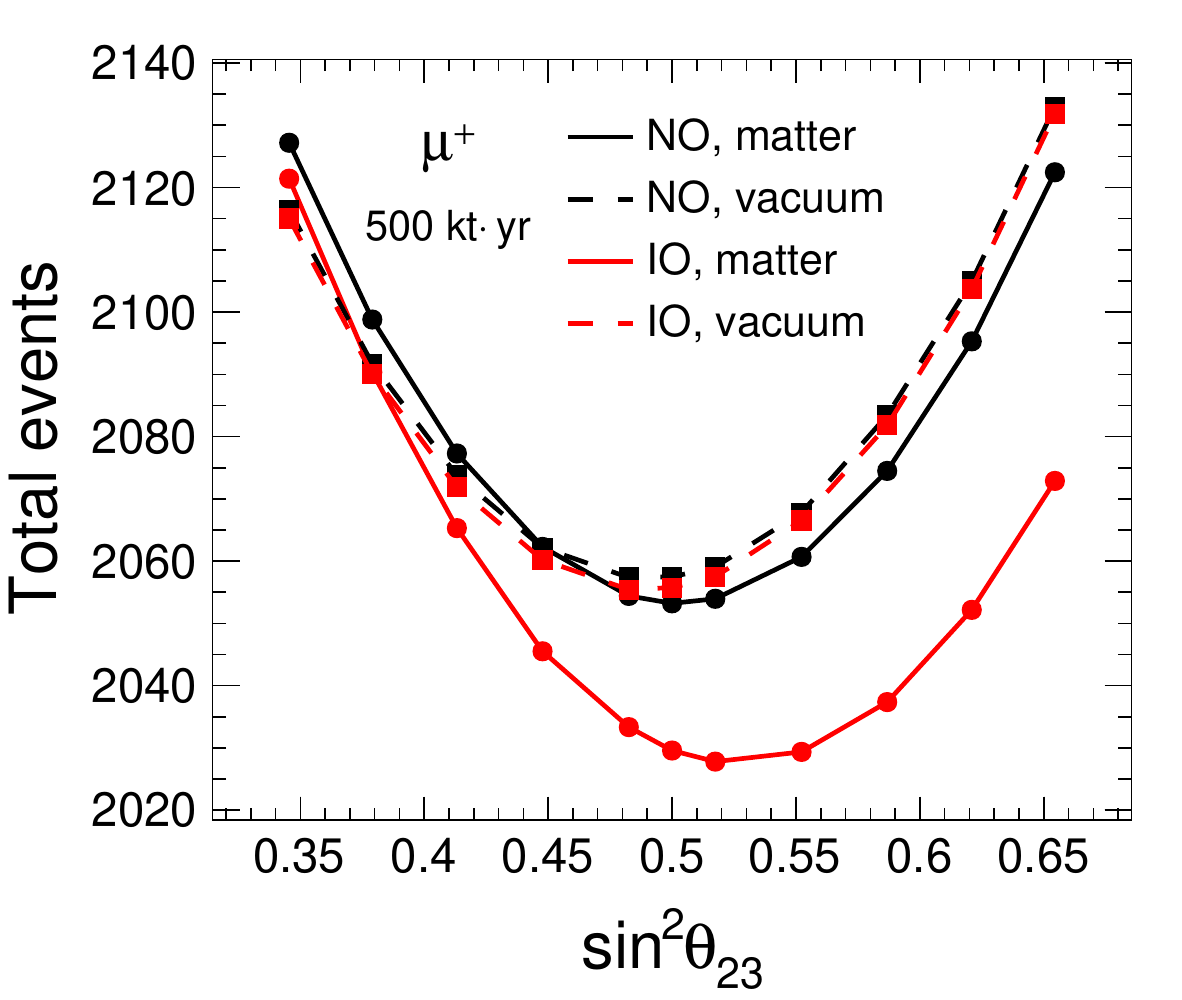}
	\caption{Total events as functions of $\sin^2\theta_{23}$ for reconstructed $\mu^-$ (left panel) and $\mu^+$ (right panel) using 500 kt$\cdot$yr exposure at ICAL. The black (red) curve denote NO (IO). Solid curves correspond to neutrino oscillation in the presence of matter with the PREM profile whereas dashed curves stand for neutrino oscillations in vacuum. Note that the scale of y-axis is different in two panels. We consider the benchmark values of other oscillation parameters given in Table~\ref{tab:osc-param-value}.}
	\label{fig:total_events_th23}
\end{figure}

Figure~\ref{fig:total_events_th23} presents total events as functions of $\sin^2\theta_{23}$ for reconstructed $\mu^-$ (left panel) and $\mu^+$ (right panel) using 500 kt$\cdot$yr exposure at ICAL. The black and red curves represent NO and IO, respectively. The solid curves correspond to the case of three-flavor neutrino oscillation in the presence of matter with the PREM profile whereas dashed curves denote the neutrino oscillations in vacuum. In both the panels of Fig.~\ref{fig:total_events_th23}, we observe that the total events are minimum for maximal mixing ($\theta_{23} = 45^\circ$) where $\sin^2\theta_{23} = 0.5$. This happens because the oscillation of $\nu_\mu$ to other flavors is proportional to  $\sin^2 2\theta_{23}$ at the leading order~\cite{Akhmedov:2004ny} and $\sin^2 2\theta_{23} = 1$ for $\theta_{23} = 45^\circ$ results into maximum oscillation of $\nu_\mu$ to other flavors.  

In both panels of Fig.~\ref{fig:total_events_th23}, total events are the same for the case of NO and IO when we consider vacuum oscillations. In the left panel of Fig.~\ref{fig:total_events_th23}, a significant difference is observed in the case of NO while considering matter effect, where the total reconstructed $\mu^-$ events in the presence of matter are less than that in vacuum. This occurs because neutrinos experience matter effects inside Earth for the case of NO, which enhances $\nu_\mu \rightarrow \nu_e$ and $\nu_e \rightarrow \nu_\mu$ oscillation probabilities. Since the initial flux of atmospheric neutrinos for $\nu_\mu$ is larger than that for $\nu_e$ as described in Sec.~\ref{sec:neutrino_flux}, the net effect is the enhanced events corresponding $\nu_\mu \rightarrow \nu_e$ oscillations. Note that this resonance in neutrino does not occur in the case of IO. Hence, the total $\mu^-$ events for IO are the same in the cases of matter and vacuum. Further, the matter effects are proportional to $\sin^2\theta_{23}$~\cite{Akhmedov:2004ny} and decrease the total $\mu^-$ events when $\theta_{23}$ is in the higher octant. On the other hand, $\sin^2 2\theta_{23}$ increases the events as $\theta_{23}$ deviates from $45^\circ$. Therefore, the minimum of the black curve gets shifted to the larger value of $\sin^2\theta_{23}$ in the higher octant.

As far as $\mu^+$ events are concerned, the right panel of Fig.~\ref{fig:total_events_th23} shows that the total events are the same for the case of NO and IO in vacuum, and the minimum events are also observed at $\sin^2\theta_{23} = 0.5$. Unlike the case of $\mu^-$, here, the matter effects appear for the case of IO as shown by the solid red curve in the right panel, whereas the events for NO are the same as that for the case of vacuum. Here, also, we can observe the shift in the minimum at a higher value of $\sin^2\theta_{23}$ in the presence of matter but for the case of IO. 

%=====================================
\subsection{Impact of $\Delta m^2_\text{eff}$ on Reconstructed Events}
\label{sec:impact_dmsqeff_events}
%=====================================

\begin{figure}
	\centering
	\includegraphics[width=0.49\linewidth]{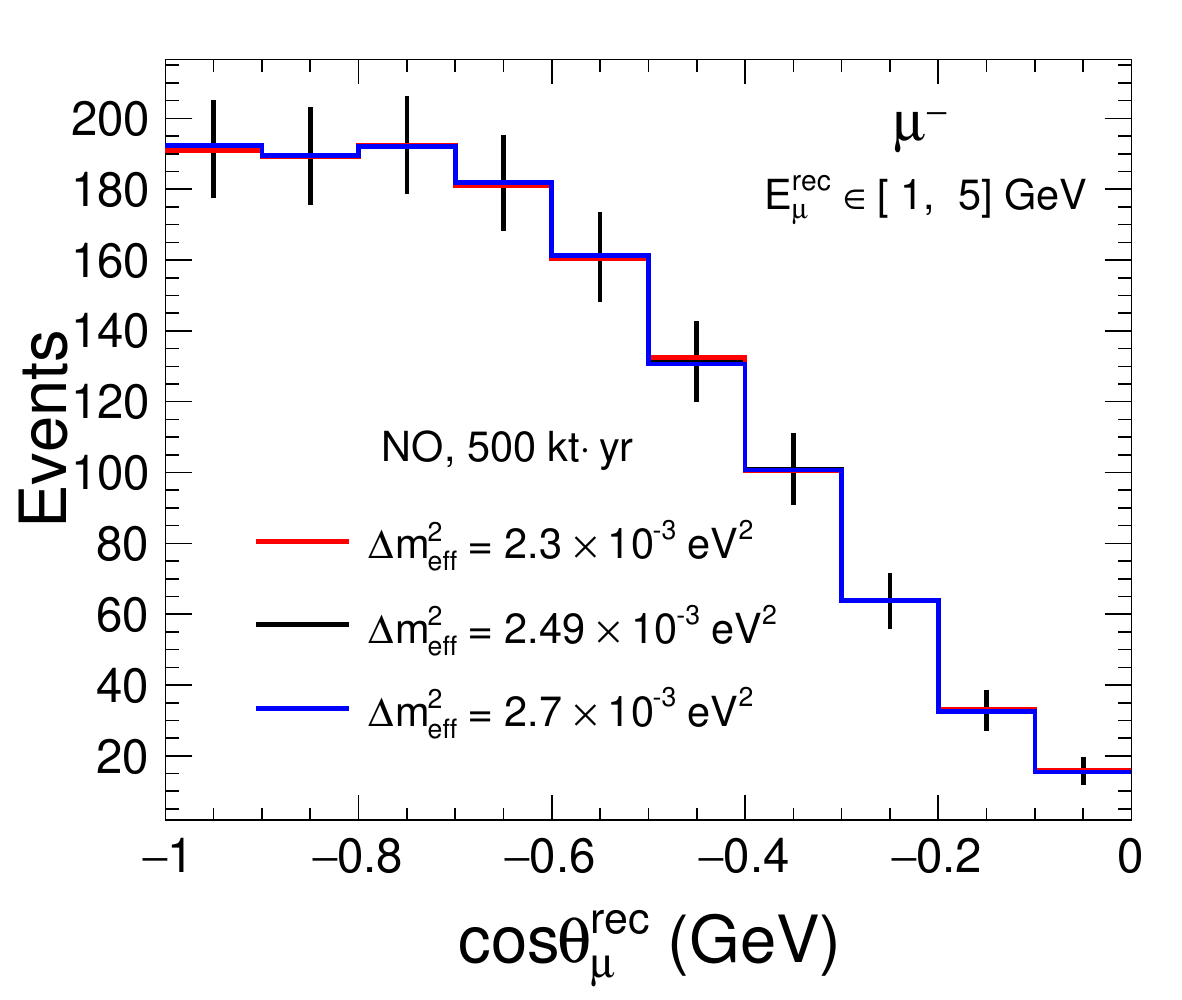}
	\includegraphics[width=0.49\linewidth]{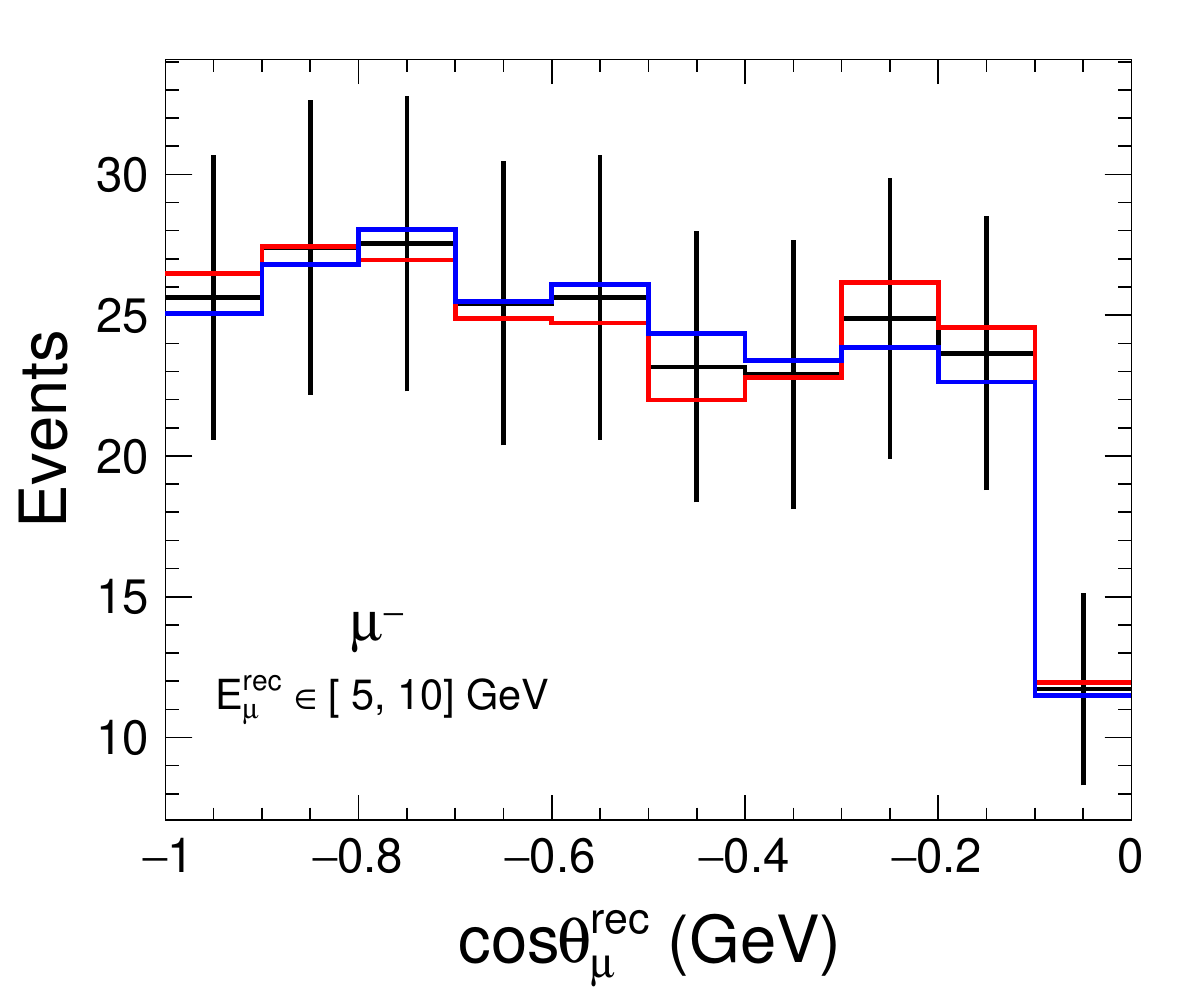}
	\includegraphics[width=0.49\linewidth]{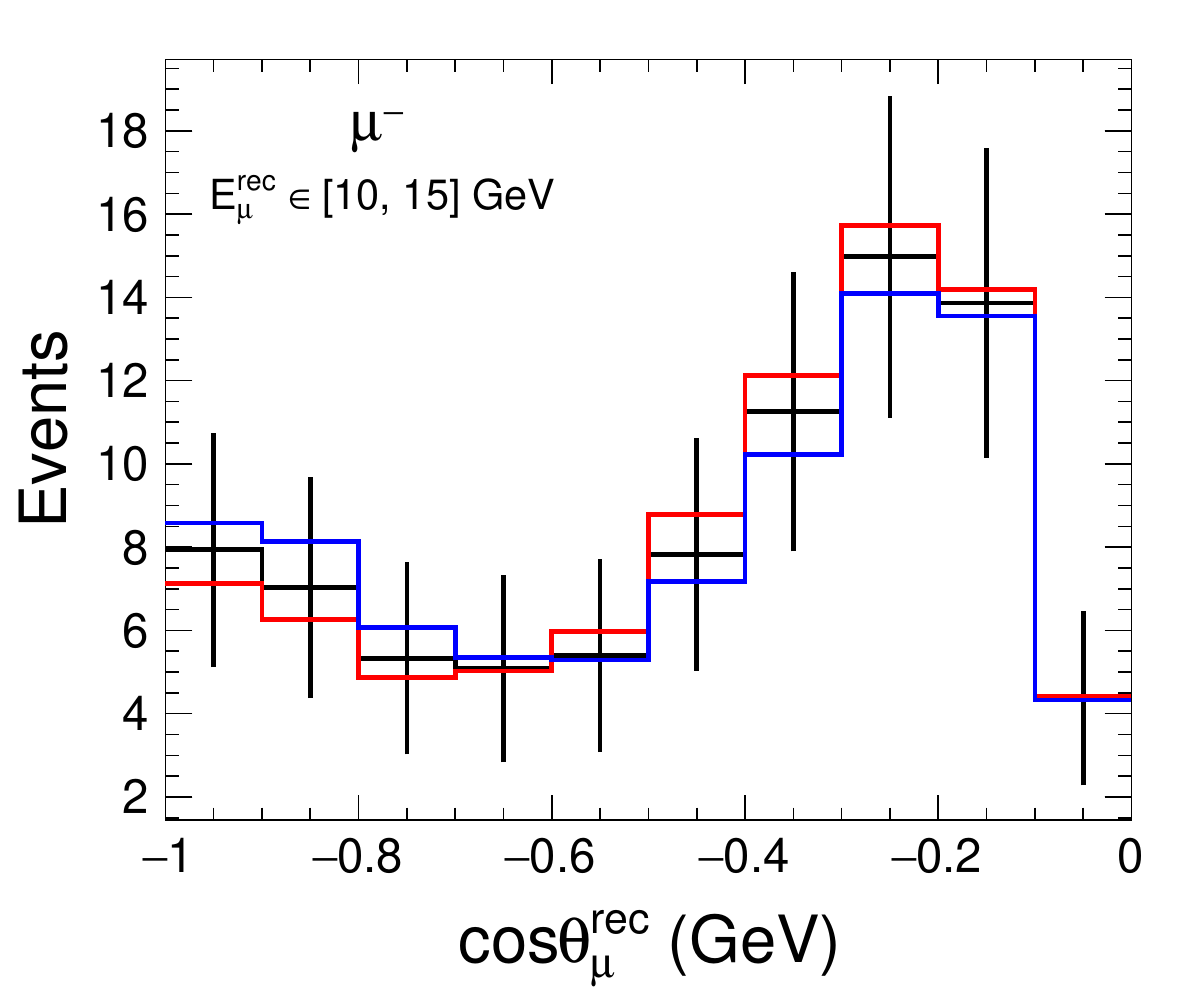}
	\includegraphics[width=0.49\linewidth]{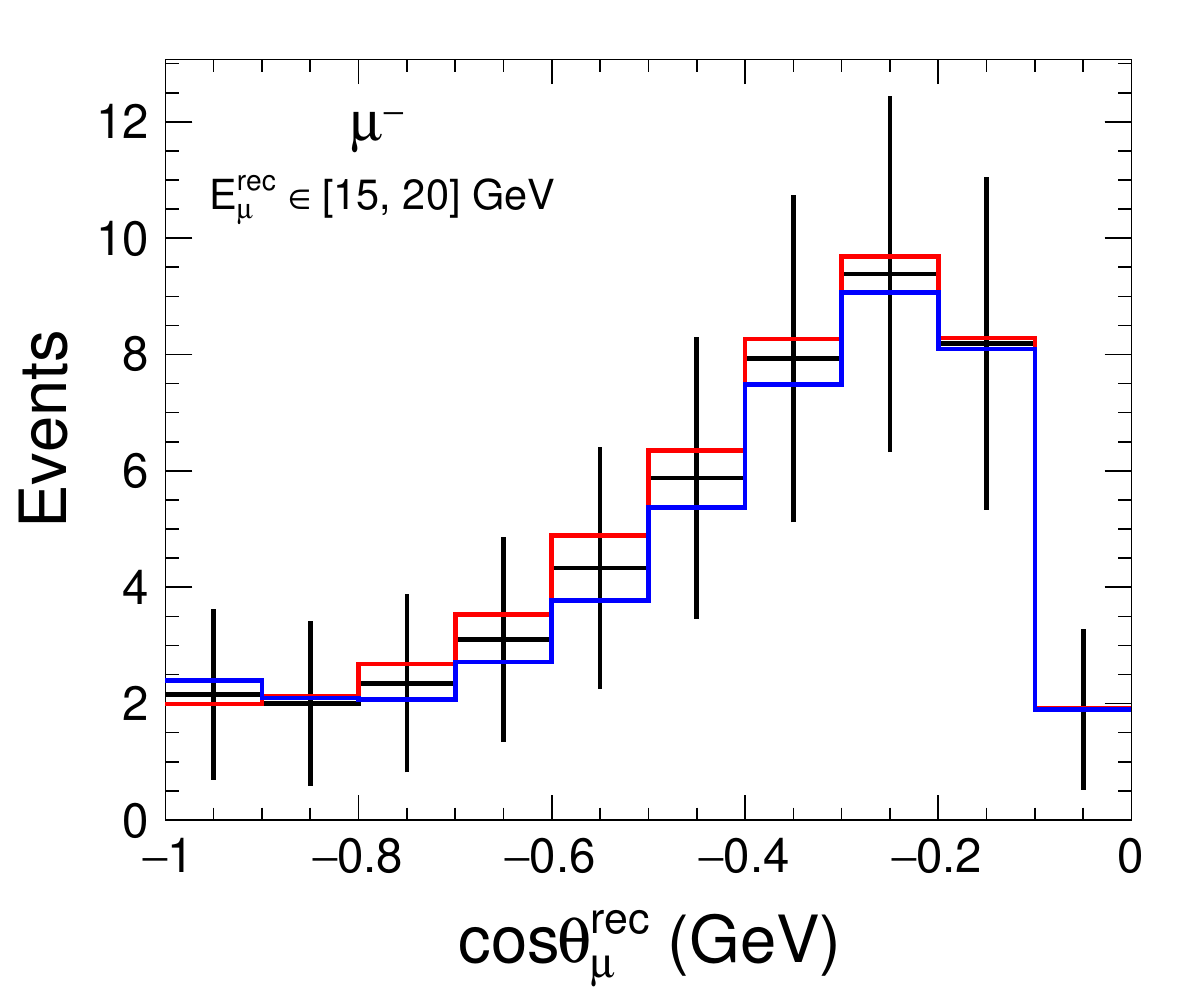}
	\caption{The distributions of reconstructed $\mu^-$ events as functions of $\cos\theta_\mu^\text{rec}$ for 500 kt$\cdot$yr exposure at ICAL. The red, black, and blue curves correspond to $\Delta m^2_\text{eff} = 2.3 \times 10^{-3}~\text{eV}^2, 2.49 \times 10^{-3}~\text{eV}^2,~\text{and}~2.7 \times 10^{-3}~\text{eV}^2$, respectively. We take energy ranges of [1, 5] GeV, [5, 10] GeV, [10, 15] GeV, and [15, 20] GeV in the top left, top right, bottom left, and bottom right panels, respectively. Note that the scales of y-axis are different in all the panels. We consider NO, and the benchmark values of other oscillation parameters given in Table~\ref{tab:osc-param-value}. }
	\label{fig:event_dist_dmsqeff_nu}
\end{figure}

\begin{figure}
	\centering
	\includegraphics[width=0.49\linewidth]{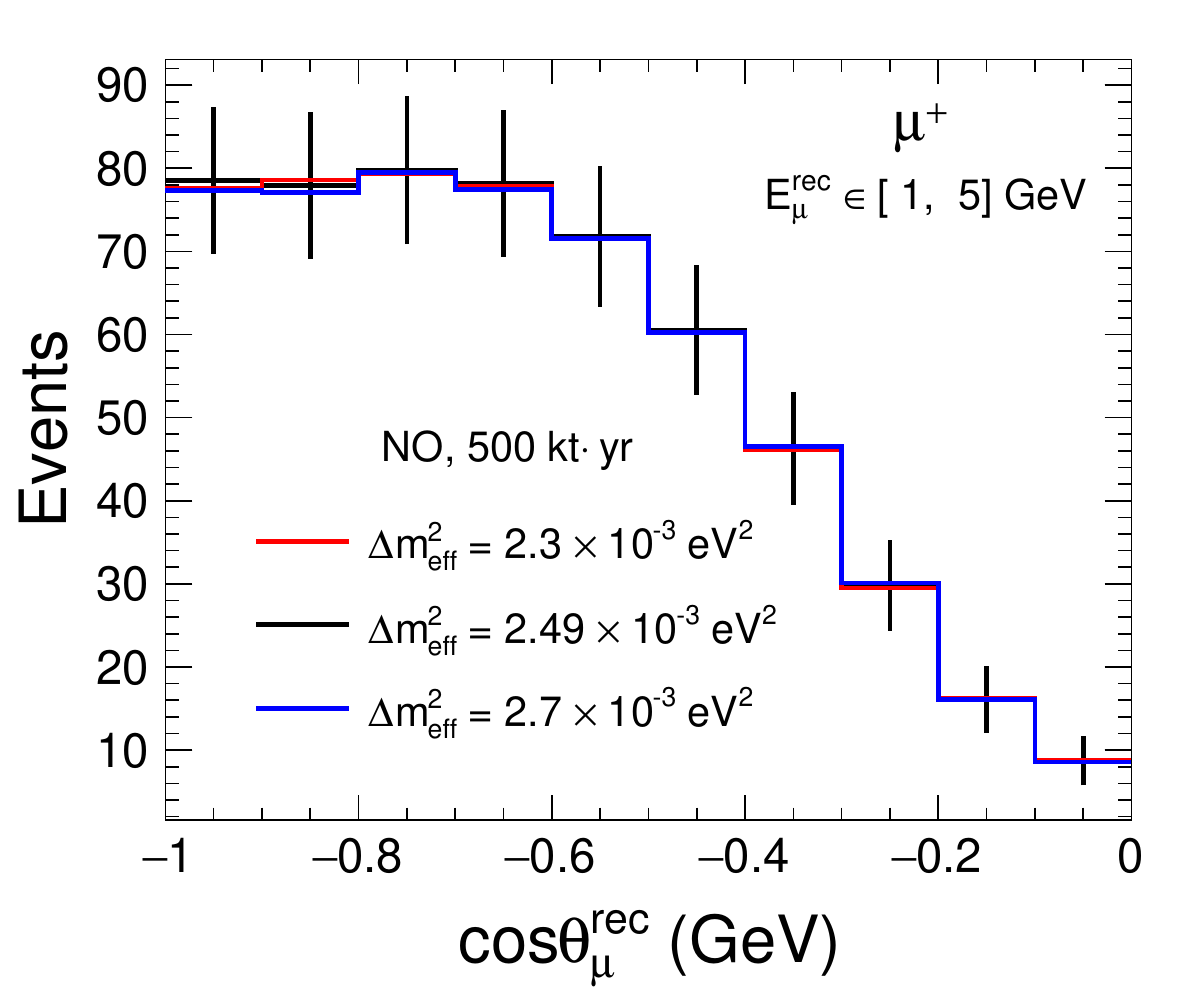}
	\includegraphics[width=0.49\linewidth]{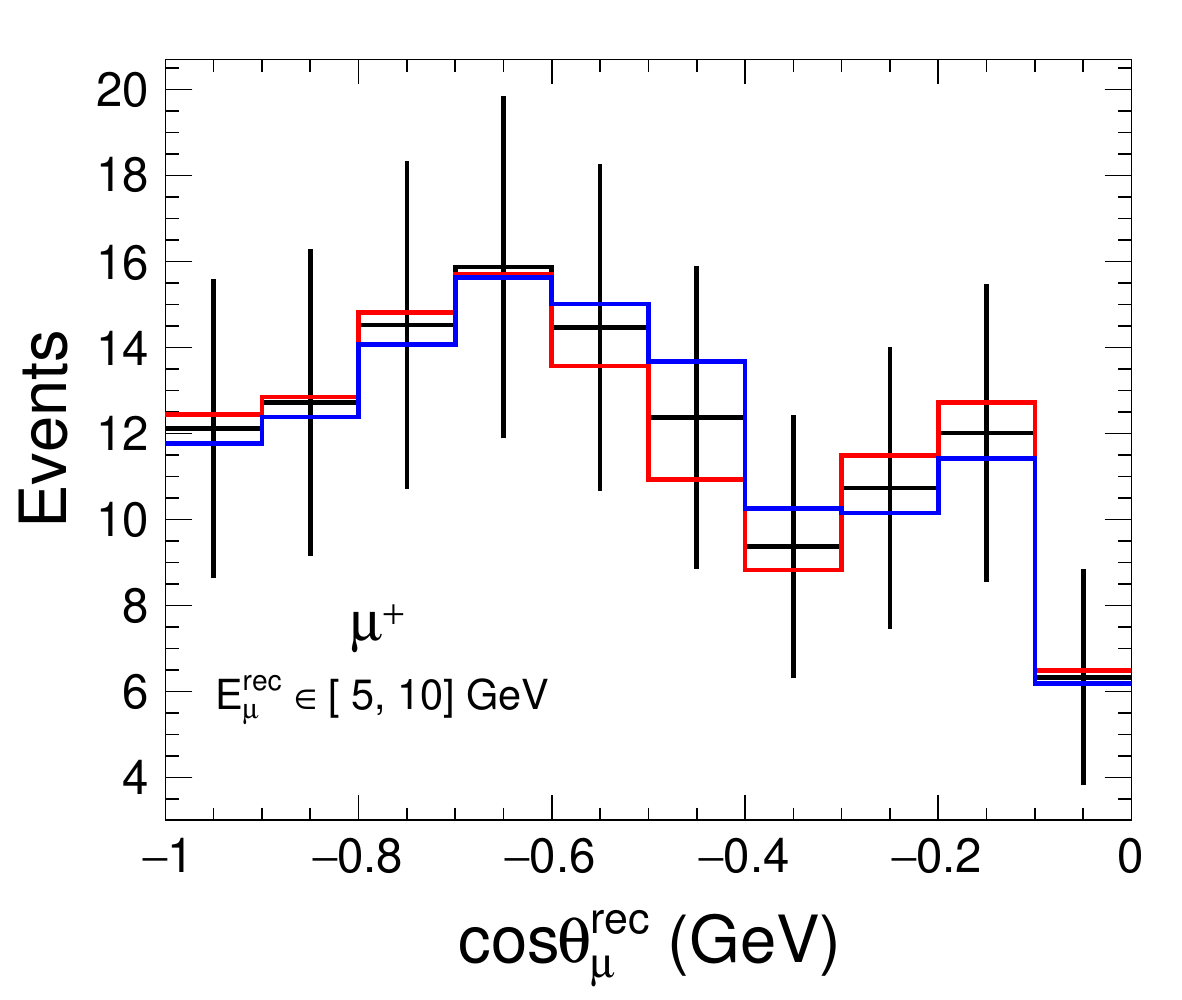}
	\includegraphics[width=0.49\linewidth]{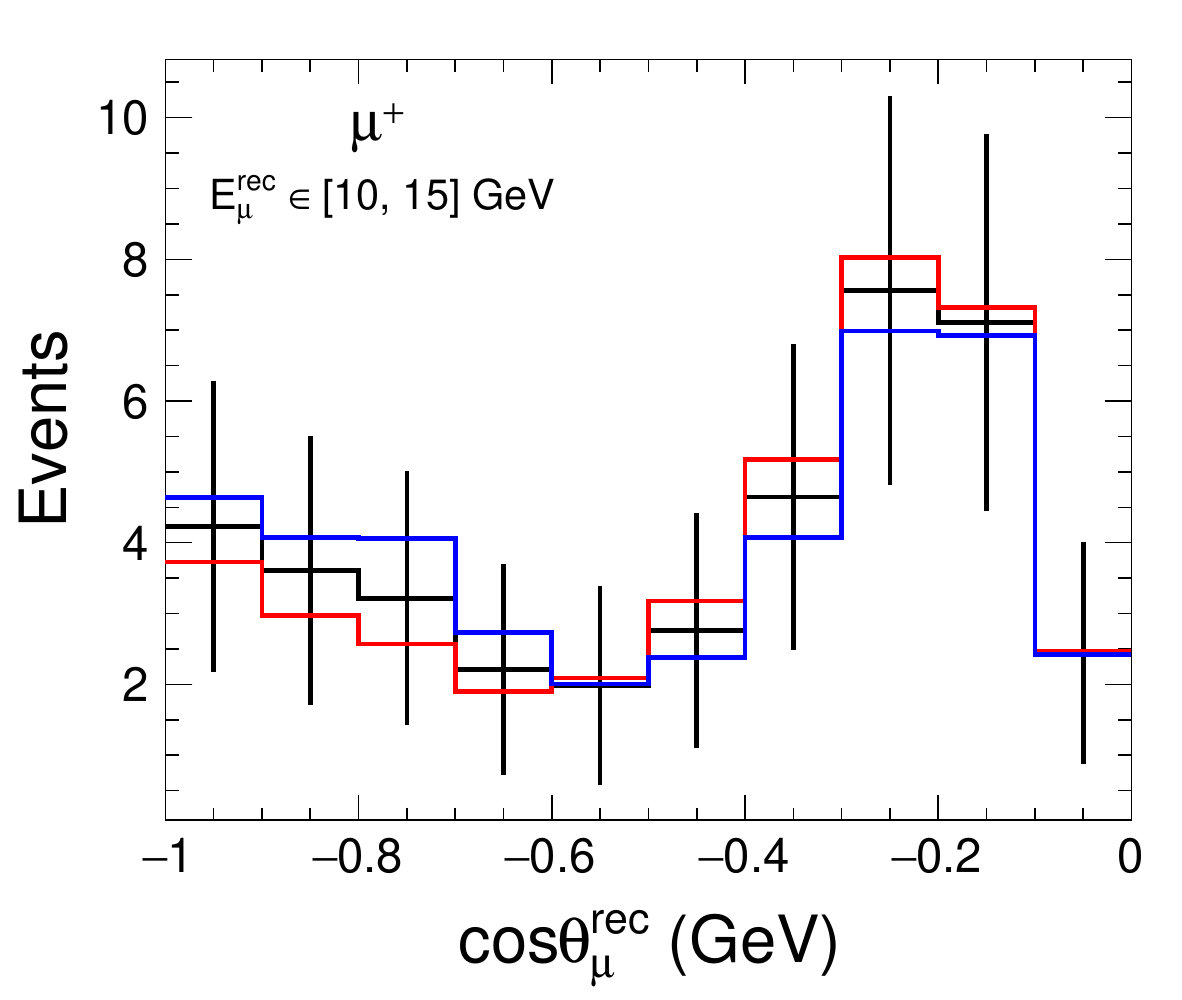}
	\includegraphics[width=0.49\linewidth]{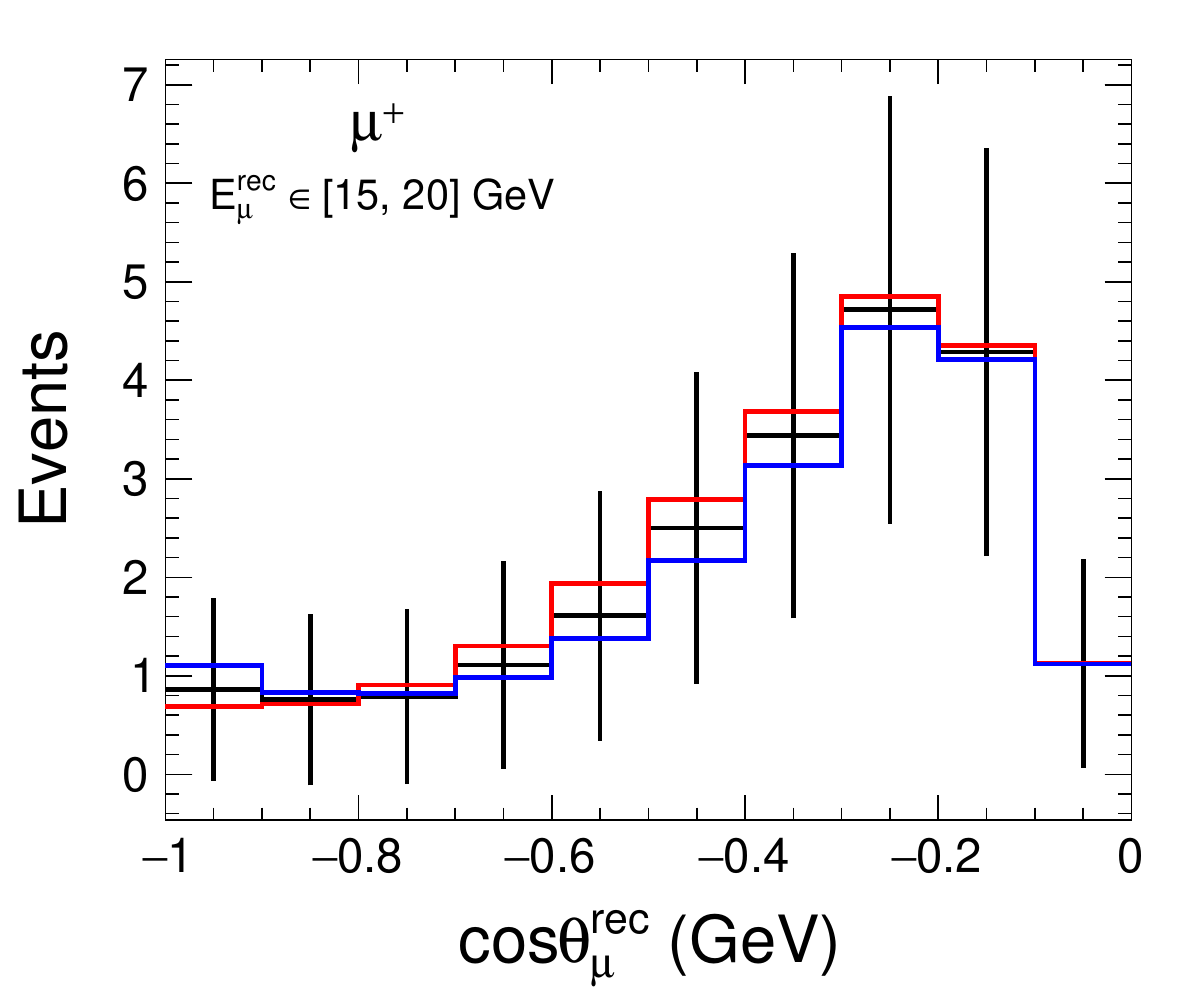}
	\caption{The distributions of reconstructed $\mu^+$ events as functions of $\cos\theta_\mu^\text{rec}$ for 500 kt$\cdot$yr exposure at ICAL. The red, black, and blue curves correspond to $\Delta m^2_\text{eff} = 2.3 \times 10^{-3}~\text{eV}^2, 2.49 \times 10^{-3}~\text{eV}^2,~\text{and}~2.7 \times 10^{-3}~\text{eV}^2$, respectively. We take energy ranges of [1, 5] GeV, [5, 10] GeV, [10, 15] GeV, and [15, 20] GeV in the top left, top right, bottom left, and bottom right panels, respectively. Note that the scales of y-axis are different in all the panels. We consider NO, and the benchmark values of other oscillation parameters given in Table~\ref{tab:osc-param-value}. }
	\label{fig:event_dist_dmsqeff_anu}
\end{figure}

So far, we have discussed the effect of atmospheric mixing angle $\theta_{23}$ on total events at ICAL.  As far as the atmospheric mass square difference $\Delta m^2_\text{eff}$ is concerned, it's effect on total number of events is not significant, instead, it alters the distribution of events. In Fig.~\ref{fig:event_dist_dmsqeff_nu}, we show how the distribution of reconstructed $\mu^-$ events as a function of $\cos\theta_{\mu}^\text{rec}$ modifies when we vary the value of $\Delta m^2_\text{eff}$. We consider 500 kt$\cdot$yr exposure at ICAL with full three-flavor neutrino oscillations in the presence of matter effect with PREM profile~\cite{Dziewonski:1981xy}. We assume normal mass ordering with benchmark values of oscillation parameters (except $\Delta m^2_\text{eff}$) given in Table~\ref{tab:osc-param-value}. The black curves correspond to the benchmark value of  $\Delta m^2_\text{eff}$ which is $2.49 \times 10^{-3}$ eV\textsuperscript{2}. In order to understand the impact of $\Delta m^2_\text{eff}$, we choose the 3$\sigma$ range obtained from ICAL analysis on precision measurements of atmospheric neutrino oscillation parameters considering muon and hadron information~\cite{Devi2014}. In Fig.~\ref{fig:event_dist_dmsqeff_nu}, this 3$\sigma$ range of $2.3 \times 10^{-3}$ eV\textsuperscript{2} and $2.7 \times 10^{-3}$ eV\textsuperscript{2} are denoted by red and blue curves, respectively. The error bars denote the statistical fluctuations for the benchmark choice shown by black curve. We show these distributions for the energy ranges of 
%1 to 5 GeV, 5 to 10 GeV, 10 to 15 GeV, and 15 to 20 GeV
[1, 5] GeV, [5, 10] GeV, [10, 15] GeV, and [15, 20] GeV in the top left, top right, bottom left, and bottom right panels, respectively. Note that the scales of y-axis are different in all these panels because the number of events decrease with energy following the power law as described in Sec.~\ref{sec:neutrino_flux}. Similarly, we present the event distributions for reconstructed $\mu^+$ events in Fig.~\ref{fig:event_dist_dmsqeff_anu}.

In Figs.~\ref{fig:event_dist_dmsqeff_nu} and \ref{fig:event_dist_dmsqeff_anu} , we can observe that the effect of $\Delta m^2_\text{eff}$ is significant for energies above 5 GeV whereas the impact is negligible for the energy range of 1 to 5 GeV. This happens because the effect of $\Delta m^2_\text{eff}$ is significant in the region of vacuum oscillations which is dominant at higher energies of 5 to 25 GeV, and is similar for both neutrinos and antineutrinos. The red and blue curves move apart from the black curve in opposite directions unlike the effect of $\theta_{23}$ in Fig.~\ref{fig:total_events_th23} where total events increase for both lower as well as higher values of  $\theta_{23}$. It is interesting to note that this effect has a zenith angle dependence which explains the dilution of this effect in total number of events. Note that though this effect of $\Delta m^2_\text{eff}$ in each bin appears small, the contribution of every bin is added while calculating the precision measurement of $\Delta m^2_\text{eff}$. Also, we consider information on only muon energy and zenith angle while plotting the event distributions in Figs.~\ref{fig:event_dist_dmsqeff_nu} and \ref{fig:event_dist_dmsqeff_anu}. The authors in Ref.~\cite{Devi:2014yaa} have demonstrated that the incorporation of hadron energy information further improves the sensitivity of ICAL towards the measurement of $\Delta m^2_\text{eff}$. 

\section{Summary}
\label{sec:ICAL_conclusion}
The upcoming 50 kt Iron Calorimeter (ICAL) detector at the India-based Neutrino Observatory (INO)~\cite{ICAL:2015stm} aims to detect atmospheric muon neutrinos and antineutrinos separately in a multi-GeV range of energies over a wide range of baselines from about 15 to 12750 km. ICAL consists of stacks of 5.6 cm thick iron layers with a gap of 4 cm for sandwiching the Resistive Plate Chambers (RPCs). The iron layers act as passive detector elements, whereas the RPCs act as active detector elements. The charged-current interactions of neutrino would produce muons and hadrons. Since muon is a minimum ionizing particle, it can pass through many layers of RPCs and leave hits in the form of a track. ICAL would be magnetized with a magnetic field of about 1.5 T which would enable ICAL to distinguish between $\mu^-$ and $\mu^+$ tracks by measuring the direction of bending. This charge identification (CID)  capability helps ICAL to differentiate muon neutrinos and antineutrinos. Since the matter effects modify the oscillation patterns differently for neutrinos and antineutrinos, CID capability plays an important role in the determination of neutrino mass ordering, which is the main goal of ICAL. As far as the hadrons are concerned, they give rise to multiple hits in the same layer of RPCs in the form of shower events. 

The muon energy resolution at ICAL is about 10 to 15\% in the muon energy range of 1 to 25 GeV~\cite{Chatterjee:2014vta}. The angular resolution of muon at ICAL is expected to be about $1^\circ$~\cite{Chatterjee:2014vta}. The reconstruction efficiency and resolutions are poor for horizontal events because horizontally-going muons are able to pass through only a few RPC layers. The ICAL detector also has some sensitivity towards hadron showers, where the energy of hadron can be measured using the number of hits in a shower, but the hadron energy resolution can be as poor as $\sim$40\%~\cite{Devi:2013wxa}. The incorporation of hadron energy information in the physics analyses had been demonstrated to improve the sensitivities of ICAL towards precision measurements of neutrino oscillation parameters~\cite{Devi:2014yaa}, the determination of neutrino mass ordering~\cite{Devi:2014yaa} and BSM physics scenarios~\cite{Khatun:2019tad,Sahoo:2021dit}. 

The analyses in this thesis have been performed using the simulated events at the ICAL detector. The neutrino interactions at ICAL are simulated using the NUANCE neutrino event generator, where the geometry of ICAL and the flux of atmospheric neutrinos at the INO site in Theni are given as inputs. The neutrino oscillations are taken into account using the reweighting algorithm. Then, the detector response is incorporated by folding the simulated event with the migration matrices for muons ~\cite{Chatterjee:2014vta} and hadrons~\cite{Devi:2013wxa} as provided by the ICAL collaboration. After applying the detector properties, the oscillation analysis can be performed using the simulated events in terms of the observable quantities which are reconstructed muon energy ($E_\mu^\text{rec}$), muon direction ($\cos\theta_\mu^\text{rec}$), and hadron energy (${E'}_\text{had}^\text{rec}$). Now, we describe the studies on response uniformity of RPC in chapter~\ref{chap:RPC_response}.
\end{refsegment}

\cleartooddpage
\chapter{Studies on Response Uniformity of RPC}
\label{chap:RPC_response}
\begin{refsegment}

The resistive plate chamber (RPC) is an active detector element at ICAL. The structure and working principle of RPCs are described in detail in Sec.~\ref{sec:RPC} where we learned that the gas mixture is confined between the resistive plates of glass or bakelite. Since the high voltage cannot be applied on the resistive plate of glass or bakelite, a coating of a conductive layer like graphite is applied on the outer surface of the resistive plate. A positive potential of about 5 kV is applied on one surface of the graphite layer and a negative potential of about - 5 kV on another surface. The potential on the graphite layer spreads uniformly due to moderate resistivity of around 1 M$\Omega$ and a uniform electric field of about 50 kV/cm is established inside the chamber. Note that the surface resistivity of the graphite layer should also be high enough to allow the induction of signals on the pickup strips.

The uniform electric field is crucial to achieve a uniform detector response. The non-uniform surface and bulk resistivity of the graphite layer present in RPC are likely to affect the detector response and dead time. In this chapter, we present the initial results of a study that is oriented towards investigating the effects of resistivity of the graphite layer on RPC signal generation. In Sec.~\ref{sec:resistivity_simulation}, we describe the framework to simulate charge transport in graphite layer. Simulation of potential buildup for uniform resistivity is given in Sec.~\ref{sec:sim_uniform_res}. Details about the experiment setup to measure surface resistivity of graphite layer is mentioned in Sec.~\ref{sec:exp_res}. The experimentally measured surface resistivity is used as an input to simulate potential buildup for non-uniform surface resistivity as given in Sec.~\ref{sec:sim_nonuniform_res}. The experimental measurement of distribution of time constant and potential are discussed in sections~\ref{sec:exp_tau} and \ref{sec:exp_potential}. Finally, we summarize out findings in Sec.~\ref{sec:rpc_summary}.

%=====================================
\section{Simulation of Charge Transport in Graphite Layer}
\label{sec:resistivity_simulation}
%=====================================

\begin{figure}
	\centering
	\includegraphics[width=0.95\linewidth]{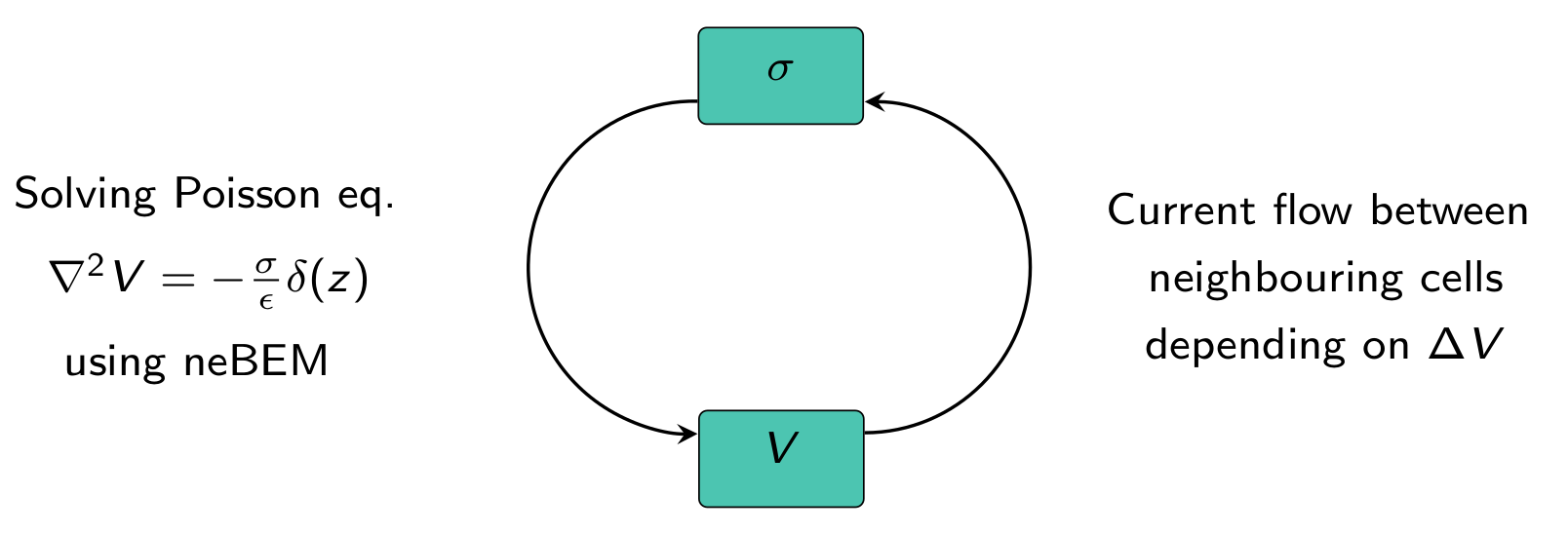}
	\caption{Procedure for simulating charge transport in a resistive layer.}
	\label{fig:chani_flowchart}
\end{figure}

We study the motion of charge in the graphite layer to understand the effect of surface resistivity on the functioning of RPC and signal generation. A ROOT-based mathematical framework has been developed where the simulation of charge transport is performed by solving the Poisson equation for a rectangular surface at each time step and by calculating the currents between subcells of the surface. A flowchart of the process is shown in Fig.~\ref{fig:chani_flowchart} where the potential ($V$) distribution is obtained from the distribution of surface charge density ($\sigma$) by solving the Poisson equation, and then, the current is calculated between neighboring cells to transport the charge at each time step. 
The Poisson equation for two-dimensional surface can be described as
\begin{align}\label{eq:Poisson}
\nabla^2 V = - \frac{\sigma}{\epsilon}\delta(z)\,,
\end{align}
where, $V$ is the potential on the surface, $\sigma$ is the surface charge density and $\epsilon$ is the electric permittivity of the material. In the present work, we solve the Poisson equation using a solver based on a nearly exact Boundary Element Method (neBEM)~\cite{mukhopadhyay:2006a,neBEM} which is available in the Garfield++ toolkit~\cite{Garfieldpp}. The neBEM solver has already been used to calculate electric field and potential in numerous electrostatic problems~\cite{Nayana:2006a,Nayana:2006b,Nayana:2006c,Mukhopadhyay:2006use,Nayana:2007,mukherjee:2007,Mukhopadhyay:2007d,Mukhopadhyay:2009}. 

To calculate the electrostatic field, the rectangular surface of width $l_x$ and length $l_y$ is divided in $n_x$ subcells along x-direction and $n_y$ subcells along y-directions. The subcells are labeled with indices from 1 to $n_x \times n_y$ starting from bottom left to top right. In the application of neBEM, Eq.~\ref{eq:Poisson} can be written in the matrix form as 
\begin{align}\label{eq:influence}
[l_{mn}][\sigma_n] = [V_m]\,.
\end{align}
where, $\sigma_n$ denotes the elements of column vector corresponding to the charge density and $V_m$ stands for elements of column vector representing the potential. Here, $l_{mn}$ is the element of influence matrix representing the potential at the center of $m^\text{th}$ subcell due to a unit charge density at the $n^\text{th}$ subcell. Since the influence matrix depends only on geometry of the system, it is calculated only once using neBEM and the same matrix can be used repeatedly for calculation of field at each time step. Now, if the charge density distribution ($\sigma$) is known, the potential distribution ($V$) can be obtain using Eq.~\ref{eq:influence}. It is also possible to obtain the charge density distribution from known potential distribution using the inverse equation of Eq.~\ref{eq:influence}
\begin{align}\label{eq:inverse_influence}
[\sigma_n] = [l_{mn}]^{-1}[V_m]\,.
\end{align}

\begin{figure}
	\centering
	\includegraphics[width=0.4\linewidth]{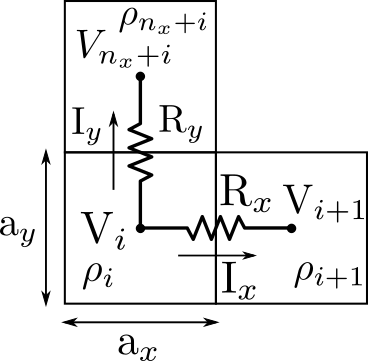}
	\caption{Effective resistance and current flow between neighboring cells due to potential difference.}
	\label{fig:currentflow}
\end{figure}

After knowing the electric potential at each cell, the transient simulation is performed by calculating the electric currents between neighboring subcells following the method described in Ref.~\cite{Chani:2012}. Note that the author in Ref.~\cite{Chani:2012} assumes the uniform surface resistivity throughout the geometry, whereas we incorporate the non-uniform surface resistivity where each cell can have a different value of surface resistivity. Figure~\ref{fig:currentflow} shows the effective resistance and the current flow between the neighboring cells due to potential difference. The effective resistance between $i^\text{th}$ cell and the cell present on its right with cell number $i + 1$, is given as 
\begin{equation}
R_x = \left[\left(\rho_i + \rho_{i+1}\right)/2\right]\times a_x/a_y.
\end{equation}
Similarly, the effective resistance between $i^\text{th}$ cell and the cell on the top with cell number $n_x +i$ is
\begin{equation}
R_y = \left[\left(\rho_i + \rho_{n_x+i}\right)/2\right] \times a_y/a_x.
\end{equation}
Now, the currents flowing between these cells can be be calculated using the Ohm's law as following
\begin{align}
(I_{x})_i = (V_i - V_{i+1})/R_x \,, \\
(I_{y})_i = (V_i - V_{n_x + i})/R_y\,,
\end{align}
here, $V_i$ is the electric potential at the $i^\text{th}$ cell.

We assume that the current flowing between cells is constant during a single time step ($\Delta t$). Hence, the amount of charge moving from one cell to another cell can be calculated as the product of the current and the time step ($\Delta t$)
\begin{equation}
\Delta q = I \times \Delta t\,.
\end{equation}
Note that there can be a connector resistance ($R_c$) which comes into picture due to resistance between the power supply and the connection points. The effective resistance along $x(y)$ direction between the connector and the cell on which the connection is made, can be given as 
\begin{equation}\label{eq:connector_res}
R_{x(y)} = R_c + (\rho/2)\times a_{x(y)}/a_{y(x)},
\end{equation}
In the present simulation, we assume the connector resistance $R_c$ to be zero for simplicity. This effective resistance in Eq.~\ref{eq:connector_res} is used to calculate the charging and discharging currents of the resistive layer following the procedure described above.

The initial conditions can be given in terms of the charge on the subcells or the applied potential, depending upon the problem under consideration. The potential, as well as the charge, at a later time are calculated by finite-difference time stepping to study the effect of non-uniformity in surface resistivity. The non-uniform distribution of input surface resistivity for these subcells is experimentally measured as described in section \ref{sec:exp_res}.

%=====================================
\section{Simulation of Potential Buildup for Uniform Resistivity}
\label{sec:sim_uniform_res}
%=====================================

%=====================================
\begin{figure}
	\centering
	\includegraphics[width=0.5\linewidth]{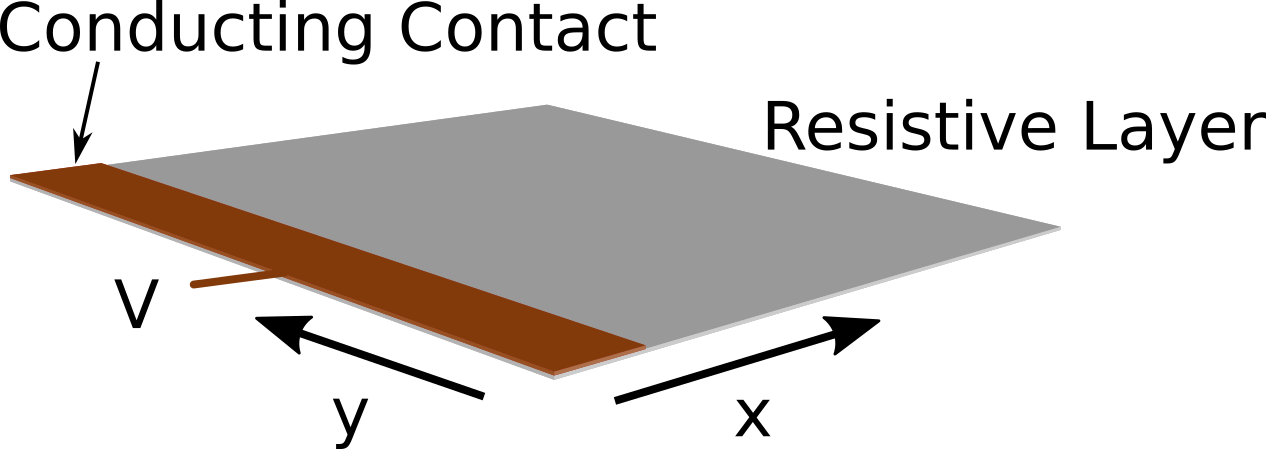}
	\caption{Simulation of potential buildup by application of a constant potential of 5000 V on the left side of the resistive layer}
	\label{fig:potential_on_resistive_layer}
\end{figure}
%=====================================

\begin{figure}
	\centering
	\includegraphics[width=0.6\linewidth]{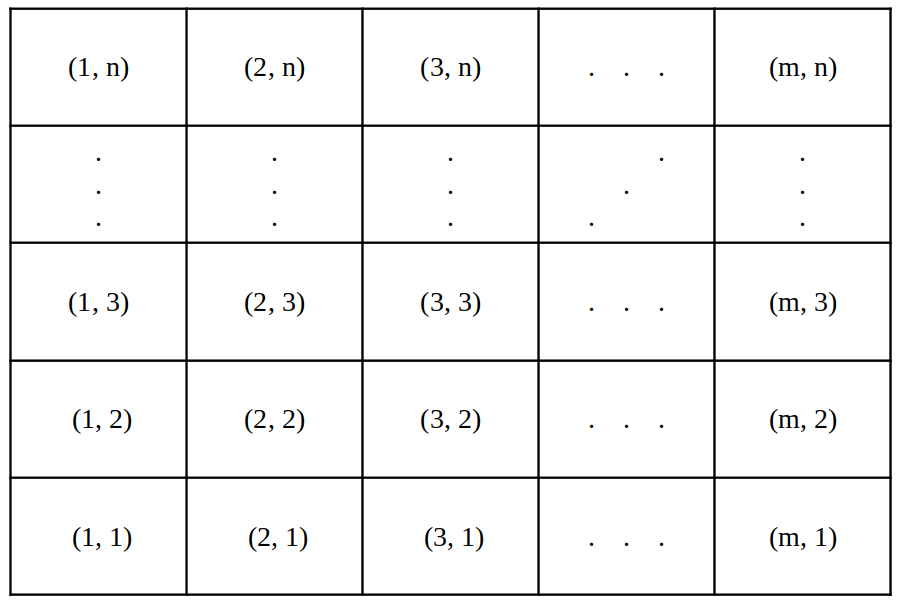}
	\caption{An $m \times n$ discretization for resistive layer. For our case, we have $m = n = 25$.}
	\label{fig:cellconvention}
\end{figure}

To study the charge transport in simulation, we apply a constant potential of 5000 V at the left side of the graphite layer as shown in Fig.~\ref{fig:potential_on_resistive_layer} and simulate potential buildup as a function of time. The size of graphite layer is $25\times25~\text{cm}^2$ which is divided into 625 cells of size $1~\text{cm}^2$ each. The nomenclature of cells follows the discretization shown in Fig.~\ref{fig:cellconvention}. The surface resistivity of the resistive layer is taken as 0.12 M$\Omega$ which is an average value of the experimentally measured surface resistivity as described in section \ref{sec:exp_res}. The relative permittivity of the graphite layer is assumed to be 10 in the current simulation. The initial charge and potential at each cell are taken as zero. The potential buildup is simulated for 60 $\mu$s, at the end of which the potential appears to reach the equilibrium.

%=====================================
\begin{figure}
	\centering
	\includegraphics[width=0.49\linewidth]{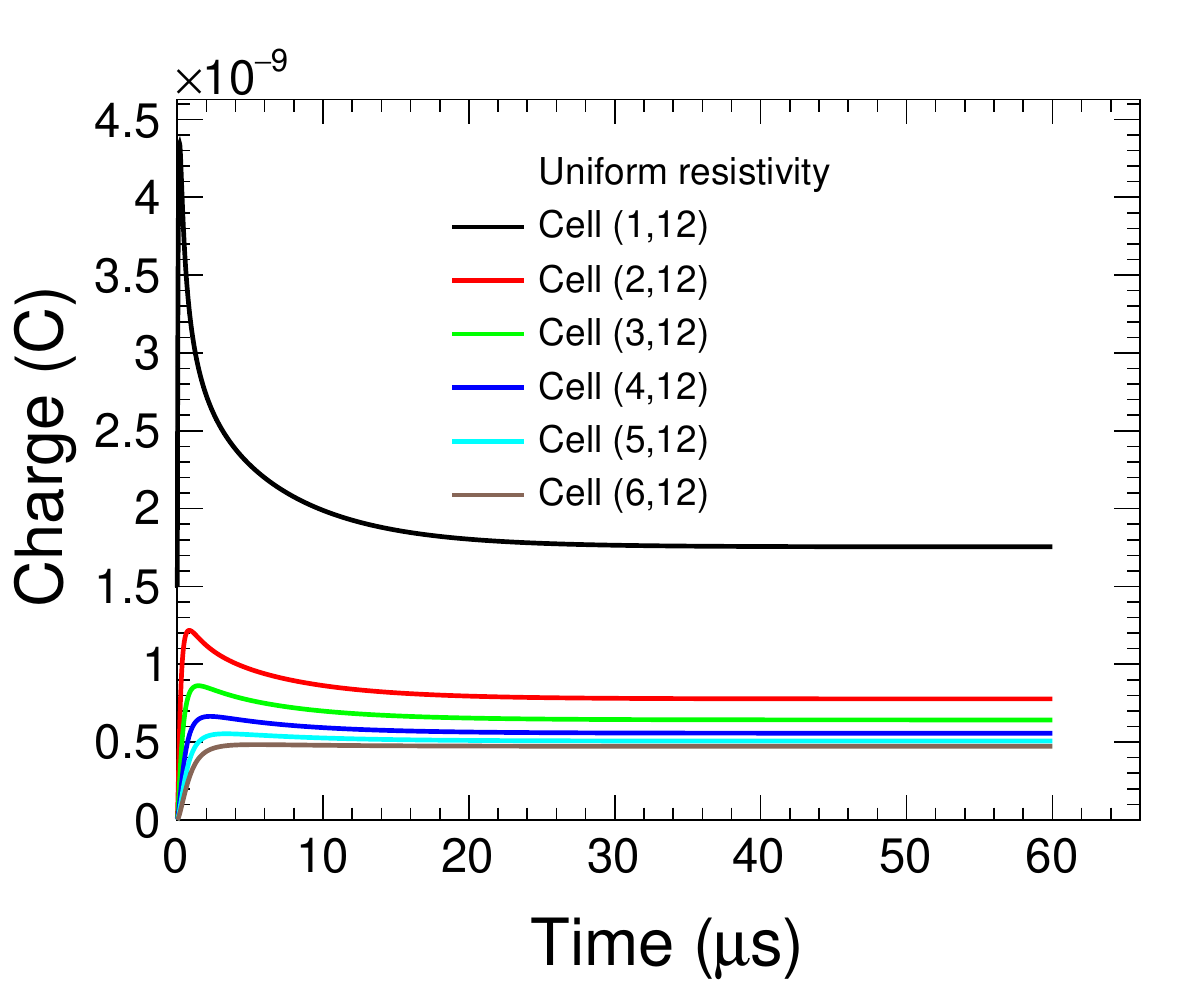}
	\includegraphics[width=0.49\linewidth]{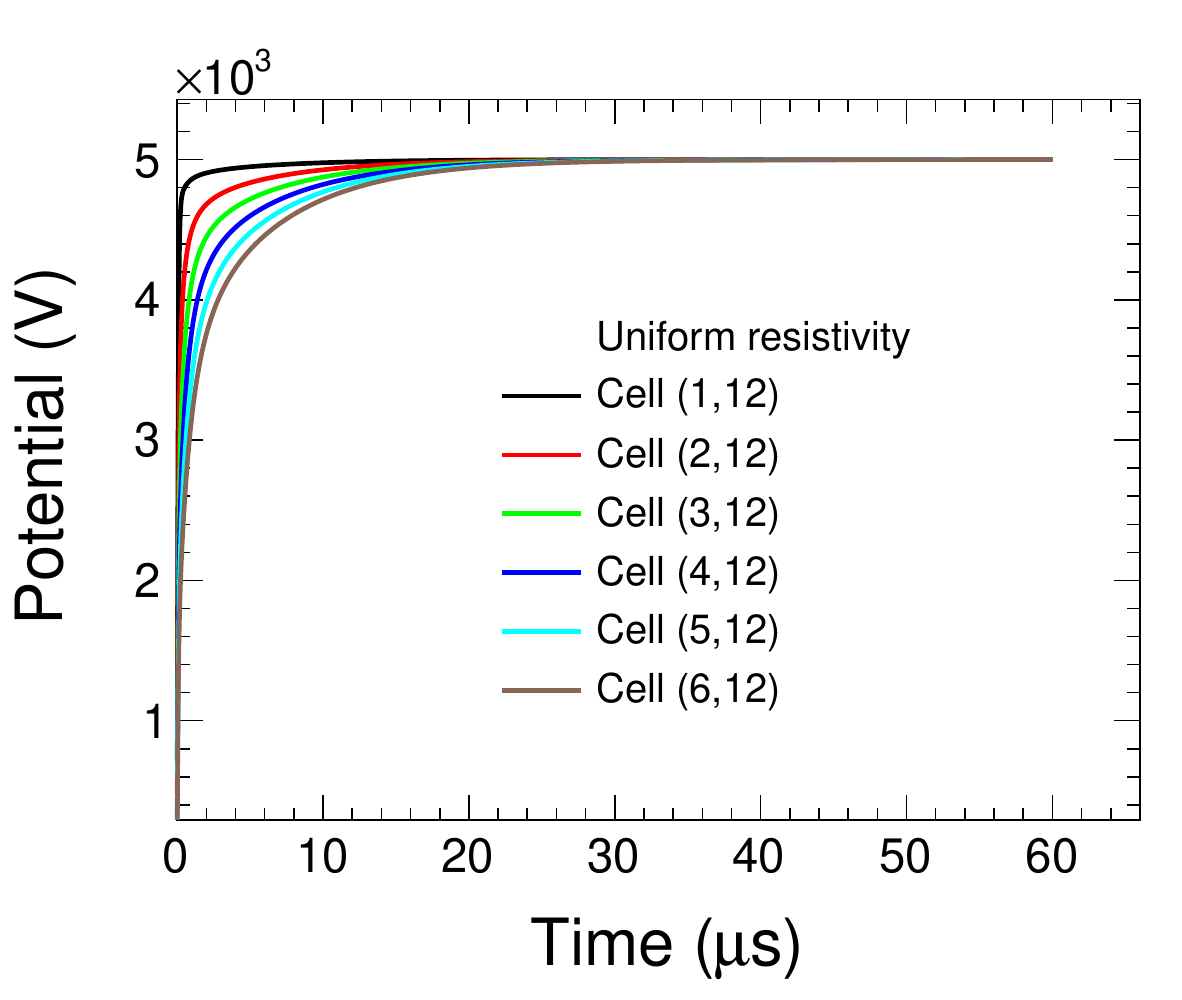}
	\caption{Simulated charge distribution (left panel) and potential distribution (right panel) as functions of time for various cells for uniform surface resistivity on the application of 5000 V at the left edge.}
	\label{fig:charge_potential_1D_uniform}
\end{figure}
%=====================================

 Figure~\ref{fig:charge_potential_1D_uniform} shows the charge (left panel) and potential (right panel) distribution as functions of time for various cells for uniform surface resistivity. We can observe in the left panel of Fig.~\ref{fig:charge_potential_1D_uniform}, that the charge in the cells first increases sharply and then saturates. A charge pile-up can also be observed in the cells near the point of application of voltage which can be due to a sudden increase in the resistivity of the layer around the point of contact. The charge pile-up decreases as we move away from the left side, where we have applied the fixed potential. The charge at saturation is also larger for the first few cells which may be due to the edge effect. 
 
 In the right panel of Fig.~\ref{fig:charge_potential_1D_uniform}, we observe that the potentials at the cells first increase sharply and then saturate to a common value of 5000 V. No pile-up is observed in the case of potential, but the rate of potential buildup decreases as we move away from the point of application of voltage. 
 
%=====================================
\begin{figure}
 	\centering
 	\includegraphics[width=0.49\linewidth]{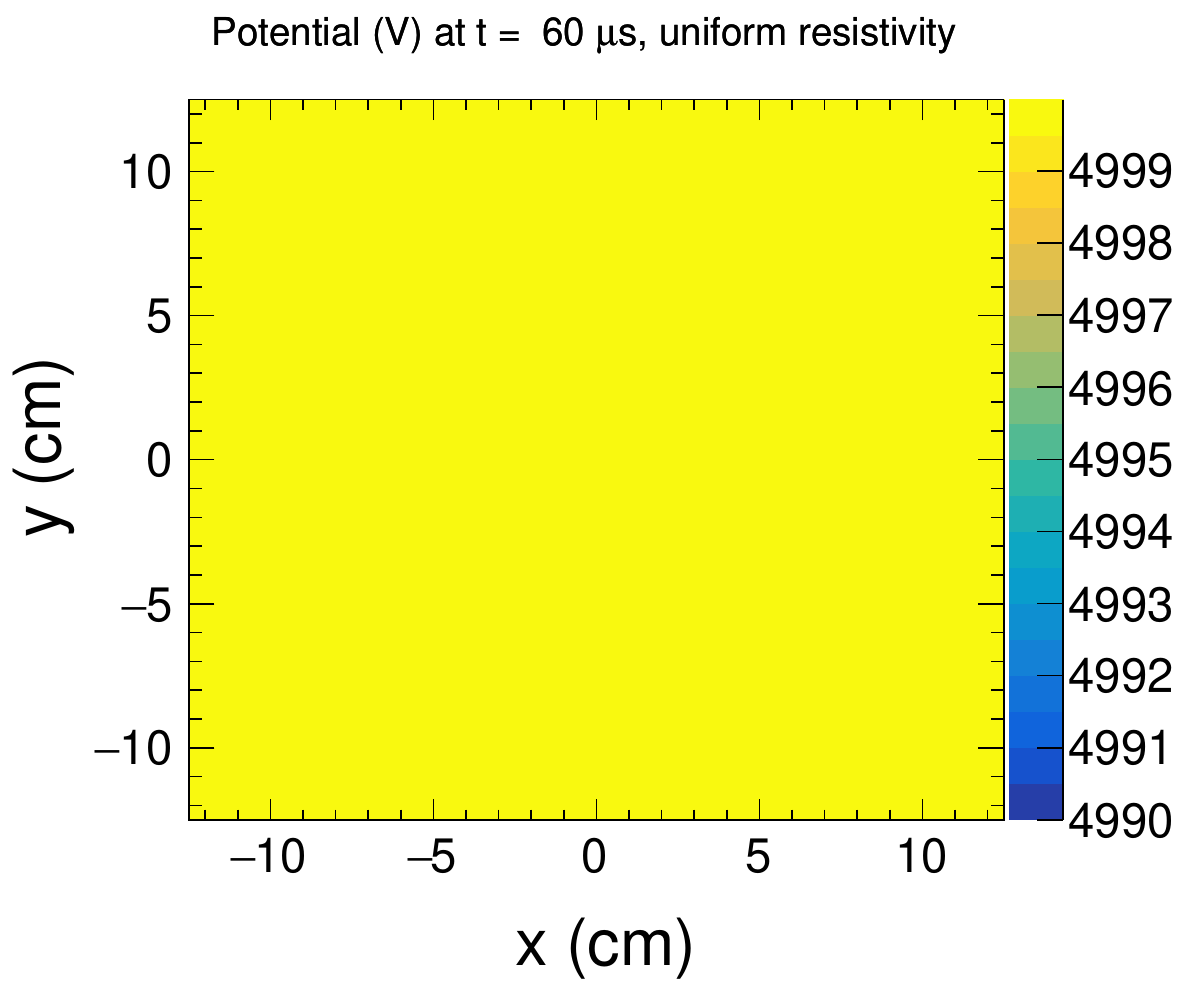}
  	\includegraphics[width=0.49\linewidth]{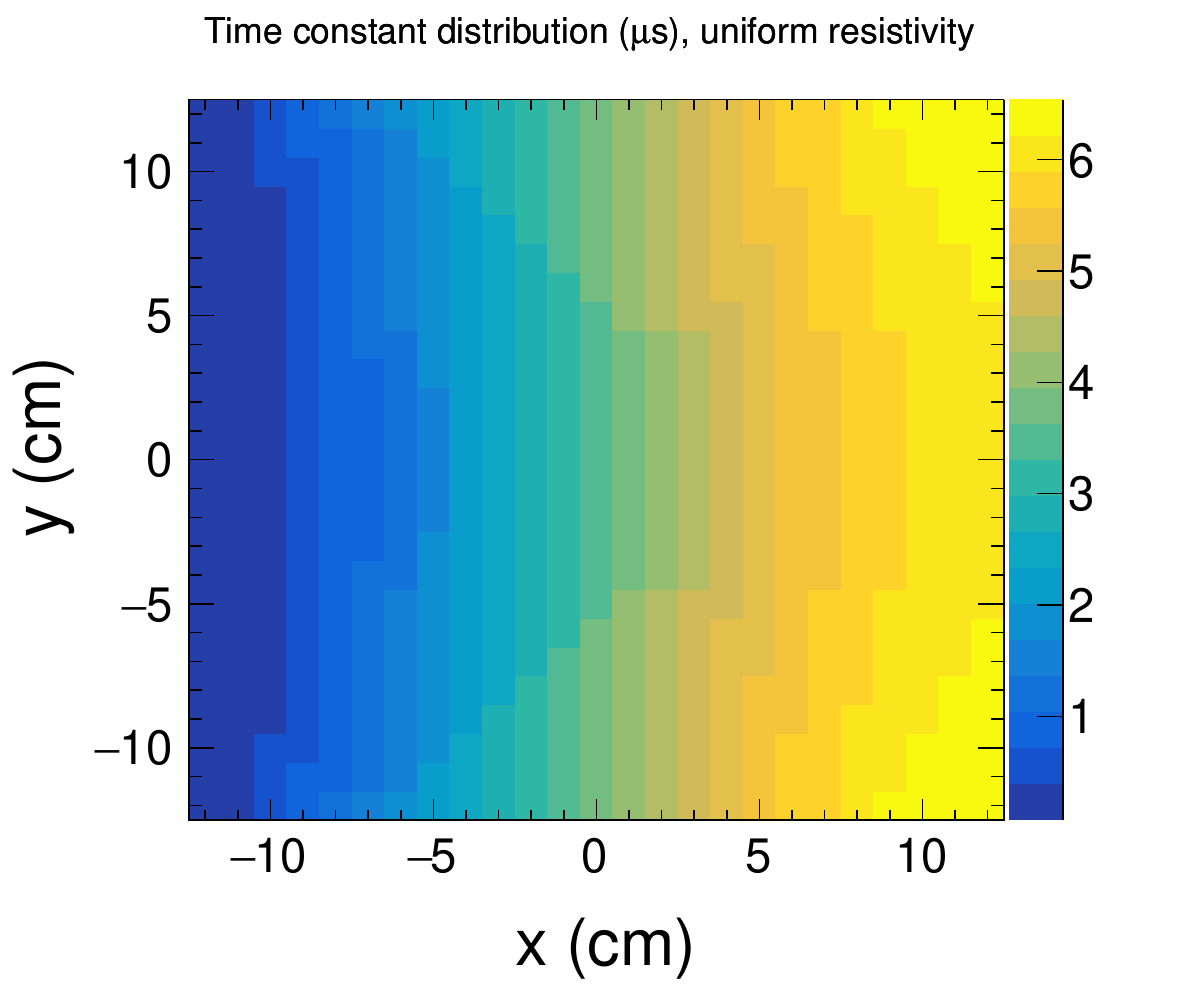}
 	\caption{Simulated 2D distribution of potential (left panel) and time constant (right panel) for uniform surface resistivity on the application of 5000 V at the left edge.}
 	\label{fig:Volt_Tau_2d_uniform}
\end{figure}
%=====================================
 
The left panel of Fig.~\ref{fig:Volt_Tau_2d_uniform} presents 2D potential distribution at the end of 60 $\mu s$, which shows that the potential at each cell reaches the same value at the end. It is interesting to note that it takes different times for every cell to saturate. We define the time constant ($\tau$) as the time required to reach a potential of $V_0(1-\exp(-t/\tau))$ where $V_0$ is the applied potential. The right panel of Fig.~\ref{fig:Volt_Tau_2d_uniform} shows the 2D distribution of the time constant, which increases as we move away from the left side along the x-direction, whereas it remains nearly constant as we move along the y-direction.

%=====================================
\begin{figure}
	\centering
	\includegraphics[width=0.49\linewidth]{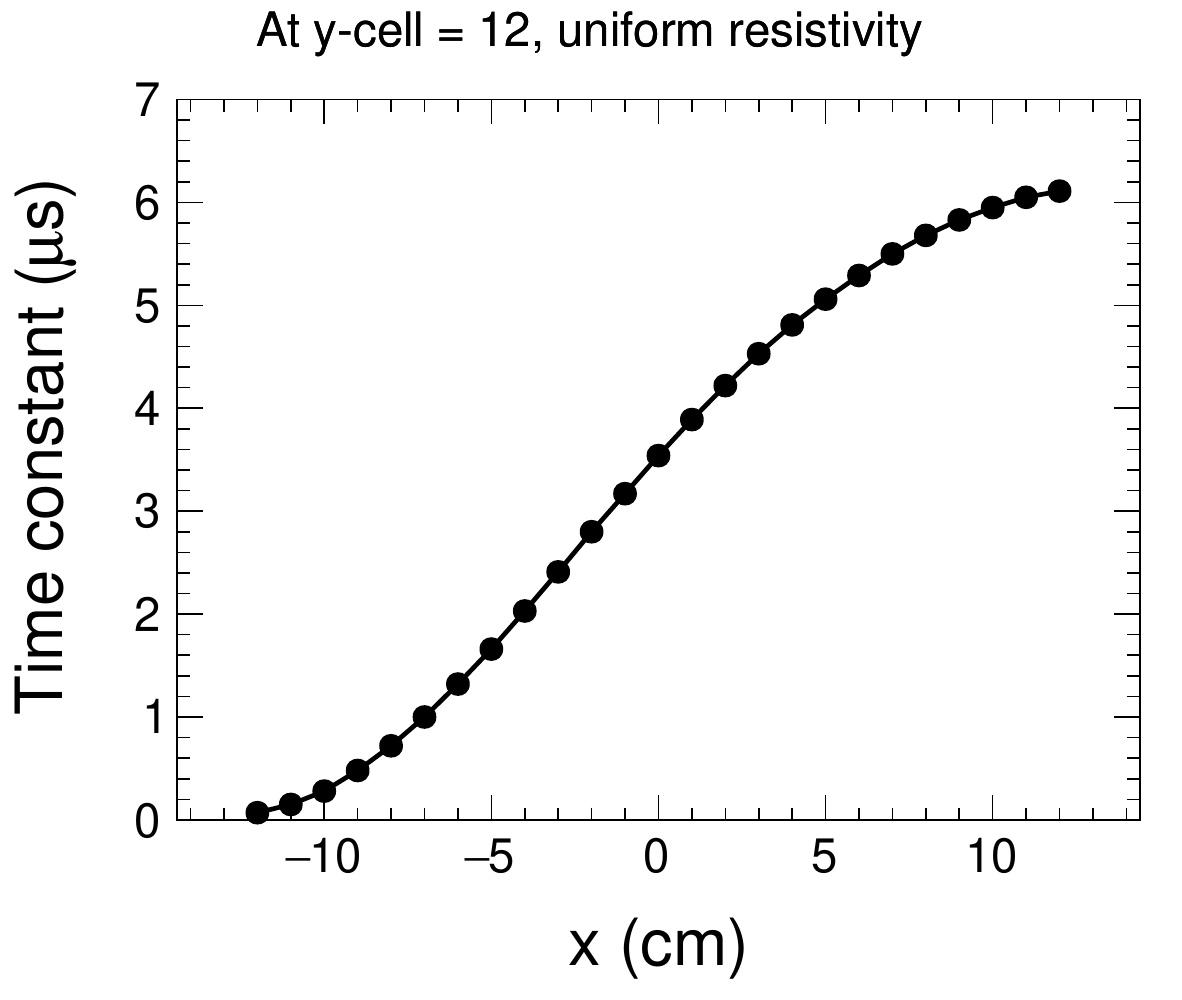}
	\includegraphics[width=0.49\linewidth]{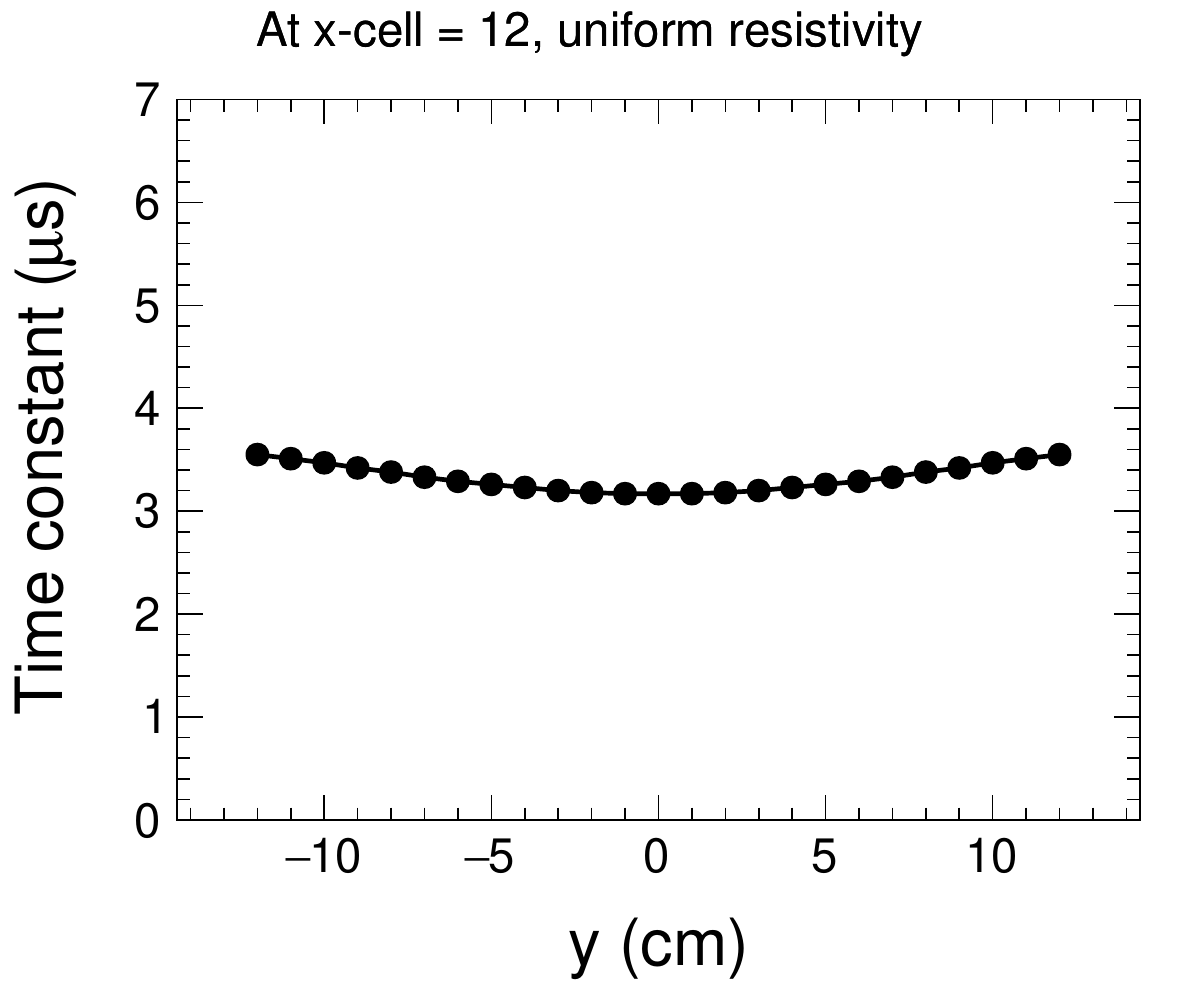}
	\caption{Simulated time constant as functions of cell positions in x (left panel) and y (right panels) directions for uniform surface resistivity on the application of 5000 V at the left edge.}
	\label{fig:Tau_Vs_xCell_yCell_uniform}
\end{figure}
%=====================================

The left panel of Fig.~\ref{fig:Tau_Vs_xCell_yCell_uniform} depicts the time constant as nearly a linear function of cell positions in the x-direction for 12\textsuperscript{th} y-cell, which is due to an increase in the effective resistance as we move away from the left side. The right panel of Fig.~\ref{fig:Tau_Vs_xCell_yCell_uniform} portrays the time constant as a function of cell positions in the y-direction for 12\textsuperscript{th} x-cell where the time constant remains nearly same for all y-cells. A slight depression can be due to variation in the effective resistance of the path caused by geometry, where the resistance along the effective path will be higher for cells around the edges ($y = \pm 12.5$) because of a longer average path from the left edge.

\begin{figure}
	\centering
	\includegraphics[width=0.49\linewidth]{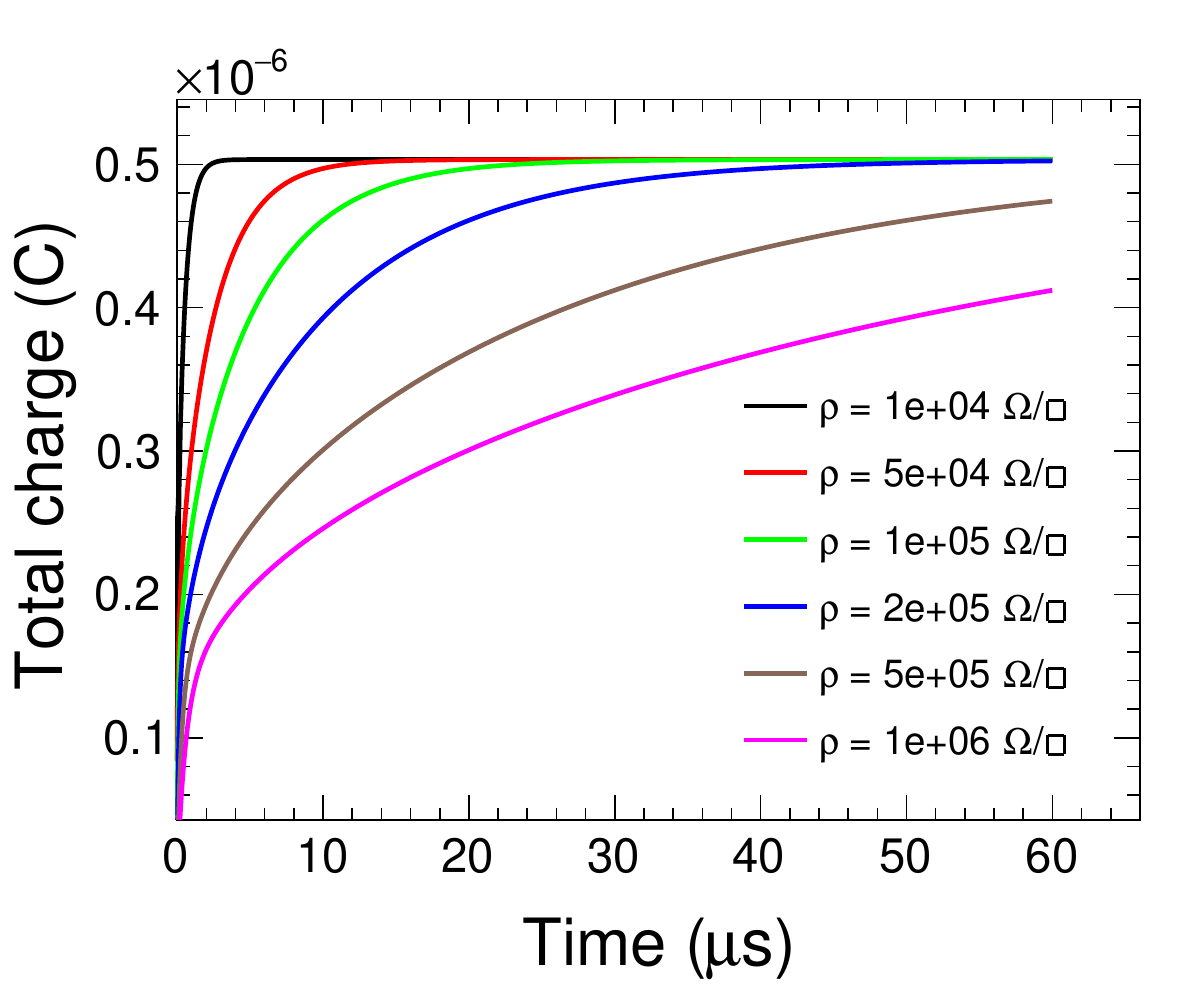}
	\includegraphics[width=0.49\linewidth]{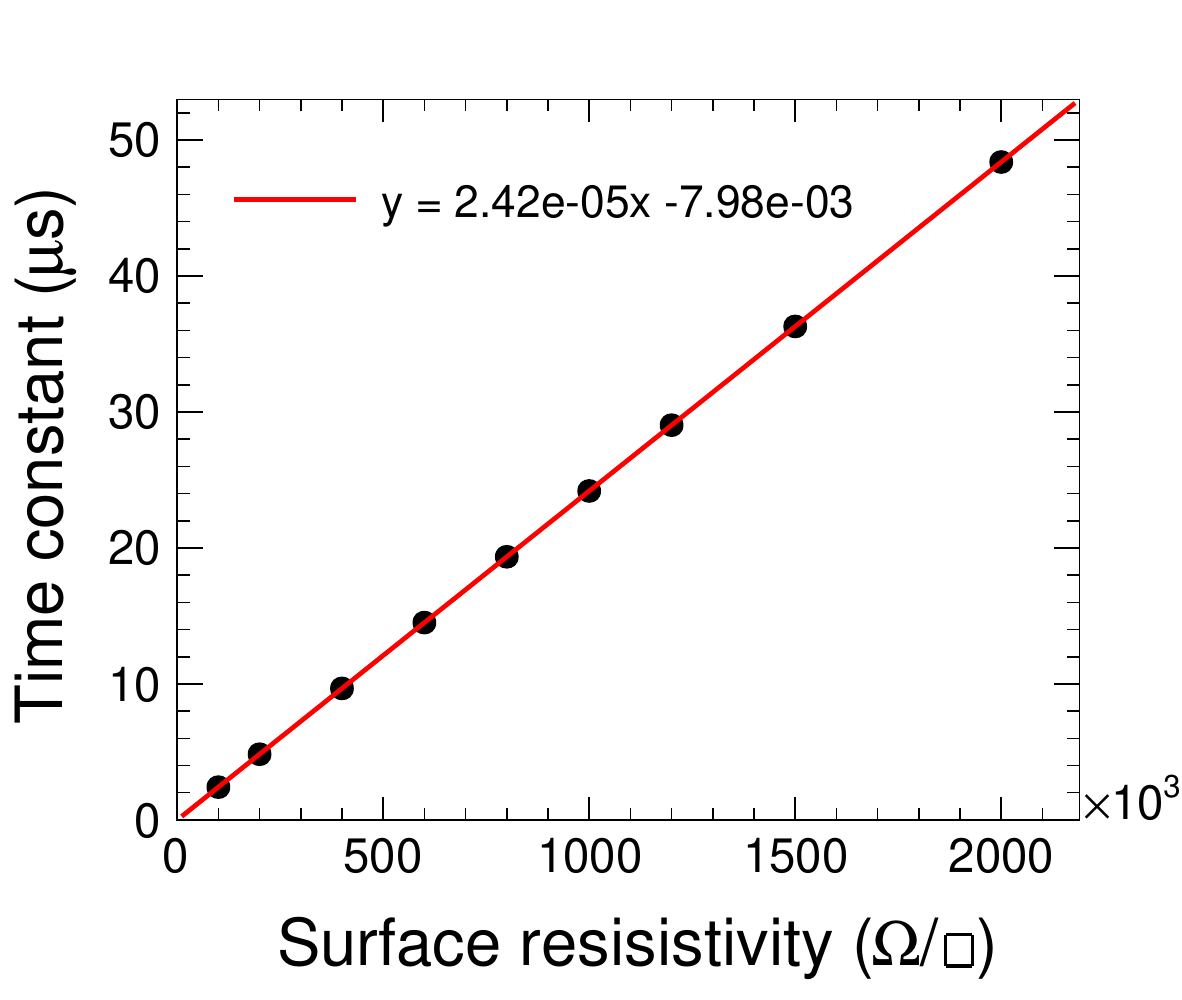}
	\caption{Simulated charging behavior for different surface resistivities in the left panel. Simulated time constant as a function of surface resistivity in the right panel.}
	\label{fig:Resistivity_variation}
\end{figure}

The right panel of Fig.~\ref{fig:Volt_Tau_2d_uniform} indicates that the time constant depends on the effective resistance. To verify that, we simulate charge buildup for various surface resistivities of the resistive layer of size $25\times 25$ cm\textsuperscript{2} and relative permittivity of 10. The left panel Fig.~\ref{fig:Resistivity_variation} shows the total charge as a function of time for various surface resistivities. We observe that the total charge at saturation is the same for all the cases of resistivities, and the charging rate decreases with the surface resistivity. 

%=====================================
%=====================================

The right panel of Fig.~\ref{fig:Resistivity_variation} shows the time constant as a linear function of surface resistivity where the time constant is defined as the time required for the total charge to reach $(1-\exp(-t/\tau))$ of total charge at saturation. Thus, we can deduce that the time constant is linearly proportional to the surface resistivity, and it should also increase with the resistance along the effective path. 

%=====================================
\section{Experimental Measurement of Surface Resistivity of Graphite Layer}
\label{sec:exp_res}
%=====================================

%=====================================
\begin{figure}
	\centering
	\includegraphics[width=0.49\linewidth]{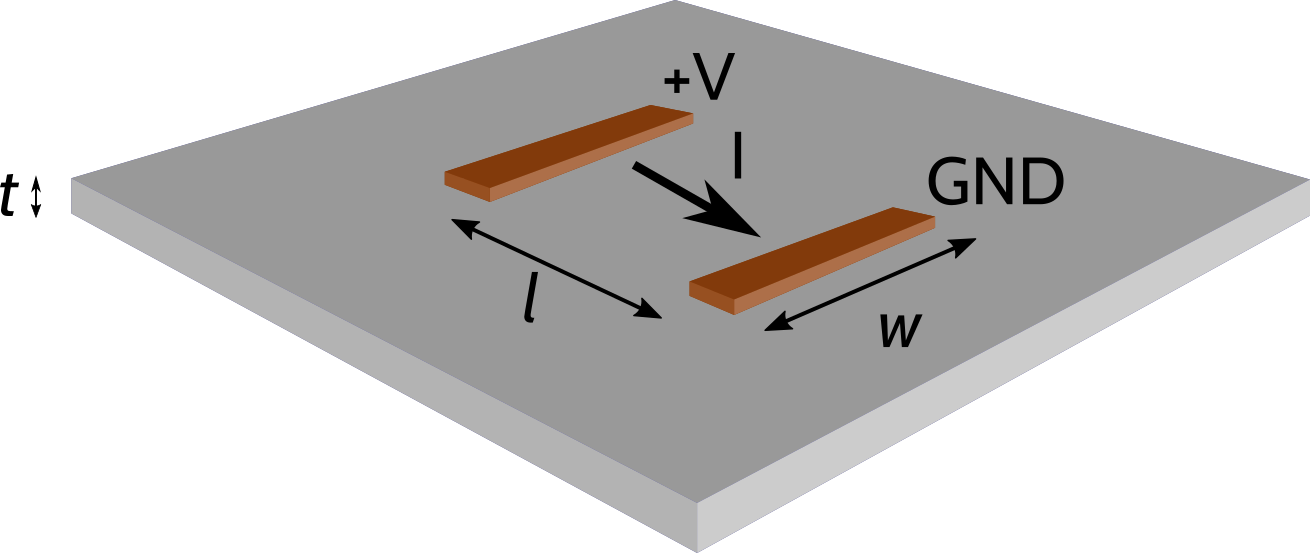}
	
	\caption{Measurement of surface resistivity of a square region of the resistive layer.}
	\label{fig:box_resistivity}
\end{figure}
%=====================================

The surface resistivity of a sheet can be obtained using surface resistance of a square region~\cite{Anant_aggarwal:2006}. Consider a sheet of thickness $t$, length $l$ and width $w$ as shown in fig. \ref{fig:box_resistivity}. The current is flowing perpendicular to $w$ on the surface. The resistance can be related to bulk resistivity as following 
\begin{align}
R = \rho \frac{l}{A} = \rho \frac{l}{w t}.
\end{align}
For a square region, $l = w$,
\begin{align}
R = \frac{\rho}{t} = R_s
\end{align}
where, $R_s$ is the surface resistivity with unit $\Omega/\Box$. Thus, the resistance measured by a square zig will be equal to the surface resistivity of the graphite layer. 

%=====================================
\begin{figure}
	\centering
	\includegraphics[width=0.46\linewidth]{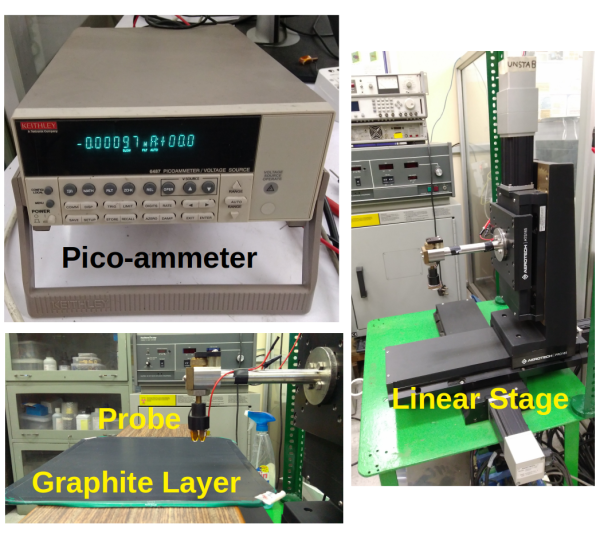}
	\includegraphics[width=0.52\linewidth]{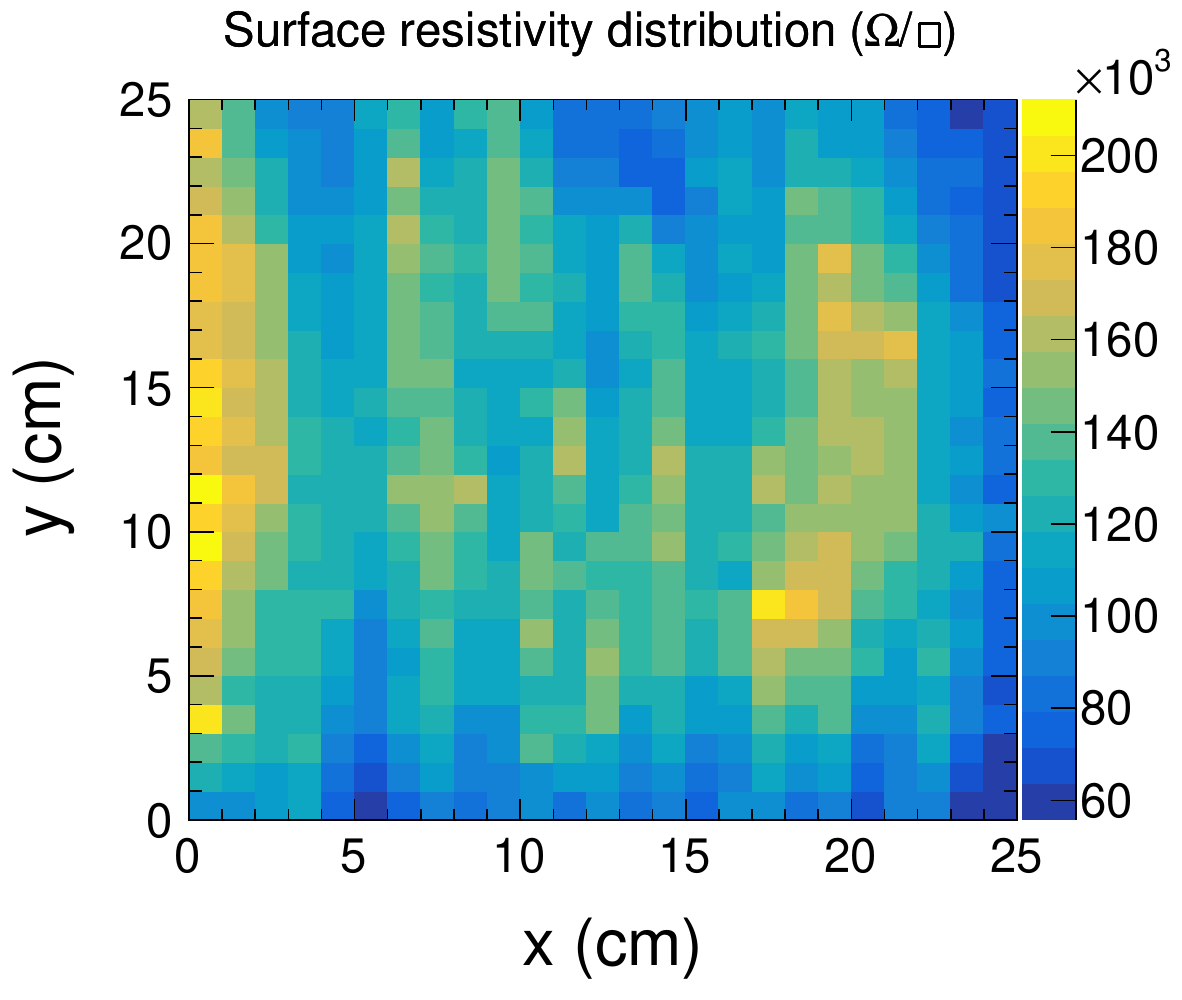}
	\caption{Experimental setup to measure surface resistivity by the application of constant potential on the resistive layer of graphite in the left panel~\cite{Kumar:2021dka}. Experimentally measured surface resistivity of graphite layer of size $25\times25~\text{cm}^2$ in the right panel.}
	\label{fig:resistivity_measurements_setup}
\end{figure}
%=====================================

We use KEITHLEY 6487 picoammeter to measure the surface resistivity where a constant potential is applied across a square zig, and a current flowing through the surface is measured. The ratio of potential and current gives us the resistance, which is equal to the surface resistivity.

The square zig probe is mounted on the AEROTECH PRO165 linear stage as shown in the left panel of Fig.~\ref{fig:resistivity_measurements_setup}. This linear stage moves in XYZ direction with high precision and can be controlled through an AEROBASIC programming language. The picoammeter and linear stage are synchronized using GPIB and serial (RS232) interface, respectively, using python program which also performs data logging as the probe moves to various cells and touches the surface.

We measure the surface resistivity of graphite layer of size $25\times25$ cm\textsuperscript{2} with 625 cells of $1 ~ \text{cm}^2$ each using the experimental setup shown in the left panel of Fig.~\ref{fig:resistivity_measurements_setup}. The graphite layer is coated on the bakelite plate of thickness 3.3 mm using spray paint. The thickness of the graphite layer in our sample is found to have a variation in the range of 10 to 30 $\mu$m. A constant potential of 10 V is applied across the square zig of size $1 ~ \text{cm}^2$. The experimentally measured 2D distribution of surface resistivity is shown in the right panel of Fig.~\ref{fig:resistivity_measurements_setup}. We can observe that the surface resistivity varies in the range of about 60 to 200 k$\Omega/\Box$. In the next section, we use this experimentally measured non-uniform surface resistivity as an input to our simulation. 

%=====================================
\section{Simulation of Potential Buildup for Non-uniform Resistivity}
\label{sec:sim_nonuniform_res}
%=====================================

Now, we describe the simulation of potential buildup, which is performed using experimentally measured non-uniform surface resistivity as input. We apply potential of 5000 V at the left edge of the resistive layer having size of $25\times25$ cm\textsuperscript{2} with 625 cells of $1 ~ \text{cm}^2$ each. The relative permittivity of the graphite layer is taken as 10. 

%=====================================
\begin{figure}
	\centering
	\includegraphics[width=0.49\linewidth]{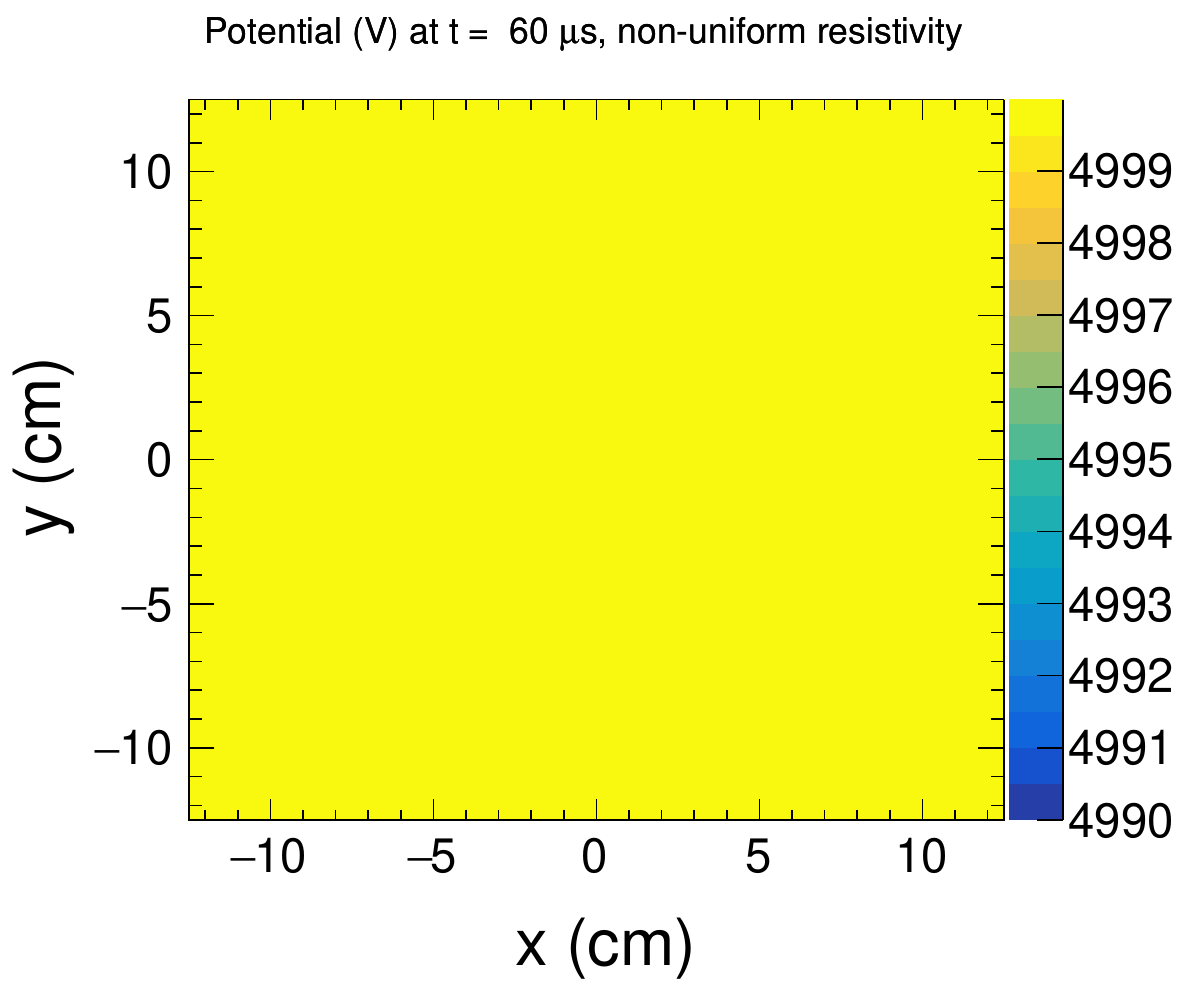}
	\includegraphics[width=0.49\linewidth]{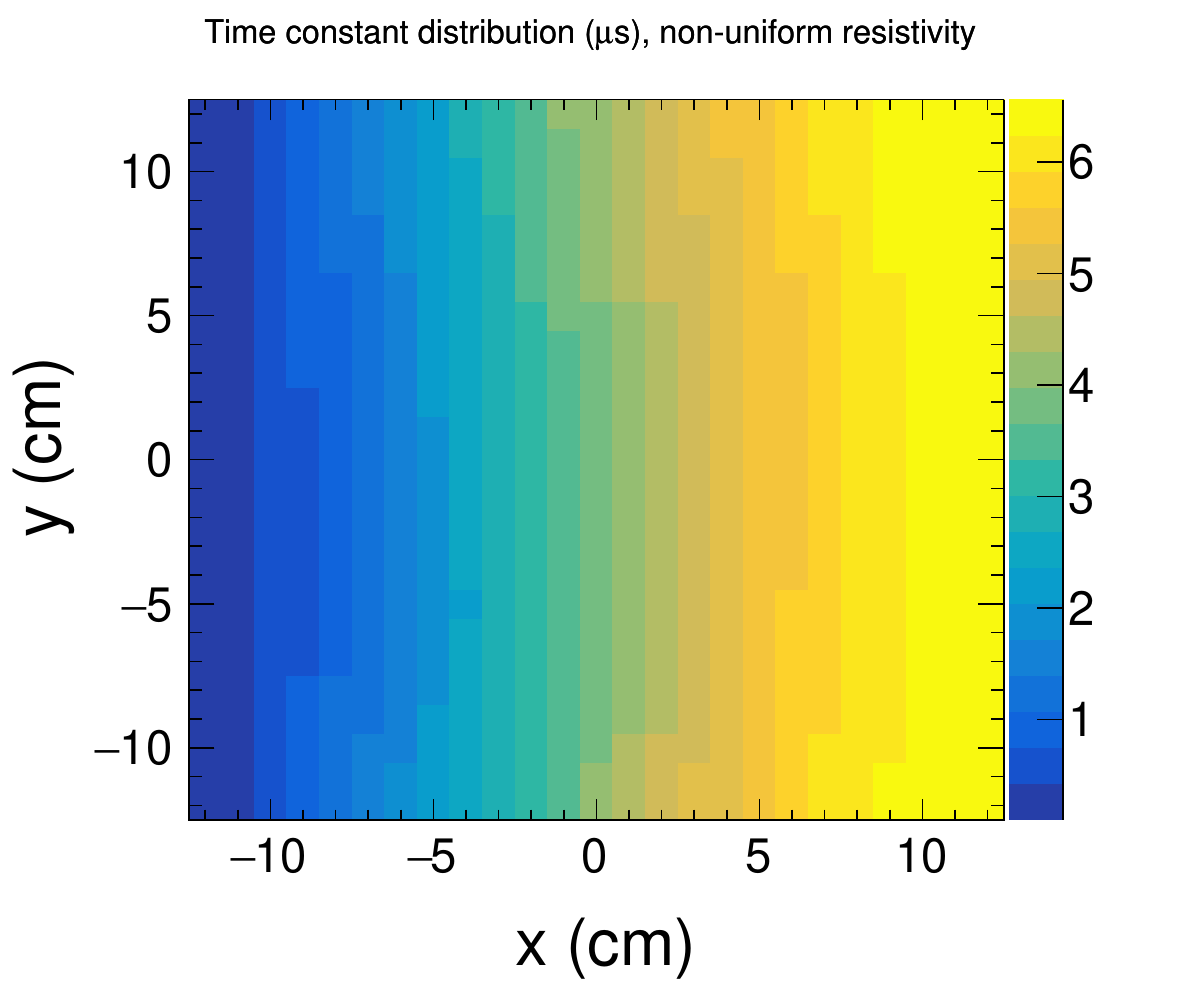}
	\caption{Simulated 2D distribution of potential (left panel) and time constant (right panel) for experimentally measured non-uniform surface resistivity on application of 5000 V at left edge.}
	\label{fig:Volt_Tau_2d_nonuniform}
\end{figure}
%=====================================

%=====================================
\begin{figure}
	\centering
	\includegraphics[width=0.49\linewidth]{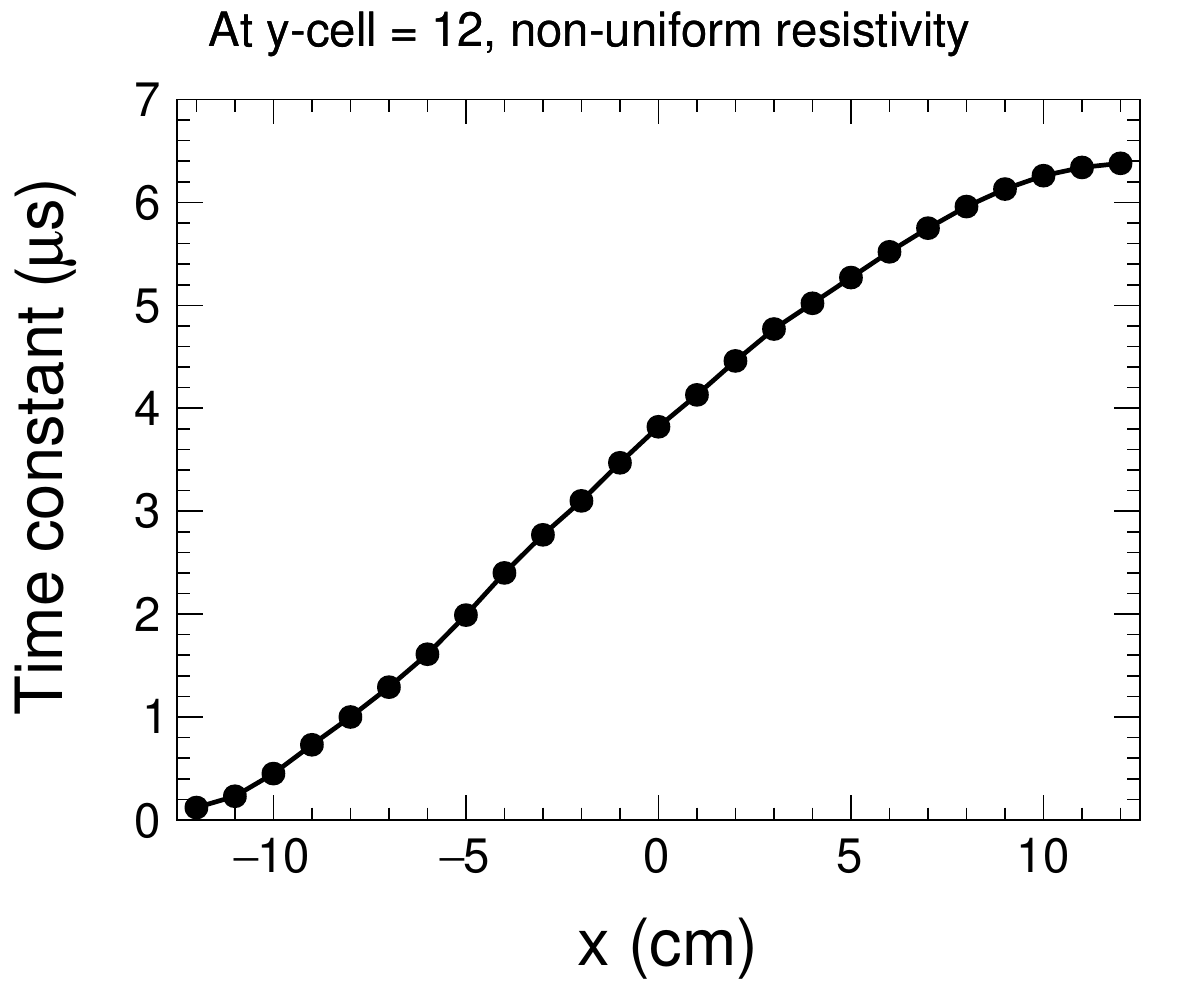}
	\includegraphics[width=0.49\linewidth]{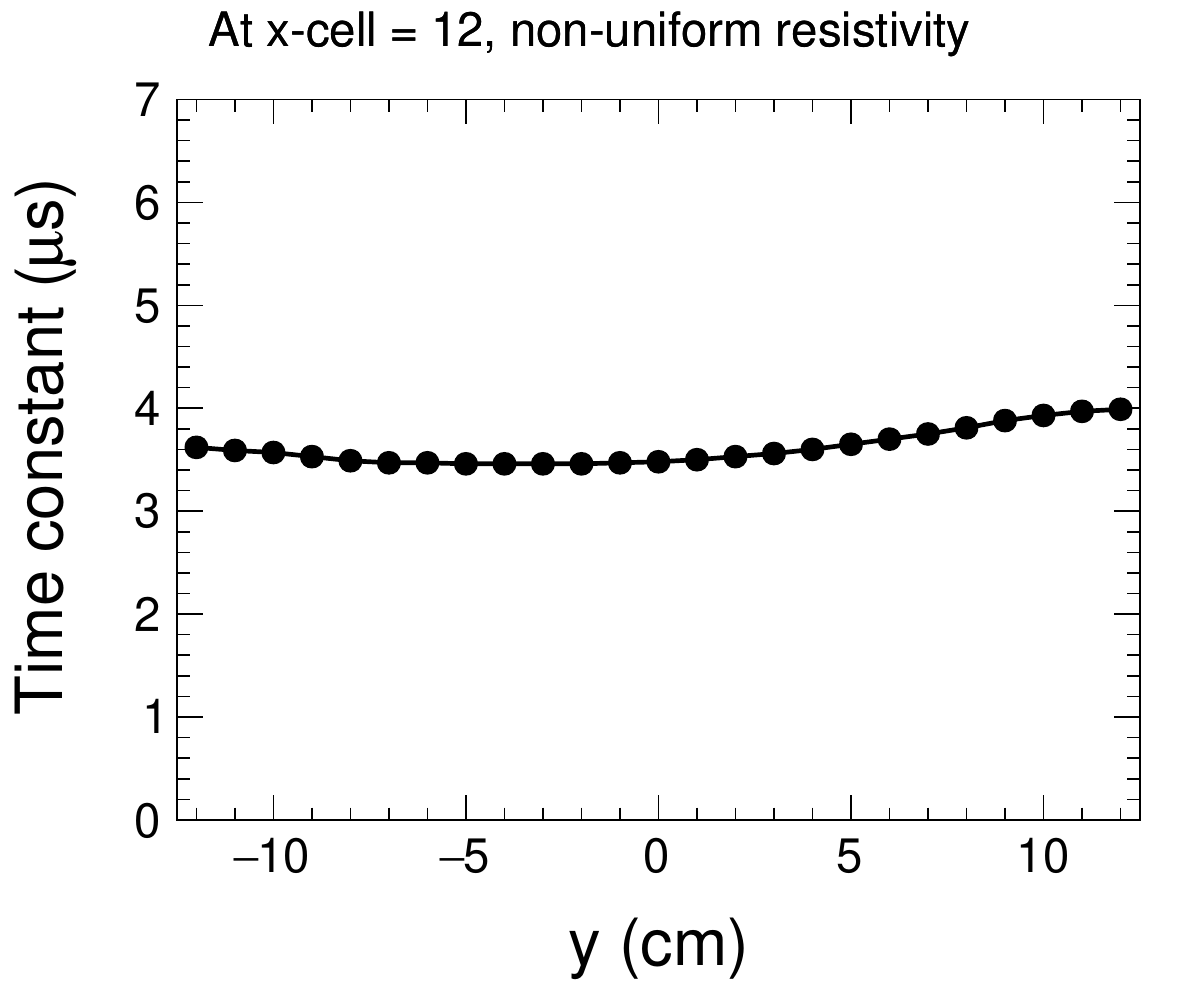}
	\caption{Simulated time constants as functions of cell locations in x (left panel) and y (right panel) directions for experimentally measured non-uniform surface resistivity on the application of 5000 V at the left edge.}
	\label{fig:Tau_Vs_xCell_yCell_nonuniform}
\end{figure}
%=====================================

The left panel of Fig.~\ref{fig:Volt_Tau_2d_nonuniform} shows the simulated 2D potential distribution for experimentally measured non-uniform surface resistivity at the end of 60 $\mu$s. We can see that the potential distribution is constant irrespective of the non-uniformity of surface resistivity. The right panel of Fig.~\ref{fig:Volt_Tau_2d_nonuniform} presents the simulated 2D distribution of time constant where small asymmetry can be observed compared to that in the case of uniform surface resistivity as shown in the right panel of Fig.~\ref{fig:Volt_Tau_2d_uniform}. The left and right panels of Fig.~\ref{fig:Tau_Vs_xCell_yCell_nonuniform} show the simulated time constants as functions of cell locations in x and y directions, respectively. Small fluctuation can be observed in both the panels of Fig.~\ref{fig:Tau_Vs_xCell_yCell_nonuniform} due to non-uniformity in surface resistivity. It can be deduced that the non-uniformity does not disturb the potential distribution, but it may introduce some fluctuations in the distribution of the time constant. In the next section, we discuss the experimental measurements of potential and time constant for the graphite layer to verify the results obtained from the simulation.

%=====================================
\section{Experimental Measurement of Time Constant}
\label{sec:exp_tau} 
%=====================================

The charging behavior of the graphite layer can be studied experimentally by applying a square pulse at the left edge and observing the response at the other cells. Since the surface resistivity of the graphite layer is comparable to the internal impedance of typical oscilloscopes (1 M$\Omega$) and multimeters (10 M$\Omega$), the measurement of potential using these probes introduces errors and results in a larger time constant. The internal impedance of the probe should be very large compared to that of the system in order to ensure accurate measurements.

\begin{figure}
	\centering
	$\vcenter{\hbox{\includegraphics[width=0.43\linewidth]{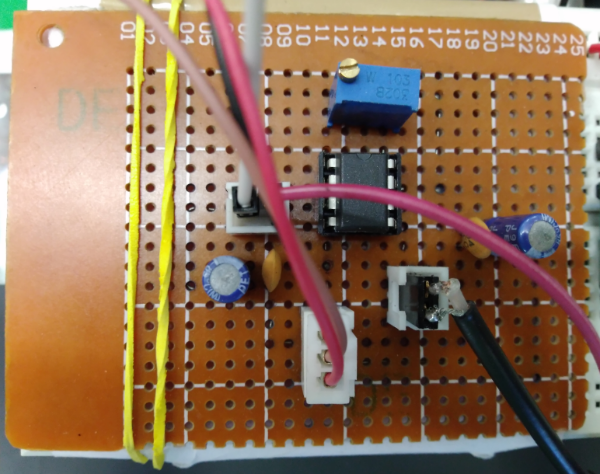}\vspace{5mm}}}$
	\caption{The high impedance circuit to experimentally measure the distribution of potential and time constant.}
	\label{fig:timeconstantexpsetup}
\end{figure}

For this purpose, we select an operational amplifier (op amp) AD845, which has a very high input impedance ($10^{11}$ k$\Omega$) and very low output impedance. The slew rate of this op amp is large enough (100 V/$\mu$s) to faithfully reproduce the input timing behavior at the output. We use this op amp as a voltage follower circuit where output follows the input and gain is unity~\cite{Anant_aggarwal:2006}. The output of this voltage follower circuit can be connected to the oscilloscope or voltmeter. Figure~\ref{fig:timeconstantexpsetup} shows this voltage following circuit for experimental measurement of potential and time constant. We have experimentally assured that the presence of op amp does not affect the input signal by confirming the agreement of a pulse generated through a function generator and measurement of its time constant.

\begin{figure}
	\centering
	\includegraphics[width=0.49\linewidth]{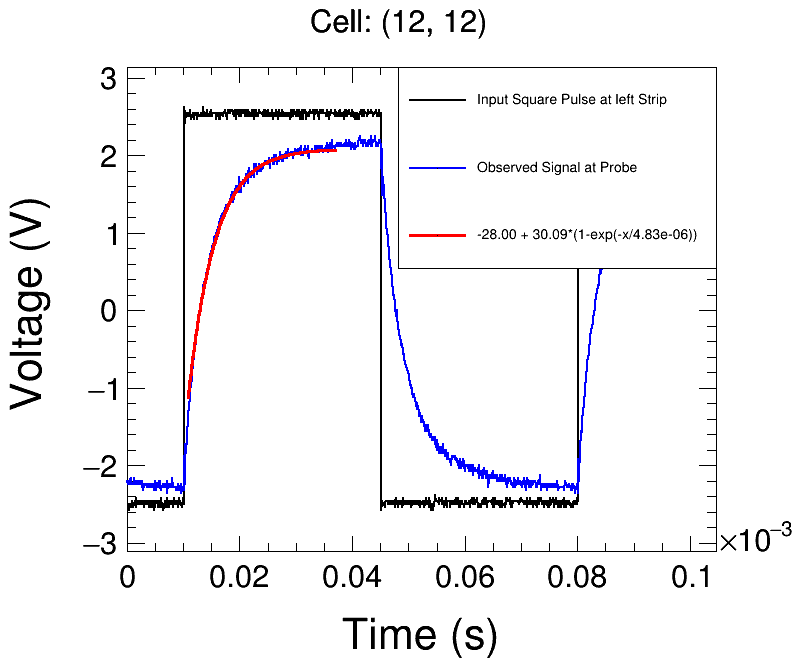}
	\includegraphics[width=0.49\linewidth]{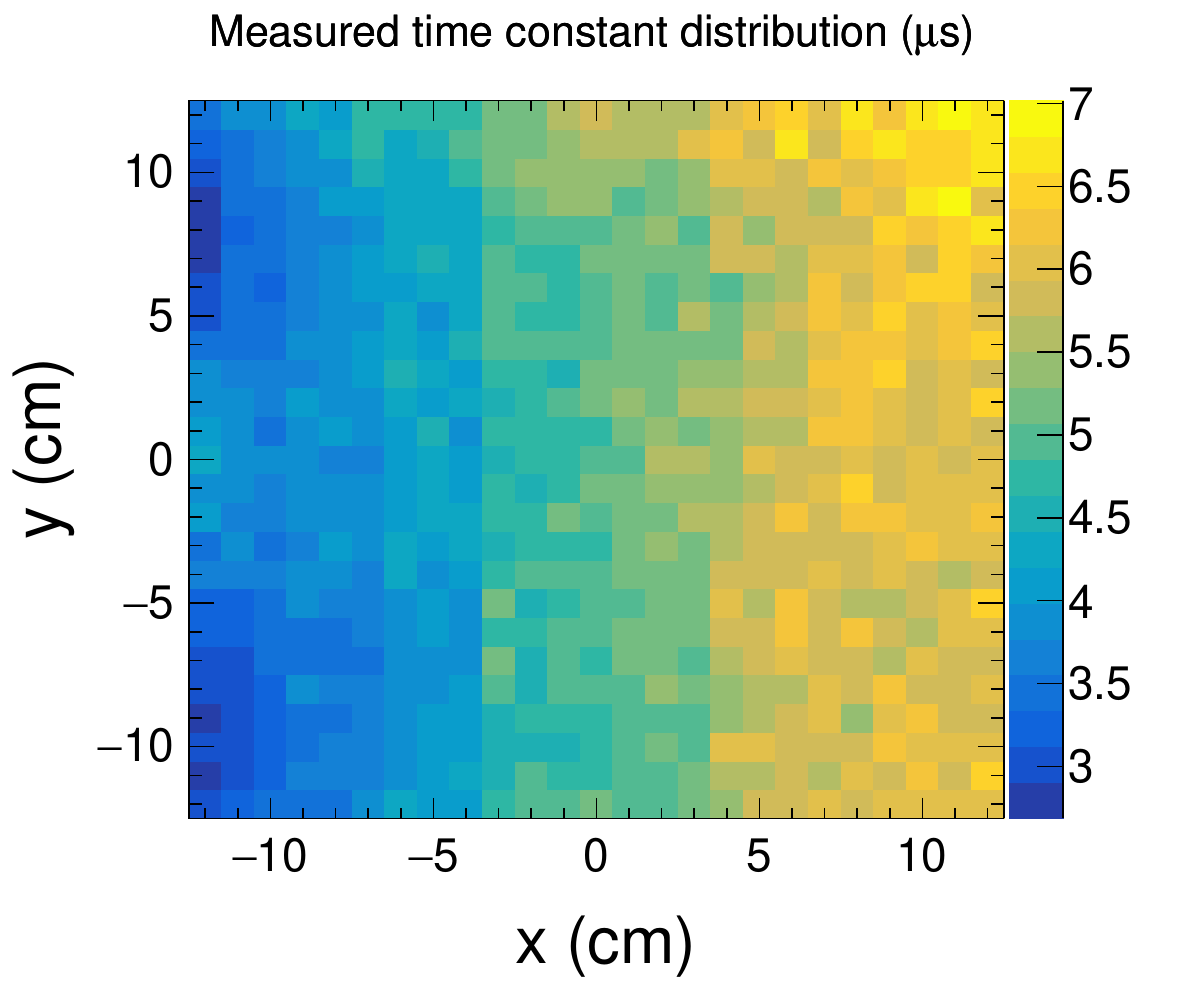}
	\caption{The left panel shows experimentally measured charging and discharging signal at cell (12,12) on the application of square pulse on the left edge of the graphite layer. The right panel shows the 2D distribution of the experimentally measured time constant.}
	\label{fig:exp_tau_25_25}
\end{figure}

The left panel of Fig.~\ref{fig:exp_tau_25_25} shows the charging and discharging potential at cell (12,12) around the center of the graphite layer on the application of square pulse on the left edge. The black and blue curves show the input pulse and the experimentally observed signal, respectively. The experimentally observed charging signal is fitted with the function $V_0(1-\exp(-t/\tau))$ as shown by the red curve where the time constant $\tau$ is obtained after fitting. The distribution of the fitted time constant for each cell has been shown in the right panel of Fig.~\ref{fig:exp_tau_25_25}. This experimentally measured time constant is distributed in the range of 3 to 7 $\mu$s, which is in agreement with the range of time constant obtained from simulation (1 to 6 $\mu$s) as shown in the right panel of Fig.~\ref{fig:Volt_Tau_2d_nonuniform}. 

\begin{figure}
	\centering
	\includegraphics[width=0.49\linewidth]{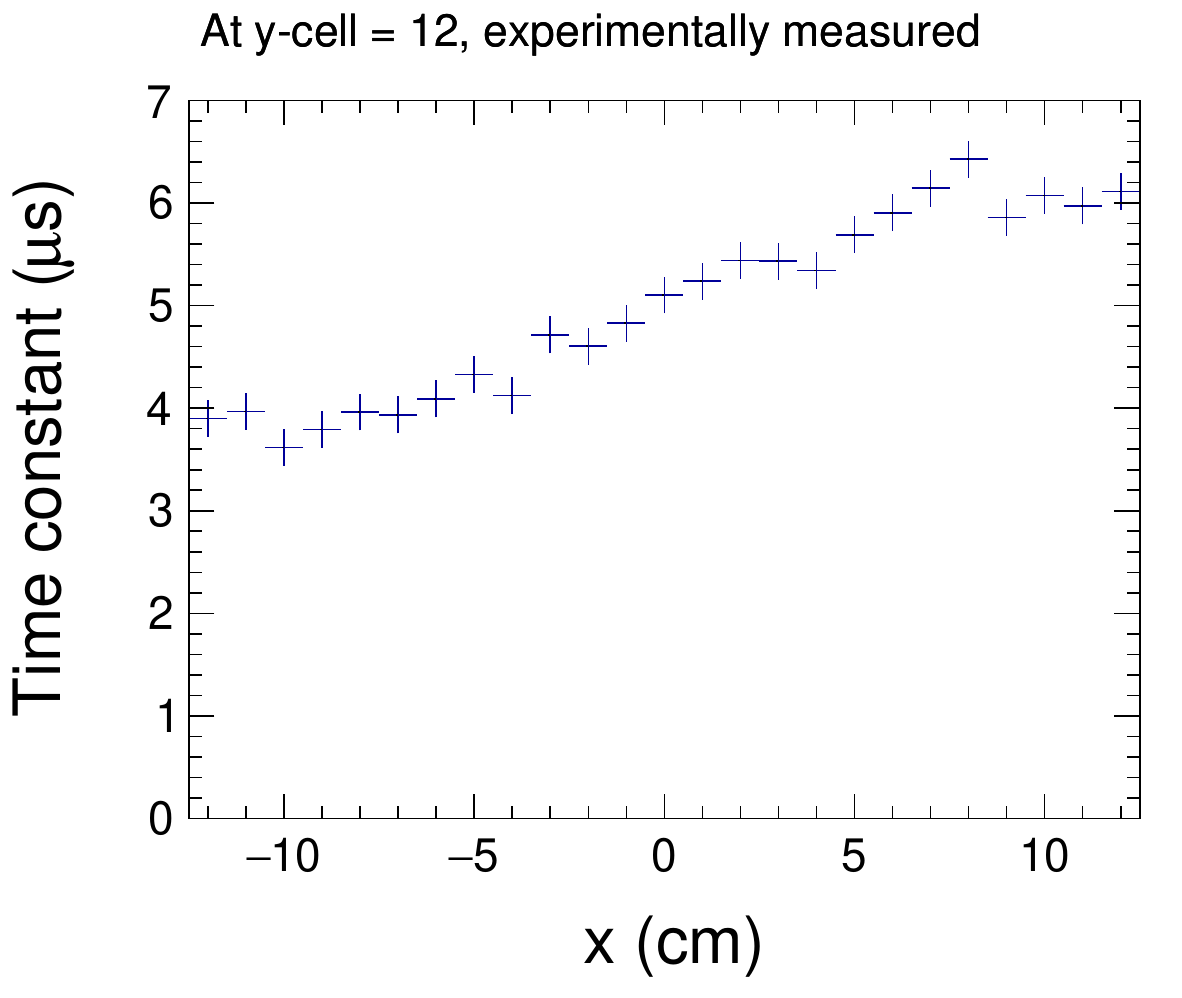}
	\includegraphics[width=0.49\linewidth]{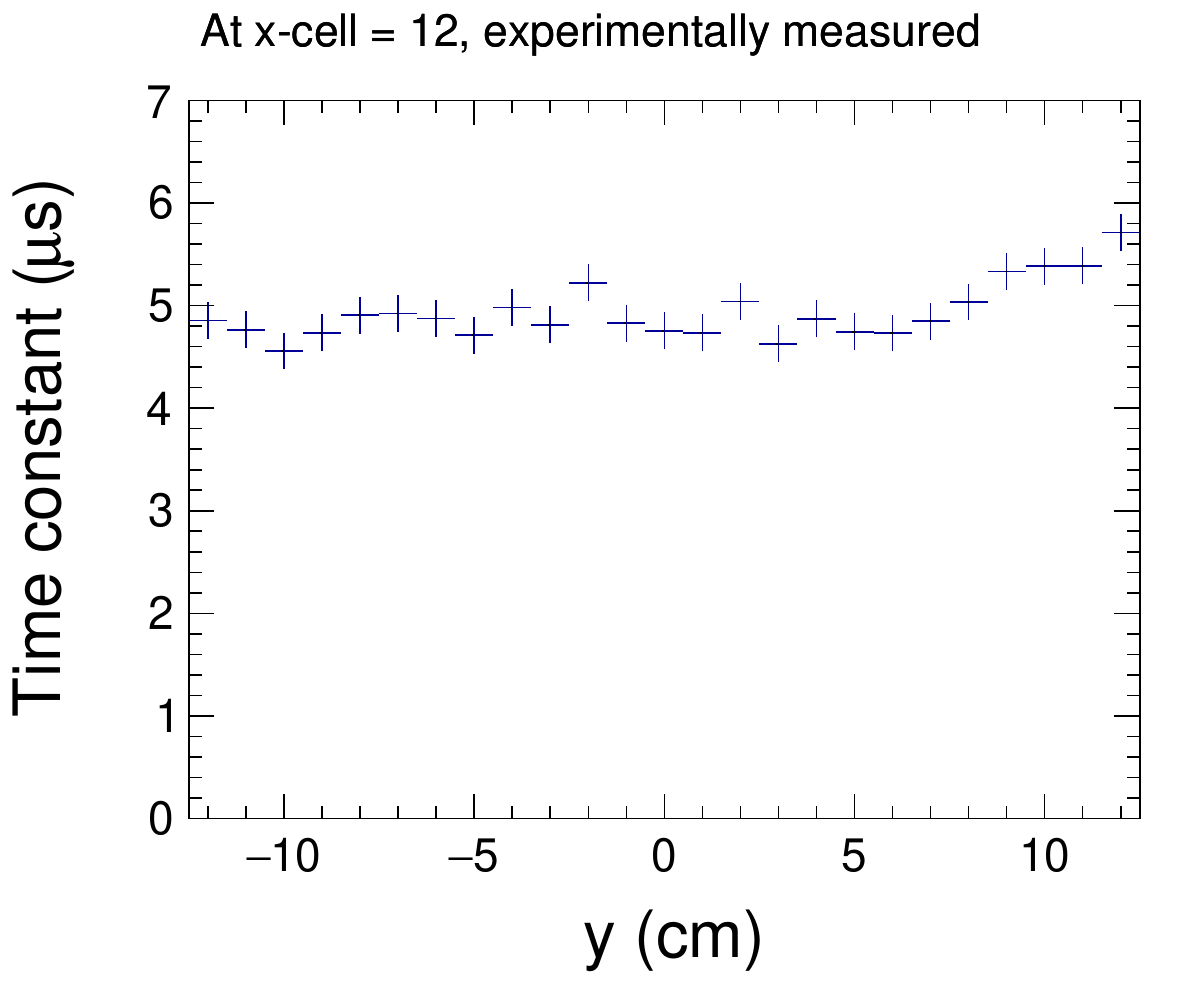}
	\caption{Experimentally measured time constants as functions of cell locations in x (left panel) and y (right panel) directions on the application of square pulse on the left edge of the graphite layer.}
	\label{fig:exp_tau_Vs_xCell_yCell}
\end{figure}

The measured time constant is also increasing linearly, as we are moving away from the region of application of square pulse along the x-direction as shown in the left panel of Fig.~\ref{fig:exp_tau_Vs_xCell_yCell}. On the other hand, the right panel of Fig.~\ref{fig:exp_tau_Vs_xCell_yCell} shows that the measured time constant is almost independent of the cell locations along the y-direction. This behavior is similar to what we observed for the simulated time constant, as shown in Fig.~\ref{fig:Tau_Vs_xCell_yCell_nonuniform}. The measured time constants for the cells around the right edge are in good agreement with those for simulation, but the measurement time constants for the cells around the left edge seem to be larger than those for simulation. The possible reason for this may be the difference in the resistance of the connector and the resistive layer.  Now, we describe the experimental measurement of the potential distribution in the next section. 

%=====================================
\section{Experimental Measurement of Potential Distribution}
\label{sec:exp_potential} 
%=====================================

\begin{figure}
	\centering
	$\vcenter{\hbox{\includegraphics[width=0.41\linewidth]{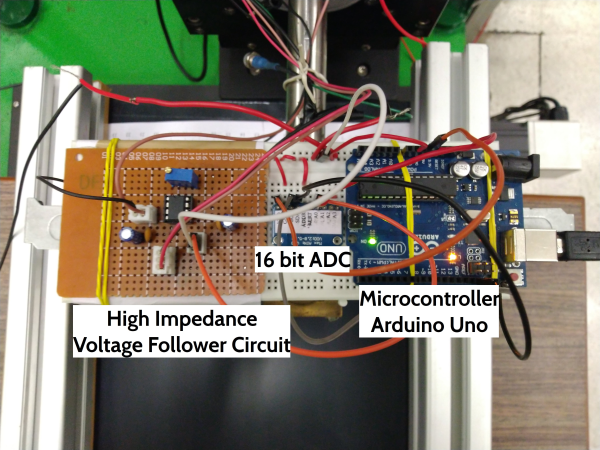}\vspace{4mm}}}$
	$\vcenter{\hbox{\includegraphics[width=0.58\linewidth]{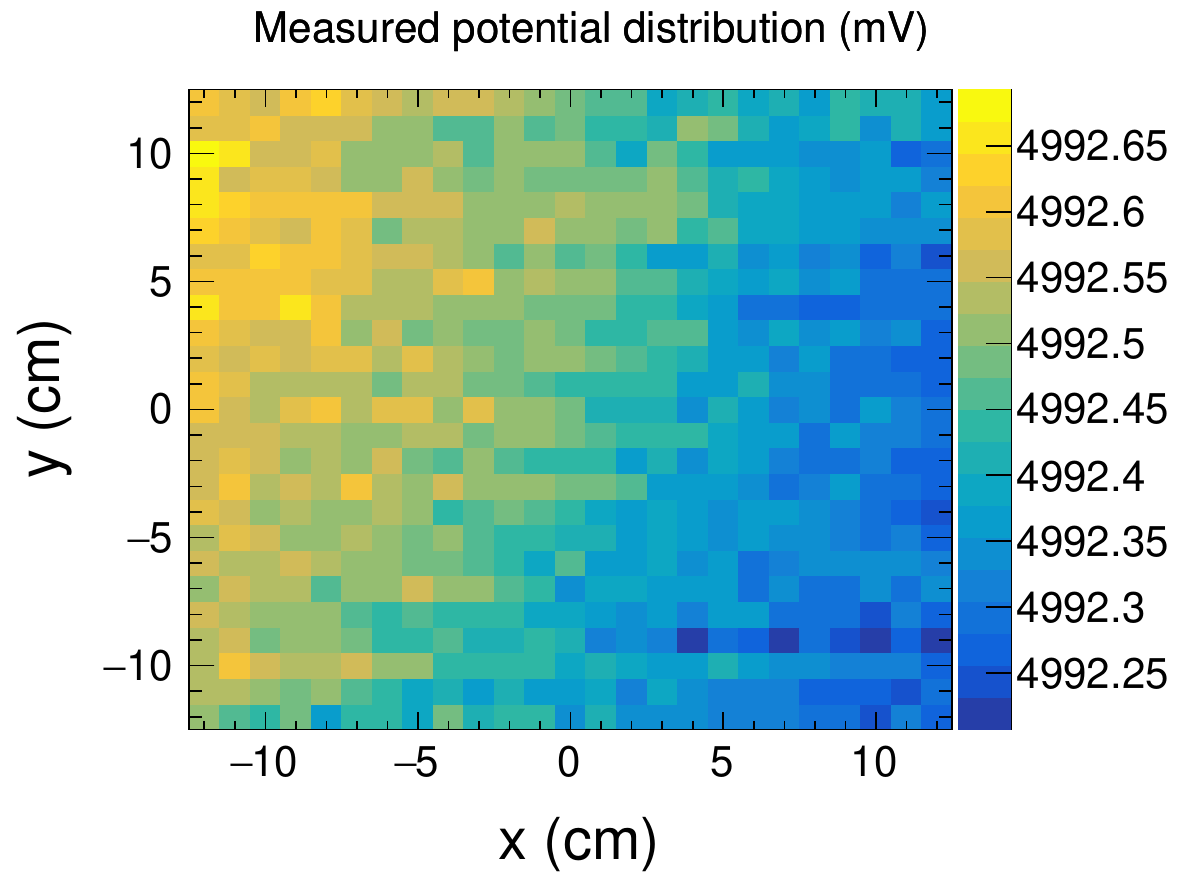}}}$
	\caption{The left panel shows the experimental circuit to measure the potential distribution. The right panel illustrates the experimentally measured potential distribution across the graphite layer on the application of 5 V on the left edge.}
	\label{fig:exp_volt_25by25_l5v}
\end{figure}

The input impedance of the multimeter (10 M$\Omega$) is comparable to the resistivity of the graphite layer, which results in the measured potential being always less than the true potential across the graphite layer. In section \ref{sec:exp_tau}, we have explained the use of op amp voltage follower circuit as a suitable solution because this has high input impedance. This circuit saturates at large input voltages (about 12 V). We use 16-bit ADC ADS1115 to measure input voltage and perform data logging using the micro-controller Arduino Uno ATmega328. This ADC has an input range of 0 to 5 V with a least count of 0.18 mV. The combination of op amp voltage follower circuit, ADC, and micro-controller acts as an effective high impedance voltage probe as shown in the left panel of Fig.~\ref{fig:exp_volt_25by25_l5v}.

%=====================================
%=====================================

This high impedance voltage probe is synchronized with the linear stage using a python program in a similar method as explained in section \ref{sec:exp_tau}. A potential of 5 V is applied on the left edge of the graphite layer, and the linear stage moves with high precision to each cell. The voltage is measured and recorded once the probe touches the surface of the graphite layer. The right panel of Fig.~\ref{fig:exp_volt_25by25_l5v} shows the experimentally measured potential distribution on graphite layer. This experimentally measured potential is distributed in the range of 4992.25 to 4992.65 mV. This fluctuation is less than 0.01\%, which indicates that the potential distribution is quite uniform, as seen for the simulated potential distribution in the left panel of Fig.~\ref{fig:Volt_Tau_2d_nonuniform}. The right panel of Fig.~\ref{fig:exp_volt_25by25_l5v} has a diagonal distribution that seems to indicate that the top left corner may be better connected than the bottom left.

%=====================================
\section{Summary}
\label{sec:rpc_summary}
%=====================================

The resistive plate chambers have the graphite coating on the resistive plate of glass or bakelite. The high electric potential applied on the graphite layer results in a uniform electric field inside the chamber. The non-uniformity of surface resistivity may affect the detector response of RPC. In this work, we explore the effect of non-uniform surface resistivity of the graphite layer on the distribution of potential and time constant. 

A ROOT-based mathematical framework using neBEM solver has been developed to simulate the charge buildup across the graphite layer of uniform and non-uniform surface resistivities. The surface resistivity of the graphite layer is measured using the linear stage and the picoammeter, which are controlled by a python program. The experimentally measured surface resistivity distribution of the graphite layer is used as an input for simulation to predict the distribution of potential and time constant. A high-impedance voltage follower circuit is designed using an operational amplifier to experimentally measure the time constant and potential distribution, which are found to be in good agreement with that of prediction from the simulation. The potential distribution is found to be uniform and independent of the fluctuations in the surface resistivity. The time constant is observed to vary in the range of 3 to 7 $\mu$s, and its distribution gets affected by the fluctuations in surface resistivity. The time constant of the resistive layer increases linearly with the surface resistivity. 

The present version of the simulation framework developed in this work can be used to study the charge flow over the surface. However, this simulation can be upgraded to simulate charge transport in the three-dimensional geometries, which will enable us to study the current flow through the bulk of the glass or bakelite of the RPC. This will allow us to simulate the seeping of charge through glass or bakelite, and the dead-time of the detector can be estimated more realistically. The effect of button spacers and side spacers can also be studied using the three-dimensional simulation. An application of time-dependent potential or charge can be used to study the propagation of the signal. We hope that this work will open a unique avenue in the simulation studies of transient current. 

\end{refsegment}

%%%%%% If you want to divide your thesis into parts %%%%%%%
%% \cleartooddpage
%% \part{Name of the part}
%%%%%%%%%%%%%%%%%%%%%%%%%%%%%%%%%%%%%%%%%%%%%%%%%%%%%%%%%%%
\cleartooddpage
\chapter{Oscillation dip and valley}
\label{chap:dip_valley}
\begin{refsegment}

It is quite remarkable to see that starting from almost no knowledge of the neutrino masses or lepton mixing parameters twenty years ago, we have been able to construct a robust, simple, three-flavor oscillation framework which successfully explains most of the data~\cite{Capozzi:2017ipn,Esteban:2018azc,NuFIT,DeSalas:2018rby}. Atmospheric neutrino experiments have contributed significantly to achieve this milestone~\cite{Learned:2019vcq,Kajita:2019bzu} by providing an avenue to study neutrino oscillations over a wide range of energies ($E_\nu$ in the range of $\sim$100 MeV to a few hundreds of GeV) and baselines ($L_\nu$ in the range of a few km to more than 12,000 km) in the presence of Earth's matter with a density varying in the range 
of 0 to 10 g/cm$^3$.

An important breakthrough in the saga of atmospheric neutrinos came in 1998 when the pioneering Super-Kamiokande (Super-K) experiment reported convincing evidence for neutrino oscillations in atmospheric neutrinos by observing the zenith angle (this angle is zero for vertically downward-going events) dependence of $\mu$-like and $e$-like events~\cite{Super-Kamiokande:1998kpq}. The data accumulated by the Super-K experiment, based on an exposure of 33 kt$\cdot$yr, showed a clear deficit of upward-going events in the zenith angle distributions of $\mu$-like events with a statistical significance of more than $6\sigma$, while the zenith angle distribution of $e$-like events did not show any significant up--down asymmetry. These crucial observations by the Super-K experiment were successfully interpreted in a two-flavor scenario assuming oscillation between $\nu_\mu$ and $\nu_\tau$, leading to the disappearance of $\nu_\mu$. In this scenario, the survival probability of $\nu_\mu$ can be expressed in the following simple fashion:
%==========================
\begin{equation}
P(\nu_\mu \rightarrow \nu_\mu) = 1 - \sin^2 2\theta_{23} 
\cdot \sin^2\left({1.27 \cdot \Delta m^2_{32} \left(\mbox{eV}^2\right) 
	\cdot {L_\nu \left(\mbox{km}\right)\over E_\nu \left(\mbox{GeV}\right)}}\right).
\label{eq:2flavor_survival}
\end{equation}
%==========================
Due to the hierarchies in neutrino mass pattern ($\Delta m^2_{21} \ll |\Delta m^2_{32}|$) and in mixing angles ($\theta_{13} \ll \theta_{12}, \theta_{23}$), the above simple two-flavor $\nu_\mu$ survival probability expression was sufficient to explain the following broad features of the Super-K atmospheric data, providing a solution of the long-standing atmospheric neutrino anomaly in terms of ``neutrino mass-induced flavor oscillations". In the limit of maximal mixing ($\theta_{23} = 45^{\circ}$), Eq.\,\ref{eq:2flavor_survival} reduces to the following simple expression:
\begin{equation}
P(\nu_\mu \rightarrow \nu_\mu) = \cos^2  \left({1.27 \cdot \Delta m^2_{32} \left(\mbox{eV}^2\right) \cdot {L_\nu \left(\mbox{km}\right)\over E_\nu \left(\mbox{GeV}\right)}}\right)\,.
\label{eq:pmumu-nu-final-omsd}
\end{equation}
%==========
\begin{itemize}	
	\item
	Around 50\% of upward-going muon neutrinos disappear after traveling long distances. For very long baselines, the above $\nu_\mu$ survival probability expression (see Eq.~\ref{eq:2flavor_survival}) approaches $1 - (1/2)\sin^22\theta_{23}$ and the observed survival probability becomes close to 0.5, suggesting that the mixing angles is close to the maximal value of 45$^\circ$.	
	\item
	The disappearance of $\nu_\mu$ events starts for the zenith angle close to the horizon indicating that the oscillation length should be around $\sim$ 400 km for neutrinos having energies close to 1 GeV. This suggests that the atmospheric mass-squared difference, $|\Delta m^2_{32}|$ should be around $\simeq 10^{-3}$ eV$^2$.	
	\item
	There is no visible excess or deficit of electron neutrinos. It suggests that the oscillations of muon neutrinos mainly occur due to $\nu_\mu \rightarrow \nu_\tau$ transitions.
	\item
	The Super-K experiment was also the first experiment to observe the sinusoidal $L/E$ dependence of the $\nu_\mu$ survival probability~\cite{Super-Kamiokande:2004orf}. A special analysis was performed by the Super-K collaboration, where they selected events with good resolution in $L/E$, largely excluding low-energy and near-horizon events. Using this high-resolution event sample, they took the ratio of the observed and expected event rates and the oscillation dip started to appear on the canvas around $L/E$ $\sim$ 100 km/GeV, with the deepest location of the dip around $L/E$ $\sim$ 500 km/GeV (see Fig.~4 in Ref.~\cite{Super-Kamiokande:2004orf}). This study was quite useful to disfavor the other alternative models which showed different $L/E$ behaviors. Note that the oscillation dip was also observed in the DeepCore data where the ratio of observed events to unoscillated Monte Carlo (MC) events is plotted against the reconstructed $L/E$ of neutrino~\cite{IceCube:2014flw}.	
\end{itemize}
%==========

The performance of ICAL is optimized for the reconstructed muon energy ($E^\text{rec}_\mu$) range of 1 GeV to 25 GeV, and reconstructed baseline ($L^\text{rec}_\mu$) from 15 km to 12000 km except around the horizon. Thus, the range of reconstructed $L^\text{rec}_\mu/E^\text{rec}_\mu$ of detected muon to which ICAL is sensitive, is from 1 km/GeV to around $10^4$ km/GeV. Therefore, we expect to see an oscillation dip (or two) as in Super-K. In this chapter, we investigate the capability of the ICAL detector to reconstruct the oscillation dip in the reconstructed  $L^\text{rec}_\mu/E^\text{rec}_\mu$  distribution of observed muon events. Moreover, it would be possible to identify the dip in $\mu^-$ and $\mu^+$ events separately, due to the muon charge identification capability of ICAL. 

In contrast to the studies in \cite{Super-Kamiokande:2004orf,IceCube:2014flw} where the ratio between data and unoscillated MC as a function of $L/E$ was used, in this chapter, we use the ratio between upward-going and downward-going events for the analysis. We exploit the fact that the downward-going atmospheric neutrinos and antineutrinos in multi-GeV energy range do not oscillate, and up-down asymmetry reflects the features mainly due to oscillations in upward-going events in these energies. The magnetic field in the ICAL detector enables us to study these distributions in neutrino and antineutrino oscillation channels separately. In this chapter, we also discuss the procedure to measure the atmospheric mass-squared difference $\Delta m^2_{32}$ and the mixing angle $\theta_{23}$ using the information from the oscillation dip, and provide the results.  

In this chapter, we point out for the first time that there is an ``oscillation valley'' feature in the two-dimensional plane of   reconstructed muon observables ($E_\mu^\text{rec}$, $\cos\theta_\mu^\text{rec}$) for $\mu^-$ and $\mu^+$ events separately  at ICAL. The presence of valley-like feature in the oscillograms of  survival probability of $\nu_\mu$ and $\bar\nu_\mu$ in the plane of neutrino energy and direction is well known. But, it is not a priori obvious that the reconstructed energy and direction of muon will serve as a good proxy for the neutrino energy and direction, and will be able to reproduce this valley. In this work, we show that the reconstructed muon observables at ICAL preserve these features, which is a non-trivial statement about the fidelity of the detector. 

The valley-like feature would be recognizable at ICAL due to its sensitivity to muons having energies in the wide range of 1 GeV to 25 GeV and excellent angular resolution at these energies as described in Sec.~\ref{sec:muon_response}. We show,  using the complete migration matrices of ICAL for muon as obtained from GEANT4 simulation~\cite{Chatterjee:2014vta} that, due to the excellent energy and direction resolution, the oscillation valley in two-dimensional ($E_\mu^\text{rec},\cos\theta_\mu^\text{rec}$) plane would appear prominently at ICAL. We also demonstrate for the first time that the mass-squared difference may be determined using the alignment of the oscillation valley.

There are studies to estimate the sensitivity of the ICAL detector for measuring the atmospheric oscillation parameters $\Delta m^2_{32}$ and $\theta_{23}$, see Refs.\,\cite{Thakore:2013xqa,Devi:2014yaa,Kaur:2014rxa,Mohan:2016gxm,Rebin:2018fdl,Chacko:2019wwm}. ICAL would be able to determine these parameters separately in the neutrino and antineutrino channels~\cite{Kaur:2017dpd,Dar:2019mnk}. The difference between these analyses and our study is that  Refs.\,\cite{Thakore:2013xqa,Devi:2014yaa,Kaur:2014rxa,Mohan:2016gxm,Rebin:2018fdl,Chacko:2019wwm,Kaur:2017dpd,Dar:2019mnk} employ the $\chi^2$ method, while the focus of our study is to identify the oscillation dip and the oscillation valley, independently in $\mu^-$ and $\mu^+$ events, and determine the oscillation parameters from them. This is possible due to the wide range of energy and baselines reconstructed at ICAL with excellent detector properties. We also include statistical fluctuations, systematic uncertainties, and errors in other oscillation parameters while determine $\Delta m^2_{32}$ and $\theta_{23}$  with 10-year data at ICAL. 

This chapter is organized in the following fashion. In Sec.~\ref{sec:dip_valley_prob}, we present the survival probabilities of neutrinos and antineutrinos as one-dimensional functions of $L_\nu/E_\nu$ and as two-dimensional oscillograms in $(E_\nu, \cos\theta_\nu)$ plane, to pinpoint the oscillation dips and the oscillation valley, respectively. Next, in Sec.~\ref{sec:dip_reco_events}, we formulate a data-driven variable, the ratio between upward-going and downward-going events (U/D), to use it as an observable for all the analyses in this chapter. We propose a novel algorithm to identify the oscillation dip and to measure its ``location'' in reconstructed $L^{\rm rec}_\mu/E^{\rm rec}_\mu$ distributions. We present the  $90\%$ C.L. range for $\Delta m^2_{32}$ expected with 500 kt$\cdot$yr exposure of ICAL using the calibration curve between the location of the dip and $\Delta m^2_{32}$.  We also estimate the allowed range for $\sin^2\theta_{23}$ at 90$\%$ C.L. using the ratio of upward-going and downward-going events (U/D). Sec.~\ref{sec:2D_E-CT} is devoted to discuss our analysis of U/D distributions in  reconstructed ($E^{\rm rec}_\mu, \cos\theta^{\rm rec}_{\mu}$) plane. For identifying the oscillation valley in   reconstructed ($E^{\rm rec}_\mu, \cos\theta^{\rm rec}_{\mu}$) plane of these distributions and to measure the ``alignment'' of the valley, we suggest a unique algorithm and demonstrate its use. We present the $90\%$ C.L. range of $\Delta m^2_{32}$ using 500 kt$\cdot$yr exposure of ICAL with the help of the calibration curve between the alignment of the valley and $\Delta m^2_{32}$. In Sec.~\ref{sec:dip_valley_conclusions}, we summarize our findings and point out the utility of our novel approach in establishing the oscillation hypothesis in atmospheric neutrino experiments.

% %=====================================
\section{Neutrino Oscillation Probabilities}
\label{sec:dip_valley_prob}
%=====================================

In this section, we discuss the survival probabilities of atmospheric $\nu_\mu$ and $\bar\nu_\mu$, reaching the detector after their propagation through the atmosphere and possibly the Earth. The oscillation probabilities are functions of energy ($E_\nu$) and zenith angle ($\theta_\nu$) of neutrino. The net distance $L_\nu$ traversed by a neutrino (its baseline) is related to its zenith angle via Eq.~\ref{eq:zen-bl-rel}.

For atmospheric neutrinos, the survival probabilities may be represented in two ways: (i) as one-dimensional functions of  $L_\nu/E_\nu$, and (ii) as two-dimensional oscillograms in the plane of energy and zenith angle of neutrinos. Below we discuss these two representations, and the major physics insights obtained via them.

%==========================
\subsection{$L_\nu/E_\nu$ dependence $\nu_\mu$ and $\bar{\nu}_\mu$ survival probabilities}
\label{sec:dip_prob}
%==========================

\begin{figure}
	\centering
	\includegraphics[width=0.49\linewidth]{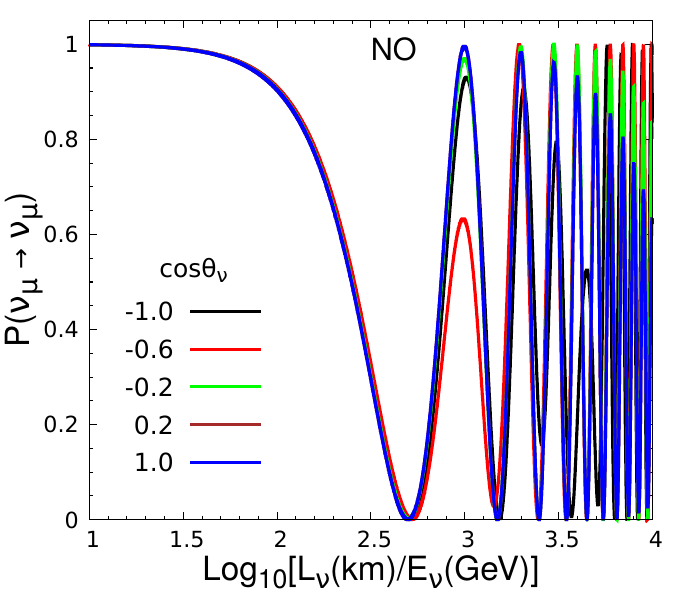}
	\includegraphics[width=0.49\linewidth]{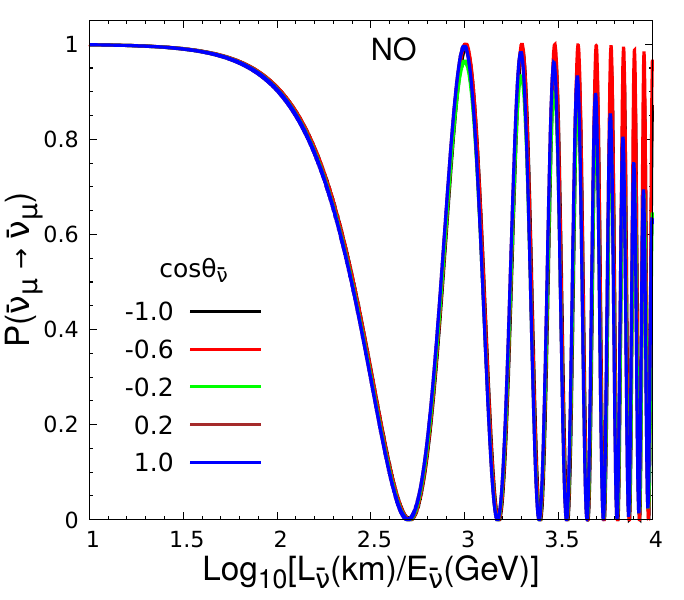}
	\caption{The $\nu_\mu$ and $\bar{\nu}_\mu$ survival probability as functions of $L_\nu/E_\nu$ in the left and right panels, respectively, in the three-flavor oscillation framework in the presence of Earth's matter using PREM profile~\cite{Dziewonski:1981xy}. The different colored curves correspond to different choices of the zenith angle ($\theta_\nu$) of neutrino (or equivalently, different neutrino baselines $L_\nu$). We use the benchmark values of oscillation parameters given in Table~\ref{tab:osc-param-value}.~\cite{Kumar:2020wgz}} 
	\label{fig:osc_dip_neutrino}
\end{figure}

Figure~\ref{fig:osc_dip_neutrino} illustrates $\nu_\mu$ and $\bar{\nu}_\mu$ survival probability as functions of $\log_{10} (L_\nu/E_\nu)$ in the left and right panels, respectively, using the benchmark values of oscillation parameters mentioned in Table~\ref{tab:osc-param-value}. The different colored curves correspond to the different zenith angles $\theta_\nu$ representing different baselines ($L_\nu$). We can observe that the first oscillation minimum appears at $\log_{10} (L_\nu/E_\nu) \approx 2.7$ for $\nu_\mu$ as well as $\bar{\nu}_\mu$. For values of $L_\nu/E_\nu$ smaller than the location of the first oscillation minimum, $\nu_\mu$ and $\bar{\nu}_\mu$ survival probabilities are almost equal and the overlapping nature of different $\cos\theta_\nu$ curves shows that there is no $\theta$-dependence beyond that coming from $L_\nu/E_\nu$. This observation indicates that the survival probabilities in this region are dominated by vacuum oscillations. However, for higher values of $L_\nu/E_\nu$, Earth's matters come into the picture and introduce an additional $\theta$-dependence beyond that coming from $L_\nu/E_\nu$ which can be prominently seen for $\nu_\mu$ survival probability in the case of NO. Since matter effects on antineutrinos are not significant in NO, the $\bar{\nu}_\mu$ survival probability does not show this feature.

%==========================
%==========================

%==========================
\subsection{Oscillograms in ($E_\nu$, $\cos\theta_\nu$) plane}
%==========================

\begin{figure}
	\centering
	\includegraphics[width=0.49\linewidth]{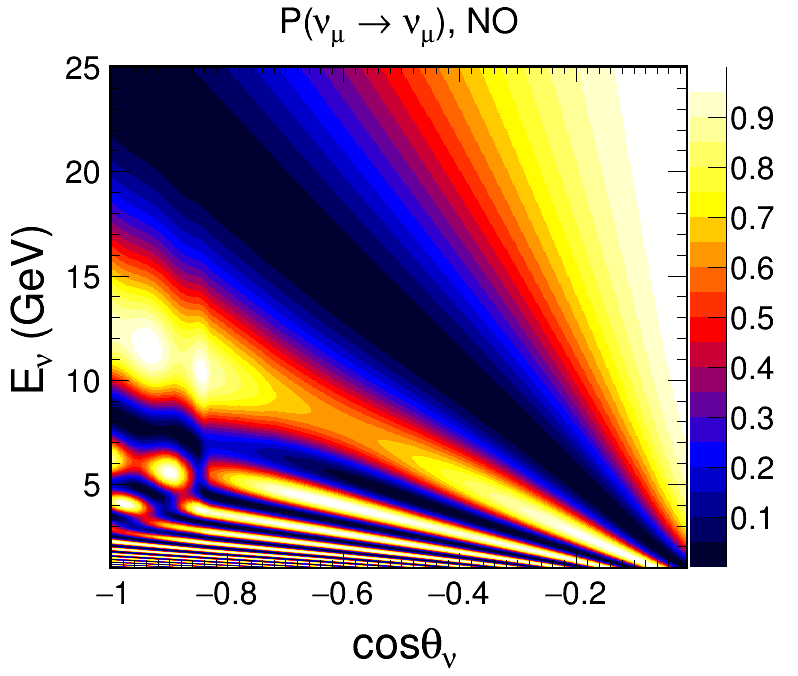}
	\includegraphics[width=0.49\linewidth]{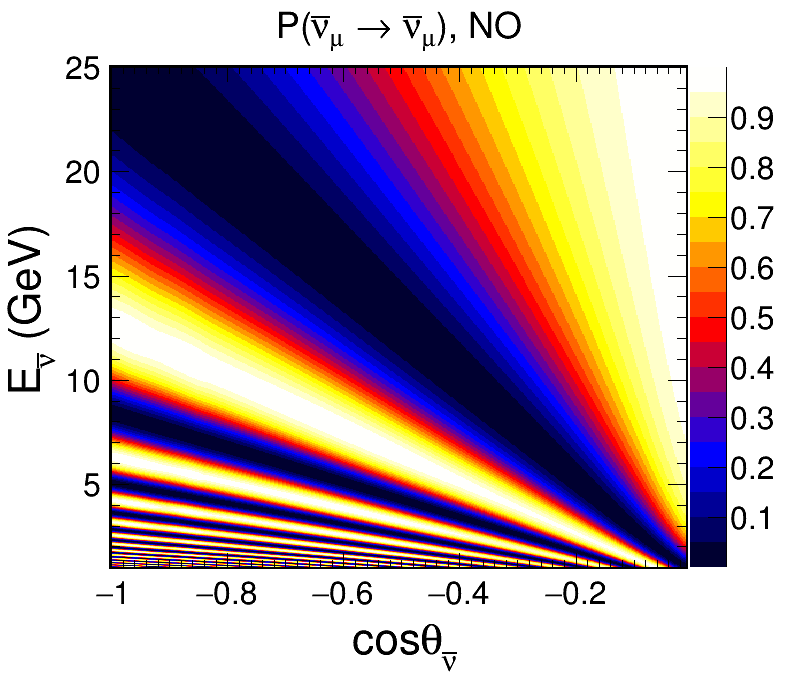}
	\caption{Oscillograms of $\nu_\mu$ and $\bar{\nu}_\mu$ survival probability in the ($E_\nu$, $\cos\theta_\nu$) plane in the left and right panels, respectively. We use three-flavor neutrino oscillations in the presence of Earth's matter considering the PREM profile~\cite{Dziewonski:1981xy}. We use the benchmark values of oscillation parameters given in Table~\ref{tab:osc-param-value}.~\cite{Kumar:2020wgz}}
	\label{fig:osc_valley_neutrino}
\end{figure}

Figure~\ref{fig:osc_valley_neutrino} shows $\nu_\mu$ and $\bar{\nu}_\mu$ survival probability oscillograms in ($E_\nu$, $\cos\theta_\nu$) plane in the left and right panels, respectively, using NO as the neutrino mass ordering. In both the panels of Fig.~\ref{fig:osc_valley_neutrino}, we observe the prominent dark blue diagonal bands, which represent the minimum survival probabilities for $\nu_\mu$ and $\bar{\nu}_\mu$. This band of oscillation minimum corresponds to the same "oscillation dip" as observed in Fig.~\ref{fig:osc_dip_neutrino} at $\log_{10} (L_\nu/E_\nu) \approx 2.7$. The nature of this band in $\nu_\mu$ and $\bar{\nu}_\mu$ survival probabilities is almost identical, which indicates that this region is dominated by vacuum oscillations. In the present work, we explore whether we can reconstruct this diagonal band of oscillation minima (an ``oscillation valley'') using the atmospheric neutrino data. The triangular region of the oscillogram below the dark blue diagonal band corresponding to lower energies and longer baselines is dominated by the matter effects\footnote{In Ref.~\cite{Akhmedov:2006hb}, the authors gave a detailed physics interpretation of the oscillograms in terms of the amplitude and phase conditions while describing various features such as MSW peaks, parametric ridges, local maxima, zeros, and saddle points.} for neutrinos (in the case of NO considered here). For the case of neutrinos, the MSW resonance~\cite{Wolfenstein:1977ue,Mikheev:1986gs,Mikheev:1986wj} can be observed as red patch around $6~\text{GeV} < E_\nu < 8~\text{GeV}$ and $-0.7 < \cos\theta_\nu < -0.5$. The neutrino oscillation length resonance~\cite{Petcov:1998su,Chizhov:1998ug,Petcov:1998sg,Chizhov:1999az,Chizhov:1999he} or parametric resonance~\cite{Akhmedov:1998ui,Akhmedov:1998xq} can be seen as yellow patches around $3~\text{GeV} < E_\nu < 6~\text{GeV}$ and $-1.0 < \cos\theta_\nu < -0.8$. The yellow triangular region above the blue band corresponds to vacuum oscillations, which is almost identical for $\nu_\mu$ and $\bar{\nu}_\mu$ survival probabilities.

%==========================
%==========================

In the next section, we demonstrate how these oscillation dip and valley features in the multi-GeV range of energies may be reflected in the reconstructed event distributions for $\mu^-$ and $\mu^+$ in an atmospheric neutrino experiment like ICAL.

%==========================
\section{Oscillation dip in $L_\mu^\text{rec}/E_\mu^\text{rec}$ distribution}
\label{sec:dip_reco_events}
%==========================

While this analysis can in principle be performed with any atmospheric neutrino detector, we use the proposed ICAL detector at INO~\cite{ICAL:2015stm} for our simulations. There are two main reasons for this (i) ICAL would be able to distinguish neutrinos from antineutrinos, and hence independent analyses for these two channels may be carried out. (ii) ICAL would be able to reconstruct the energy as well as the direction of $\mu^-$ and $\mu^+$ (produced in the CC interactions of $\nu_\mu$ and $\bar\nu_\mu$, respectively) with a good precision, which will be crucial for the analysis. These are also the reasons due to which ICAL would be able to provide a direct measurement of neutrino mass ordering by observing the Earth's matter effects separately in neutrino and antineutrino channels in the multi-GeV range of energies~\cite{ICAL:2015stm}. To perform the present analyses, we use the method described in Sec.~\ref{sec:event_simulation} to simulate the reconstructed muon events at the ICAL detector. 

In this work, we use the ratios of upward-going and downward-going reconstructed muons in various energy and direction bins. Note that the upward-going vs. downward-going events may lead to some ambiguity when the events are near-horizon, i.e., $-0.2 < \cos\theta_{\mu}^\text{rec} < 0.2$ (or $73 ~ \text{km} < L_\mu^\text{rec} < 2621 ~ \text{km}$). For these events, there is a significant change in the neutrino path length for a small variation in the estimated neutrino arrival direction. However, the detector response of ICAL to such events is anyway very poor, owing to the horizontally places RPCs and iron layers. It is observed that these events (corresponding to $\log_{10} [L_\mu^{\text{rec}}/E_\mu^{\text{rec}}]$ in the range of 1.5 -- 2.0) do not affect the analysis\footnote{Whenever we mention the value of $\log_{10} [L/E]$ in this chapter, we take $L$ and $E$ in the units of km and GeV, respectively.}. Note that the Super-K collaboration selected only the ``high-resolution'' events in their $L/E$ analysis~\cite{Kajita:2014koa, Itow:2013zza}. In particular, their analysis rejected neutrino events near the horizon, as well as low-energy events where the large scattering angles would have led to large errors in the reconstruction of neutrino direction. The low efficiency of ICAL for the near-horizontal event, and our analysis threshold of 1 GeV for muons, automatically incorporates both of these filters. However, a difference between our analysis and the $L/E$ analysis of Super-K~\cite{Kajita:2014koa} also needs to be pointed out. While the analysis in~\cite{Kajita:2014koa} is in terms of the inferred $L_\nu$ and $E_\nu$ of neutrinos, our analysis is in terms of the $L_\mu^\text{rec}$ and $E_\mu^\text{rec}$ of muons. The energy deposited at the detector by hadrons in the final state of neutrino interaction can be reconstructed at ICAL~\cite{Devi:2013wxa}, but we do not use the hadron energy information in this study.

Now, we explore the expected event distributions at ICAL for $\mu^-$ and $\mu^+$ as functions of reconstructed\footnote{The observable $L_\mu^\text{rec}$, which is the effective baseline, is related to the reconstructed muon direction $\theta_\mu^\text{rec}$ in the following fashion,
\begin{equation}
L_\mu^\text{rec} = \sqrt{(R+h)^2 - (R-d)^2\sin^2\theta_\mu^\text{rec}} \,-\, (R-d)\cos\theta_\mu^\text{rec} \,.
\label{eq:zen-bl-mu-rel}
\end{equation}
Note that $L_\mu^\text{rec}$ is an observable associated with the reconstructed muon direction, and not to be related directly with the neutrino direction.} $L_\mu^\text{rec}/E_\mu^\text{rec}$, with an MC event sample corresponding to 1000 years and simulated event samples corresponding to 10 years.  The 1000-year MC sample represents the expected observations for the scenario where unlimited statistics is available, which will help us understand how much information on oscillation features remains intact after the detection process if statistics is not the limitation. The analysis using such a large statistics will guide our algorithms. It will also be used for the calibrations of oscillation parameters like $\Delta m^2_{32}$ and $\sin^2\theta_{23}$. On the other hand, the analysis with the 10-year simulated sample would help us estimate the effects due to statistical fluctuations, which is an important consideration for low-statistics experiments such as those with atmospheric neutrinos. To this end, we analyze 100 statistically independent sets of the 10-year simulated samples.

The expected number of reconstructed $\mu^-$ and $\mu^+$ events at the 50 kt ICAL detector in 10 years are already presented in Table~\ref{tab:total_events} where, U stands for the number of muon events in the upward-going direction ($\cos\theta_\mu^\text{rec} > 0$) whereas D represent the number of muons events in the downward-going direction ($\cos\theta_\mu^\text{rec} < 0$). The U/D ratio for reconstructed muon is defined as
%==========================
\begin{equation}\label{eq:U/D_def}
\text{U/D} (E_\mu^\text{rec}, \cos\theta_\mu^\text{rec}) \equiv
\frac{N(E_\mu^\text{rec}, -|\cos\theta_\mu^\text{rec}|)}{N(E_\mu^\text{rec}, +|\cos\theta_\mu^\text{rec}|)} \; ,
\end{equation}
%==========================
where $N(E_\mu^\text{rec}, \cos\theta_\mu^\text{rec})$ is the number of muon events in the bins with reconstructed energy $E_\mu^\text{rec}$ and the reconstructed zenith angle $\theta_\mu^\text{rec}$. We associate the U/D ratio as a function of reconstructed $ \cos\theta_\mu^\text{rec}$ with the upward-going muon events, i.e., with $\cos\theta_\mu^\text{rec} < 0$. We can also associate this U/D asymmetry with the $L_\mu^\text{rec}$ corresponding to these upward-going muon events, which is given in terms of $\cos\theta_\mu^\text{rec}$ by Eq.~\ref{eq:zen-bl-mu-rel}.

The U/D ratio can be taken as a proxy for the survival probability $P(\nu_\mu \rightarrow \nu_\mu)$ because about 98\% events at the ICAL detector are contributed from $\nu_\mu \rightarrow \nu_\mu$ survival channel, and only about 2\% of total events come from $\nu_e \rightarrow \nu_\mu$ appearance channel for benchmark values of oscillation parameters given in Table~\ref{tab:osc-param-value}. The use of this ratio also minimizes the dependence of our analysis on those systematic uncertainties which are symmetric for upward and downward-going events. Now, we shall analyze the distributions of events and U/D ratios as the functions of $L_\mu^\text{rec}/E_\mu^\text{rec}$ for $\mu^-$ and $\mu^+$ events separately.

%=================================
\subsection{Events and U/D ratio using 1000-year Monte Carlo simulation }
\label{sec:LbyE-1000yrs}
%=================================

%==========================
\begin{figure}
	\centering
	\includegraphics[width=0.45\linewidth]{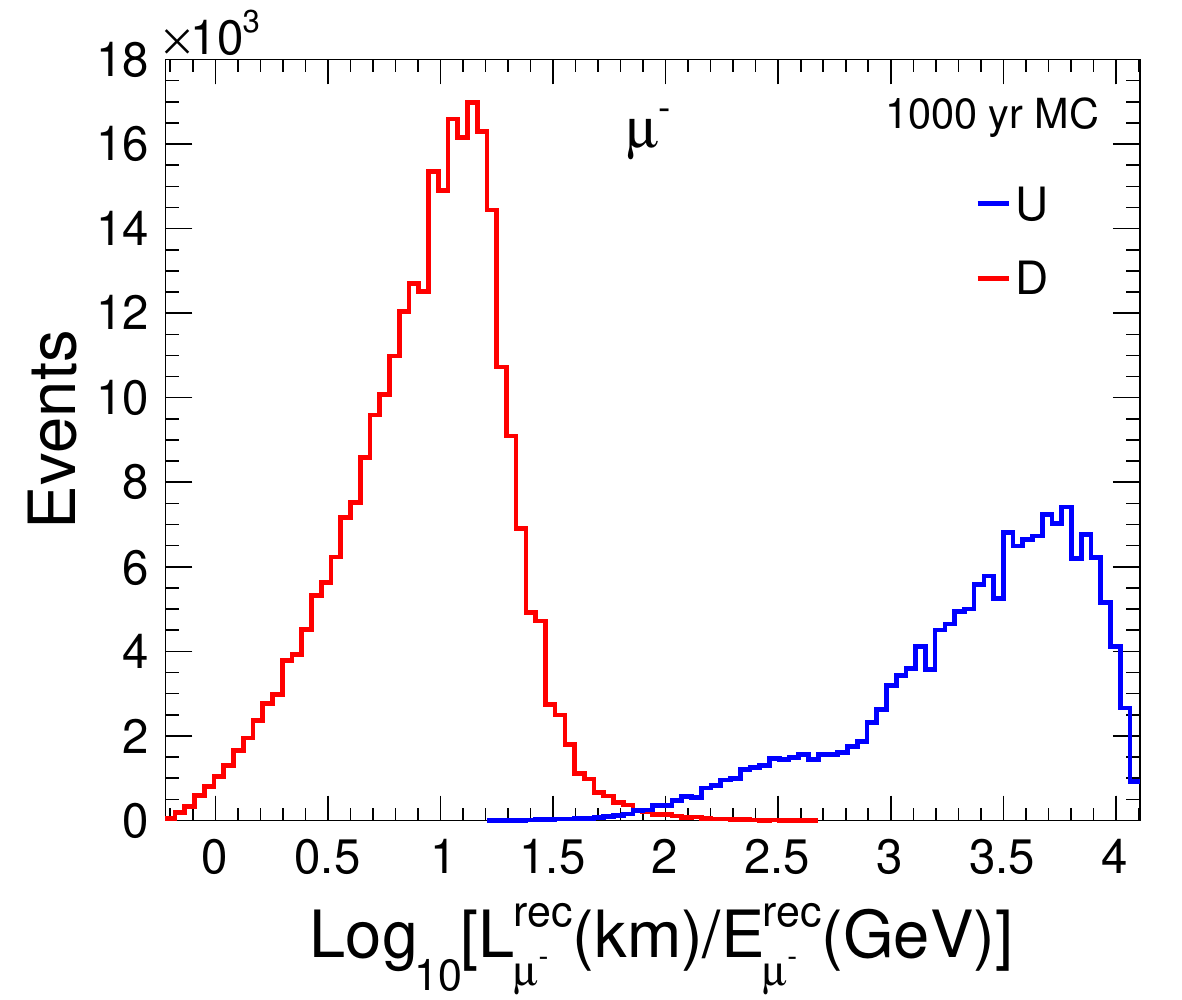}
	\includegraphics[width=0.45\linewidth]{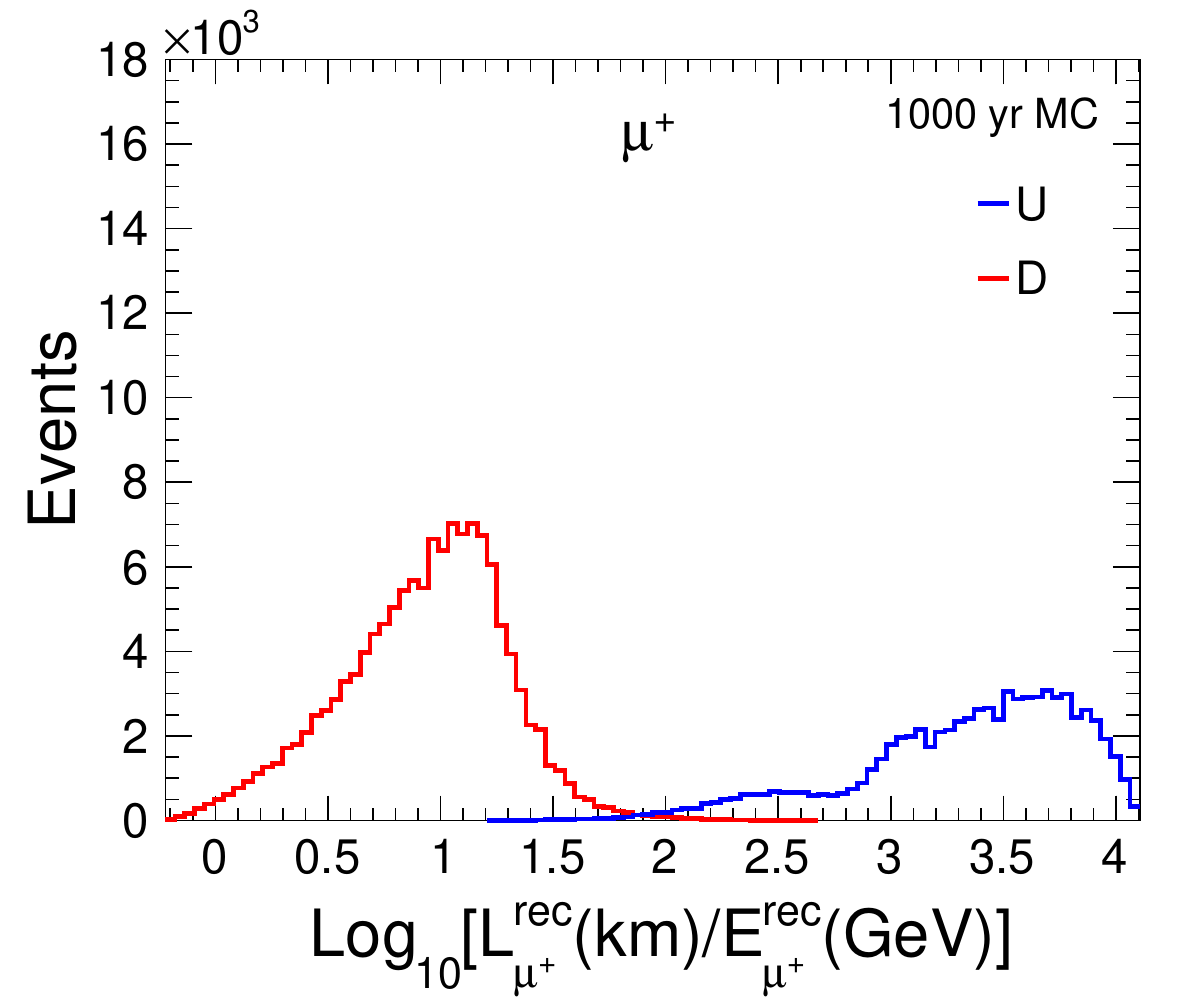}
	\includegraphics[width=0.45\linewidth]{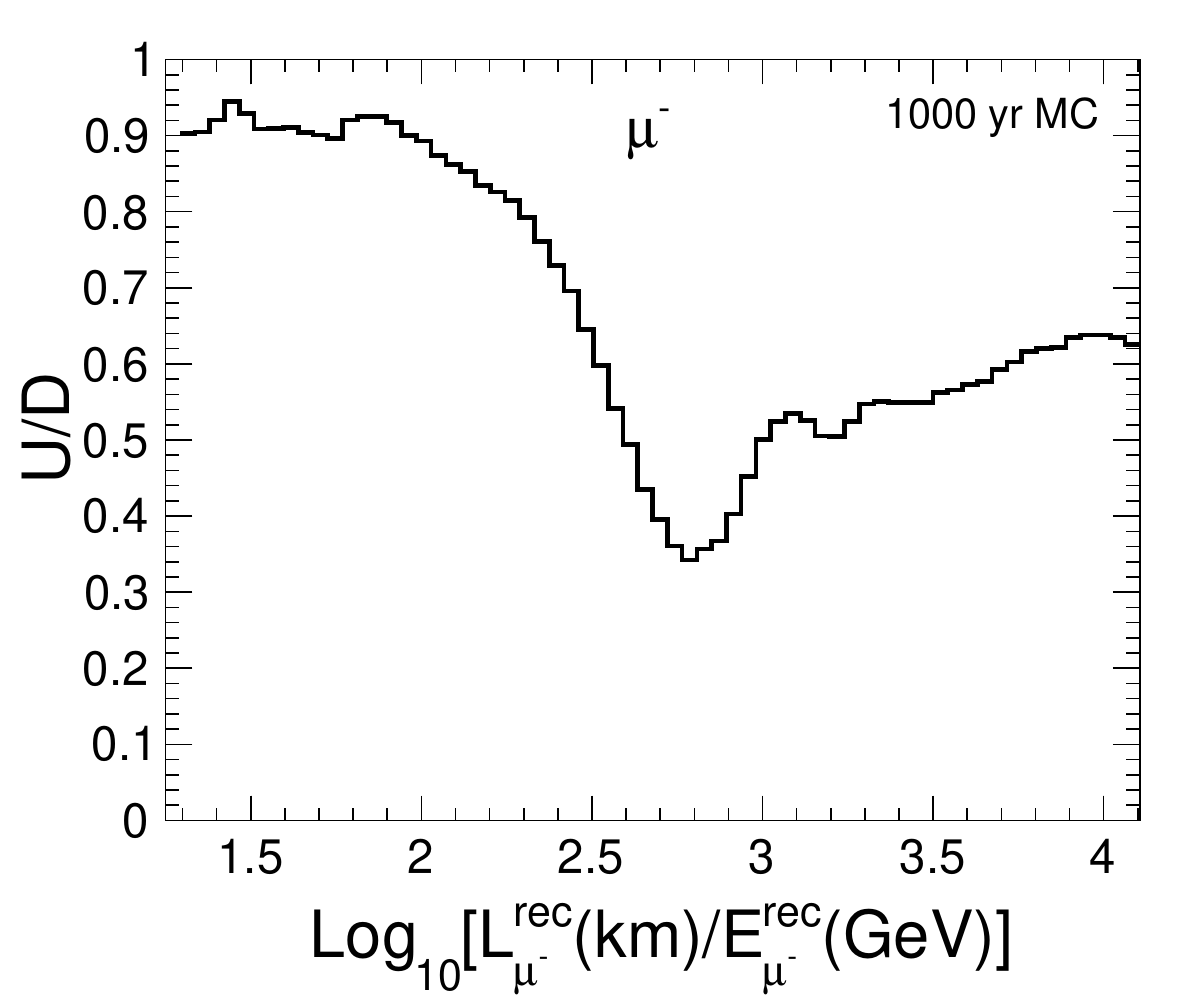}
	\includegraphics[width=0.45\linewidth]{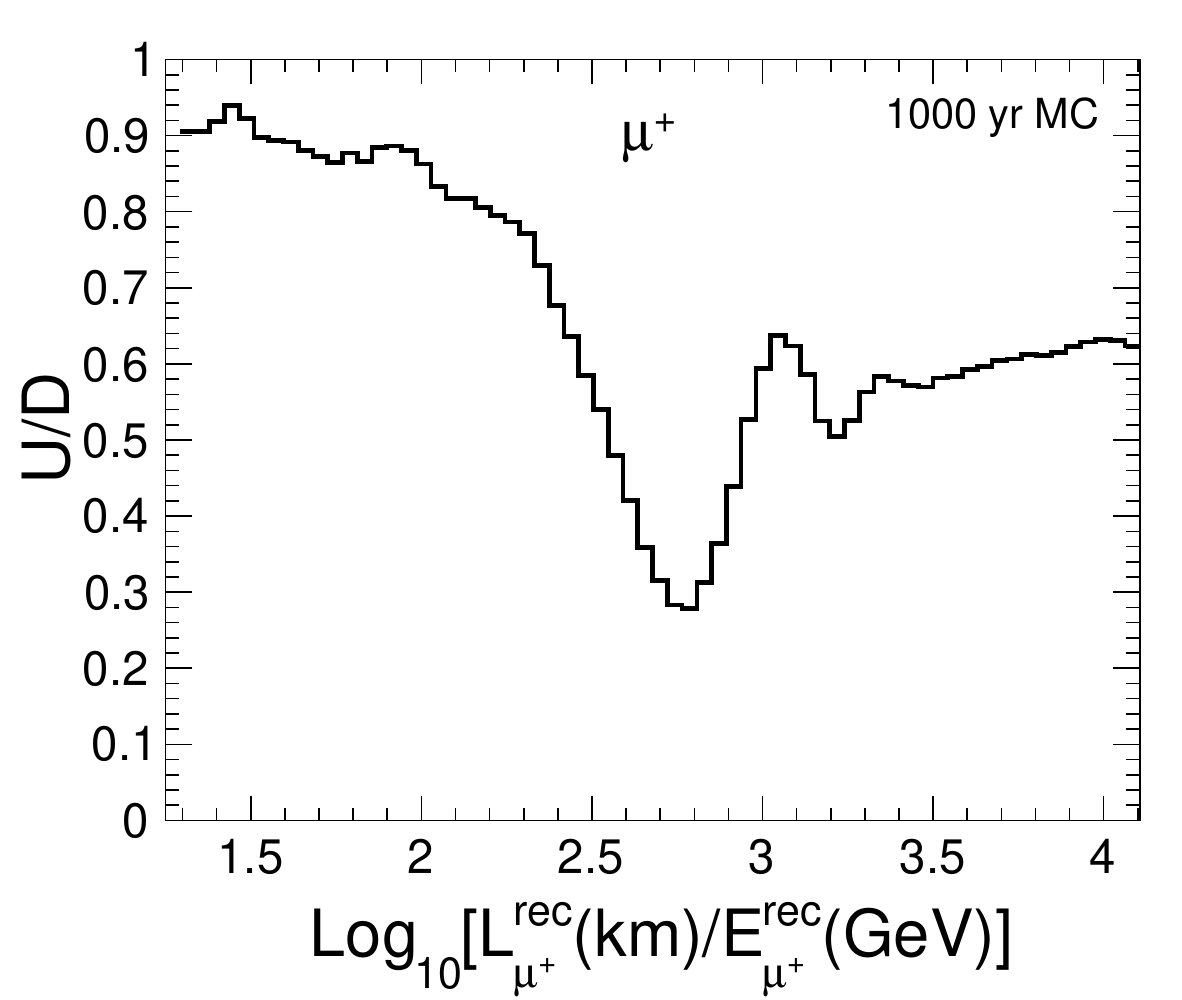}
	\caption{The top panels show the expected distributions of reconstructed $\mu^-$ (left panel) and $\mu^+$ (right panel) events as functions of $\log_{10}[L_\mu^\text{rec}/E_\mu^\text{rec}]$. The blue and red curves represent upward-going (U) and downward-going (D) events, respectively. The bottom panels show the distributions of U/D ratios as functions of $\log_{10}[L_\mu^\text{rec}/E_\mu^\text{rec}]$ for $\mu^-$ (left panel) and $\mu^+$ (right panel). These results are obtained using the 1000-year MC data sample. Here, $E_\mu^\text{rec}$ lies in the energy range of 1 to 25 GeV. We use the benchmark values of oscillation parameters given in Table~\ref{tab:osc-param-value}.~\cite{Kumar:2020wgz}}
	\label{fig:osc_dip_1000yr}
\end{figure}
%==========================

The distribution of reconstructed $\mu^-$ (left panels) and $\mu^+$ (right panels) events for 1000-year MC simulation is shown in the upper panels of Fig.~\ref{fig:osc_dip_1000yr}. We consider the reconstructed muon events in the energy range of $(E_\mu^\text{rec})_\text{min} = 1$ GeV to $(E_\mu^\text{rec})_\text{max} = 25$ GeV. Since $L_\mu^\text{rec}$ varies in the range of 15 to 12757 km for atmospheric neutrinos, the minimum and maximum values of $\log_{10}[L_\mu^\text{rec}/E_\mu^\text{rec}]$ are given as $\log_{10}[(L_\mu^\text{rec})_\text{min}/(E_\mu^\text{rec})_\text{max}]$ = 0.22 and $\log_{10}[(L_\mu^\text{rec})_\text{max}/(E_\mu^\text{rec})_\text{min}] = 4.1$, respectively. Since the minimum value of $L_\mu^\text{rec}$ for upward-going events is about 437 km, the minimum value of $\log_{10}[L_\mu^\text{rec}/E_\mu^\text{rec}]$ turns out to be 1.2 for these events. We bin the data in 100 uniform bins of $\log_{10}[L_\mu^\text{rec}/E_\mu^\text{rec}]$. The blue and red curves in the upper panels of Fig.~\ref{fig:osc_dip_1000yr} represent upward and downward-going reconstructed muon events, respectively. We can observe that the number of upward-going events are less than that of downward-going events because the upward-going $\nu_\mu$ have traveled larger distances and have had a higher chance of oscillating to neutrinos of other flavors. The number of $\mu^+$ events are less than that of $\mu^-$ events because the interactions cross-section of antineutrino is smaller than that of neutrinos by a factor of approximately 2 at these energies.

In the lower panels of Fig.~\ref{fig:osc_dip_1000yr}, we divide the range for upward-going muons ($\cos\theta_\mu^\text{rec} < 0$), i.e. $\log_{10}[L_\mu^\text{rec}/E_\mu^\text{rec}] = 1.2 - 4.1$, in 66 uniform bins. A downward-going event with a given $\cos\theta_\mu^\text{rec} > 0$ is then assigned to the bin corresponding to $-\cos\theta_\mu^\text{rec}$. We calculate the U/D ratio in each bin by dividing the number of upward-going muon events by that of downward-going muon events, and plot with the corresponding value of $\log_{10}[L_\mu^\text{rec}/E_\mu^\text{rec}]$. In both $\mu^-$ and $\mu^+$ channels, we can observe two major dips in the U/D ratio at $\log_{10}[L_\mu^\text{rec}/E_\mu^\text{rec}] \sim 2.75$ and $\sim3.2$, respectively. In the present study, we focus our attention on the observation of the first oscillation dip.

Note that, while the two dips described above correspond to the first two oscillation minima in the survival probability as shown in Fig.~\ref{fig:osc_dip_neutrino}, they are not exactly at the same location. This happens because of the crucial difference that, while Fig.~\ref{fig:osc_dip_neutrino} uses the true values of neutrino energy and zenith angle of the atmospheric neutrino, Fig.~\ref{fig:osc_dip_1000yr} uses the reconstructed value of energy and zenith angle of the muon, which is produced from the CC interaction of that neutrino. In the CC deep-inelastic scattering of neutrino, the hadrons in the final state take away some fraction of incoming neutrino energy, which results in $E_\mu^\text{rec} < E_\nu$. Due to the decrease in  $E_\mu^\text{rec}$, the dips shift towards higher values of $L_\mu^\text{rec}/E_\mu^\text{rec}$. This effect is described quantitatively in terms of the inelasticity of an event which is defined as $y = 1 - E_\mu / E_\nu$. Since in the multi-GeV range of energy, the average inelasticity for antineutrino events ($\langle y_{\bar{\nu}} \rangle \approx 0.3$)  are smaller than that in the neutrino events($\langle y_{\nu} \rangle \approx 0.45$)~\cite{Devi:2014yaa}, the shift in the location of dip is smaller for the case of $\mu^+$ than $\mu^-$.

Note that the dips in Fig.~\ref{fig:osc_dip_1000yr} are shallower and broader than those in Fig.~\ref{fig:osc_dip_neutrino}, where the dips reach all the way to the bottom, that is $P(\nu_\mu \rightarrow \nu_\mu) \approx 0$, for $\sin^2\theta_{23} = 0.5$. This smearing occurs due to the difference between the momenta of the parent neutrino and resulting muon, as well as the limitation on the reconstruction of muon momentum in the detector. It can also be observed that the smearing is more in reconstructed $\mu^-$ events as compared to reconstructed $\mu^+$ events in Fig.~\ref{fig:osc_dip_1000yr}. This difference is caused by the broader spread in the inelasticities of interactions for neutrinos that for antineutrinos. Due to this spread, the dips in $\mu^-$ events undergo more smearing, and hence become shallower, as compared to the dips in $\mu^+$ events.

%==========================
\subsection{Events and U/D ratio using 10-year simulated data}
\label{sec:LbyE-10yrs}
%==========================

In this section, we describe the expected distribution of reconstructed muon events as functions of $L_\mu^\text{rec}/E_\mu^\text{rec}$ with an exposure of 500 kt$\cdot$yr (i.e. 10 years) at the ICAL detector. Due to limited statistics of atmospheric neutrinos, the statistical fluctuations are expected to have a significant effect on the accuracy of the results. We incorporate statistical fluctuations by taking 100 independent simulated sets and calculating the mean and root-mean-square (rms) deviations of relevant quantities.

%==========================
\begin{table}[htb!]
	\centering
	\begin{tabular}{|c|c|c|c c|}
		\hline
		Observable & Range & Bin width & \multicolumn{2}{c|}{Number of bins} \\
		\hline 
		\multirow{7}{*}{ $\log_{10}\left[\frac{L_\mu^\text{rec}(\text{km})}{E_\mu^\text{rec}(\text{GeV})}\right]$} & [0, 1] & 0.2 & 5 & \rdelim\}{7}{7mm}[34] \\ 
		& [1, 1.6] &  0.06 & 10  & \\
		& [1.6, 1.7]& 0.1& 1  & \\
	    & [1.7, 2.3]&0.3 & 2  & \\
		& [2.3, 2.4]& 0.1& 1  & \\
		& [2.4, 3.0] &0.06 & 10  &\\
		& [3, 4] & 0.2 & 5  & \\
		\hline
	\end{tabular}
	\caption{The binning scheme considered for $\log_{10}[L_\mu^\text{rec}/E_\mu^\text{rec}]$ for reconstructed $\mu^-$ and $\mu^+$ events of 10-year simulated data.~\cite{Kumar:2020wgz}}
	\label{tab:binning-1D-10years}
\end{table}
%==========================

\begin{figure}
	\centering
	\includegraphics[width=0.45\linewidth]{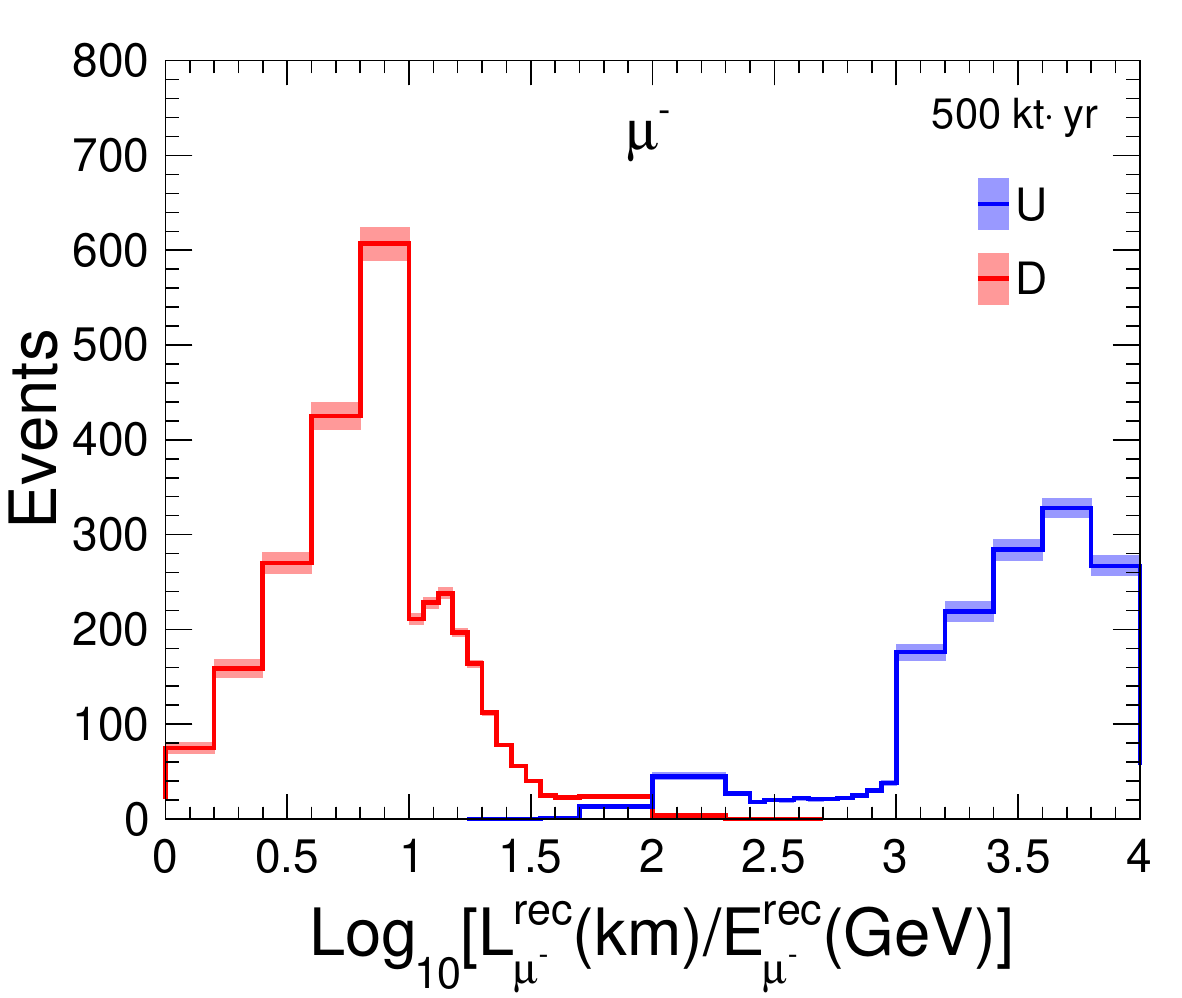}
	\includegraphics[width=0.45\linewidth]{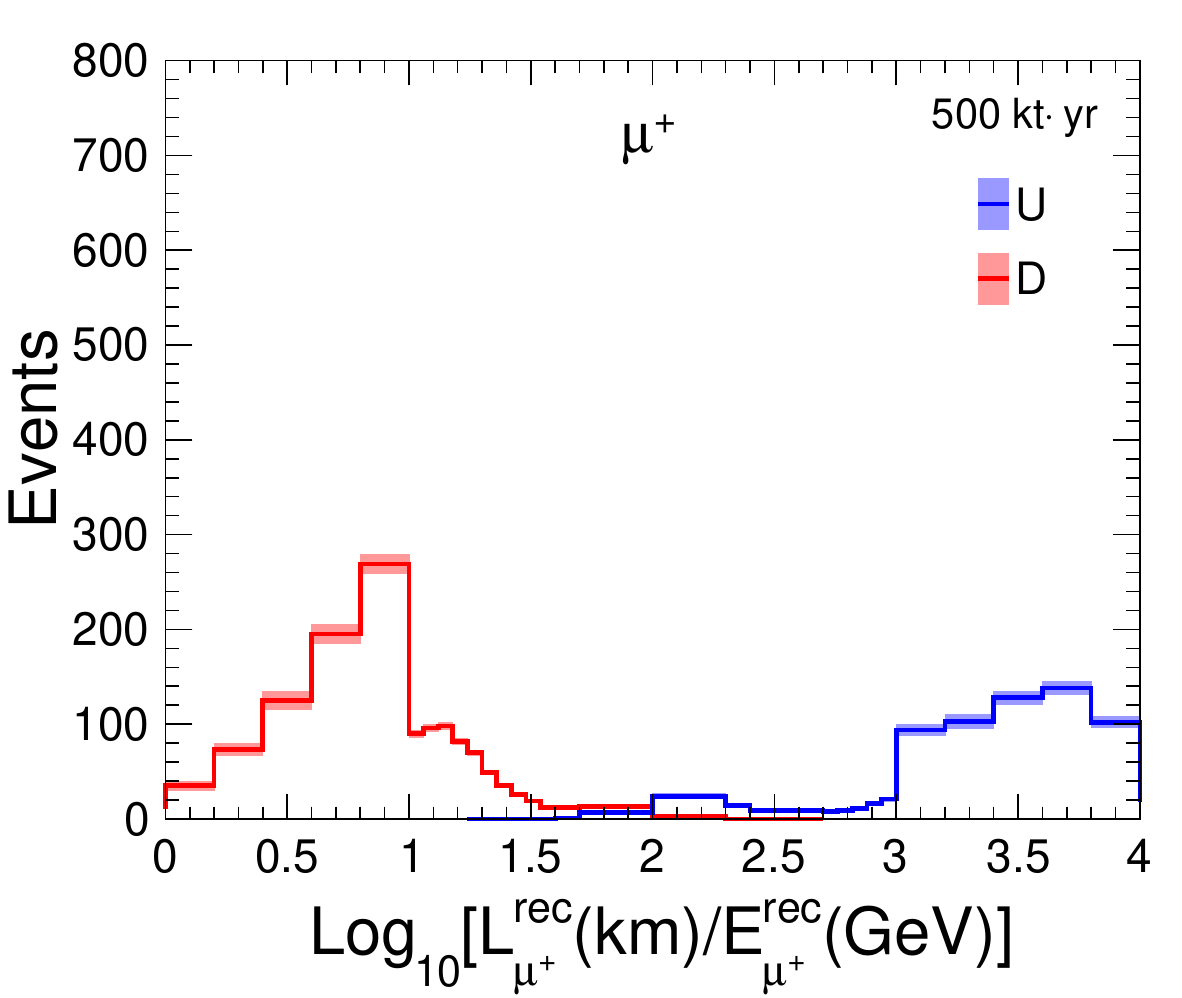}
	\includegraphics[width=0.45\linewidth]{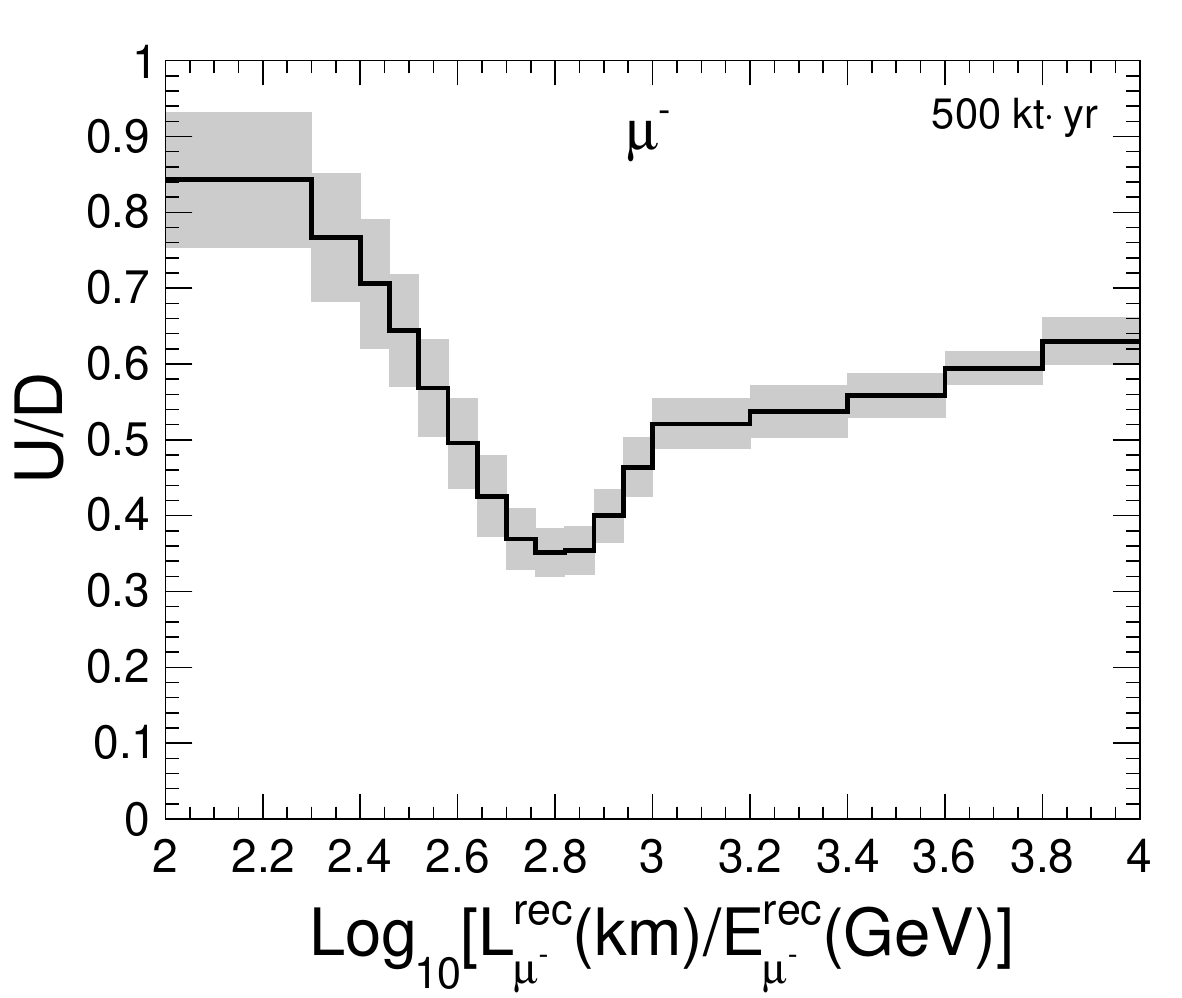}
	\includegraphics[width=0.45\linewidth]{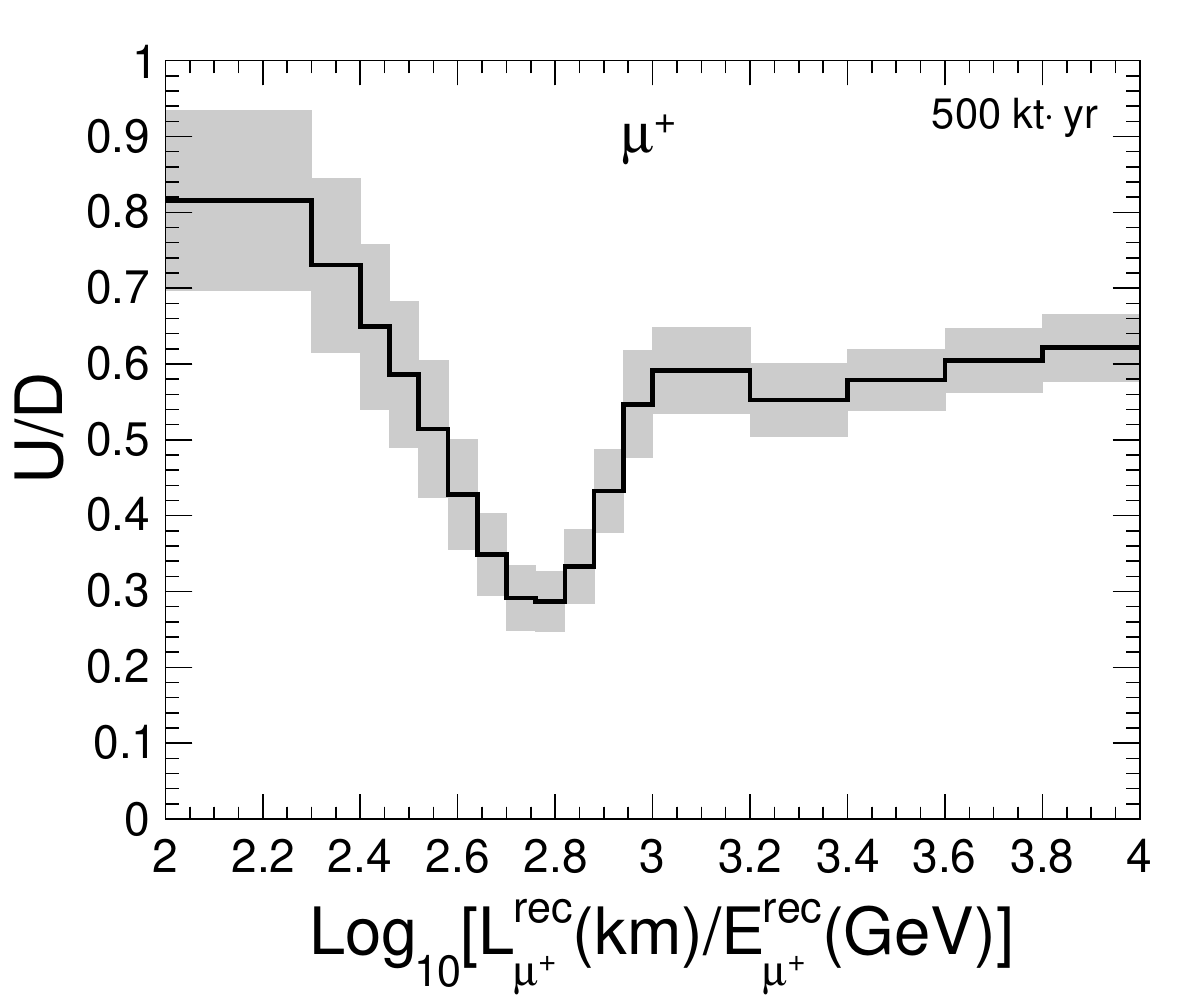}
	\caption{The top panels show the expected distributions of reconstructed $\mu^-$ (left panel) and $\mu^+$ (right panel) events as functions of $\log_{10}[L_\mu^\text{rec}/E_\mu^\text{rec}]$. The blue and red curves represent upward-going and downward-going events, respectively. The shaded boxes correspond to the statistical uncertainties. The bottom panels show the distributions of U/D ratios as functions of $\log_{10}[L_\mu^\text{rec}/E_\mu^\text{rec}]$. The heights of gray boxes correspond to expected statistical uncertainties in the U/D ratio. Here, we use 100 independent simulated sets of data for 10 years at 50 kt ICAL, i.e., 500 kt$\cdot$yr. We use the benchmark values of oscillation parameters given in Table~\ref{tab:osc-param-value}.~\cite{Kumar:2020wgz}}
	\label{fig:osc_dip_10yr}
\end{figure}

We bin the muon events in reconstructed $\log_{10}[L_\mu^\text{rec}/E_\mu^\text{rec}]$ as described in Sec.~\ref{sec:LbyE-1000yrs}. However, since the number of events in 10 years are much smaller than that for 1000 years, we choose a non-uniform binning scheme shown in Table~\ref{tab:binning-1D-10years} such that typically, we have at least 10 down-going events in each bin. We consider total 34 bins for $\log_{10}[L_\mu^\text{rec}/E_\mu^\text{rec}]$ in the range of 0 to 4. The distributions of reconstructed $\mu^-$ (left panel) and $\mu^+$ (right panel) events in these bins are shown in the top panels of Fig.~\ref{fig:osc_dip_10yr}. The blue and red curves in the top panels of Fig.~\ref{fig:osc_dip_10yr} represent upward, and downward-going reconstructed muon events, respectively. The colored boxes denote the statistical uncertainties, which are estimated using rms deviation from 100 independent simulated sets of data for 10 years.

%==========================
%==========================

The U/D ratio in the reconstructed $\log_{10}[L_\mu^\text{rec}/E_\mu^\text{rec}]$ bins are shown in the bottom panels of Fig.~\ref{fig:osc_dip_10yr}, following the procedure mentioned in Sec.~\ref{sec:LbyE-1000yrs}. We consider $\log_{10}[L_\mu^\text{rec}/E_\mu^\text{rec}]$ in the range of 2 to 4, which is the range of our interest. The statistical fluctuations shown in the figure with gray boxes are the rms deviations calculated from the distributions of the U/D ratio in 100 independent sets of simulated data for 10 years. We can observe that, while the first oscillation dip is quite prominent, the second dip is lost due to statistical fluctuations and the broad binning scheme that we have to consider owing to low statistics in 10 years. While this does not rule out the possibility of identifying the second oscillation dip using more efficient algorithms, in the present work, we shall focus on the first oscillation dip. Henceforth, whenever we mention the dip position, we refer to the position of the first dip. We can also observe that the dip for the case of $\mu^+$ is deeper than that for $\mu^-$ analogous to the 1000-year MC sample as described in Sec.~\ref{sec:LbyE-1000yrs}.

%==========================
\subsection{Identifying the dip with 10-year simulated data}
\label{sec:dip_fitting}
%==========================

\begin{figure}
	\centering
	\includegraphics[width=0.45\linewidth]{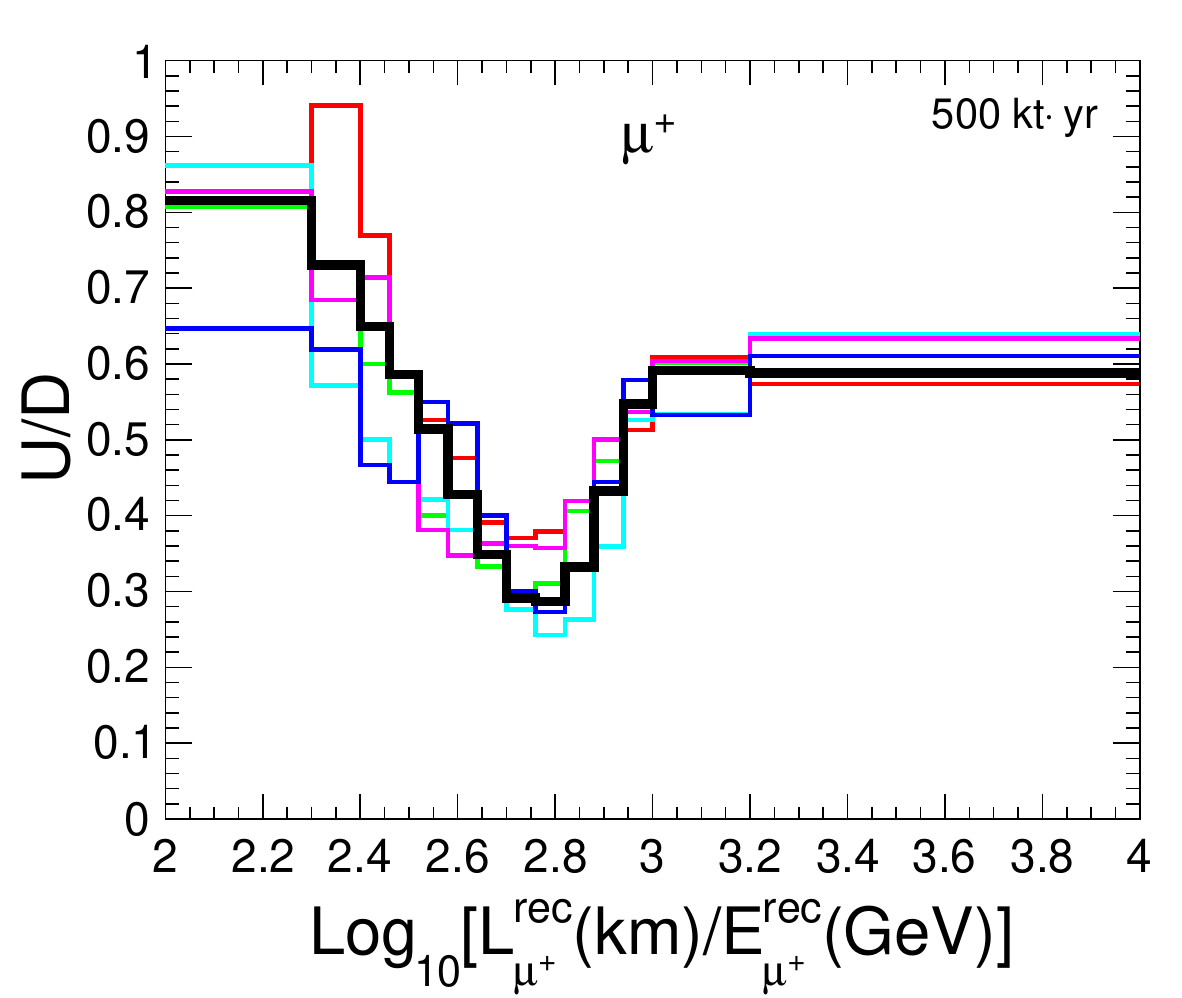}
	\includegraphics[width=0.45\linewidth]{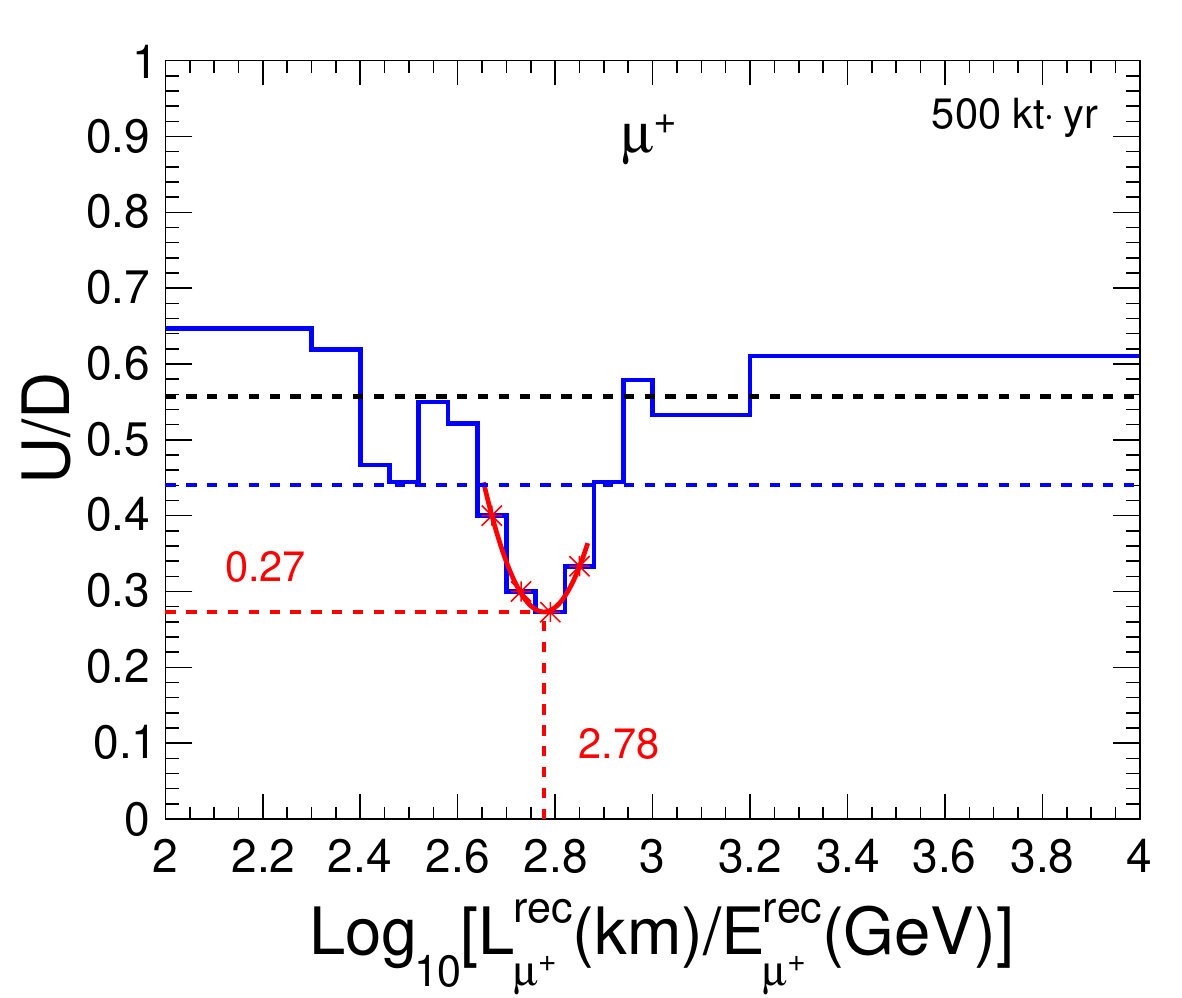}
	\caption{The left panel shows the U/D ratios as functions of reconstructed $\log_{10}[L_\mu^\text{rec}/E_\mu^\text{rec}]$ for five independent simulated data sets with 500 kt$\cdot$yr exposure at ICAL by thin colored lines. The solid black curve represents the mean of 100 such data sets. The red parabola in the right panel represents the fit to the dip in one of the simulated data sets (blue curve), obtained after employing the dip-identification algorithm described in the text. The dashed black line denotes the initial U/D ratio threshold used in the algorithm, while the dashed blue line represents the U/D ratio threshold after the identification of the dip.~\cite{Kumar:2020wgz}}
	\label{fig:osc_dip_fitting}
\end{figure}

In the left panel of Fig.~\ref{fig:osc_dip_fitting}, the U/D ratios are shown as functions of reconstructed $\log_{10}[L_\mu^\text{rec}/E_\mu^\text{rec}]$ of muons for five statistically independent simulated data sets with 500 kt$\cdot$yr exposure at ICAL. Clearly, the statistical fluctuations are significant and may result in the misidentification of the location of the dip. An identification of the dip location should not correspond simply to the minimum value of the U/D ratio, instead it should also be guided by the values of the U/D ratios in the neighboring bins. In order to accomplish this, we propose a dip-identification algorithm as described below.

%==========================
%==========================

We start by considering the region corresponding to $\log_{10}[L_\mu^\text{rec}/E_\mu^\text{rec}]$ in the range of 3.2 to 4.0 as a single bin because in this region, the oscillations are expected to be quite rapid, which leads to the averaging of the U/D ratio to a constant value. Let us consider $R_0$ to be the measured value of the U/D ratio in this bin. From 100 independent simulated sets, we find that the statistical fluctuations in data sets result in the rms deviations of $\Delta R_0 \approx 0.02$ and $0.03$ in the U/D ratio for $\mu^-$ and $\mu^+$, respectively. We take this fluctuation into account, and start with an initial ratio threshold of $R_\text{th} \equiv R_0 - 2\Delta R$, shown by the dashed black line  in the right panel of Fig.~\ref{fig:osc_dip_fitting}. All the bins with the measured values of U/D ratio less than $R_\text{th}$ form the initial candidates for the dip position. However, all these bins need not be a part of the actual dip, due to statistical fluctuations. To identify the actual bins forming the dip, we try to find the cluster of consecutive bins such that all of which have the U/D ratio lower than that in all the other bins. This is accomplished by decreasing the value of $R_\text{th}$ till all the bins with the U/D ratio less than $R_\text{th}$ are contiguous. This final value of $R_\text{th}$ is shown as the dashed blue line in the right panel of Fig.~\ref{fig:osc_dip_fitting}.

After the identification of a cluster of contiguous bins as the dip region using dip-identification algorithm, we fit the U/D ratios in these bins of dip region with a parabola as shown in the right panel of Fig.~\ref{fig:osc_dip_fitting}. The value of reconstructed $\log_{10}[L_\mu^\text{rec}/E_\mu^\text{rec}]$ corresponding to the minimum U/D ratio obtained from this fit, denoted by $x_\text{min}$, would be identified with the ``location'' of the dip.

%==========================
\subsection{Measurement of $\Delta m^2_{32}$ and $\theta_{23}$ from the dip and U/D ratio}
\label{sec:Calib-LbyE}
%==========================

\begin{figure}
	\centering
	\includegraphics[width=0.45\linewidth]{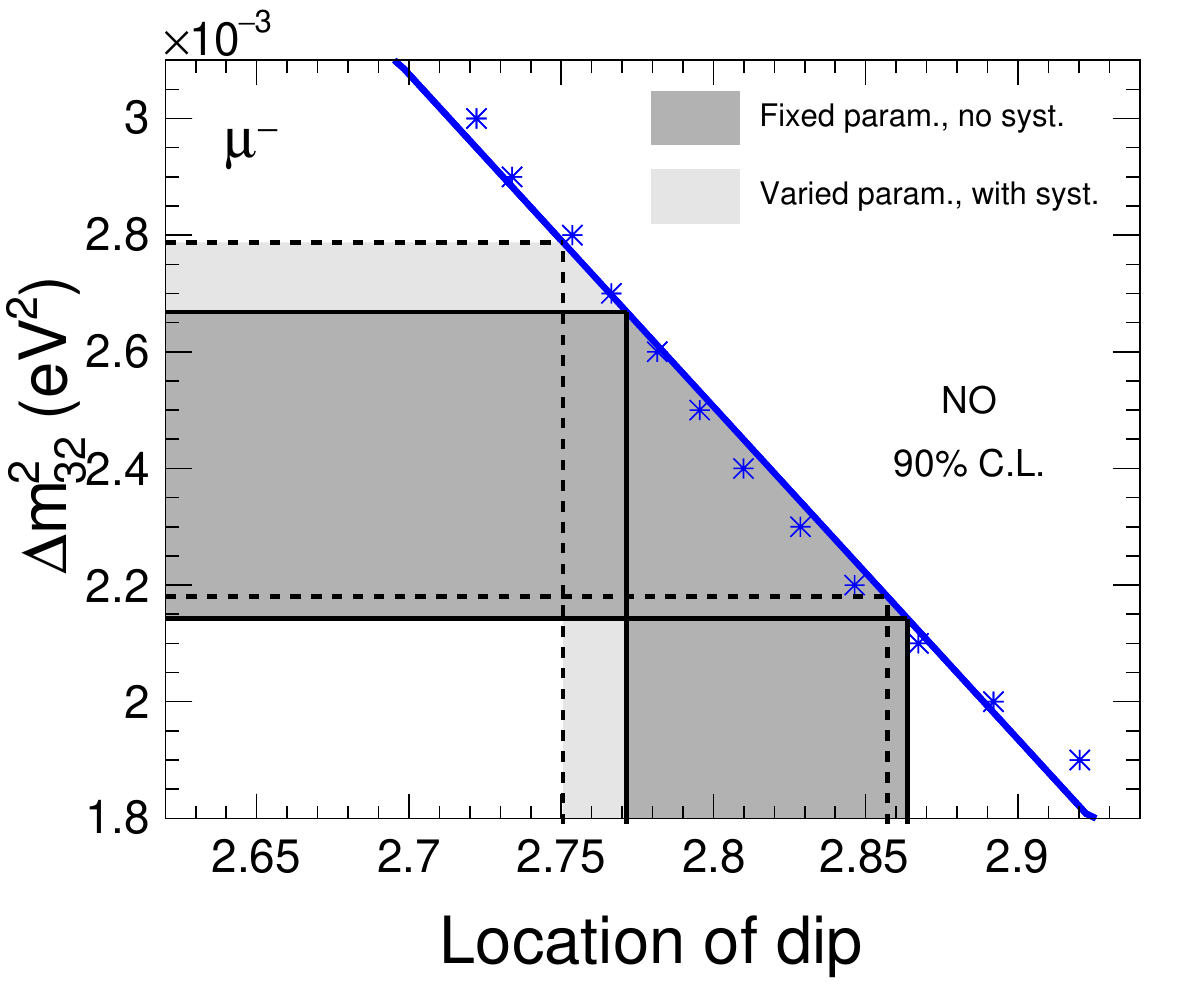}
	\includegraphics[width=0.45\linewidth]{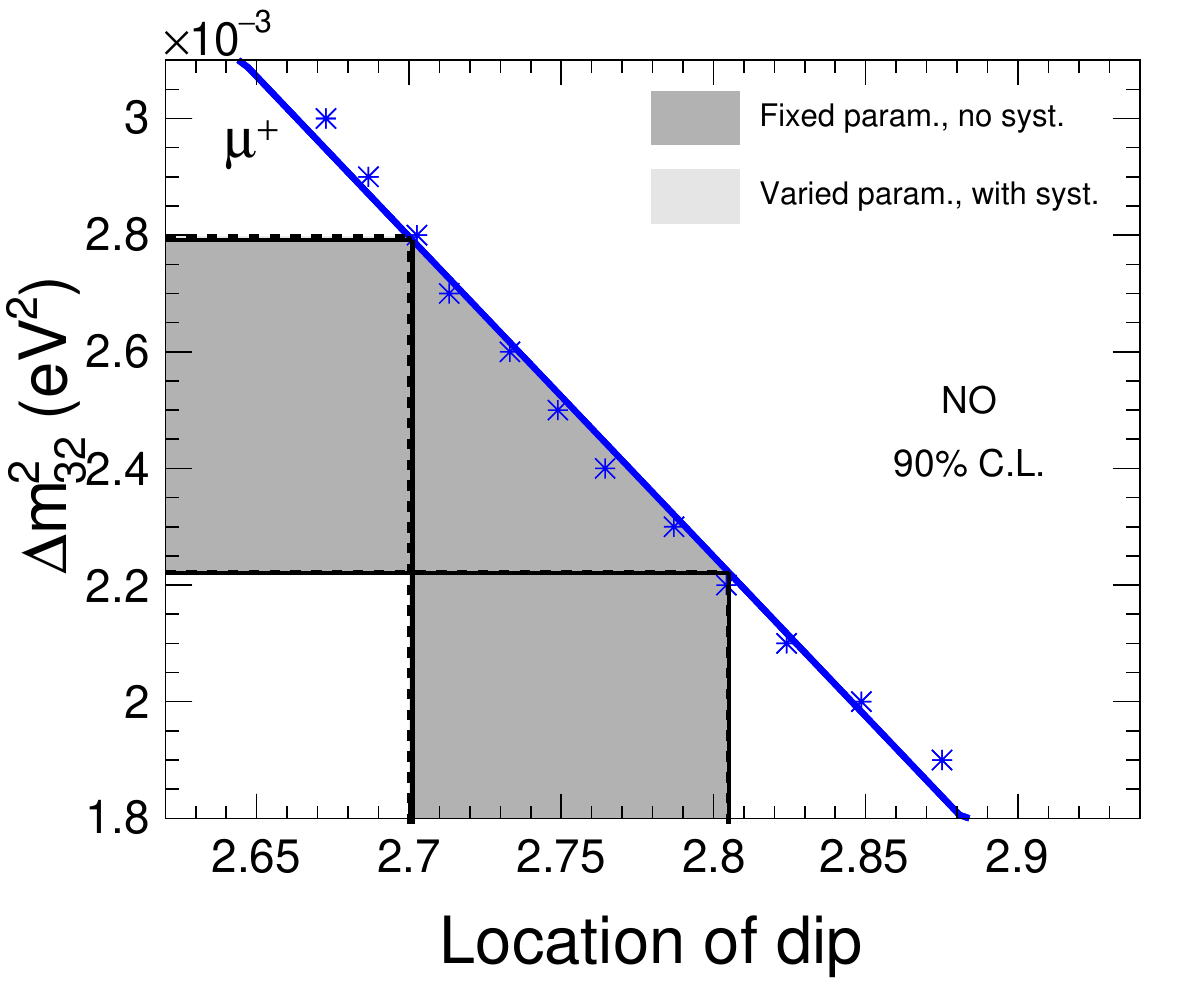}
	\includegraphics[width=0.45\linewidth]{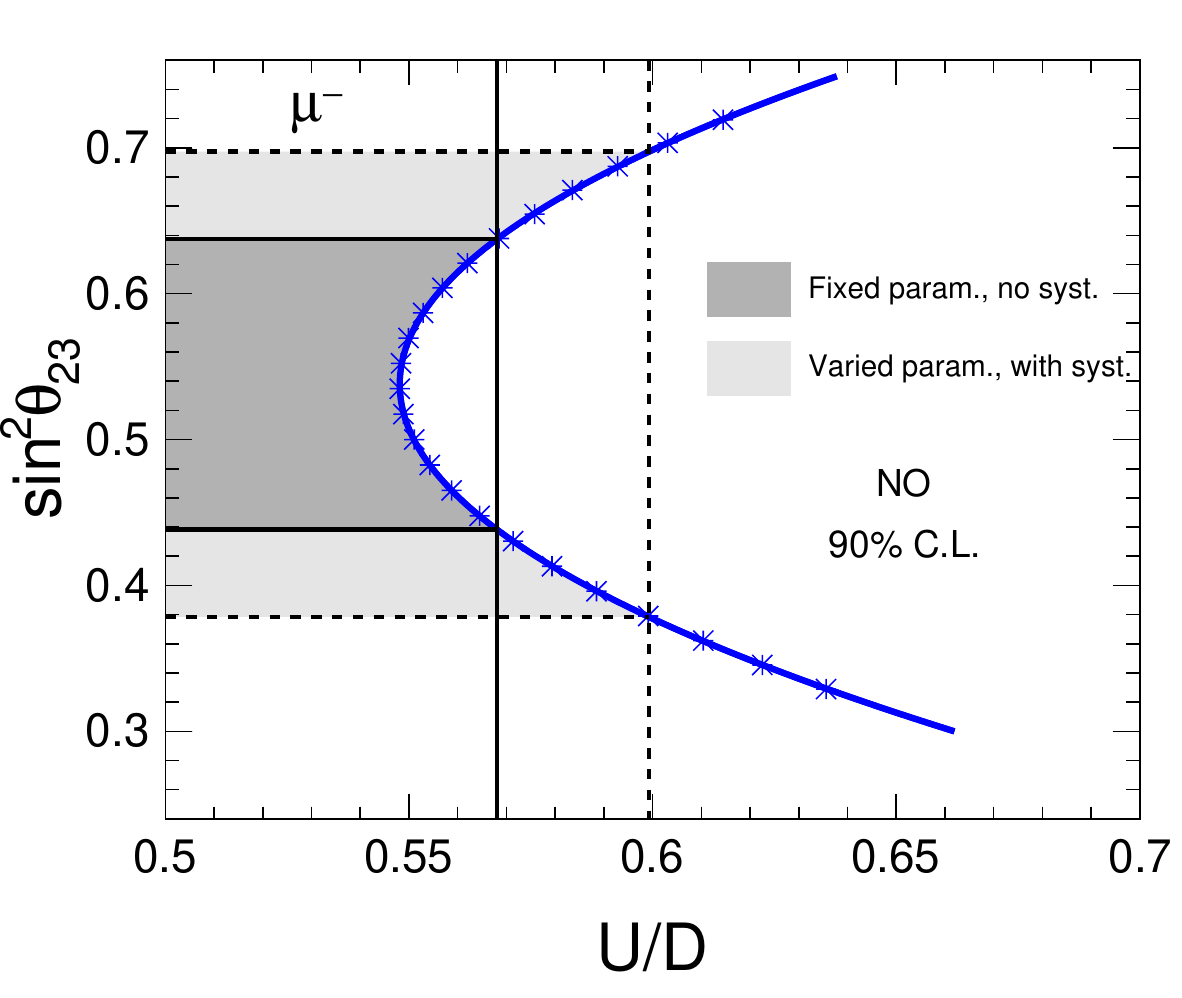}
	\includegraphics[width=0.45\linewidth]{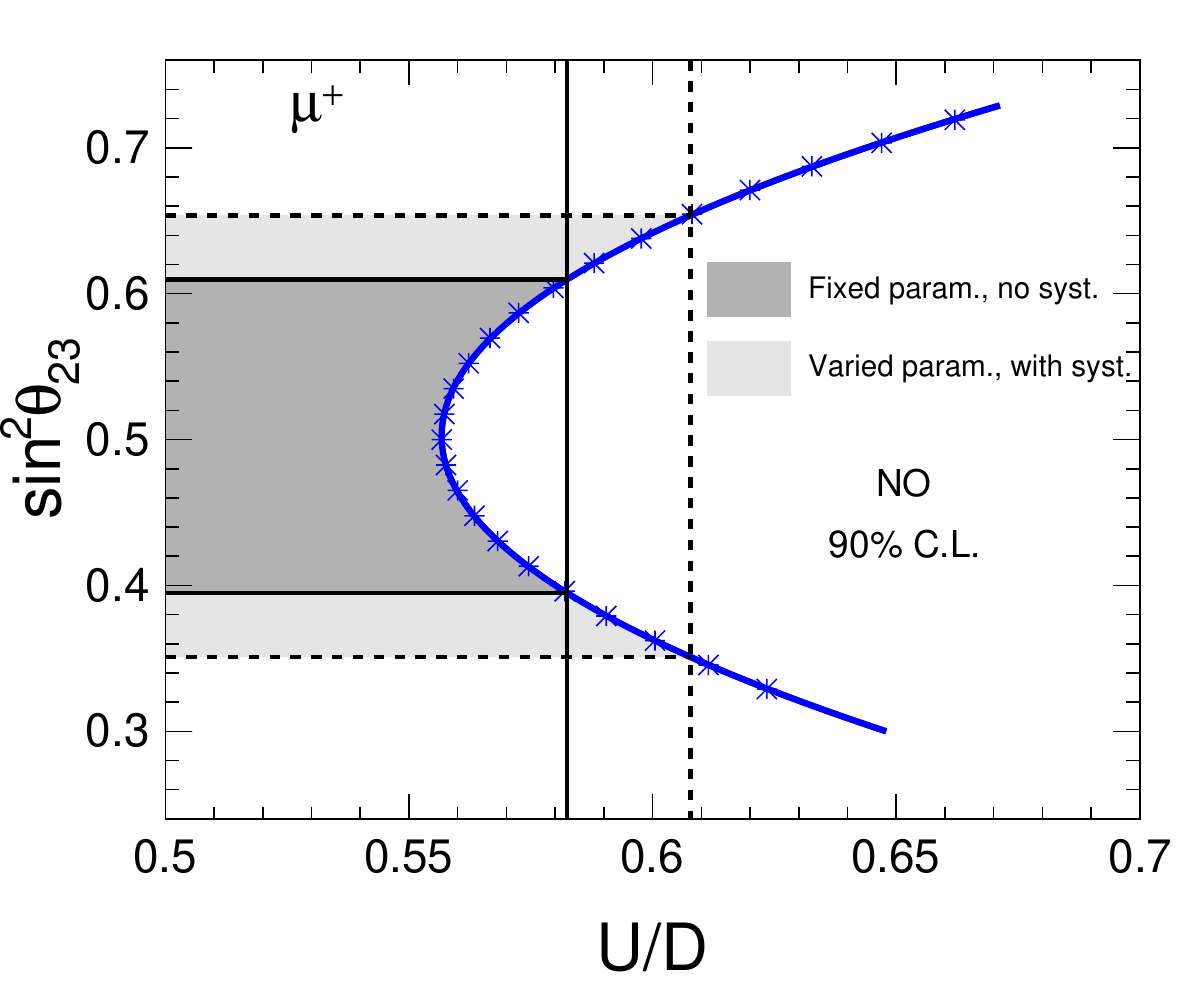}
	\caption{Upper panels: The blue points and the blue line correspond to
		the calibration of actual $\Delta m^2_{32}$ with the location of the
		dip, obtained using the 1000-year MC data sample. The light (dark) gray bands represent the 90\% C.L. regions of the location of the dip, and hence the inferred $\Delta m^2_{32}$ through calibration, for $\Delta m^2_{32} \,(\text {true}) = 2.46\times 10^{-3}$ eV$^2$,  with (without) systematic errors and uncertainties in the other oscillation parameters.  
		Lower panels: The blue points and the blue line correspond to the
		calibration of actual $\sin^2\theta_{23}$ with the total U/D ratio
		across all bins. The light (dark) gray bands represent the 90\% C.L. allowed ranges of the U/D ratio, and hence the inferred $\sin^2\theta_{23}$
		through calibration, for $\sin^2\theta_{23} \,(\text {true}) = 0.5$,  with (without) systematic errors and uncertainties in the other oscillation parameters. For all the confidence regions, we use 10-year exposure of ICAL. The results obtained  from $\mu^-$ and $\mu^+$ events are shown in the left and right panels, respectively. For the fixed-parameter analysis, we use the benchmark oscillation parameters given in Table~\ref{tab:osc-param-value}, while for the inclusion of systematic uncertainties and variation of oscillation parameters, we use the procedure described in Sec.~\ref{sec:Calib-LbyE}.~\cite{Kumar:2020wgz}}
	\label{fig:osc_dip_results}
\end{figure}

The first oscillation dip in the $\nu_\mu$ survival probability $P(\nu_\mu \rightarrow \nu_\mu)$, in the two-flavor approximation in vacuum oscillation, appears at $L_\nu(\text{km})/E_\nu(\text{GeV}) = \pi/(2.54\,\Delta m^2_{32}\, (\text{ eV}^2))$ (see Eq.~\ref{eq:2flavor_survival}). As discussed in Sec.~\ref{sec:LbyE-1000yrs}, the location of the dip $x_\text{min}$ in the distribution of $L_\mu^\text{rec}/E_\mu^\text{rec}$ would have a complicated dependence on the three-flavor neutrino oscillations, Earth's matter effects, neutrino fluxes, CC interaction cross-sections, inelasticities of events, and detector responses. However, we can calibrate $\Delta m^2_{32}$ with respect to the location of the dip, using the 1000-year MC sample. The binning scheme used for the calibration is the same as in Sec.~\ref{sec:dip_fitting}. For obtaining the calibration curves, we keep the neutrino oscillation parameters (except the one to be calibrated) fixed at the benchmark values mention in Table~\ref{tab:osc-param-value}. In the upper panel of Fig.~\ref{fig:osc_dip_results}, the blue points denote the values of $x_\text{min}$ obtained from the distribution of U/D ratio for different values of $\Delta m^2_{32}$. It can be observed that these points lie close to a straight line, and we can draw a calibration curve (blue line) that would enable us to infer the actual value of $\Delta m^2_{32}$, given the $x_\text{min}$ value determined from the data. Note that this calibration curve also depends on the analysis procedure, including the binning scheme and bin-identification algorithm.

%==========================
%==========================

While the calibration curve provides a one-to-one mapping between the actual value of $\Delta m^2_{32}$  and the location of the dip for sufficiently large data (such as 1000 years of exposure), practically speaking, the available data is going to be limited. The statistical fluctuations introduced due to this will result in uncertainties in the determination of $\Delta m^2_{32}$ using this method. To estimate these uncertainties, we generate 100 statistically independent simulated data sets with an exposure of 500 kt$\cdot$yr each, and determine the location of the dip in each of them. The resulting distribution of the location of dip allows us to determine the allowed regions for the dip location at 90\% C.L., and hence the 90\% C.L. allowed regions for the calibrated $\Delta m^2_{32}$ values. These allowed regions are shown as dark gray bands in the top panels of Fig.~\ref{fig:osc_dip_results} where we have kept all the oscillation parameters fixed at the benchmark values mentioned in Table~\ref{tab:osc-param-value} and have incorporated any systematic uncertainty. The figure indicates that the allowed range for $\Delta m^2_{32}$ at 90\% C.L. is $(2.14 - 2.67) \times 10^{-3}~\text{eV}^2$ from $\mu^-$ data and $(2.22 - 2.79) \times 10^{-3}~\text{eV}^2$ from $\mu^+$ data, while considering NO as mass ordering.

We also investigate the effects on the measurements of $\Delta m^2_{32}$ and $\theta_{23}$ due to uncertainties in the other oscillation parameters. For this, we vary the values of the other oscillation parameters in 100 statistically independent unoscillated data sets. For each of these data sets, 20 random choices of oscillation parameters are taken following the Gaussian distributions:
\begin{align}
\Delta m^2_{21} &= (7.4 \pm 0.2)\times 10^{-5} \,\text{eV}^2, \nonumber\\
\Delta m^2_{32} &= (2.46\pm 0.03)\times 10^{-3} \,\text{eV}^2\,,\nonumber \\
\sin^2 2\theta_{12} &= 0.855 \pm 0.020, \nonumber \\
\sin^2 2\theta_{13} &= 0.0875 \pm 0.0026,\nonumber \\
\sin^2\theta_{23} &= 0.50\pm 0.03,  
\end{align}
which are consistent with the present neutrino global fit results~\cite{NuFIT,Esteban:2020cvm,deSalas:2020pgw,Marrone:2021}. This way, the variations of the results over a large number (2000) of different combinations of values of the other oscillation parameters are effectively taken into account.  Note that the value of the oscillation parameter to be determined is kept fixed at benchmark value as given in Table~\ref{tab:osc-param-value}. We keep $\delta_\text{CP}$ fixed at zero because its effect on $\nu_\mu$ and $\bar\nu_\mu$ survival probabilities is negligible in the multi-GeV energy range.

We do not see any major changes in the measurement of $\Delta m^2_{32}$ caused by the uncertainties in the other oscillation parameters, viz. $\theta_{12}$, $\theta_{23}$, $\theta_{13}$, and $\Delta m^2_{21}$. This is expected, because (a) the mixing angle $\theta_{23}$ does not change the dip-locations of $\nu_\mu$ and $\bar\nu_\mu$ survival probabilities, (b) the mixing angle $\theta_{13}$ is already precisely measured, (c) the solar oscillation parameters $\Delta m^2_{21}$ and $\theta_{12}$ have negligible impact on $\nu_\mu$ and $\bar\nu_\mu$ survival probabilities in the multi-GeV energy range.    

In addition, we take into account the five major systematic uncertainties in the neutrino fluxes and cross sections that are used in the standard ICAL analyses~\cite{ICAL:2015stm}. These five uncertainties are (i) 20\% in overall flux normalization, (ii) 10\% in cross sections, (iii) 5\% in the energy dependence, (iv) 5\% in the zenith angle dependence, and (v) 5\% in overall systematics. For each of the 2000 simulated data sets, we modify the number of events in each $(E_\mu^\text{rec}, \cos\theta_\mu^\text{rec})$  bin as
\begin{equation}
N = N^{(0)} (1 + \delta_1) (1 + \delta_2) 
(E_\mu^\text{rec}/E_0)^{\delta_3}
(1 + \delta_4 \cos \theta_\mu^\text{rec}) (1 + \delta_5)~,
\end{equation}
where $N^{(0)}$ is the theoretically predicted number of events, and $E_0 = 2$ GeV. Here, $(\delta_1, \delta_2, \delta_3, \delta_4, \delta_5)$ is an ordered set of random numbers, generated separately for each simulated data set, with the Gaussian distributions centered around zero and the $1\sigma$ widths given by $(20\%, 10\%, 5\%, 5\%, 5\%)$. The normalization uncertainties and energy tilt uncertainty are expected to be canceled in the U/D ratio, while the zenith angle distribution uncertainty affects the upward-going and downward-going events differently, and hence would be expected to affect the U/D ratio. We explicitly check this and indeed find this to be true.

After incorporating systematic uncertainties and varying oscillation parameters, we get the 90$\%$ C.L. allowed range for $\Delta m^2_{32}$ from $\mu^-$ events as $(2.18 - 2.79)\times 10^{-3}$ eV$^2$, and from $\mu^+$ data as $(2.22 - 2.80)\times10^{-3}$ eV$^2$, for $\Delta m^2_{32} \,(\text {true}) = 2.46\times 10^{-3}$ eV$^2$. These results are shown with light gray bands in the upper panels of Fig.~\ref{fig:osc_dip_results}. Note that  the change in the measurement of $\Delta m^2_{32}$ from $\mu^+$ events after inclusion of systematic uncertainties and errors in other oscillation parameters is quite small, whereas that from $\mu^-$ events is larger. This is because the dip in $\mu^+$ data is quite sharp, owing to the lower inelasticity in $\bar\nu_\mu$ CC interactions.

We can follow the same procedure to determine the value of the mixing angle $\theta_{23}$ from the data. In principle, this is related to the depth of the dip, and such a calibration could be obtained using the 1000-year MC events. However, the statistical fluctuations in the 500 kt$\cdot$yr simulated data are observed to give rise to large uncertainties. On the other hand, it is found that the total U/D ratio of all events with $L_\mu^\text{rec}/E_\mu^\text{rec}$ in the range of 2.2 -- 4.1 gives a much better estimate of $\theta_{23}$. In the lower panels of Fig.~\ref{fig:osc_dip_results}, we show in blue the calibration for $\sin^2\theta_{23}$ with the measured total U/D ratio, obtained using the 1000-year MC sample for several $\sin^2\theta_{23} \,(\text {true})$ values. We also show the 90\% C.L. allowed values of $\sin^2\theta_{23}$ inferred using the 100 independent simulated data sets with 500 kt$\cdot$yr exposure each, keeping all the oscillation parameters fixed at benchmark values as given in Table \ref{tab:osc-param-value}, with dark gray bands. The expected allowed range for $\sin^2\theta_{23}$ at $90\%$ C.L. is (0.44 -- 0.64) from $\mu^-$ events, and (0.39 -- 0.61) from $\mu^+$ events. With the systematic uncertainties and the variation of  oscillation parameters $\Delta m^2_{32}$, $\theta_{13}$, $\theta_{12}$, and $\Delta m^2_{21}$ as mentioned above, we get the 90$\%$ C.L. allowed range for $\sin^2\theta_{23}$ from $\mu^-$ events as (0.38 -- 0.70), and from $\mu^+$ data as (0.35 -- 0.65), for $\sin^2\theta_{23} \,(\text {true}) = 0.5$. These results are shown with light gray bands in the lower panels of Fig.~\ref{fig:osc_dip_results}.

Note that we do not claim or demand that our analysis gives better results than those obtained with the complete $\chi^2$ analysis done for ICAL in Refs.~\cite{Thakore:2013xqa,Devi:2014yaa}. The $\chi^2$ analyses take into account the complete event spectra, while our focus is on the identification of the dip, a feature that would reconfirm the oscillation paradigm for neutrinos. The observation that the inferred $\Delta m^2_{32}$ range is comparable to the range expected from the complete $\chi^2$ analysis points to the conclusion that most of the information about the $\Delta m^2_{32}$ is concentrated in the location of the U/D dip.

%=====================================
\section{Oscillation valley in the $(E_\mu^\text{rec}, \cos\theta_\mu^\text{rec})$ plane}
\label{sec:2D_E-CT}
%=====================================

The oscillation dip discussed in the last section was a dip in the U/D ratio as a function of $L_\mu^\text{rec}/E_\mu^\text{rec}$. The dependence on $L_\mu^\text{rec}/E_\mu^\text{rec}$ was motivated from the approximate form of the neutrino oscillation probability (with two flavors in vacuum). However, if the detector can reconstruct both $L_\mu$ (hence $\cos\theta_\mu$) and $E_\mu$ accurately, then one can go a step ahead and look at the distribution of the U/D ratio in the $(E_\mu, \cos\theta_{\mu})$ plane. We perform such an analysis in this section, and find that such a distribution can have many interesting features with physical significance, which may be identifiable with sufficient data. In particular, we point out an ``oscillation valley'' corresponding to the dark diagonal band in Fig.~\ref{fig:osc_valley_neutrino}, whose nature can provide stronger tests for the oscillation hypothesis, albeit only with a large amount of data. On the other hand, we show that the identification of the oscillation valley, and the determination of $\Delta m^2_{32}$ based on its ``alignment'', is possible with simulated data of 500 kt$\cdot$yr exposure at ICAL.

%=================================
\subsection{Events and U/D ratio using 1000-year Monte Carlo simulation}
\label{sec:2D-dist-1000yr}
%=================================

In this section, we discuss the distributions of events and the U/D ratio, in the plane of reconstructed  energy $E_{\mu}^\text{rec}$ and zenith angle $\cos\theta_{\mu}^\text{rec}$ of the $\mu^-$ and $\mu^+$ events, for a 1000-year MC sample. The main reason for considering such a huge exposure here is not to miss those features, which survive in spite of our not using $E_\nu$ and $\cos\theta_\nu$ directly, but which may disappear due to the limitation of statistics.

%=================================
\begin{figure}
	\centering
	\includegraphics[width=0.45\linewidth]{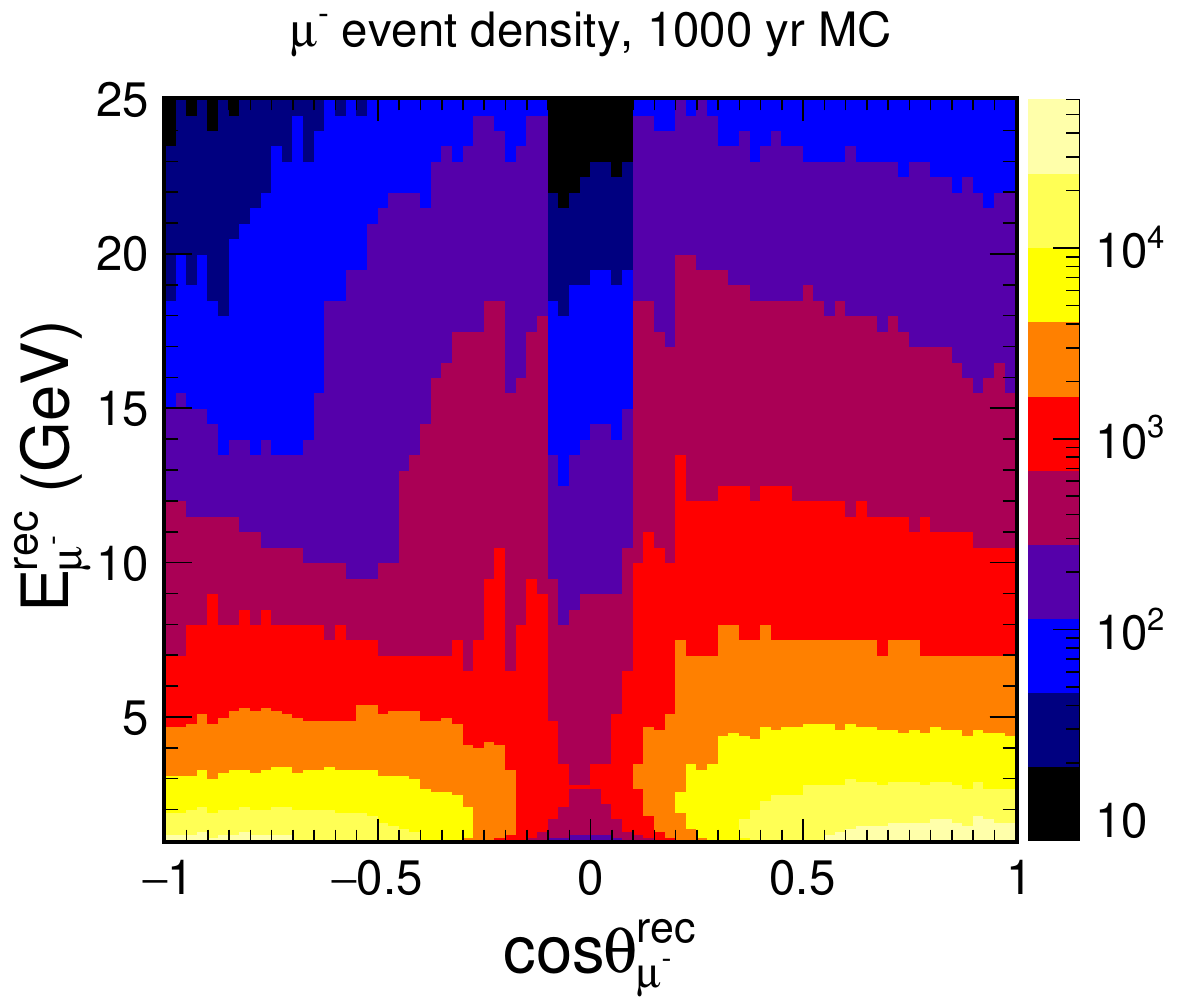}
	\includegraphics[width=0.45\linewidth]{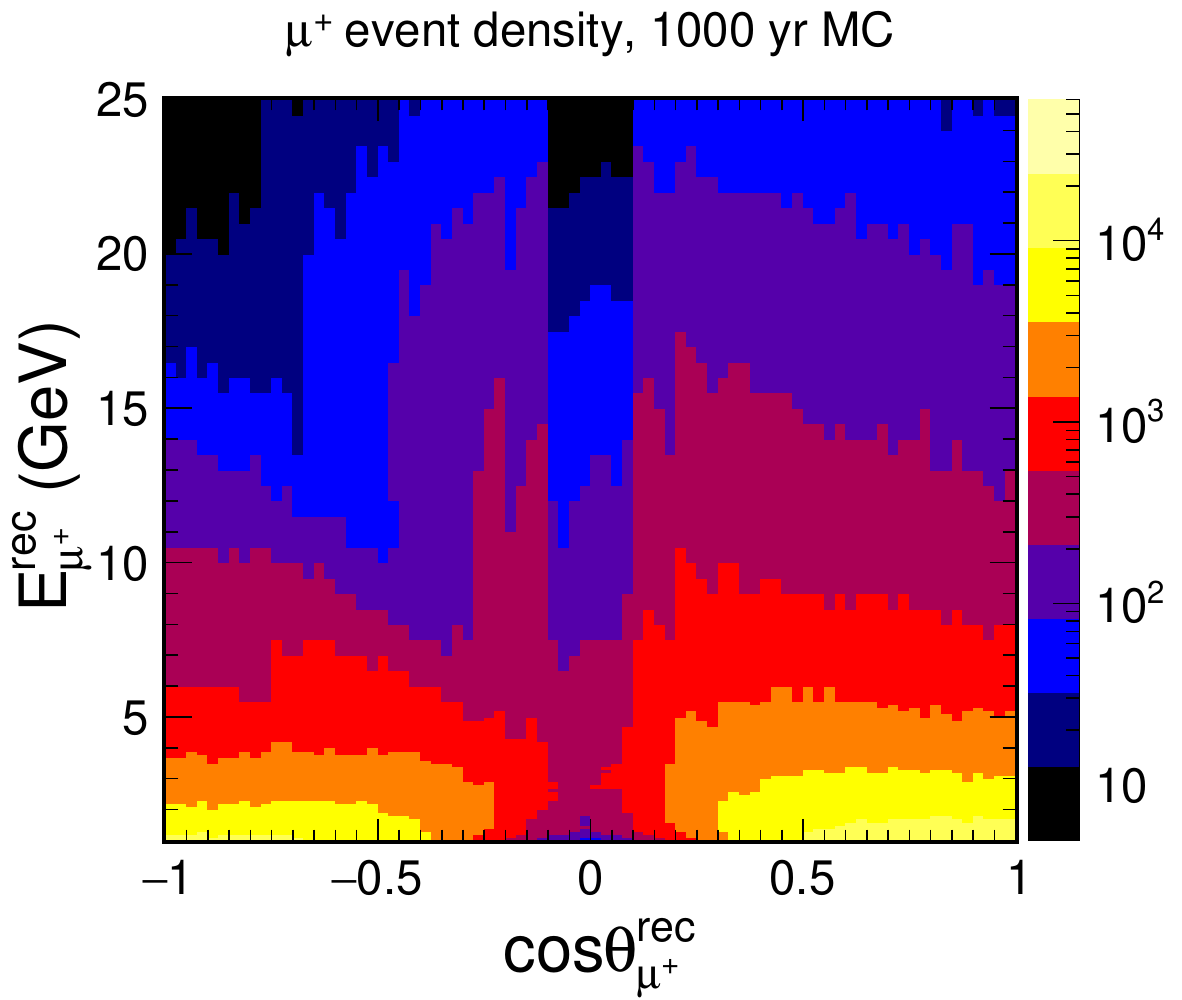}
	\includegraphics[width=0.45\linewidth]{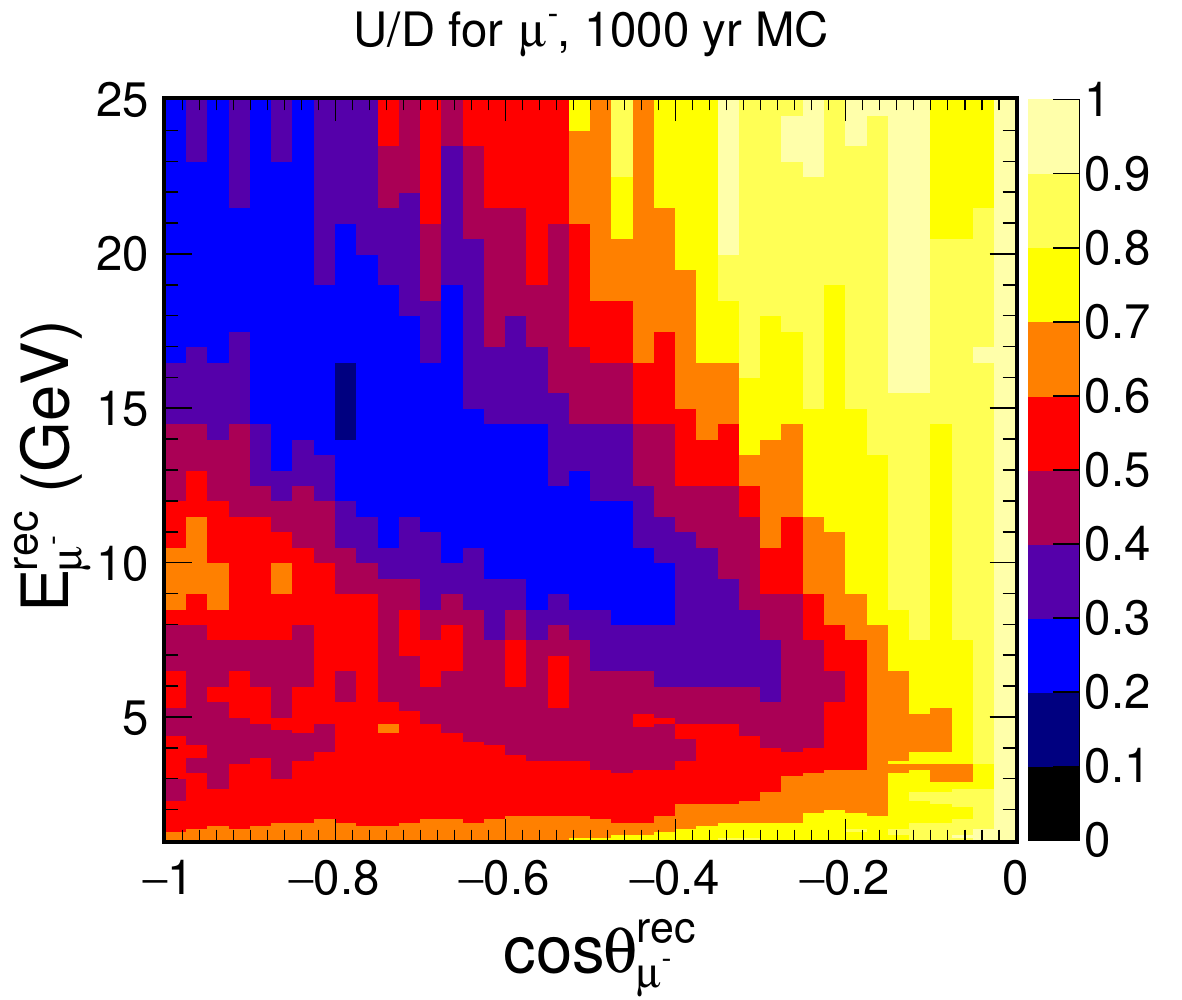}
	\includegraphics[width=0.45\linewidth]{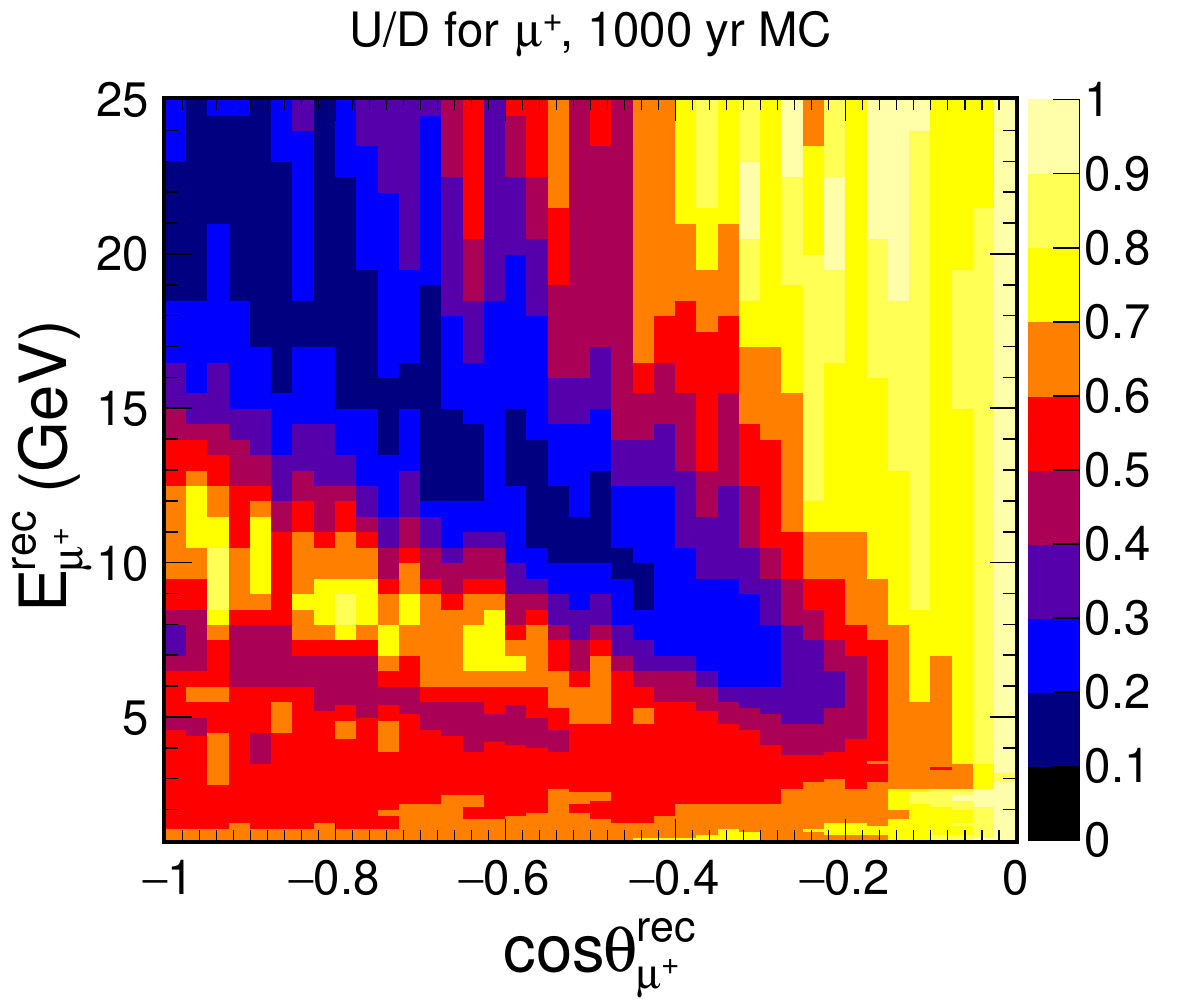}
	\caption{The distributions of event densities (upper panels) and U/D ratios (lower panels), for $\mu^{-}$ (left panels) and $\mu^{+}$ (right panels), in the ($E_\mu^\text{rec},\cos\theta_\mu^\text{rec}$) plane, obtained using the 1000-year MC sample. We use the oscillation parameters given in Table~\ref{tab:osc-param-value}. Note that the event density in the upper panels has units of [GeV$^{-1}$sr$^{-1}$], and is obtained by dividing the number of events in each bin by ($2\pi~\times$ the bin area), i.e. by ($2\pi \Delta E_{\mu}^\text{rec} \Delta \cos\theta_{\mu}^\text{rec}$). Here $\Delta E_{\mu}^\text{rec}$ and  $\Delta \cos\theta_{\mu}^\text{rec}$ are the height and the width of the bin, respectively. Note that the U/D($E_\mu^\text{rec}, \cos\theta_\mu^\text{rec}$) is defined only for $\cos\theta_\mu^\text{rec} < 0$ (see Eq.~\ref{eq:U/D_def}).~\cite{Kumar:2020wgz}}
	\label{fig:osc_valley_1000yr}
\end{figure}
%=================================

The quantities shown in Fig.~\ref{fig:osc_valley_1000yr} are binned based on the reconstructed values of $E_{\mu}^\text{rec}$ and $\cos\theta_{\mu}^\text{rec}$. The $\cos\theta_\mu^\text{rec}$ range of -1.0 to 1.0 has been divided into 80 bins of equal width. We have a total of 84 bins in $E_{\mu}^\text{rec}$ in the range of 1 GeV to 25 GeV. The bin width is not uniform in $E_\mu^\text{rec}$, since at higher energies the number of events decreases rapidly. The first 45 $E_\mu^\text{rec}$ bins, in the range of 1.0 GeV to 5.5 GeV, have a width of 0.1 GeV each, and the last 39 bins, in the range of 5.5 GeV to 25 GeV, have a width of 0.5 GeV each. The upper panels of Fig.~\ref{fig:osc_valley_1000yr} present the distributions of $\mu^-$ (left panel) and $\mu^+$ (right panel) event densities. The abrupt lowering of the event density for $-0.2<\cos\theta_{\mu}^\text{rec}<0.2$ is due to the poor reconstruction efficiency of the ICAL detector for the events coming near the horizontal direction. Note that this feature was absent in Fig.~\ref{fig:osc_dip_1000yr} since it was distributed over a large $L_\mu^\text{rec}/E_\mu^\text{rec}$ range.

In lower panels of Fig.~\ref{fig:osc_valley_1000yr}, we show the U/D ratio in each bin, in the ($E_\mu^\text{rec}, \cos\theta_\mu^\text{rec}$) plane.  Note that the U/D($E_\mu^\text{rec}, \cos\theta_\mu^\text{rec}$) is defined only for $\cos\theta_\mu^\text{rec} < 0$ (See Eq.~\ref{eq:U/D_def}). This also makes the features in Fig.~\ref{fig:osc_valley_1000yr} resemble those in Fig.~\ref{fig:osc_valley_neutrino}, thus enabling their understanding in terms of the survival probabilities shown in the oscillograms. The lower panels may be observed to have many interesting features. The most prominent one is of course the light/dark blue band, corresponding to U/D $\lesssim 0.25$, extending diagonally from $(E_\mu^\text{rec}, \cos\theta_{\mu}^\text{rec}) \approx (5 ~\text{GeV},-0.3)$ to $(E_\mu^\text{rec}, \cos\theta_{\mu}^\text{rec}) \approx (25 ~\text{GeV},-1.0)$. This is the band with the lowest values of U/D ratio in this plane, and we shall henceforth refer to it as the ``oscillation valley''. Clearly, this valley is deeper for $\mu^+$, since the number of dark blue bins with U/D $< 0.15$ is seen to be larger for $\mu^+$. This is consistent with the observation of a deeper oscillation dip in $\mu^+$, as discussed in Sec.~\ref{sec:LbyE-1000yrs}. The main reason for this is the smaller inelasticity of antineutrino CC events. The differences observed between the left and right panels may be attributed to the Earth matter effects, which affect the survival probabilities of neutrinos and antineutrinos differently for a given neutrino mass ordering.

%=================================
\subsection{Events and U/D ratio using 10-year simulated data}
\label{sec:2D-dist-10yr}
%=================================

\begin{figure}
	\centering
	\includegraphics[width=0.45\linewidth]{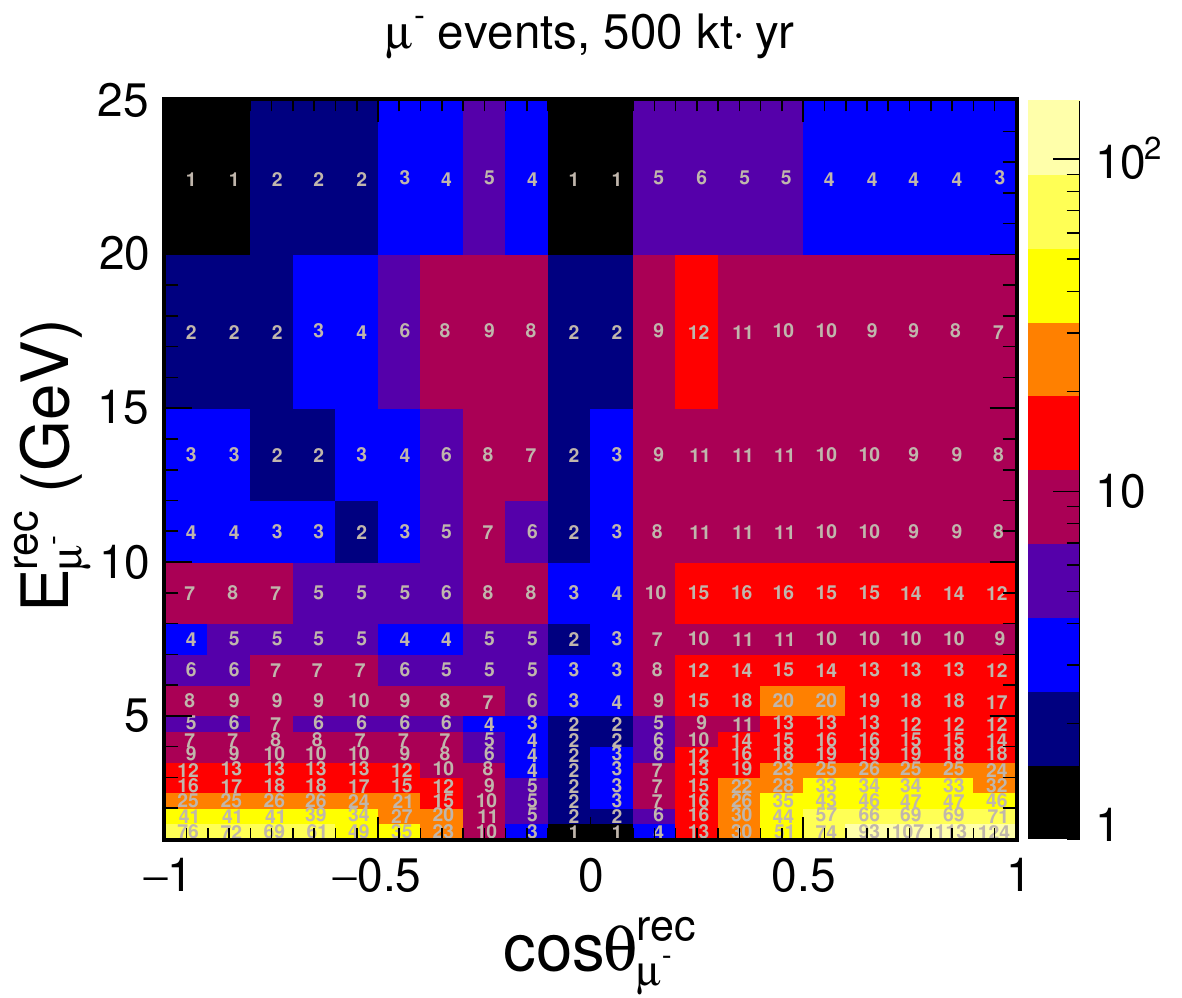}
	\includegraphics[width=0.45\linewidth]{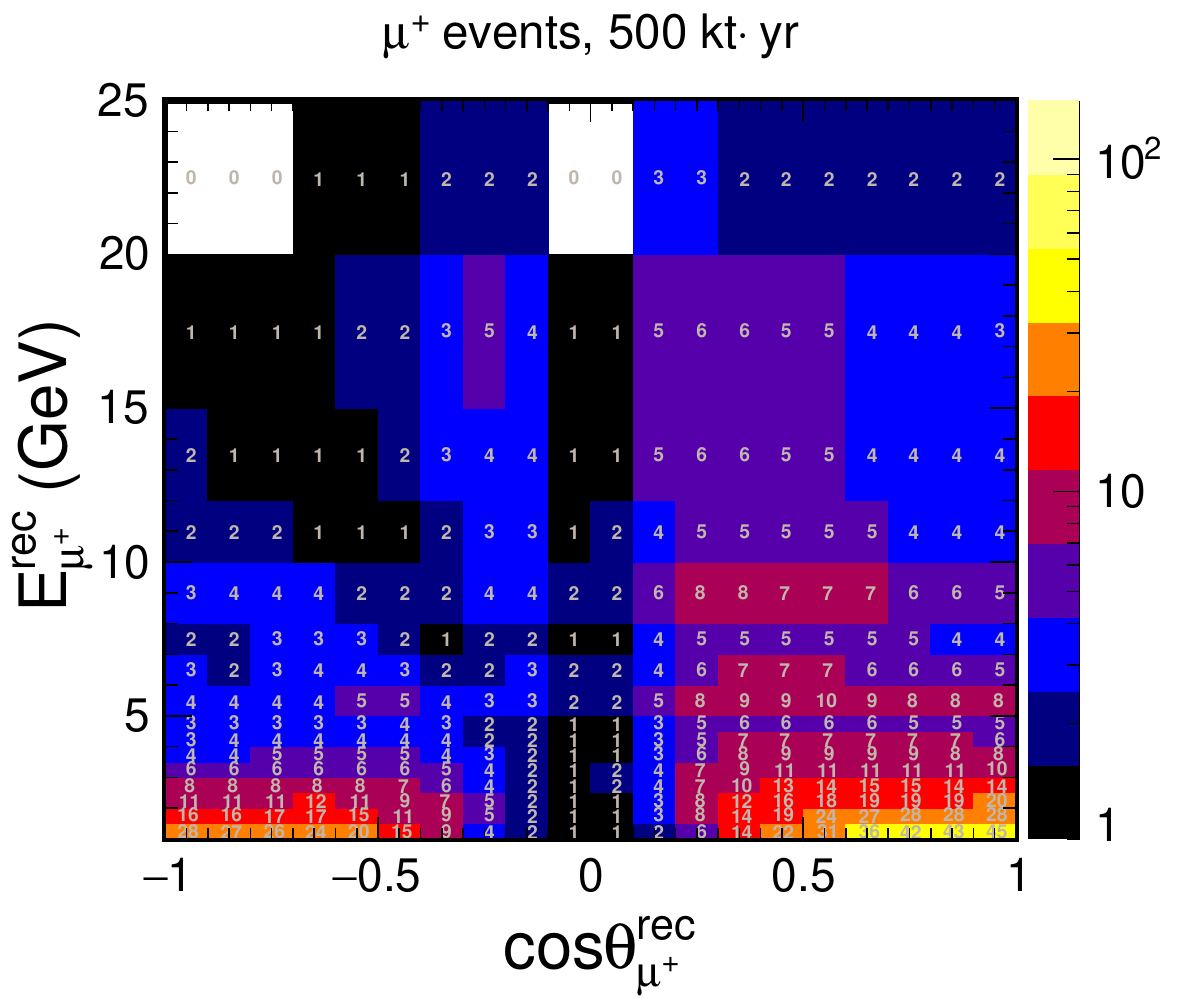}
	\includegraphics[width=0.45\linewidth]{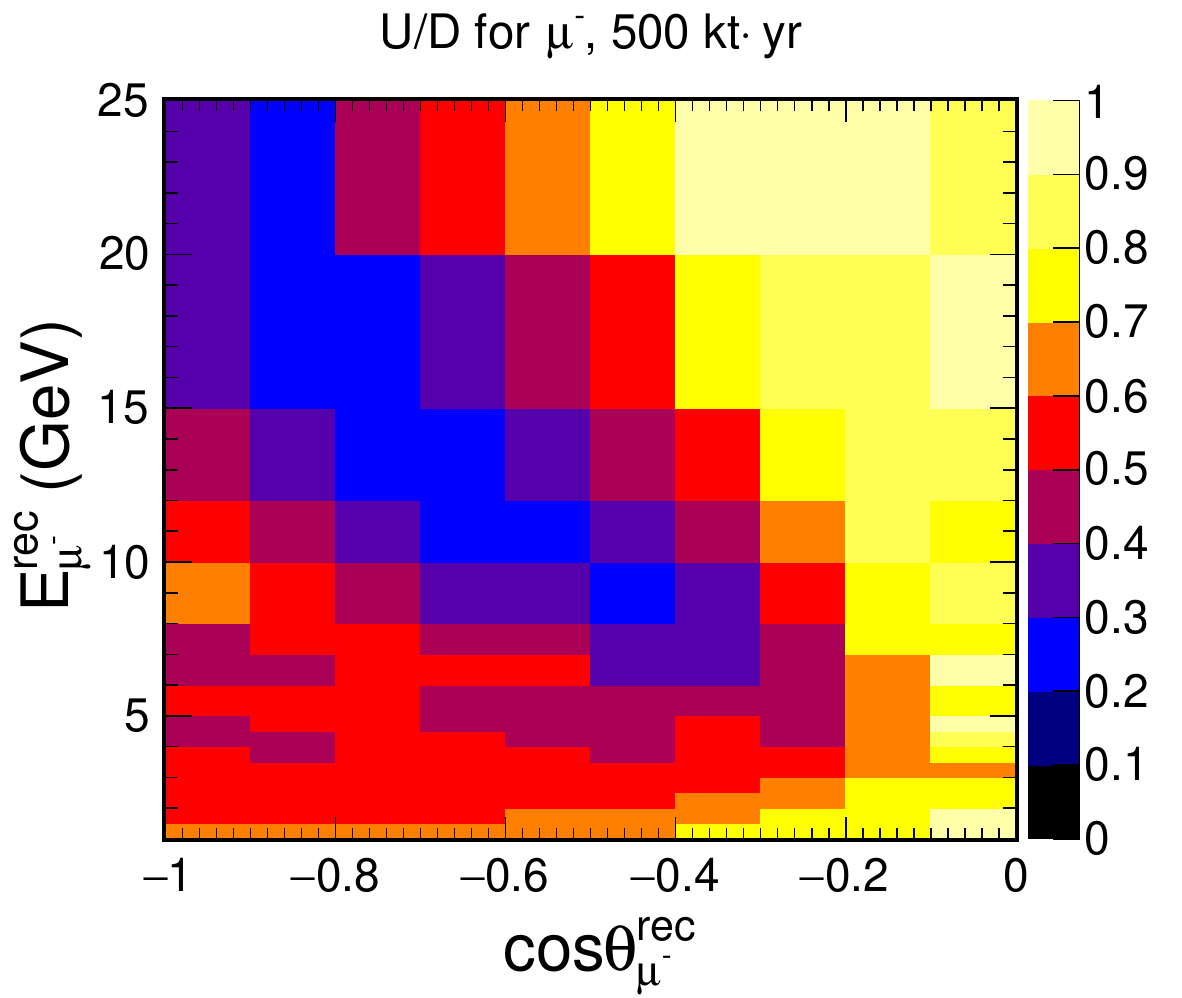}
	\includegraphics[width=0.45\linewidth]{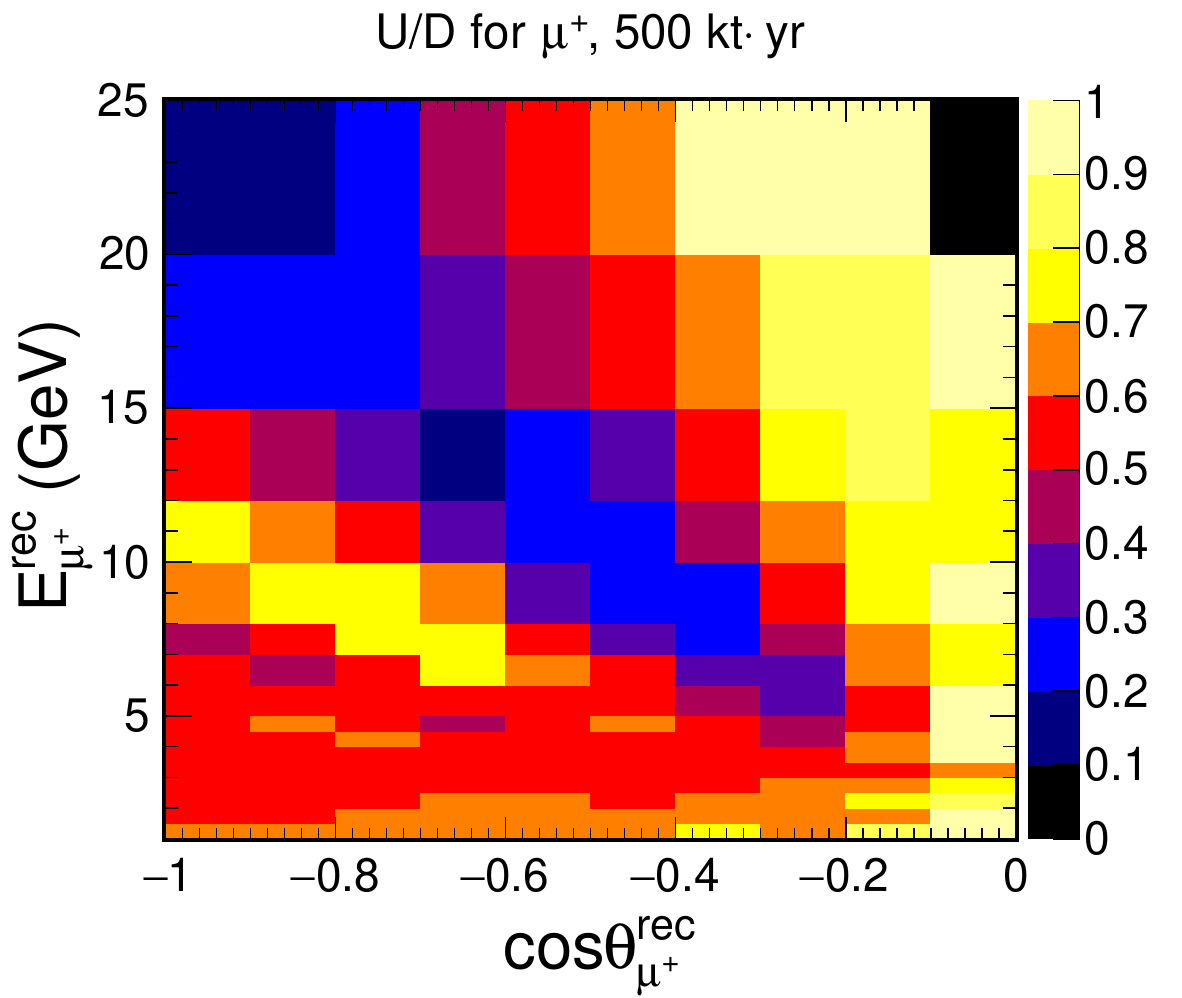}
	\caption{The distributions of mean number of events (upper panels)
		and mean U/D ratios (lower panels) for $\mu^{-}$ (left panels) and
		$\mu^{+}$ (right panels) in the plane of $\cos\theta_{\mu}^\text{rec}$
		and $E_{\mu}^\text{rec}$. The mean is calculated from 100 independent
		data sets of exposure 500 kt$\cdot$yr each at ICAL.
		The number of events written in white are rounded to the
		nearest integers. 
		We use the oscillation parameters given in Table~\ref{tab:osc-param-value}.  Note that the U/D($E_\mu^\text{rec},  \cos\theta_\mu^\text{rec}$) is defined only for $\cos\theta_\mu^\text{rec} < 0$ (see Eq. \ref{eq:U/D_def}).~\cite{Kumar:2020wgz}}
	\label{fig:osc_valley_10yr}
\end{figure}

In the upper panels of Fig.~\ref{fig:osc_valley_10yr}, we show the expected event distributions of $\mu^-$ (left panel) and $\mu^+$ (right panel) events, for 500 kt$\cdot$yr exposure at ICAL, in the plane of reconstructed muon energy ($E_{\mu}^\text{rec}$) and reconstructed muon zenith angle ($\cos\theta_{\mu}^\text{rec}$). The $\cos\theta_{\mu}^\text{rec}$ range of -1.0 to 1.0 is divided into 20 uniform bins of width 0.1 each. The binning in $E_{\mu}^\text{rec}$ is non-uniform, and is such that there are at least a few number of events in each bin (except in the largest energy bins, where this may not be possible due to the smaller neutrino flux at higher energies). The $E_{\mu}^\text{rec}$ range of 1 GeV to 25 GeV has been divided into 16 non-uniform bins, as given in Table \ref{tab:binning-2D-10years}. 

%------------------------------------------
\begin{table}[htb!]
	\centering
	\begin{tabular}{|c|c|c|c c|}
		\hline
		Observable & Range & Bin width & \multicolumn{2}{c|}{Number of bins} \\
		\hline 
		\multirow{4}{*}{ $E_\mu^\text{rec}$ } & [1, 5] &  0.5 & 8 & \rdelim\}{5}{7mm}[16] \\ 
		& [5, 8] & 1 & 3  & \\
		\multirow{2}{*}{ (GeV) }  & [8, 12] & 2 & 2  & \\
		& [12, 15] & 3 & 1  & \\
		& [15, 25] & 5 & 2  & \\
		\hline 
		$\cos\theta_\mu^\text{rec}$ & [-1.0, 1.0] & 0.1 & 20  & \cr
		\hline
	\end{tabular}
	\caption{The binning scheme considered for $E_\mu^\text{rec}$ and $\cos\theta_\mu^\text{rec}$ for $\mu^-$ and $\mu^+$ events of 10-year simulated data while we show event distribution and U/D plots in the ($E_\mu^\text{rec}$, $\cos\theta_\mu^{\text{rec}}$) plane. This binning scheme is used in analysis of oscillation valley as well.~\cite{Kumar:2020wgz} }
	\label{tab:binning-valley-10years}
\end{table}
%-----------------------------------------------

In order to take care of the statistical fluctuations that will clearly be significant with 500 kt$\cdot$yr data, we generate 100 independent data sets, each with this exposure. In the upper panels of Fig.~\ref{fig:osc_valley_10yr}, we show the mean number of events obtained from these 100 data sets. Similarly, in the lower panels of Fig.~\ref{fig:osc_valley_10yr}, we calculate the U/D ratios in each bin for the above 100 sets, and present their mean values in the figure, using appropriate colors. Note that the U/D($E_\mu^\text{rec}, \cos\theta_\mu^\text{rec}$) is defined only for $\cos\theta_\mu^\text{rec} < 0$ (see Eq. \ref{eq:U/D_def}).

Clearly, while many of the features corresponding to matter effects seem to get diluted due to the coarser bins and statistical fluctuations, the oscillation valley along the diagonal survives. Note that a good reconstruction of the oscillation valley would be an evidence for the fidelity of the ICAL detector. In the following sections, we shall provide a procedure for identifying this oscillation valley and extracting information about the oscillation parameters from it.

%=================================
\subsection{Identifying the valley with 10-year simulated data}
\label{sec:2D-fit-procedure}
%=================================

We now describe the methodology adopted for the identification of the oscillation valley in the $(E_\mu^\text{rec}, \cos\theta_\mu^\text{rec})$ plane, and the determination of its alignment. To get the alignment of the oscillation valley, we fit the U/D distribution for $\mu^-$ or $\mu^+$ independently with a functional form
\begin{equation}
F_0(E_\mu^\text{rec}, \cos\theta_\mu^\text{rec}) = Z_0 + N_0 \cos^2
\left(m_0 \frac{\cos\theta_\mu^\text{rec}}{E_\mu^\text{rec}} \right),
\label{eq:fitting-func-sm}
\end{equation}
where $Z_0$, $N_0$, and $m_0$ are the independent parameters to be determined from the fitting of U/D distributions. The parameter $Z_0$ quantifies the minimum depth of the fitted U/D ratio, $N_0$ is the normalization constant, whereas $m_0$ is the slope of the oscillation valley. Since more than $95\%$ of the events at ICAL are contributed from the $\nu_\mu \to \nu_\mu$ and $\bar\nu_\mu \to \bar\nu_\mu$ survival probabilities, we expect that the function $F_0$, that resembles Eq.~\ref{eq:pmumu-nu-final-omsd}, would be suitable for fitting the oscillation valley in the $(E_\mu^\text{rec}$, $\cos\theta_\mu^\text{rec})$ plane. Here, the slope $m_0$ can be used directly to calibrate $\Delta m^2_{32}$. The parameters $N_0$ and $Z_0$ also contain information about the atmospheric mixing parameters, however, they cannot be connected easily to the determinations of these parameters.

\begin{figure}
	\centering
	\includegraphics[width=0.45\linewidth]{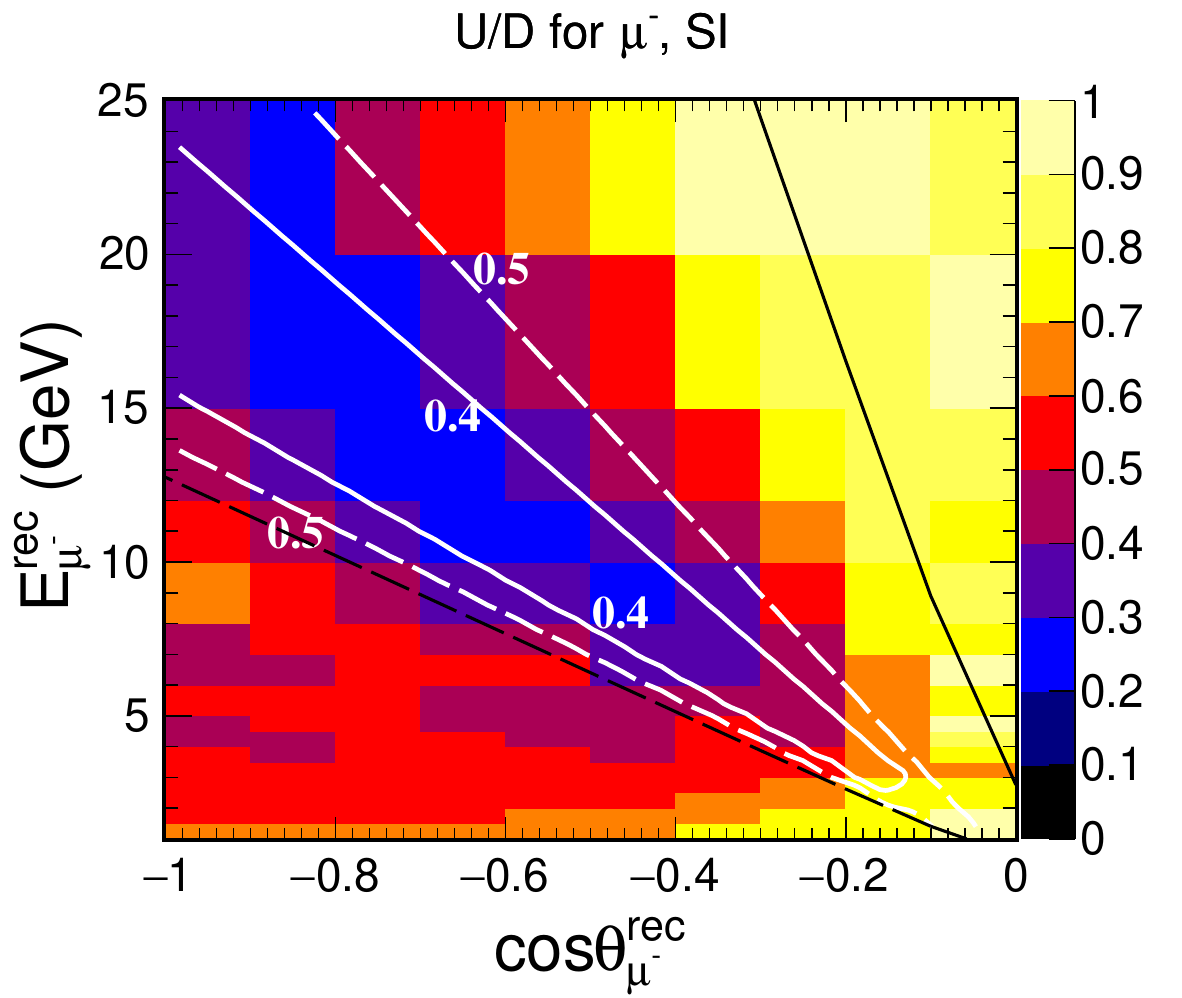}
	\includegraphics[width=0.45\linewidth]{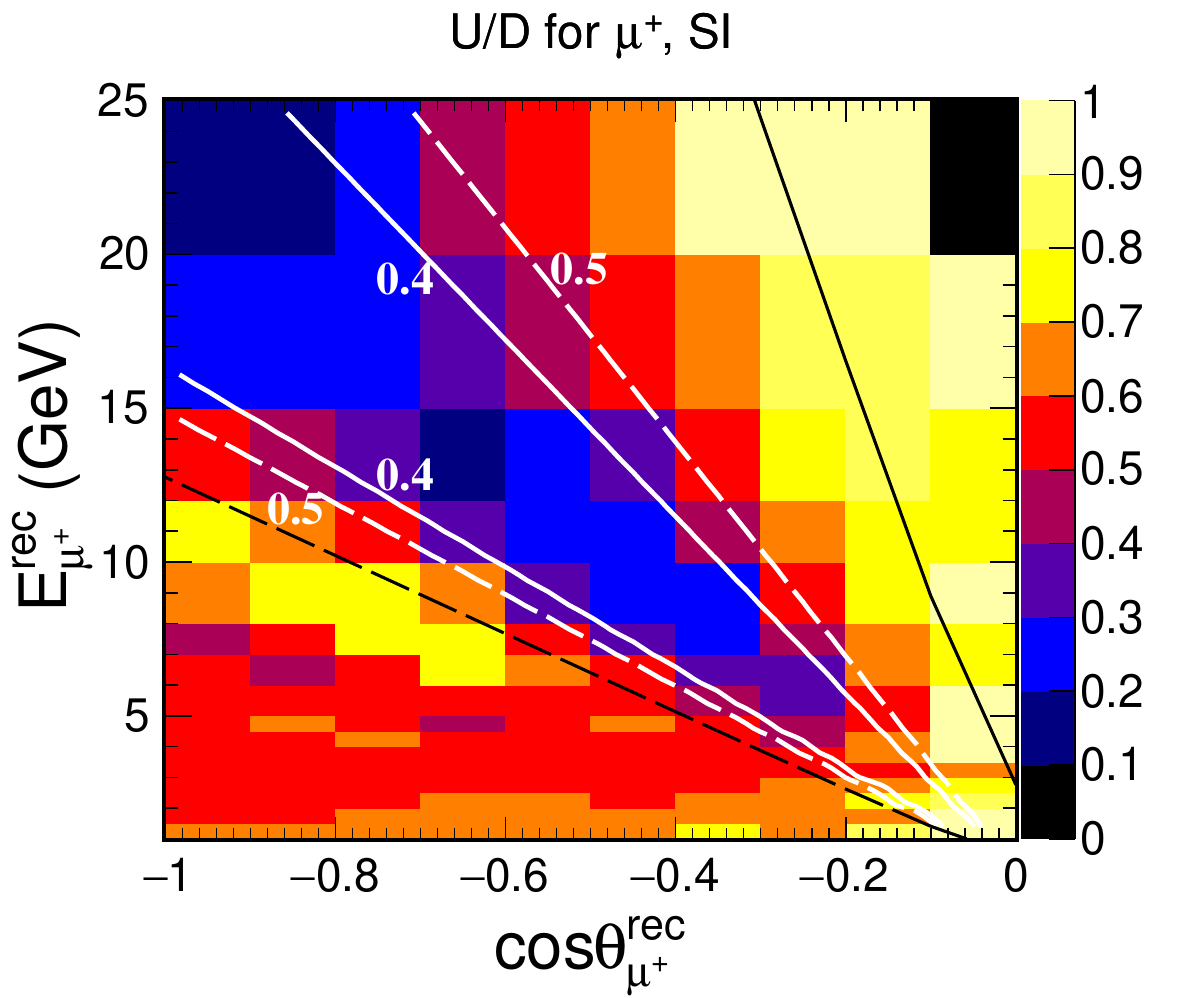}
	\caption{Demonstration of the fitting for identifying the valley and determining its alignment. The distributions of U/D ratios in ($E_\mu^\text{rec}$, $\cos\theta_\mu^\text{rec}$) plane, for $\mu^-$ and $\mu^+$ events, in left and right panels, respectively. This is the average of 100 independent simulated data sets, each for 10-year exposure at ICAL. The white solid and dashed lines correspond to the contours with fitted U/D ratio equal to 0.4 and 0.5, respectively, obtained after the fitting of oscillation valley with the function $F_0$ as given in Eq.~\ref{eq:fitting-func-sm}. The black solid and dashed lines correspond to $\log_{10}[L_\mu^\text{rec}/E_\mu^\text{rec}] =$ 2.2 and 3.0, respectively. We consider normal mass ordering, and the benchmark value of oscillation parameters given in Table~\ref{tab:osc-param-value}.~\cite{Kumar:2021lrn}}
	\label{fig:osc_valley_fitting}
\end{figure}

Figure~\ref{fig:osc_valley_fitting} also shows the contour lines for the fitted U/D ratio of 0.4 and 0.5. The contour lines for $\mu^+$ are seen to reproduce the data better than those for $\mu^-$. The reason behind this is the matter effects, which appear in $\mu^-$ data since the mass ordering is assumed to be NO here. Note that the fitting function $F_0$ is designed for only the vacuum oscillation, and we focus only on the region where vacuum oscillation is expected to dominate. Since the events with $\log_{10} [L_\mu^{\text{rec}}/E_\mu^{\text{rec}}] > 3.1 $ are susceptible to significant matter effect and rapid oscillations, the information from these bins cannot be directly connected to the dip corresponding to the vacuum oscillations. As far as the region $\log_{10} [L_\mu^{\text{rec}}/E_\mu^{\text{rec}}] < 2.2$ is concerned, the oscillation are not developed yet. Therefore, we only choose the region $2.2 < \log_{10} [L_\mu^{\text{rec}}/E_\mu^{\text{rec}}] < 3.1$ for our analysis, where vacuum oscillations are expected to dominate, and to be resolvable. The cut of $\log_{10} [L_\mu^{\text{rec}}/E_\mu^{\text{rec}}] < 2.2$ also gets rid of most of the events near horizon, where the distances traveled by the neutrinos have large errors. In order to minimize the data from the bins with clearly large fluctuations, we use another cut on the maximum value of the U/D ratio. This cut is taken to be U/D $< 0.9$ for both $\mu^-$ and $\mu^+$ events.

Our analysis is rather simple, focused only on the identification of the valley and on determining its alignment. We show here that 500 kt$\cdot$yr of exposure would be enough to locate the valley and quantify its alignment even with our simplified treatment. In the following section, we shall also show that the value of $\Delta m^2_{32}$ may be calibrated using the alignment of the valley.

%=================================
\subsection{Measurement of $\Delta m^2_{32}$ from the alignment of the valley}
\label{sec:Calib-valley}
%=================================

%=================================
\begin{figure}
	\centering
	\includegraphics[width=0.45\linewidth]{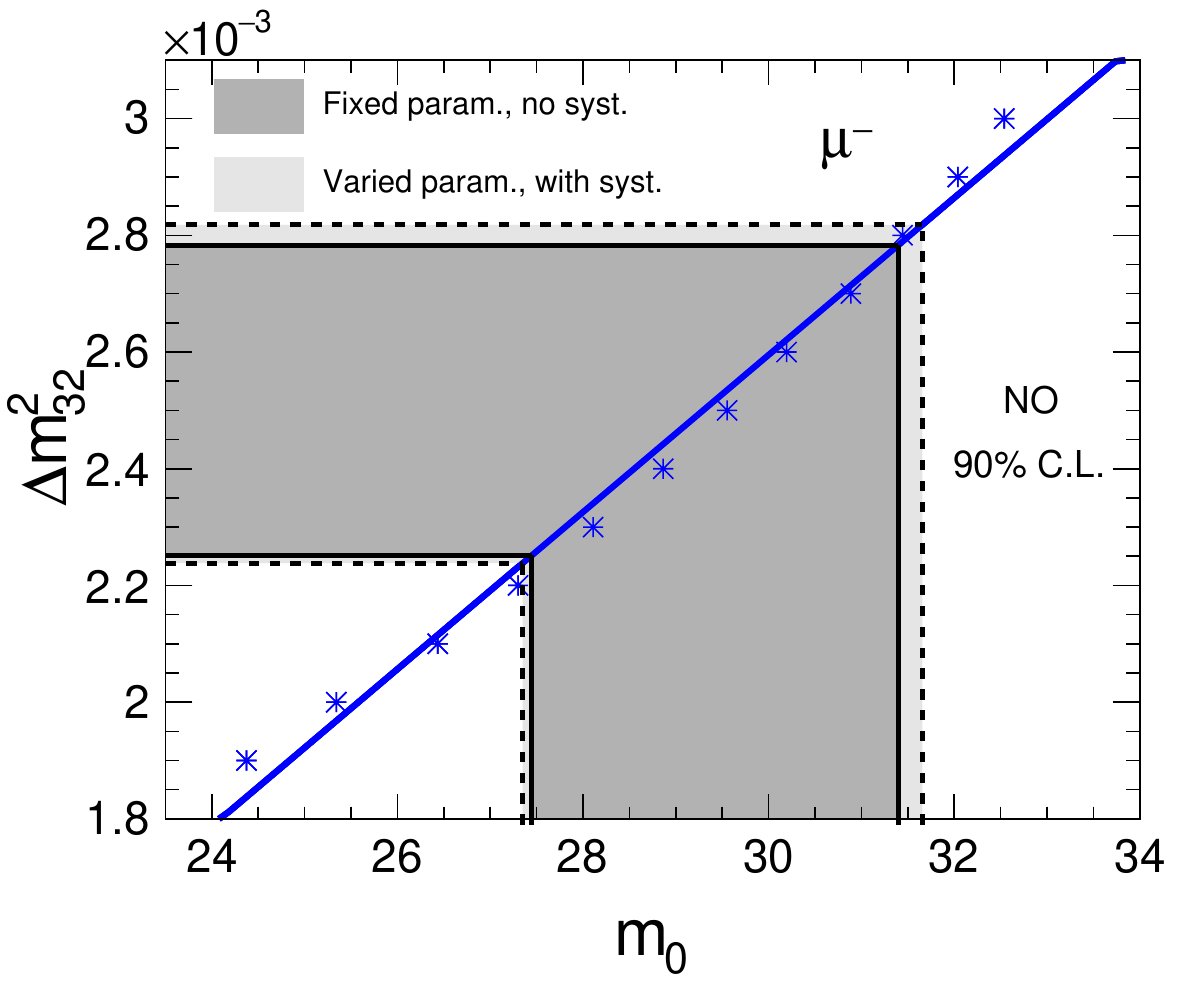}
	\includegraphics[width=0.45\linewidth]{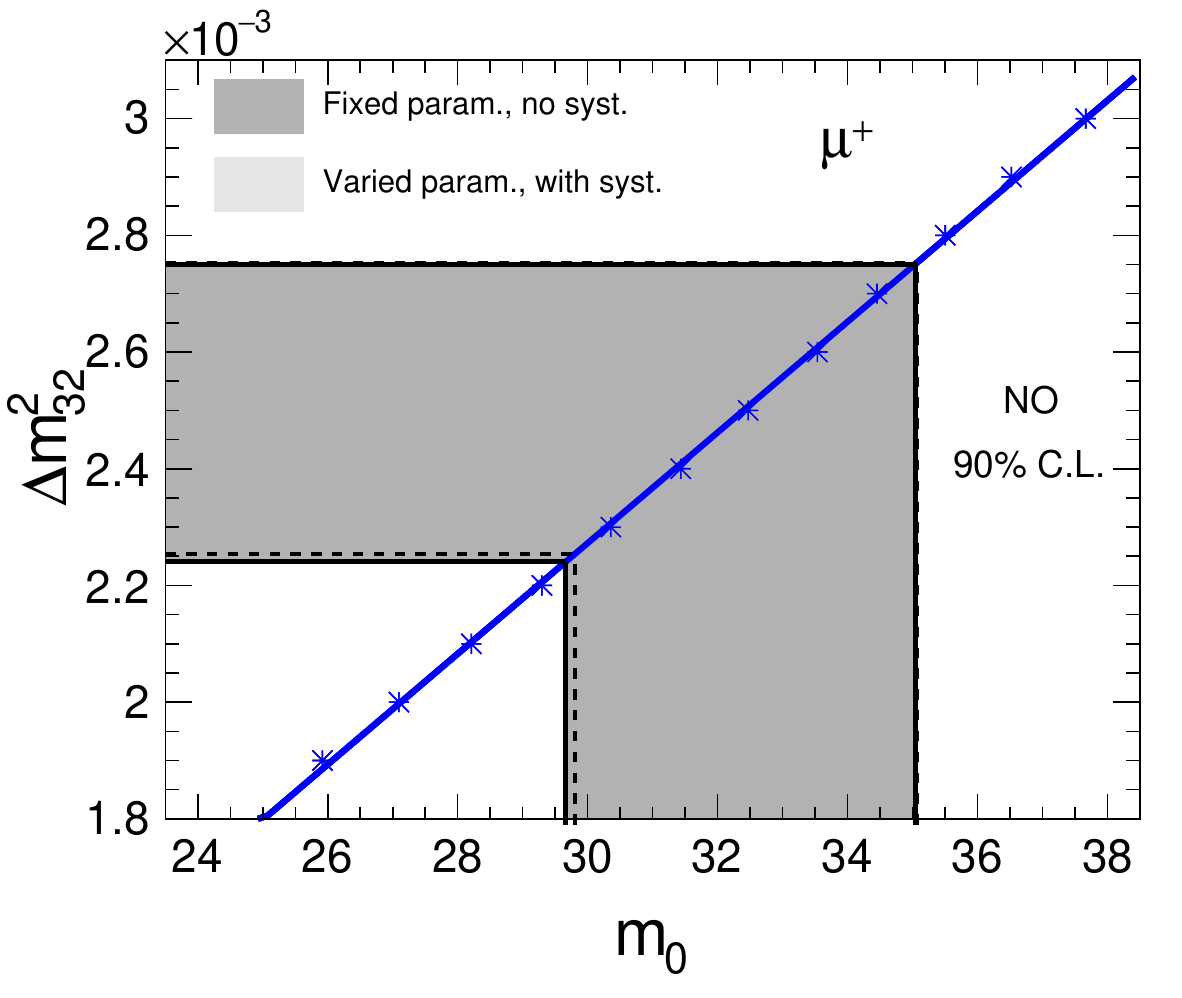}
	\caption{The blue stars and the blue lines indicate the calibration curves of $\Delta m^2_{32}$ against $m_0$, obtained using 1000-year MC data with normal mass ordering. The gray bands represent the 90\% C.L. allowed ranges of $m_0$ (vertical bands), and hence of $\Delta m^2_{32}$ (horizontal bands), with a given input value of $\Delta m^2_{32}$ = 2.46$\times 10^{-3}$ eV$^2$, for an exposure of 10 years at ICAL. The results for $\mu^-$ and $\mu^+$ events are shown separately, in the left and right panels, respectively. The light (dark) gray  bands show the ranges when the errors on other oscillation parameters and the impact of systematic uncertainties are included (excluded), as discussed in Sec.~\ref{sec:Calib-LbyE}. For fixed-parameter case, we use the benchmark value of the oscillation parameters given in Table~\ref{tab:osc-param-value}.~\cite{Kumar:2021lrn}}
	\label{fig:osc_valley_results}
\end{figure}
%=================================

For the same reasons described in Sec.~\ref{sec:Calib-LbyE}, we expect the alignment of the valley, as determined through the procedure outlined in the previous section, to lead to the value of $\Delta m^2_{32}$. Indeed, for upward-going events, $L_\mu^\text{rec}$ is approximately proportional to $\cos \theta_\mu^\text{rec}$, so that the alignment of the valley, i.e. $E_\mu^\text{rec} (\text{GeV})/\cos\theta_\mu^\text{rec}$, is approximately  $(5.08/\pi)\,\Delta m^2_{32} (\text{eV}^2)\, R(\text{km})$.

We calibrate for $\Delta m^2_{32}$ by fitting the U/D ratio of 1000-year MC data with input $\Delta m^2_{32}$ values in the range $(1.9 - 3.0)\times 10^{-3}$ eV$^2$, and obtaining the corresponding  value of $m_0$. The statistical fluctuations are estimated by simulating 100 independent data sets of 10 years for a given input value of $\Delta m^2_{32}$ $= 2.46\times 10^{-3}$ eV$^2$, and fitting for the U/D ratio independently with Eq.~\ref{eq:fitting-func-sm}. The 100 values of $m_0$ thus obtained provide the expected uncertainties on $\Delta m^2_{32}$. The dark gray bands in Fig.~\ref{fig:osc_valley_results} present the 90\% C.L. allowed values for $m_0$, and hence the 90\% C.L. allowed values of the calibrated $\Delta m^2_{32}$, with all the oscillation parameters  fixed at benchmark values as given in Table~\ref{tab:osc-param-value}.  The figure indicates that the 90$\%$ C.L. allowed range for $\Delta m^2_{32}$ from $\mu^-$ events is $(2.25 - 2.78)\times10^{-3}~\text{eV}^2$, while that from $\mu^+$ data is $(2.24 - 2.75)\times10^{-3}~\text{eV}^2$, for $\Delta m^2_{32} \,(\text {true}) = 2.46\times 10^{-3}$ eV$^2$. The light gray bands in Fig.~\ref{fig:osc_valley_results} shows the (small) deterioration in the $90\%$ C.L. range of $\Delta m^2_{32}$ due to the incorporation of the present uncertainties in the oscillation parameters ($\theta_{12}$, $\theta_{23}$, $\theta_{13}$, and $\Delta m^2_{21}$), and all the five systematic errors, as discussed in Sec.~\ref{sec:Calib-LbyE}. With systematic uncertainties and error in other oscillation parameters, we get the expected $90\%$ C.L. allowed range for $\Delta m^2_{32}$ from $\mu^-$ events as $(2.24 - 2.82)\times 10^{-3}~\text{eV}^2$ and from $\mu^+$ events as $(2.25 - 2.75)\times 10^{-3}~\text{eV}^2$.

%=================================
\section{Summary and Concluding Remarks}
\label{sec:dip_valley_conclusions}
%=================================

Atmospheric neutrino experiments access a large range of energies and baselines for neutrino oscillations, hence they are capable of testing basic features of the neutrino flavor conversion framework over a large parameter space. In this work, we study the potential of the ICAL detector to visualize the $L/E$ dependence of survival probabilities in neutrino and antineutrino channels separately, by distinguishing $\mu^-$ and $\mu^+$ events. For this, we use the reconstructed muon energy and zenith angle, as opposed to the inferred neutrino energy and zenith angle. Further, we consider the ratio of upward-going (U) and downward-going (D) events, instead of the ratio of observed oscillated events and simulated unoscillated events. The reconstructed $L_\mu^\text{rec}/E_\mu^\text{rec}$ distribution of the U/D ratio for 1000-yr MC indicates that the feature of the first oscillation dip in this ratio in muon neutrinos (and antineutrinos) is still preserved in the detected $\mu^-$ (and $\mu^+$). We demonstrate that this first oscillation dip can be clearly identified at ICAL with an exposure of 500 kt$\cdot$yr, i.e. 10-year running of the 50 kt ICAL.

We develop a novel dip-identification algorithm to identify the oscillation dip and measure the value of $\Delta m^2_{32}$, using the information from the first oscillation dip in the $L_\mu^\text{rec}/E_\mu^\text{rec}$ distributions of the U/D ratio, in $\mu^-$ and $\mu^+$ events separately. This algorithm finds a contiguous set of $L_\mu^\text{rec}/E_\mu^\text{rec}$ bins with the lowest U/D values, and uses the combined information in all of them to determine the dip location precisely. The value of $\Delta m^2_{32}$ is then calibrated against the location of the dip using the 1000-yr MC, and the statistical uncertainties expected with the 10-year data are estimated using simulations of 100 independent data sets. The expected 90\% C.L. ranges of $\Delta m^2_{32}$ obtained after taking into account the systematic uncertainties and varying the other oscillation parameters over their currently allowed ranges, are summarized in Table~\ref{tab:dip_valley_result_summary}. We also calibrate $\theta_{23}$, however, here simply the ratio of the total number of U and D events is found to be a better choice to calibrate against, rather than the depth of the dip. We find that, when the true value of $\sin^2\theta_{23}$ is 0.50, the U/D ratio would determine the 90\% C.L. range of its value to be (0.38 -- 0.70) with $\mu^-$ events, and (0.35 -- 0.65) with $\mu^+$ events, taking into account systematic uncertainties and errors in the other oscillation parameters.

\begin{table}[htb!]
	\begin{center}
		\begin{tabular}{|c|c|c|}
			\hline
			\multirow{2}{3cm}{\centering Reconstructed observables} &
			\multicolumn{2}{c|}{$90\%$ C.L. range}   \\
			\cline{2-3}
			& $\mu^-$ & $\mu^+$ \\
			\hline
			& & \\
			$L_\mu^\text{rec}/E_\mu^\text{rec}$ & $(2.18~\text{-}~2.79)\times 10^{-3}$ eV$^2$
			& $(2.22 ~\text{-}~ 2.80)\times10^{-3}$ eV$^2$ \\
			& &  \\
			\hline
			\hline
			& & \\
			$(E^\text{rec}_\mu$, $\cos\theta^\text{rec}_\mu)$   & $(2.24 ~\text{-}~ 2.82)\times 10^{-3}$ eV$^2$ & $(2.25 ~\text{-}~ 2.75)\times 10^{-3}$ eV$^2$  \\
			& & \\
			\hline
		\end{tabular}
		\caption{The expected $90\%$ C.L. allowed ranges for $\Delta m^2_{32}$, from the analyses using $L_\mu^\text{rec}/E_\mu^\text{rec}$ and $(E_\mu^\text{rec}, \, \cos\theta_\mu^\text{rec})$ distributions of the U/D ratio, with 500 kt$\cdot$yr exposure. The true value of $\Delta m^2_{32}$ is taken to be $2.46 \times 10^{-3}$ eV$^2$, and the other oscillation parameters $\theta_{12}$, $\theta_{23}$, $\theta_{13}$, and $\Delta m^2_{21}$ are varied over their currently allowed ranges, and systematics are taken into account.~\cite{Kumar:2020wgz,Kumar:2021lrn}}
		\label{tab:dip_valley_result_summary}
	\end{center}
\end{table}

In this chapter, we point out for the first time that the identification of an ``oscillation valley'' feature is possible, in the distribution of the U/D ratio in the plane of $(E_\mu^\text{rec}, \, \cos \theta_\mu^\text{rec}$). The oscillation dip observed in the one-dimensional $L_\mu^\text{rec}/E_\mu^\text{rec}$ analysis above now appears as an oscillation valley in the two-dimensional $(E_\mu^\text{rec},\cos\theta_\mu^\text{rec}$) plane. We go on to fit the oscillation valley for finding the alignment of the valley in the $(E_\mu^\text{rec},\cos\theta_\mu^\text{rec}$) plane, and show that it serves as a good proxy for $\Delta m^2_{32}$. As in the oscillation dip analysis, we calibrate $\Delta m^2_{32}$ against the alignment of the valley using 1000-yr MC data, and estimate the uncertainties with 10-year data using 100 independent data sets. The expected 90\% C. L. ranges of $\Delta m^2_{32}$ with systematic uncertainties and errors in other oscillation parameters are summarized in Table~\ref{tab:dip_valley_result_summary}.

Note that the aim of our approach is not to compare the precisions achieved by different methods -- indeed, this study does not aim to compete with the conventional $\chi^2$ method for precision. Our aim is to present an approach (``oscillation dip'') that has proved useful in the past for establishing the oscillation hypothesis and rejecting some alternative hypotheses, and to go one step beyond it, by performing the ``oscillation valley'' analysis in the two-dimensional $(E_\mu^\text{rec}, \, \cos \theta_\mu^\text{rec}$) plane. An important feature of our analysis is that the reconstruction of dip and valley and their identification processes are data-driven. Of course, to retrieve the information on oscillation parameters using the calibration curve, we use MC simulations.

Note that the precisions on $\Delta m^2_{32}$, obtainable from our dip and valley analyses, are very similar to each other. This is expected, since the L/E dependence is assumed and the two-dimensional analysis offers no advantages for the determination of $\Delta m^2_{32}$ in the context of the SM.  However, our focus has been on confirming that such two-dimensional reconstruction of the valley survives in the detector environment, i.e. the ICAL characteristics are sufficient to reconstruct the valley feature. This would also act as the fidelity test for the detector. The power of the valley analysis would become evident when L/E dependence is not assumed, where it would open up avenues for testing the oscillation framework from more angles -- for example, by looking for effects of new physics on the shape, alignment, width, or depth of the valley. In the context of non-standard interactions (NSI) of neutrinos, this will be addressed in the next chapter.

Though we have presented our results for ICAL, the same procedure can be used for any present or upcoming atmospheric neutrino experiment that has access to a large range of energies and baselines. An atmospheric neutrino experiment can perform the analysis based on the oscillation dip as well as the oscillation valley. Of course, if the detector cannot identify the muon charge, the data from neutrino and antineutrino channels will have to be combined, which may dilute some of the observable features. However, the principle of performing the oscillation valley analysis in the $(E_\mu^\text{rec}, \, \cos \theta_\mu^\text{rec}$) plane -- going one step ahead of the $L_\mu^\text{rec}/E_\mu^\text{rec}$ analysis -- may be adopted in any atmospheric neutrino experiment. Note that while the fixed-baseline experiments may in principle be able to identify the $L_\nu/E_\nu$ oscillation dip, they do not have access to the $(E_\nu, \, \cos \theta_\nu$) plane, and hence they cannot perform the oscillation valley analysis.

Even then, certain features of ICAL make it uniquely suited for such an analysis. It will likely be the first detector to identify the oscillation dip separately for the neutrino and antineutrino channels, which will also be a test for new physics that may affect the neutrino and antineutrino differently. Further, as far as the oscillation valley analysis goes, a crucial requirement for the capability to identify the alignment of the valley is the enough number of points to perform the fitting as shown in the Fig.~\ref{fig:osc_valley_fitting}. This needs a large energy range to which the detector is sensitive (large number of y-bins), and an excellent angular resolution (large number of x-bins). ICAL is thus uniquely suited for the oscillation valley analysis, for providing an orthogonal approach to establish the nature of neutrino oscillations, and hence for making the neutrino oscillation picture more robust.

\end{refsegment}

\cleartooddpage
\chapter{Probing Non-standard Interactions}
\label{chap:NSI}
\begin{refsegment}
Neutrino oscillations are the consequences of mixing among different neutrino flavors and non-degenerate values of neutrino masses, with at least two neutrino masses nonzero. However, neutrinos are massless in the Standard Model (SM) of particle physics, and therefore, physics beyond the SM (BSM) is needed to accommodate nonzero neutrino masses and mixing. Many models of BSM physics suggest new non-standard interactions (NSI) of neutrinos~\cite{Wolfenstein:1977ue}, which may affect neutrino production, propagation, and detection in a given experiment. The possible impact of these NSI on neutrino oscillation experiments have been studied extensively, for example see Refs.~\cite{Valle:1987gv,Guzzo:1991hi,Guzzo:2000kx,Huber:2001zw,Gago:2001si,Escrihuela:2011cf,Gonzalez-Garcia:2011vlg,Ohlsson:2012kf,Gonzalez-Garcia:2013usa,Farzan:2017xzy,Miranda:2015dra,Dev:2019anc}. In this chapter, we propose a new method for identifying NSI at atmospheric neutrino experiments which can reconstruct the energy, direction, as well as charge of the muons produced in the detector due to charged-current interactions of $\nu_\mu$ and $\bar\nu_\mu$.

In this work, we shall focus on the neutral-current NSI, which may be described at low energies via effective four-fermion dimension-six operators as~\cite{Wolfenstein:1977ue}   
\begin{equation}
{\mathcal L}_{\text{NC-NSI}} = -2\sqrt 2 \, G_F \,\varepsilon^{f}_{\alpha\beta, C} 
\, (\bar\nu_{\alpha}\gamma^\rho P_L\nu_{\beta}) \, 
(\bar f \gamma_\rho P_C f) \, ,
\end{equation}
where $G_{F}$ is the Fermi constant. The dimensionless parameters $\varepsilon^{f}_{\alpha \beta, C}$ describe the strength of NSI, where the superscript $f \in \{e,u,d\}$ denotes the matter fermions ($e$: electron, $u$: up-quark, $d$: down-quark), and the indices $\alpha, \beta \in \{e,\mu,\tau \}$ refer to the neutrino flavors. The subscript $C \in \{L,R \}$ represents the chiral projection operator $P_L = (1 - \gamma_5)/2$ or $P_R = (1 + \gamma_5)/2$. The hermiticity of the interactions demands $\varepsilon^{f}_{\beta\alpha, C} = (\varepsilon^{f}_{\alpha \beta, C})^*$.

The effective NSI parameter relevant for the neutrino propagation through matter is 
\begin{equation}
\varepsilon_{\alpha\beta} \equiv \sum_{f=e,u,d} \left(\varepsilon_{\alpha\beta,L}^{f}
+ \varepsilon_{\alpha\beta,R}^{f}\right) \,\,\frac{N_f}{N_e}
\equiv \sum_{f=e,u,d} \varepsilon_{\alpha\beta}^{f}\,\, \frac{N_f}{N_e}\,,
\label{eq:eps-definition}
\end{equation}
where $N_f$ is the number density of fermion $f$. In the approximation of a neutral and isoscalar Earth, the number densities of electrons, protons, and neutrons are identical, which implies $N_u \approx N_d \approx 3 N_e$. Thus,
\begin{equation}
\varepsilon_{\alpha\beta} \approx \varepsilon^e_{\alpha\beta}
+ 3 \, \varepsilon^u_{\alpha\beta} + 3 \, \varepsilon^d_{\alpha\beta}\,. 
\end{equation}
In the presence of NSI, the modified effective Hamiltonian for neutrino propagation through matter is 
\begin{equation}
H_\text{eff} = \frac{1}{2E} \, U \left(\begin{array}{ccc}
0 & 0 & 0 \\
0 & \Delta m^2_{21} & 0 \\
0 & 0 & \Delta m^2_{31}\end{array}
\right) U^\dag + V_\text{CC}\left(\begin{array}{ccc}
1 + \varepsilon_{ee} & \varepsilon_{e\mu} & \varepsilon_{e\tau} \\
\varepsilon_{e\mu}^* & \varepsilon_{\mu\mu} & \varepsilon_{\mu\tau} \\
\varepsilon_{e\tau}^* & \varepsilon_{\mu\tau}^* & \varepsilon_{\tau\tau}
\end{array}\right)\,.
\end{equation}
The quantity $V_\text{CC} \equiv \sqrt{2} G_F N_e$ is the effective matter potential due to the coherent elastic forward scattering of neutrinos with electrons inside the medium via the SM gauge boson $W$. Thus, the effective potential due to NSI would be ${(V_\text{NSI})}_{\alpha\beta} = \sqrt{2} G_F N_e \, \varepsilon_{\alpha\beta}$. For antineutrinos, $V_\text{CC} \rightarrow -V_\text{CC}$, $U \rightarrow U^*$, and $\varepsilon_{\alpha\beta} \to \varepsilon_{\alpha\beta}^*$. 

In the present study, we suggest a novel approach to unravel the presence of flavor-changing neutral-current NSI parameter $\varepsilon_{\mu\tau}$, via its effect on the propagation of multi-GeV atmospheric neutrinos and antineutrinos through Earth matter. We choose $\varepsilon_{\mu\tau}$ as the NSI parameter to focus on, since it can significantly affect the evolution of the mixing angle $\theta_{23}$ and the mass-squared difference $\Delta m^2_{32}$ in matter~\cite{Agarwalla:2021zfr}, which in turn would alter the survival probabilities of atmospheric muon neutrinos and antineutrinos substantially~\cite{Gonzalez-Garcia:2004pka,Mocioiu:2014gua}. Here, we consider $\varepsilon_{e\mu} = \varepsilon_{e\tau} = \varepsilon_{\mu\mu} - \varepsilon_{\tau\tau} = 0$. In general, $\varepsilon_{\mu\tau}$ can be complex, i.e. $\epsmutau \equiv |\epsmutau| e^{i\phi_{\mu\tau}}$. However, in the disappearance channels $\nu_\mu \to \nu_\mu$ and $\bar\nu_\mu \to  \bar\nu_\mu$ that dominate in our analysis\footnote{The appearance channels $\nu_e \to \nu_\mu$ and $\bar\nu_e \to \bar\nu_\mu$ also contribute to the muon events in our analysis, however these channels are not affected by $\epsmutau$ to leading order~\cite{Kopp:2007ne}.}, $\epsmutau$ appears only as $|\epsmutau| \cos\phi_{\mu\tau}$ at the leading order~\cite{Kopp:2007ne}. Thus, a complex phase only changes the effective value of $\epsmutau$ to a real number between $-|\epsmutau|$ and $+|\epsmutau|$ at the leading order. We take advantage of this observation, and restrict ourselves to real values of $\epsmutau$ in the range $-0.1 \leq \epsmutau \leq 0.1$. From the arguments given above, this covers the whole range of complex values of $\epsmutau$ with $|\epsmutau| \leq 0.1$.

Based on the global neutrino data analysis in Ref.~\cite{Esteban:2018ppq}, where the possible contributions to NSI from only up and down quark have been included, the bound on $|\epsmutau|$ turns out to be $|\varepsilon_{\mu\tau}|<$ 0.07 at 2$\sigma$ confidence level. A phenomenological study to constrain $\epsmutau$ using preliminary IceCube and DeepCore data  has been performed in~\cite{Esmaili:2013fva}. Recently, the IceCube collaboration has placed strong constraints on $\epsmutau$ using eight years of atmospheric muon neutrino data at TeV scale~\cite{IceCube:2022ubv}. The existing bounds on $\varepsilon_{\mu\tau}$ from various neutrino oscillation experiments are listed in Table~\ref{tab:existing-limit}. An important point to note is the energies of neutrinos involved in these measurements: the IceCube results are obtained using high energy events ($>300$ GeV)~\cite{Salvado:2016uqu}, the energy threshold of DeepCore is around 10 GeV~\cite{IceCube:2017zcu}, while the Super-K experiment is more efficient in the sub-GeV energy range~\cite{Super-Kamiokande:2011dam}.

%%%%%%%%%%%%%%%%%%%%%%%%%%%%%%%%%%%%%%%%%%%%%%%%%%%%%%%%%%%%%%%%%%%% 
\begin{table}[h!]
	\begin{center}
		\begin{tabular}{|c|c|c|}
			\hline \hline 
			\multirow{2}{*}{Experiment}    & \multicolumn{2}{c|}{$90\%$ C.L. bounds} \\
			\cline{2-3}
			& Convention in~\cite{Salvado:2016uqu,IceCube:2017zcu,IceCube:2022ubv,Super-Kamiokande:2011dam} & Our convention~\cite{Ohlsson:2012kf,Gonzalez-Garcia:2013usa,Choubey:2015xha,Farzan:2017xzy,Khatun:2019tad}\\ 
			\hline
			IceCube \cite{Salvado:2016uqu} & 
			$-0.006 < \tilde{\varepsilon}_{\mu\tau} < 0.0054 $ &
			$-0.018<\varepsilon_{\mu\tau}<0.0162 $ \\
			DeepCore \cite{IceCube:2017zcu} &
			$-0.0067<\tilde{\varepsilon}_{\mu\tau}< 0.0081$  & 
			$-0.0201<\varepsilon_{\mu\tau} <0.0243 $ \\ 
			IceCube \cite{IceCube:2022ubv} & 
			$-0.0041 < \tilde{\varepsilon}_{\mu\tau} < 0.0031 $ &
			$-0.0123<\varepsilon_{\mu\tau}<0.0093 $ \\
			Super-K \cite{Super-Kamiokande:2011dam}& 
			$|\tilde{\varepsilon}_{\mu\tau}| < 0.011$ & 
			$|\varepsilon_{\mu\tau}| < 0.033$ \\
			\hline \hline
		\end{tabular}
	\end{center}
	\caption{Existing bounds on $\varepsilon_{\mu\tau}$ at $90\%$ confidence level. Note that the bounds presented in~\cite{Salvado:2016uqu,IceCube:2017zcu,IceCube:2022ubv,Super-Kamiokande:2011dam} are on $\tilde{\varepsilon}_{\mu\tau}$ that is defined according to the convention $V_\text{NSI} = \sqrt{2} G_F N_d\, \tilde{\varepsilon}_{\mu\tau}$, while we use the convention  $V_\text{NSI} = \sqrt{2} G_F N_e \,\epsmutau$ ($\epsmutau$ is defined in Eq.~\ref{eq:eps-definition}). Since $N_d \approx 3 N_e$ in Earth, the bounds in \cite{Salvado:2016uqu,IceCube:2017zcu,Super-Kamiokande:2011dam} on $\tilde{\varepsilon}_{\mu\tau}$ have been converted to the bounds on $\epsmutau$, using $\epsmutau = 3\, \tilde{\varepsilon}_{\mu\tau}$, as shown in the third column.~\cite{Kumar:2021lrn}}
	\label{tab:existing-limit}
\end{table}
%%%%%%%%%%%%%%%%%%%%%%%%%%%%%%%%%%%%%%%%%%%%%%%%%%%%%%%%%%%%%%%%%%%

The ICAL detector would be sensitive to multi-GeV neutrinos, since it can efficiently detect muons in the energy range of 1--25 GeV. Note that the NOLR~\cite{Petcov:1998su,Chizhov:1998ug,Petcov:1998sg,Chizhov:1999az,Chizhov:1999he} / parametric resonance~\cite{Akhmedov:1998ui,Akhmedov:1998xq} and the  MSW resonance~\cite{Mikheev:1986wj,Mikheev:1986gs,Wolfenstein:1977ue} due to Earth matter take place for neutrino energies around 3--6 GeV and 6--10 GeV, respectively, so ICAL would also be in a unique position to detect any interplay between the matter effects and NSI. Another important feature is that ICAL can explore physics in neutrinos and antineutrinos separately, unlike Super-K and IceCube/DeepCore. The studies of physics potential of ICAL for detecting NSI have shown that, using the reconstructed muon momentum, it would be possible to obtain a bound of $|\varepsilon_{\mu\tau}| < 0.015$ at 90$\%$ C.L.~\cite{Choubey:2015xha} with 500 kt$\cdot$yr exposure. When information on the reconstructed hadron energy in each event is also included, the expected 90$\%$ C.L. bound improves to $|\varepsilon_{\mu\tau}|<0.010$~\cite{Khatun:2019tad}. The results in~\cite{Choubey:2015xha,Khatun:2019tad} are obtained using a $\chi^2$ analysis with the pull method~\cite{Huber:2002mx,Fogli:2002au,Gonzalez-Garcia:2004pka}.

The wide range of neutrino energies and baselines available in atmospheric neutrino experiments offer an opportunity to study the features of ``oscillation dip'' and ``oscillation valley'' in the reconstructed $\mu^-$ and $\mu^+$ observables, as demonstrated in the chapter~\ref{chap:dip_valley} (based on Ref.~\cite{Kumar:2020wgz}). These features can be clearly identified in the ratios of upward-going and downward-going muon events at ICAL. If the muon neutrino disappearance is solely due to non-degenerate masses and non-zero mixing of neutrinos, then the valley in both $\mu^-$ and $\mu^+$ is approximately a straight line. The location of the dip, and the alignment of the valley, can be used to determine $\Delta m^2_{32}$ (see chapter~\ref{chap:dip_valley}). These features may undergo major changes in the presence of NSI, and can act as smoking gun signals for NSI. 

The novel approach, which we propose in this chapter, is to probe the  NSI parameter $\varepsilon_{\mu\tau}$ based on the elegant features associated with the oscillation dip and valley, both of which arise from the same physics phenomenon, viz. the first oscillation minimum in the muon neutrino survival probability. For the oscillation dip feature, we note that non-zero $\epsmutau$ shifts the oscillation dip location in opposite directions for $\mu^-$ and $\mu^+$. We demonstrate that this opposite shift in dip location due to NSI can be clearly seen in the $\mu^-$ and $\mu^+$ data by reconstructing $L_\mu^\text{rec}/E_\mu^\text{rec}$ distributions, thanks to the excellent energy and direction resolutions for muons at ICAL. We develop a whole new analysis methodology to extract the information on $\varepsilon_{\mu\tau}$ using the dip locations. For this, we define a new variable exploiting the contrast between the shifts in reconstructed dip locations, which eliminates the dependence of our results on the actual value of $\Delta m^2_{32}$. For the oscillation valley feature, we notice that the valley becomes curved in the presence of non-zero $\varepsilon_{\mu\tau}$, and the direction of this bending is opposite for neutrino and antineutrino. We then demonstrate that this opposite bending can indeed be observed in expected $\mu^-$ and $\mu^+$ events. We propose a methodology to extract the information on the bending of the valley in terms of the reconstructed muon variables, and use it for identifying NSI. 

In Sec.~\ref{sec:NSI_prob}, we discuss the oscillation probabilities of neutrino and antineutrino in the presence of non-zero $\varepsilon_{\mu\tau}$, and discuss the shifts in the dip locations as well as the bending of the oscillation valleys in the survival probabilities of $\nu_\mu$ and $\bar\nu_\mu$. In Sec.\,\ref{sec:evt-ical}, we investigate the survival of these two striking features in the  reconstructed $L_\mu^{\rm rec}/E_\mu^{\rm rec}$ distributions and in the ($E_\mu^{\rm rec},\,\cos\theta_\mu^{\rm rec}$) distributions of $\mu^-$ and $\mu^+$ events separately at ICAL. In Sec.~\ref{sec:reco-dip}, we propose a novel variable for identifying the NSI, which is based on the contrast in the shifts of dip locations in $\mu^-$ and $\mu^+$. This variable leads to the calibration of $\varepsilon_{\mu\tau}$, and is used to find the expected bound on $\varepsilon_{\mu\tau}$ froma 500 kt$\cdot$yr exposure of ICAL. In Sec.~\ref{sec:reco-valley}, we measure the contrast in the curvatures of the oscillation valleys in $\mu^-$ and $\mu^+$ in the presence of NSI, and use it for determining the expected bound on $\varepsilon_{\mu\tau}$ from the valley analysis at ICAL. Finally, in Sec.~\ref{sec:NSI_conclusion}, we summarize our
findings and offer concluding remarks.

%=======================================
\section{Oscillation Dip and Valley in the Presence of NSI}
\label{sec:NSI_prob}
%=======================================

In the limit of $\theta_{13} \to 0$, and the approximation of one mass scale dominance scenario [$\Delta m^2_{21} L/(4E) \ll \Delta m^2_{32} L/(4E)$] and constant matter density, the survival probability of $\nu_\mu$ when traveling a distance $L_\nu$ is given by\,\cite{Coleman:1998ti,Gonzalez-Garcia:2004pka}
\begin{equation}
P(\nu_\mu\rightarrow\nu_\mu) = 1 - \sin^2 2\theta_\text{eff} \,
\sin^2\left[ \xi \,\frac{\Delta m^2_{32} L_\nu}{4 E_\nu} \right]\,,
\label{eq:pmumu-nsi-mutau-omsd}
\end{equation}
where
\begin{equation}
\sin^2 2\theta_\text{eff} =
\frac{|\sin 2\theta_{23} +  2 \beta \, \eta_{\mu\tau}|^2}{\xi^2}\,,
\label{eq:pmumu-thetaeff-omsd}
\end{equation}
\begin{equation}
\xi = \sqrt{|\sin 2\theta_{23} + 2 \beta \, \eta_{\mu\tau}|^2
	+ \cos^2 2\theta_{23}}\,,
\label{eq:pmumu-xi-omsd}
\end{equation}
and
\begin{equation}
\eta_{\mu\tau} = \frac{ 2 E_\nu \,V_\text{CC} \,\varepsilon_{\mu\tau}}{|\Delta m^2_{32}|}
\,,
\end{equation}
where $\beta \equiv \text{sgn}(\Delta m^2_{32})$. That is, $\beta = +1$ for normal mass ordering, and $\beta=-1$ for inverted mass ordering.

In the limit of maximal mixing ($\theta_{23} = 45^{\circ}$), Eq.\,\ref{eq:pmumu-nsi-mutau-omsd} reduces to the following simple expression\,\cite{Mocioiu:2014gua}:
\begin{equation}
P(\nu_\mu\rightarrow\nu_\mu) = \cos^2 \left[L_\nu \left(
\frac{\Delta m^2_{32}}{4E_\nu} + \varepsilon_{\mu\tau} V_\text{CC} \right)\right]\,.
\label{eq:NSI-pmumu-nu-final-omsd}
\end{equation}
We further find that the correction due to the deviation of $\theta_{23}$ from maximality, $\chi \equiv \theta_{23}-\pi/4$, is second order in the small parameter $\chi$, and hence can be neglected. Here, one notes that the NSI parameter $\epsmutau$ primarily modifies the wavelength of neutrino oscillations. It also comes multiplied with $V_\text{CC}$, which increases at higher baselines inside the  Earth's matter.  Thus, the modification in $\nu_\mu$ survival probabilities due to NSI varies with the baseline $L_\nu$ (or neutrino zenith angle $\cos\theta_\nu$).

%%%%%%%%%%%%%%%%%%%%%%%%%%%%%%%%%%%%%%%%%%%%%%%%%%%%%%%%%%%%%%%%%
\subsection{Effect of $\epsmutau$ on the oscillation dip in the $L_\nu/E_\nu$}
\label{subsec:osc-dip}
%%%%%%%%%%%%%%%%%%%%%%%%%%%%%%%%%%%%%%%%%%%%%%%%%%%%%%%%%%%%%%%%

\begin{figure}
	\centering
	\includegraphics[width=0.45\linewidth]{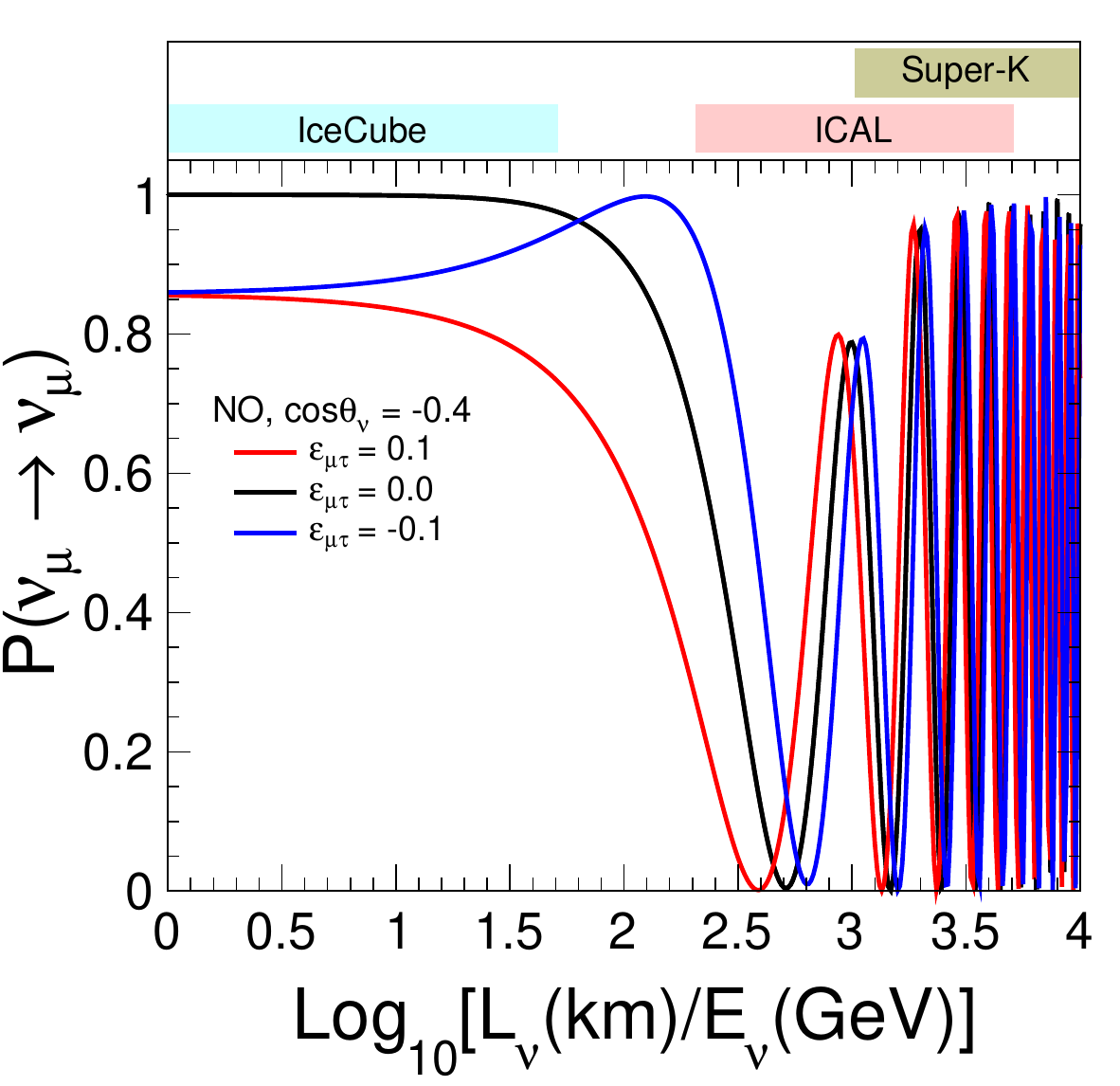}
	\includegraphics[width=0.45\linewidth]{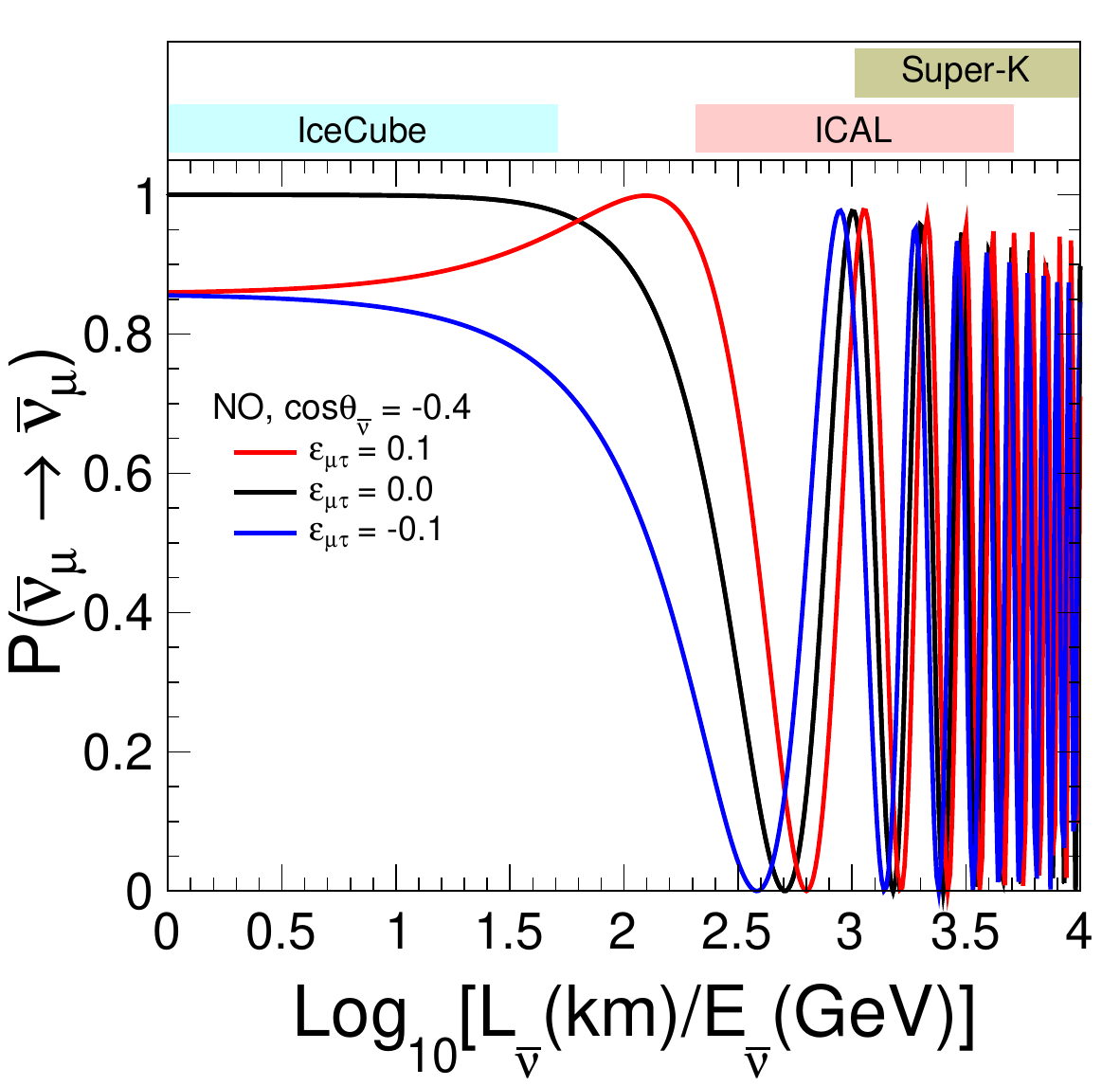}
	\includegraphics[width=0.45\linewidth]{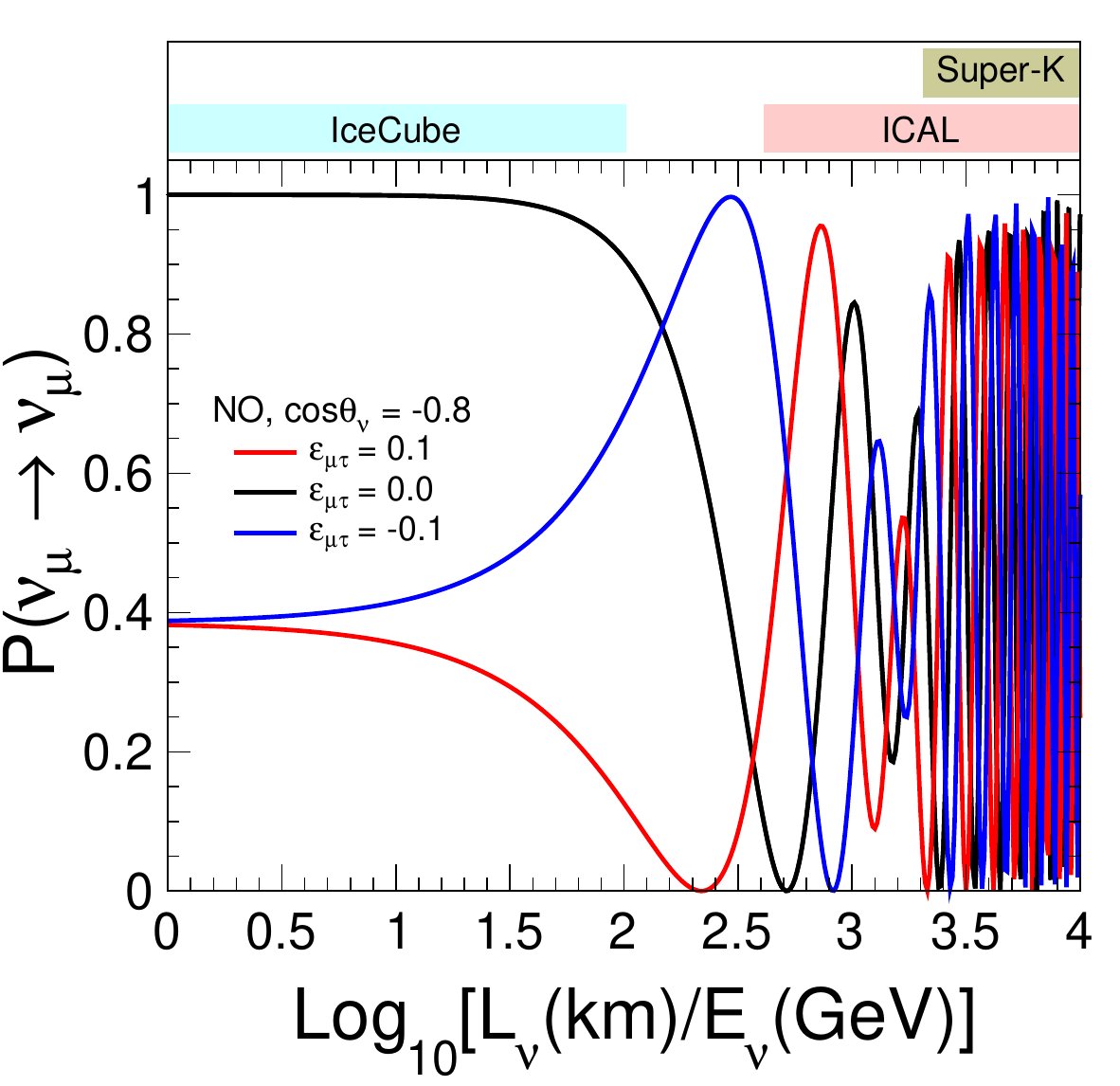}
	\includegraphics[width=0.45\linewidth]{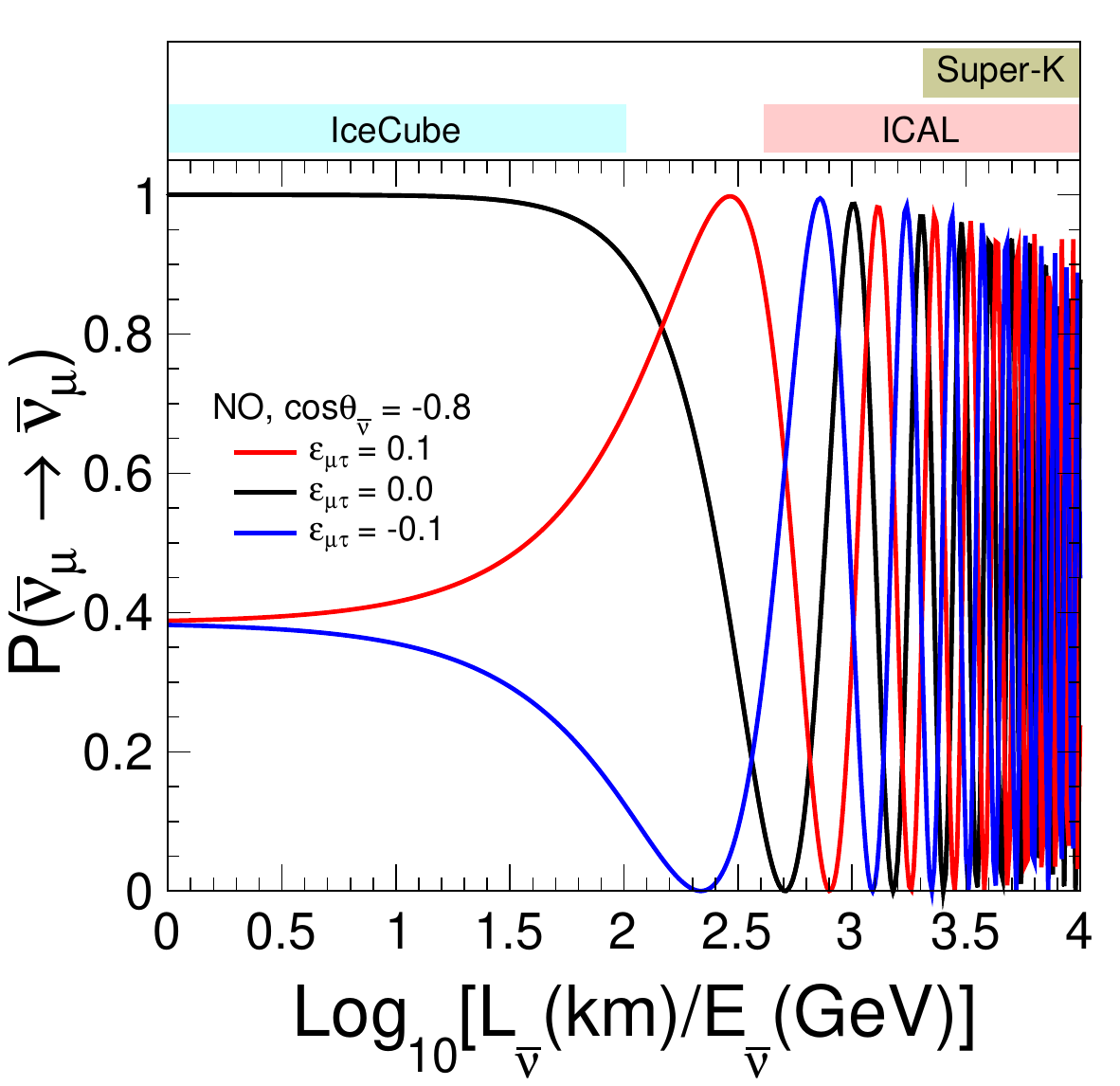}
	\caption{The three-flavor survival probabilities of $\nu_\mu$ and $\bar\nu_\mu$ in the presence of matter with PREM profile as functions of $\log_{10}(L_\nu/E_\nu)$ in left and right panels, respectively. The black lines correspond to $\varepsilon_{\mu\tau} = 0$ (standard interactions: SI), whereas red and blue lines are with $\varepsilon_{\mu\tau} = 0.1$, and $-0.1$, respectively. The neutrino direction taken in the upper (lower) panels is $\cos\theta_\nu = -0.4$ ($-0.8$). The horizontal bands shown here are the indicative $L_\nu/E_\nu$ ranges that these detectors are well-suited for. We consider normal mass ordering, and the benchmark values of oscillation parameters given in Table~\ref{tab:osc-param-value}.~\cite{Kumar:2021lrn}}
	\label{fig:osc_dip_neutrino_NSI}
\end{figure}

In Fig.\,\ref{fig:osc_dip_neutrino_NSI}, we present the three-flavor survival probabilities of $\nu_\mu$ and $\bar\nu_\mu$ in the presence of matter with PREM profile as functions of $L_\nu/E_\nu$, for two fixed values of $\cos\theta_\nu$ (i.e. $\cos \theta_\nu = -0.4, -0.8$), and for three values of the NSI parameter $\varepsilon_{\mu\tau}$ (i.e. $\epsmutau= +0.1, 0.0, -0.1$). The benchmark values of the oscillation parameters are given in Table~\ref{tab:osc-param-value}.

Figure\,\ref{fig:osc_dip_neutrino_NSI} also indicates the sensitivity ranges for the atmospheric neutrino experiments Super-K, ICAL, and IceCube. Note that these ranges are different for $\cos\theta_\nu=-0.4$ and $-0.8$, which correspond to $L_\nu$ around 5100 km and 10200 km, respectively. While calculating these $L_\nu/E_\nu$ ranges, the energy ranges chosen are those for which the detectors perform very well: we use the $E_\nu$ range of 100 MeV -- 5 GeV for Super-K~\cite{Super-Kamiokande:2005mbp}, 1--25 GeV for ICAL~\cite{ICAL:2015stm}, and 100 GeV -- 10 PeV for IceCube~\cite{IceCube:2013dkx}. Note that these ranges are only indicative. The following observations may be made from the figure.

\begin{itemize}
	
	\item For $\log_{10}[L_\nu/E_\nu]$ in the range of [0 -- 1.5], the survival probabilities of both $\nu_\mu$ and $\bar\nu_\mu$ are observed to be suppressed in the presence of non-zero $\varepsilon_{\mu\tau}$. This is because, although the oscillations due to neutrino mass-squared difference do not develop for such small value of $L_\nu/E_\nu$, i.e. $\Delta m^2_{32}L_\nu / E_\nu \ll 1$, the disappearance of $\nu_\mu$ is possible due to the $L_\nu \varepsilon_{\mu\tau} V_\text{CC}$ term, which can be high for large baselines (for core-passing neutrinos, $V_\text{CC}$ is large). For example, log$_{10} [L_\nu/E_\nu] = 1$ may correspond to $L_\nu=5000$ km and $E_\nu = 500$ GeV, so that $\Delta m^2_{32}L_\nu/E_\nu \approx 0.025$. However, for this baseline, the average density of the Earth is $\rho \approx 3.9$ g/cc, and hence for $\epsmutau=0.1$, we have $L_\nu \epsmutau V_\text{CC} \approx 0.93$, which takes the oscillation probability away from unity. The effect of NSI at such small $L_\nu/E_\nu$ is energy-independent, and can be seen at detectors like IceCube~\cite{Salvado:2016uqu} due to its better performance at high energy. Note, however, that in the high energy limit, the $\nu_\mu$ survival probability depends only on the magnitude of $\epsmutau$ (Eq.~\ref{eq:pmumu-nu-final-omsd}). As a result, it may be difficult for IceCube to determine sgn($\epsmutau$), if indeed $|\epsmutau|$ turns out to be nonzero. 
	
	\item As we go to higher value of log$_{10} [L_\nu/E_\nu]$ ($>2$), the term containing neutrino mass splitting becomes comparable to the $L \varepsilon_{\mu\tau} V_\text{CC}$ term in Eq.\,\ref{eq:pmumu-nu-final-omsd}, and the competition between these two terms leads to an energy dependence. When the oscillations due to the mass splitting are the dominating contribution, the change in oscillation wavelength due to non-zero $\varepsilon_{\mu\tau}$ results in a shift of the oscillation minima towards left or right side (lower or higher values of $L_\nu/E_\nu$, respectively), depending on the amplitude of $\varepsilon_{\mu\tau}$ and its sign. The direction of shift in the dip location depends on whether it is neutrino or antineutrino (since the sign of $V_\text{CC}$ for them are opposite), and on the neutrino mass ordering. This effect on the shift in the dip location is discussed in next paragraph in detail. This region is relevant for Super-K, INO, and most of the long-baseline experiments.
	
\end{itemize}

We now discuss the modification of the oscillation dip due to non-zero $\varepsilon_{\mu\tau}$. First, using the approximate expression of $\nu_\mu$ survival probability from Eq.~\ref{eq:NSI-pmumu-nu-final-omsd}, we obtain the value of $L_\nu/E_\nu$ at the first oscillation minimum (or the dip):  
\begin{equation}
\left. \frac{L_\nu}{E_\nu} \right|_\text{dip} = \frac{2\pi}{|\Delta m^2_{32}
	+ 4\varepsilon_{\mu\tau} V_\text{CC} E_\nu|}\,.
\label{eq:osc-min-lbye}
\end{equation}
The above expression may be written by expressing $V_\text{CC}$ in terms of the line-averaged matter density
\footnote{We can write $V_\text{CC}$ approximately as a function of matter density $\rho$
\begin{equation}
V_\text{CC} \approx \pm 7.6\times Y_{e} \times \frac{\rho}{10^{14}\,
	\text{g}/\text{cm}^{3}}\,\, \text{eV}\,.
\label{eq:vcc}
\end{equation}
Here, $Y_e = \frac{N_e}{N_p + N_n}$ is the relative electron number density. In an electrically neutral and  isoscalar medium, $Y_e=0.5$. The positive (negative) sign is for neutrino (antineutrino).}
 $\rho$ and taking care of units, 
\begin{equation}
\left. \frac{L_\nu [\text{km]}}{E_\nu [\text{GeV}]} \right|_\text{dip}
= \frac{\pi}{ \left| 2.54 \cdot \Delta m^2_{32} [\text{eV}^2]
	\pm 7.7\times 10^{-4} \cdot \rho [\text{g}/\text{cm}^3] \cdot Y_{e}
	\cdot\varepsilon_{\mu\tau} \cdot   E_\nu [\text{GeV}] \right|  }\,.
\label{eq:osc-min-lbye-units}
\end{equation}
Here, a positive sign in the denominator corresponds to neutrinos, whereas a negative sign is for antineutrinos. This approximation is useful for understanding the shift of dip position with non-zero $\varepsilon_{\mu\tau}$. Let us take the case of neutrino with normal ordering. With positive $\varepsilon_{\mu\tau}$, the denominator of Eq.\,\ref{eq:osc-min-lbye} increases, thus the oscillation minimum appears at a lower value of $L_\nu/E_\nu$ than that for the SI. On the other hand, with negative $\varepsilon_{\mu\tau}$, the oscillation dip would occur at a higher value of $L_\nu/E_\nu$. These two features can be seen clearly in the left panels of Fig.\,\ref{fig:osc_dip_neutrino_NSI}. For antineutrinos and the same mass ordering, since the matter potential has the opposite sign, the shift of oscillation dip is in the opposite direction as compared to that for neutrinos, given the same $\varepsilon_{\mu\tau}$.

Expanding Eq.\,\ref{eq:osc-min-lbye-units} to first order in $\epsmutau$, one can write
\begin{equation}
\left. \frac{L_\nu [\text{km}]}{E_\nu [\text{GeV}]} \right|_\text{dip}
= \frac{\pi}{\left| 2.54\times \Delta m^2_{32} [\text{eV}^2] \right|}
\mp \frac{ \pi \times 1.19 \times 10^{-4}\cdot \rho [\text{g}/\text{cm}^3]
	\cdot Y_{e} \cdot   E_\nu [\text{GeV}]}{\beta \,
	(\Delta m^2_{32})^2 [\text{eV}^4]} \cdot \varepsilon_{\mu\tau}  \,,
\label{eq:lbye-dip-calibration}
\end{equation}
where $\beta \equiv \text{sgn}(\Delta m^2_{32})$, as defined earlier. Here, the negative sign corresponds to neutrinos, whereas the positive sign is for antineutrino. This indicates that, for small values of $\epsmutau$, the shift in the dip location will be linear in $\epsmutau$, and will have opposite sign for neutrinos and antineutrinos, as well as for the two mass orderings. We have checked that for $E_\nu<25$ GeV and typical line-averaged density of the Earth, indeed $\varepsilon_{\mu\tau}<0.1$ is small enough for the above approximation to hold.

%%%%%%%%%%%%%%%%%%%%%%%%%%%%%%%%%%%%%%%%%%%%%%%%%%%%%%%%%%%%%%%%%%%%%%%
\subsection{Effect of $\epsmutau$ on the oscillation valley in the
	($E_\nu$, $\cos\theta_\nu$) plane}
\label{subsec:osc-valley}
%%%%%%%%%%%%%%%%%%%%%%%%%%%%%%%%%%%%%%%%%%%%%%%%%%%%%%%%%%%%%%%%%%%%%%

\begin{figure}
	\centering
	\includegraphics[width=0.45\linewidth]{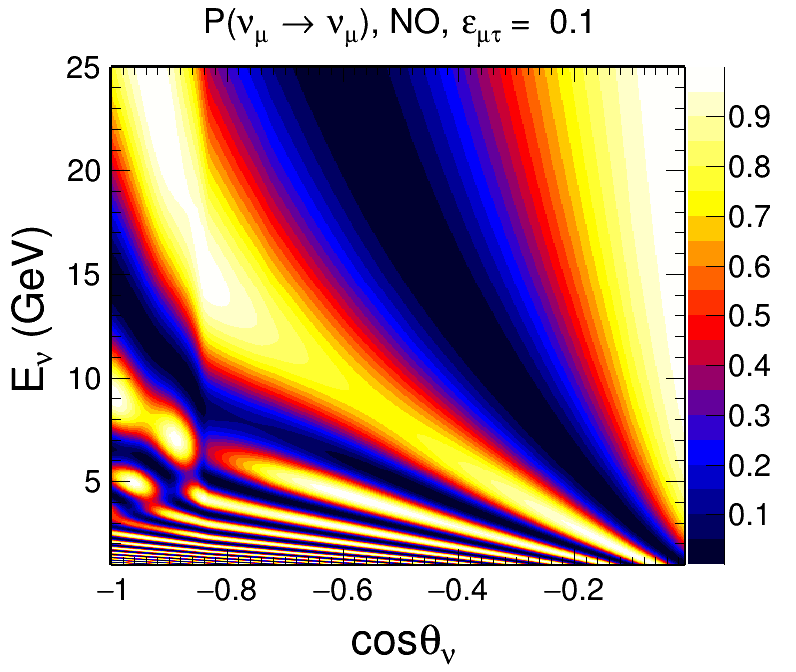}
	\includegraphics[width=0.45\linewidth]{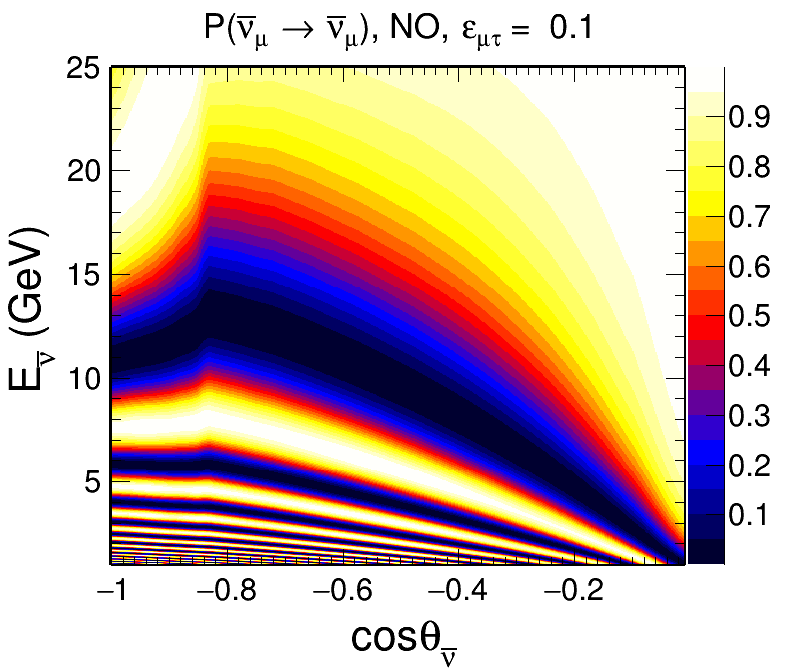}
	\includegraphics[width=0.45\linewidth]{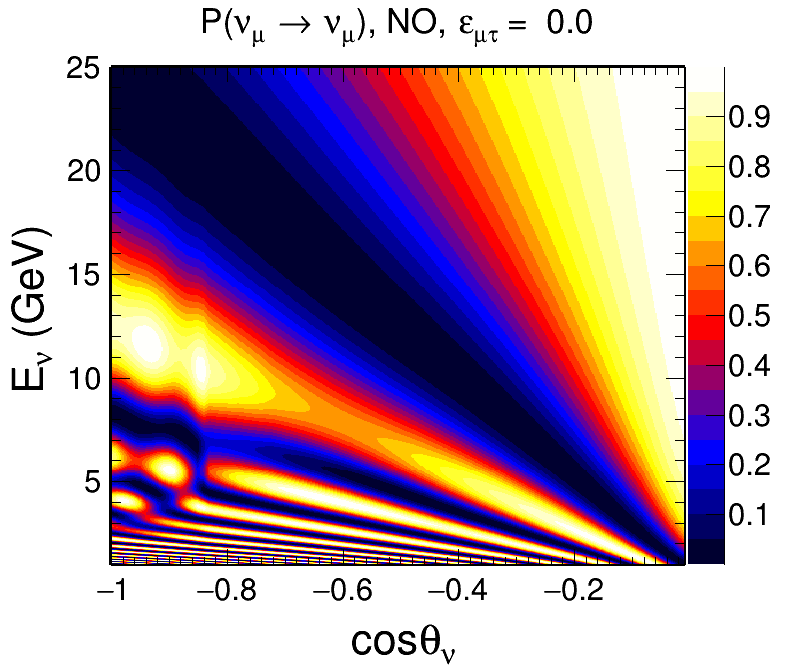}
	\includegraphics[width=0.45\linewidth]{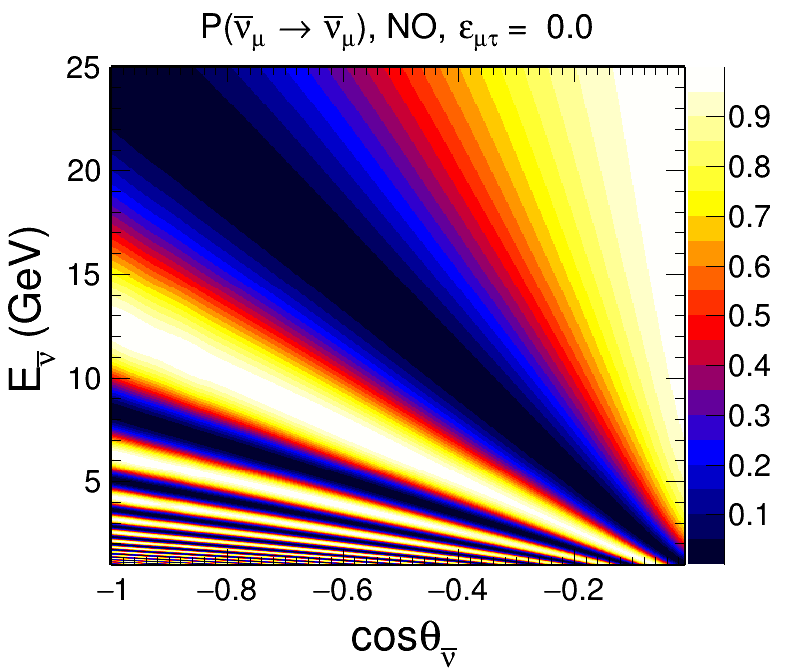}
	\includegraphics[width=0.45\linewidth]{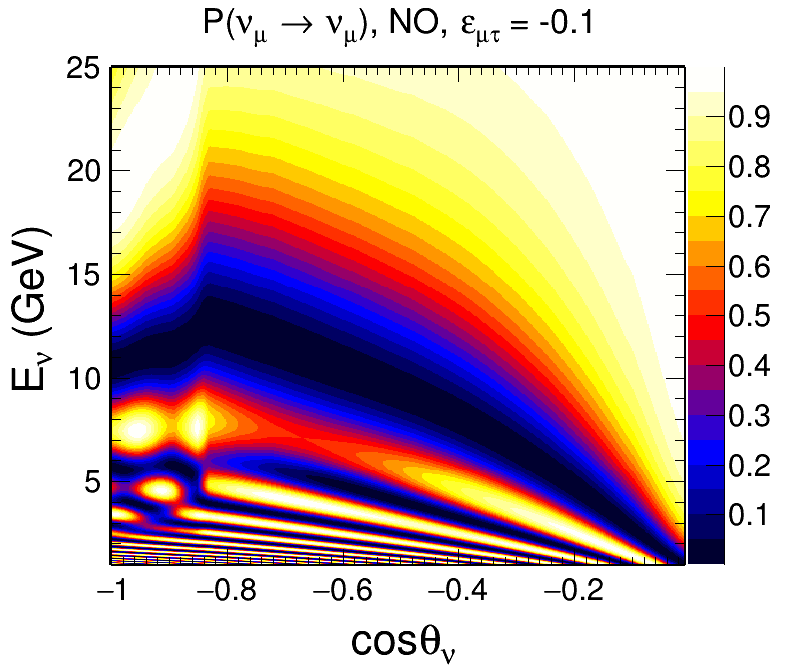}
	\includegraphics[width=0.45\linewidth]{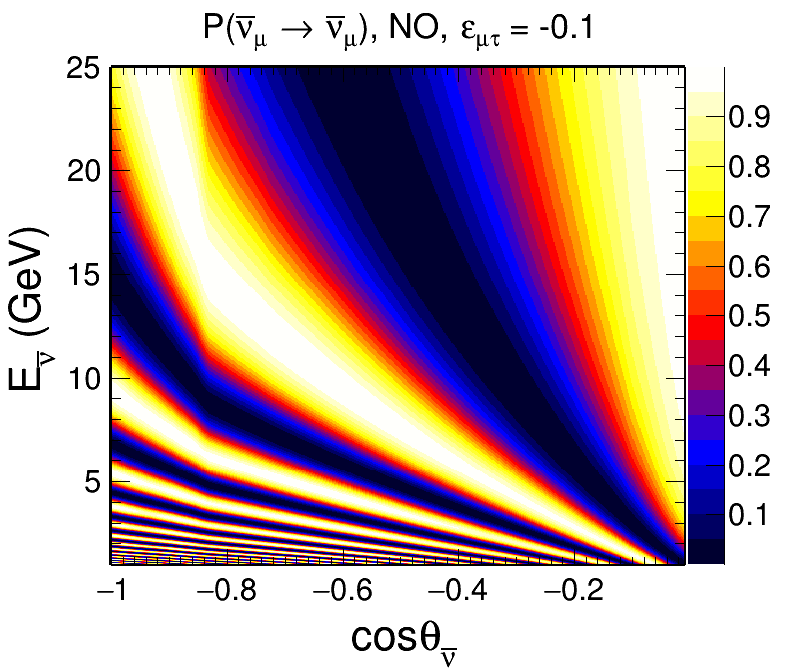}
	\caption{The oscillograms for the survival probabilities in ($E_\nu,\cos\theta_\nu$) plane for neutrino and antineutrino in the left and right panels, respectively, with normal mass ordering and the benchmark values of oscillation parameters from Table~\ref{tab:osc-param-value}. The value of $\varepsilon_{\mu\tau}$ is taken as +0.1, 0 (SI), and -0.1 in the top, middle and bottom panels, respectively.~\cite{Kumar:2021lrn}}
	\label{fig:osc_valley_neutrino_NSI}
\end{figure}

In Fig.\,\ref{fig:osc_valley_neutrino_NSI}, we show the oscillograms of survival probabilities for $\nu_\mu$ and $\bar\nu_\mu$, in the plane of neutrino energy and cosine of neutrino zenith angle ($\cos\theta_\nu$), with NO as the neutrino mass ordering. We show only upward-going neutrinos ($\cos \theta_\nu <0$), since for downward-going neutrinos ($\cos\theta_\nu>0$), the baseline $L_\nu$ is very small and neutrino oscillations do not develop, making $P_{\mu\mu} \approx 1.0$. The differences observed among the upper, middle, and lower panels are due to different values of $\varepsilon_{\mu\tau}$. A major impact of NSI is observed on the feature corresponding to the first oscillation minima, seen in the figure as a broad blue/black diagonal band. We refer to this broad band as the oscillation valley~\cite{Kumar:2020wgz}. In the SI case, the oscillation valley appears like a triangle with straight edges, while in the presence of NSI, its edges seem to acquire a curvature. The sign of this curvature is opposite for neutrinos and antineutrinos, as can be seen by comparing the left and right plots of upper and lower panels of Fig.\,\ref{fig:osc_valley_neutrino_NSI}. 

The modification in the oscillation valley due to NSI may be explained from the following relation between $E_\nu$ and $\cos\theta_\nu$ at the first oscillation minima. We rewrite Eq.\,\ref{eq:osc-min-lbye} with $L_\nu \approx 2R |\cos\theta_\nu|$:
\begin{equation}
E_\nu |_\text{valley} \approx \frac{| \Delta m^2_{32}| }{
	(\pi / |R \cos\theta_\nu|) \;  
	- 4 \beta \, \varepsilon_{\mu\tau} V_\text{CC}}\, .
\label{eq:osc-min-Ecostheta-wounit}
\end{equation}
Taking care of units and expressing $V_\text{CC}$ in terms of line-averaged constant matter density $\rho$, the above expression may be written as 
\begin{equation}
E_\nu[\text{GeV}]|_\text{valley} = \frac{|\Delta m^2_{32} [\text{eV}^2]|}{
	(\pi /| 5.08 \cdot R [\text{km}] \cdot \cos\theta_\nu|) \; 
	\mp 3.02 \times 10^{-4} \cdot \beta \cdot \varepsilon_{\mu\tau} \cdot Y_e
	\cdot \rho [\text{g}/\text{cm}^3]  }  \,.
\label{eq:osc-min-Ecostheta}
\end{equation}
Here, the negative sign in the denominator corresponds to neutrinos, whereas the positive sign is for antineutrinos. Putting $\varepsilon_{\mu\tau}=0$ in Eq.\,\ref{eq:osc-min-Ecostheta} gives the condition of the first oscillation minima to be $E_\nu= | (5.08/\pi)\cdot \Delta m^2_{32}\cdot R [\text{km}] \cdot \cos\theta_\nu|$, and this relation  clearly shows that in the SI case, the minimum of the oscillation valley is a straight line in the ($E_\nu\,,\cos\theta_\nu$) plane.

Now in the normal mass ordering scenario, if $\varepsilon_{\mu\tau} >0$ in Eq.\,\ref{eq:osc-min-Ecostheta}, then for neutrinos, $E_\nu$ increases for a given $\cos\theta_\nu$ as compared to the SI case. As a result, the oscillation valley (the broad blue/black band) bends towards higher values of energies in top left panel. On the other hand, for antineutrino, the oscillation valley tilts towards lower energies for $\varepsilon_{\mu\tau}>0$, as can be seen in the top right panel. For negative values of  $\varepsilon_{\mu\tau}$, the oscillation valley bends in the opposite direction, for both neutrinos and antineutrinos, as shown in the bottom panels of  Fig.\,\ref{fig:osc_valley_neutrino_NSI}. In the inverse mass ordering scenario, the bending of the oscillation valley will be in the direction opposite to that in the normal mass ordering scenario.

From Eqs.~\ref{eq:lbye-dip-calibration} and~\ref{eq:osc-min-Ecostheta}, the shift in the oscillation minima and the bending in the oscillation valley are in opposite directions for normal and inverted mass orderings. The effect of $\epsmutau$, therefore, depends crucially on the mass ordering. In this chapter, we present our analysis in the scenario where the mass ordering is known to be NO. The analysis for IO may be performed in an exactly analogous manner. Our results are presented with the exposure corresponding to 10 years of ICAL data-taking. It is expected that the mass ordering will be determined from the neutrino experiments (including ICAL) by this time.

%%%%%%%%%%%%%%%%%%%%%%%%%%%%%%%%%%%%%%%%%%%%%%%%%%%%%%%%%%%%%%%
%%%%%%%%%%%%%%%%%%%%%%%%%%%%%%%%%%%%%%%%%%%%%%%%%%%%%%%%%%%%%%
\section{Impact of NSI on the Event Distribution at ICAL}
\label{sec:evt-ical}
%%%%%%%%%%%%%%%%%%%%%%%%%%%%%%%%%%%%%%%%%%%%%%%%%%%%%%%%%%%%%
%%%%%%%%%%%%%%%%%%%%%%%%%%%%%%%%%%%%%%%%%%%%%%%%%%%%%%%%%%%%%%

While the effect of NSI on the neutrino and antineutrino survival probabilities was discussed in the last section, it is important to confirm whether the features present at the probability level can survive in the observables at a detector, and if they can be reconstructed. Here is where the response of the detector plays a crucial role. Neutrinos cannot be directly detected in an experiment, however the charged leptons produced from their charged-current interactions in the detector contains information about their energy, direction, and flavor, which can be recovered depending on the nature of the detector. We simulate the reconstructed muon events at the ICAL detector following the procedure mentioned in Sec.~\ref{sec:event_simulation}.

%%%%%%%%%%%%%%%%%%%%%%%%%%%%%%%%%%%%%%%%%%%%%%%%%%%%%%%%%%%%%%%%%%
\subsection{The $L_\mu^\text{rec}/E_\mu^\text{rec}$ Distributions}
\label{sec:events-L/E}
%%%%%%%%%%%%%%%%%%%%%%%%%%%%%%%%%%%%%%%%%%%%%%%%%%%%%%%%%%%%%%%%%

\begin{figure}
	\centering
	\includegraphics[width=0.45\linewidth]{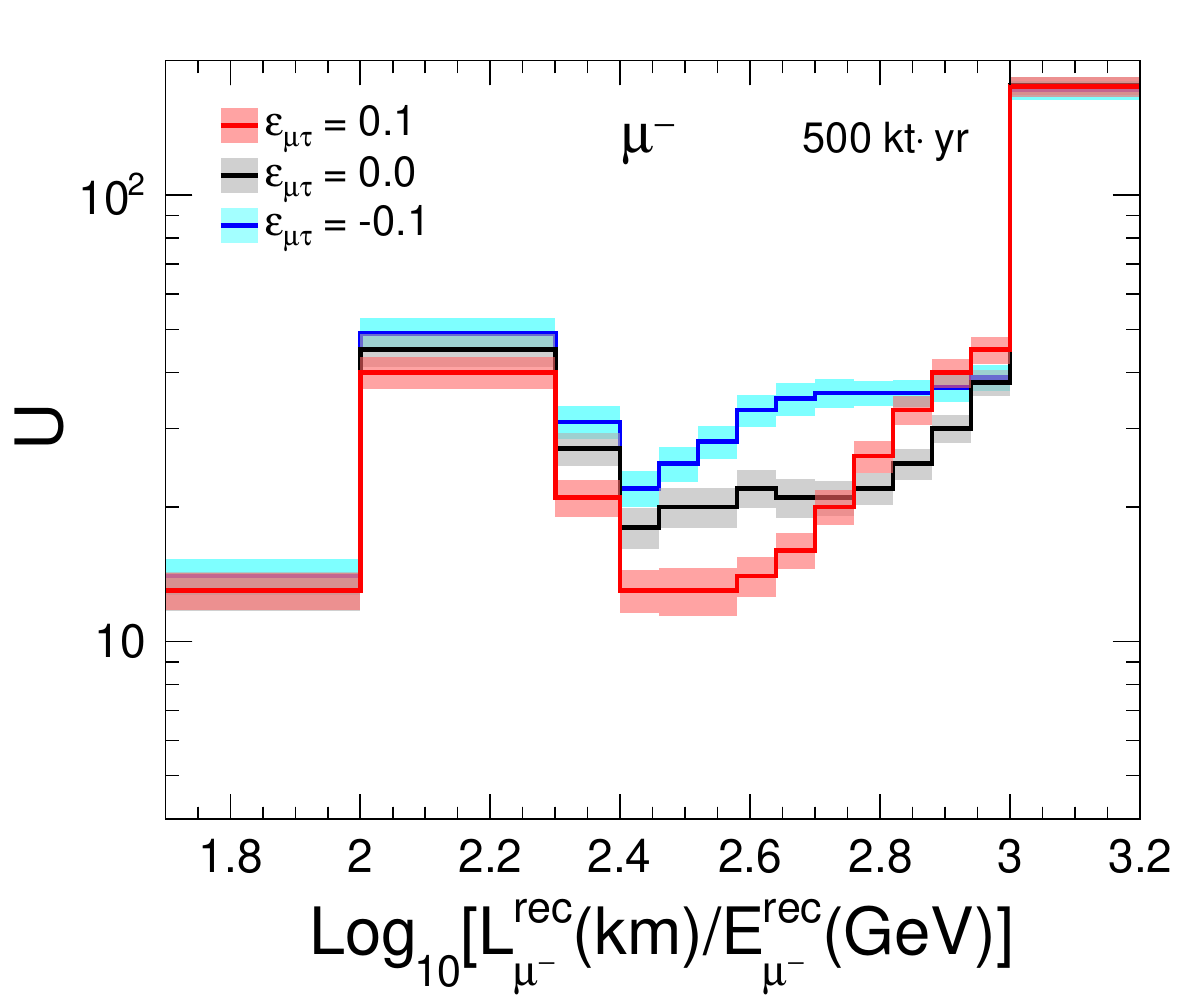}
	\includegraphics[width=0.45\linewidth]{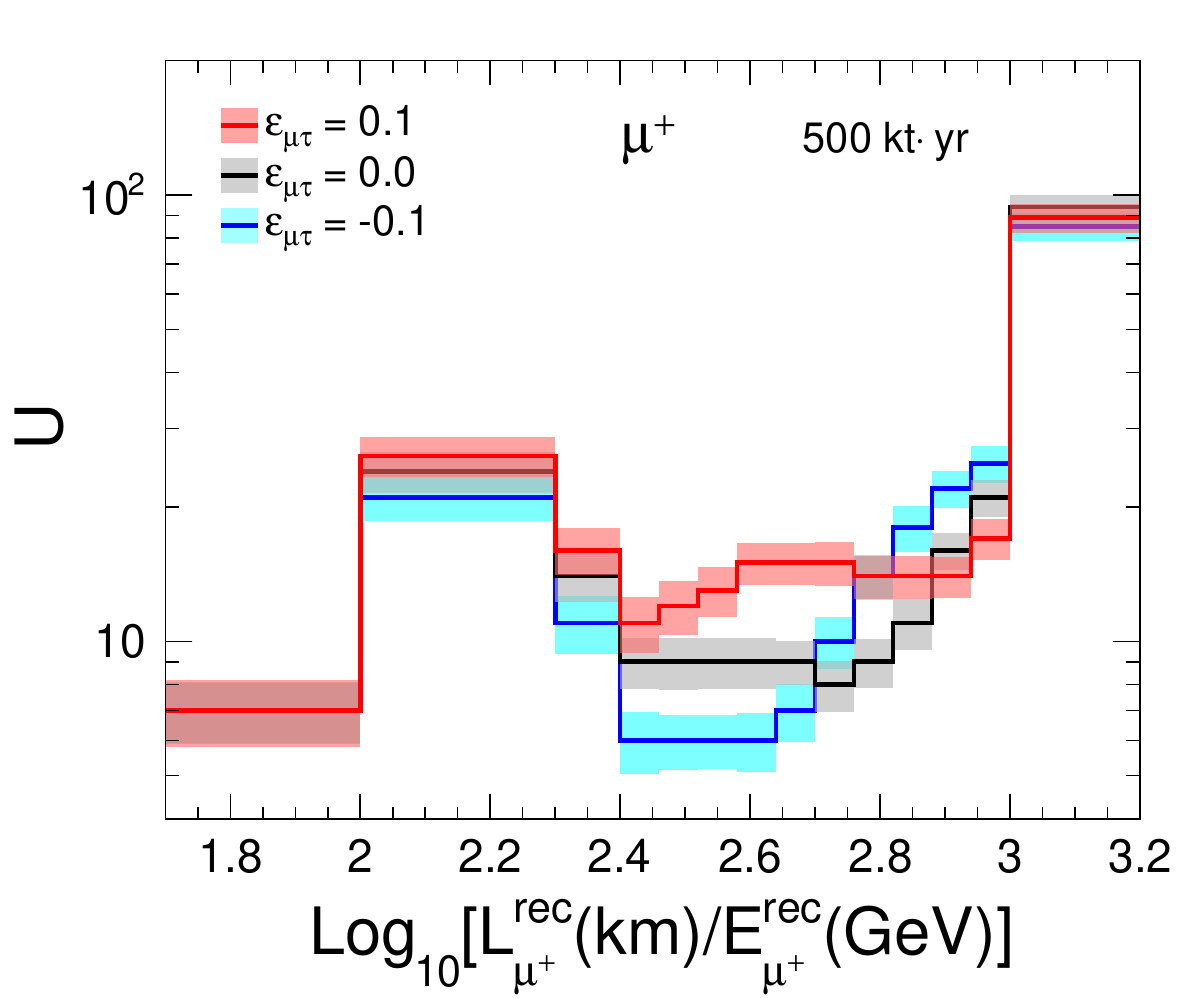}
	\caption{The $\log_{10}[L_\mu^\text{rec}/E_\mu^\text{rec}]$ distributions of $\mu^-$ and $\mu^+$ events in the left and right panels, respectively, expected in 10 years at ICAL. The black curves correspond to SI ($\varepsilon_{\mu\tau}=0$), whereas red and blue curves are for $\varepsilon_{\mu\tau} = 0.1$ and -0.1, respectively. The statistical uncertainties shown by shaded boxes are the root-mean square (rms) fluctuations of 100 independent 10-year data sets. The solid lines show the mean of these 100 distributions. We consider normal mass ordering, and the benchmark values of oscillation parameters given in Table~\ref{tab:osc-param-value}.~\cite{Kumar:2021lrn}}
	\label{fig:events_1D_10yr_NSI}
\end{figure}

Fig.\,\ref{fig:events_1D_10yr_NSI} presents the $\log_{10}[L_\mu^\text{rec}/E_\mu^\text{rec}]$ distributions of the upward-going $\mu^-$ and $\mu^+$ events (U, $\cos\theta_\mu^\text{rec} <0$) with SI ($\varepsilon_{\mu\tau} = 0$) and with SI + NSI ($\varepsilon_{\mu\tau} = \pm 0.1$) expected with 10-year exposure at ICAL. We include the statistical fluctuations in the number of events by simulating 100 independent data sets, and calculating the mean and root-mean-square deviation for each bin. For $\log_{10}[L_\mu^\text{rec}/E_\mu^\text{rec}]$, we use the same binning scheme as shown in Table~\ref{tab:binning-1D-10years}. We have a total of 34 non-uniform bins of $\log_{10}[L_\mu^\text{rec}/E_\mu^\text{rec}]$ in the range 0 to 4. The downward-going events (D, $\cos\theta_\mu^\text{rec} >0$) are not affected significantly by oscillations or by additional NSI interactions. The $\log_{10}[L_\mu^\text{rec}/E_\mu^\text{rec}]$ distribution of downward-going $\mu^-$ and $\mu^+$ events is identical to the one shown in the upper panels of Fig.~\ref{fig:osc_dip_10yr} in chapter~\ref{chap:dip_valley}, and we do not repeat it here.

Fig.\,\ref{fig:events_1D_10yr_NSI} clearly shows that the $\log_{10}[L_\mu^\text{rec}/E_\mu^\text{rec}]$ range of 2.4 to 3.0 would be important for NSI studies, since non-zero $\varepsilon_{\mu\tau}$ has the largest effect in this range. With positive $\varepsilon_{\mu\tau}$, the number of $\mu^-$ events is lower as compared to that of SI case, whereas the number of $\mu^+$ events is higher. If $\varepsilon_{\mu\tau}$ is negative, then the modifications of the number of $\mu^-$ and $\mu^+$ events are the other way around. As a consequence, if a detector is not able to distinguish between the $\mu^-$ and $\mu^+$ events, the difference between SI and NSI would get diluted substantially. The charge identification capability of the magnetized ICAL detector would be crucial in exploiting this observation.

%%%%%%%%%%%%%%%%%%%%%%%%%%%%%%%%%%%%%%%%%%%%%%%%%%%%%%%%%%%%%%%%%%%%%%%%%
\subsection{Distributions in the ($E_\mu^\text{rec}$, $\cos\theta_\mu^\text{rec}$) plane}
\label{sec:events-2d}
%%%%%%%%%%%%%%%%%%%%%%%%%%%%%%%%%%%%%%%%%%%%%%%%%%%%%%%%%%%%%%%%%%%%%%%%%

\begin{figure}
	\centering
	\includegraphics[width=0.45\linewidth]{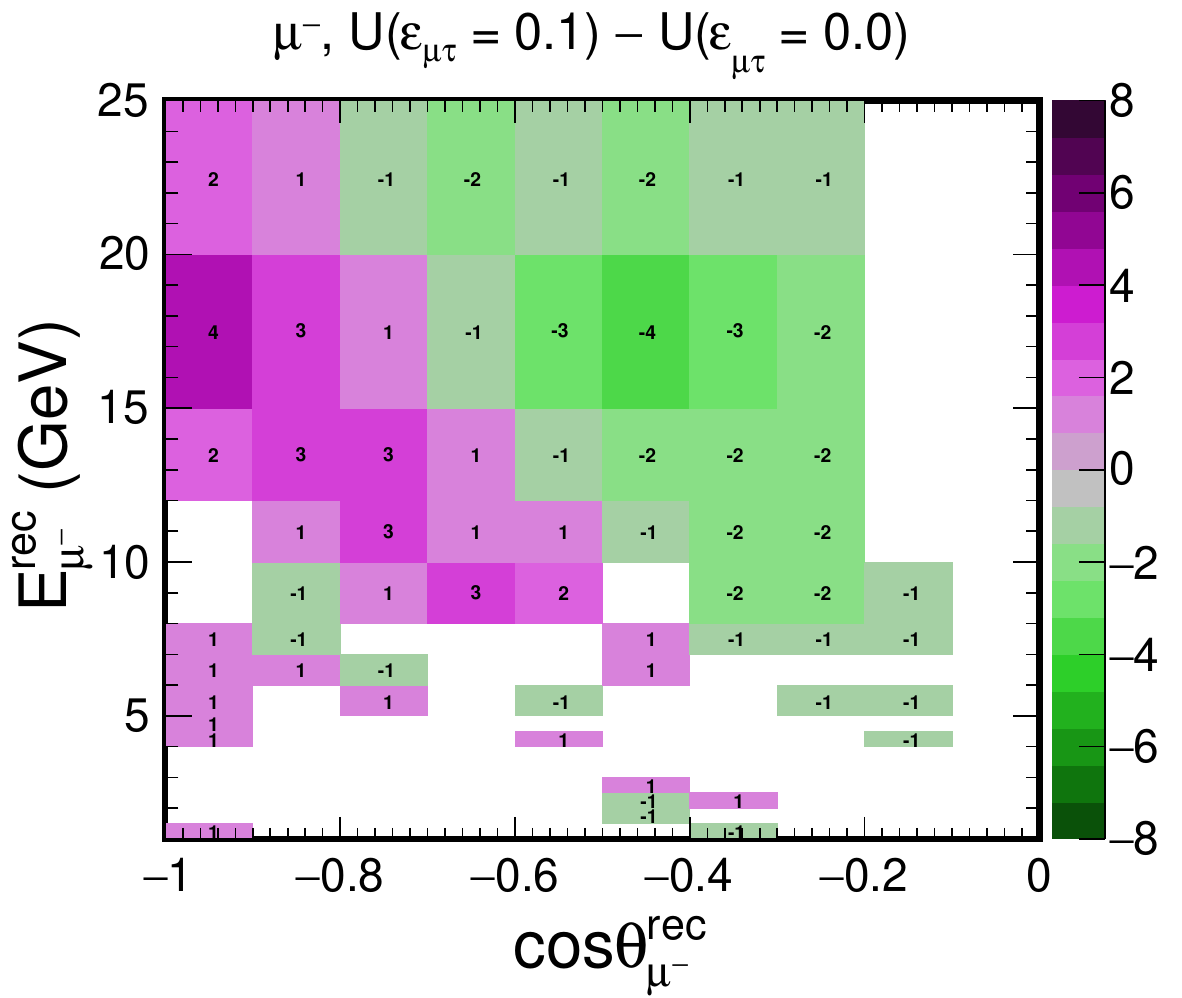}
	\includegraphics[width=0.45\linewidth]{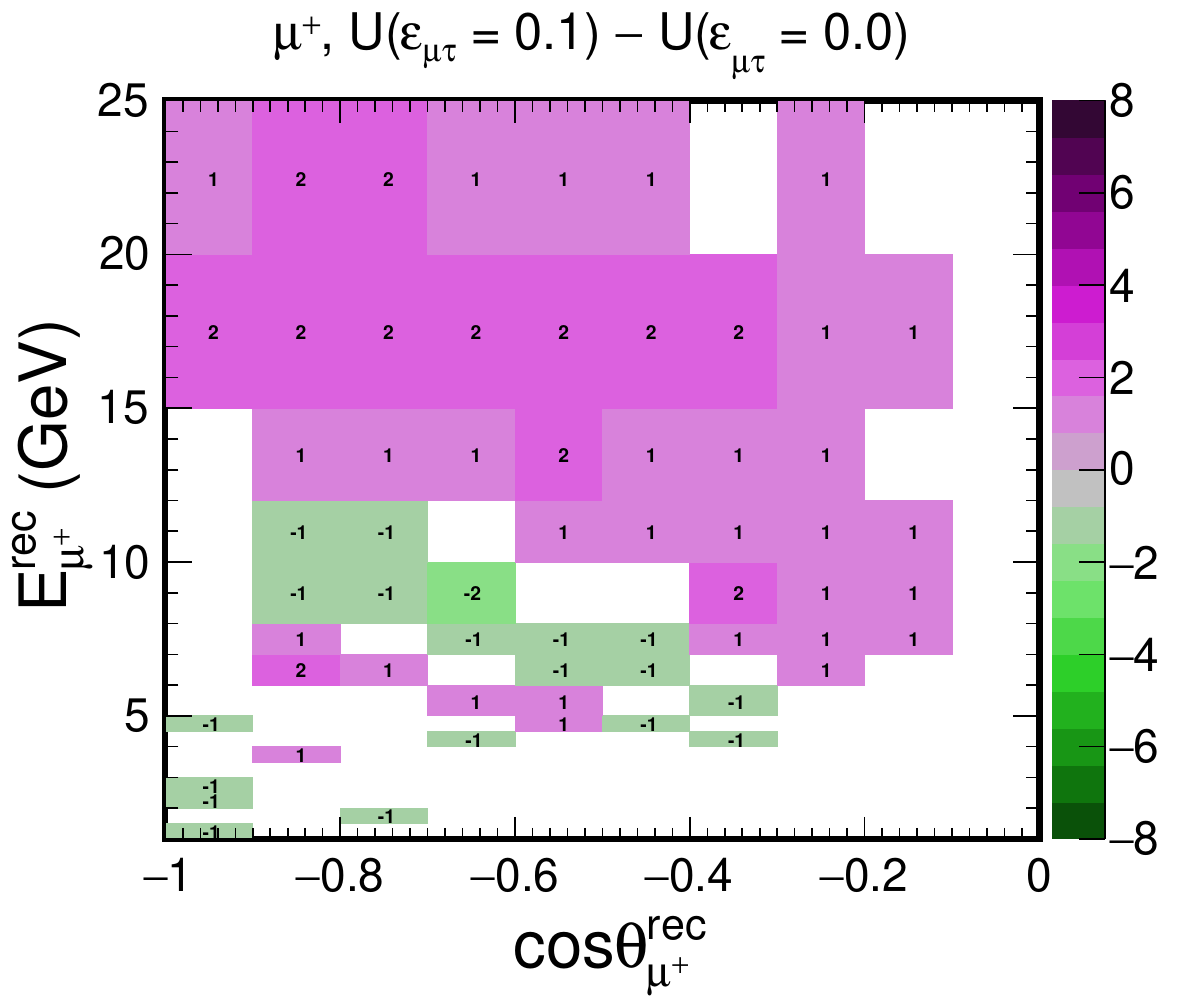}
	\includegraphics[width=0.45\linewidth]{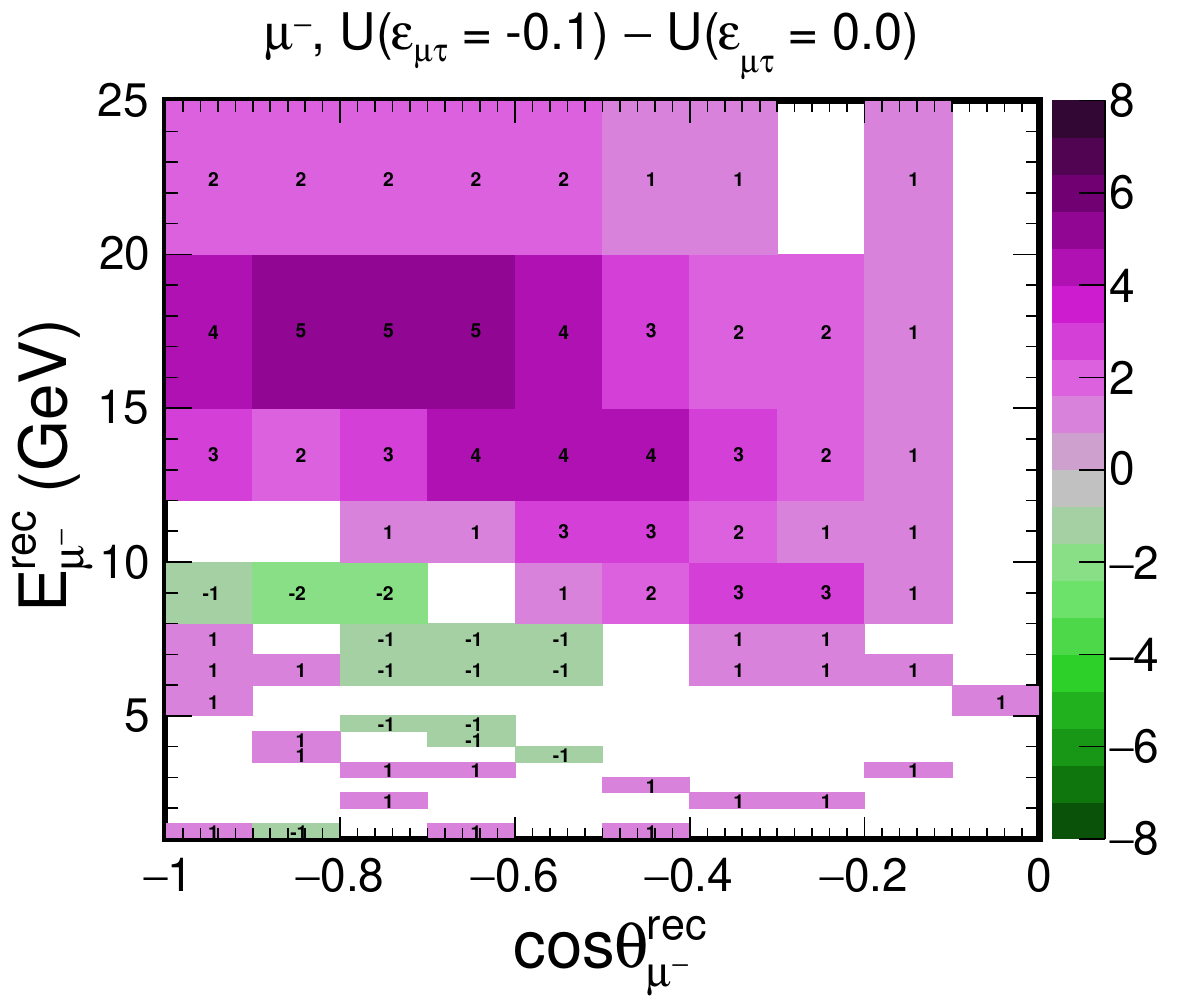}
	\includegraphics[width=0.45\linewidth]{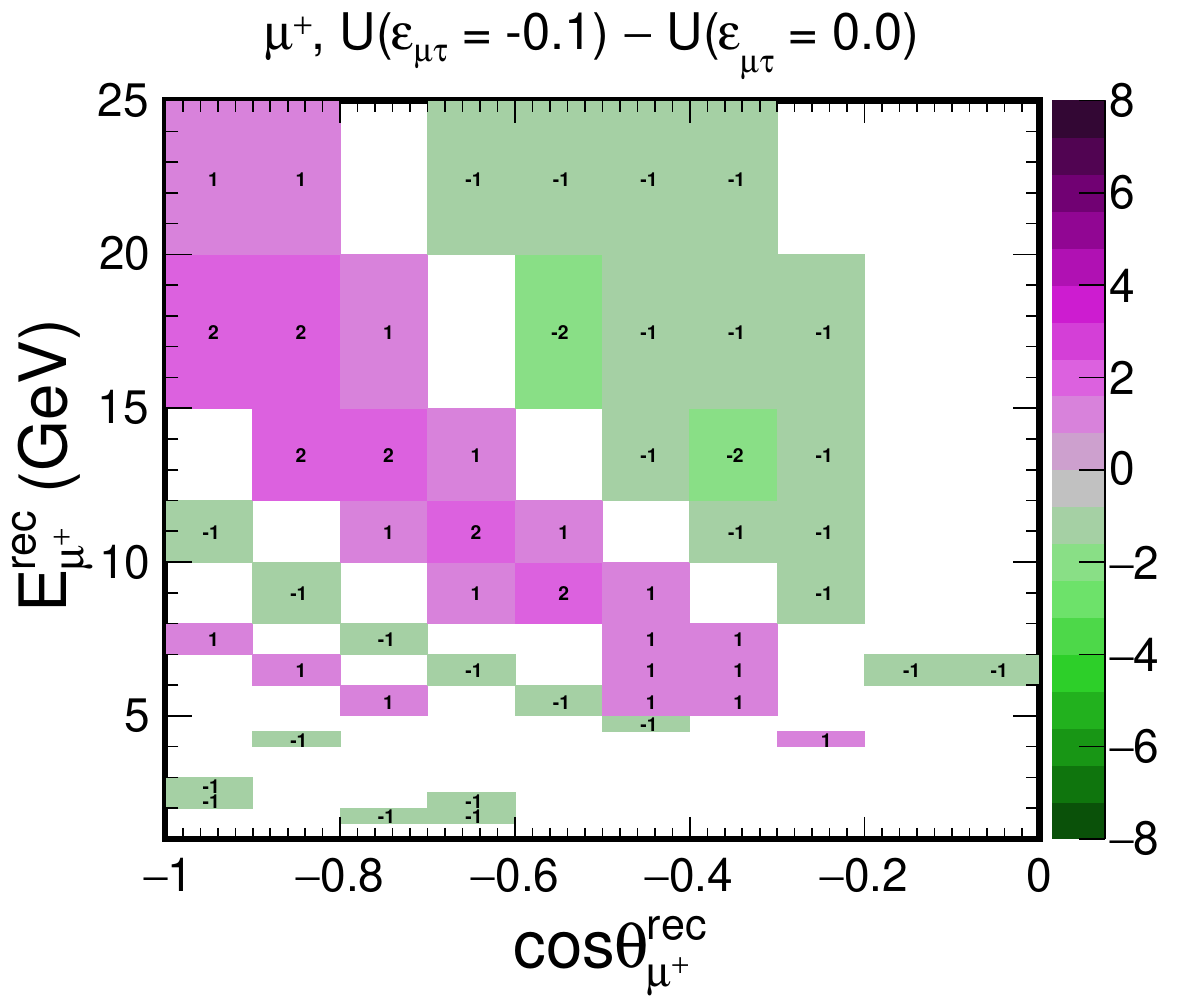}
	\caption{The $(E_\mu^\text{rec},\,\cos\theta_\mu^\text{rec})$ distributions of difference of events between SI with non-zero NSI and only SI ($\varepsilon_{\mu\tau} = 0$) expected with 500 kt$\cdot$yr of ICAL. The Left and right panels are for $\mu^-$ and $\mu^+$ events, respectively, whereas the upper and lower panels correspond to $\varepsilon_{\mu\tau}= 0.1$ and $-0.1$, respectively. We consider normal mass ordering, and the benchmark values of oscillation parameters given in Table~\ref{tab:osc-param-value}.~\cite{Kumar:2021lrn}}
	\label{fig:events_2D_10yr_NSI}
\end{figure}

In order to study the effect of NSI on the distribution of events in the ($E_\mu^\text{rec}$, $\cos\theta_\mu^\text{rec}$) plane, we bin the data in reconstructed observables, $E_\mu^\text{rec}$ and $\cos\theta_\mu^\text{rec}$. We have a total of 16 non-uniform $E_\mu^\text{rec}$ bins in the range 1 -- 25 GeV, and the whole range of $-1 \leq \cos\theta_\mu^\text{rec}\leq 1$  is divided into 20 uniform bins. The reconstructed $E_\mu^\text{rec}$ and $\cos\theta_\mu^\text{rec}$ are binned with the same binning scheme as shown in Table~\ref{tab:binning-valley-10years}. For demonstrating event distributions, we scale the 1000-year MC sample to an exposure of 10 years. The difference in the number of events with SI + NSI ($|\varepsilon_{\mu\tau}| = 0.1$) and the number of events with SI, for $\mu^-$ and $\mu^+$ events are shown in Fig.\,\ref{fig:events_2D_10yr_NSI}. Note that, in Fig.~\ref{fig:events_2D_10yr_NSI}, we choose the range corresponding to upward-going events ($\cos\theta_\mu^\text{rec} < 0$) only.

From this figure, we can make the following observations.
\begin{itemize}
	
	\item The events in the bins with $E_\mu^\text{rec} > 5$ GeV and $\cos\theta_\mu^\text{rec}<-0.2$ are particularly useful for NSI searches at ICAL. This is true for both $\mu^-$ and $\mu^+$ events.
	
	\item There is almost no difference in the number of events between SI+NSI and SI at $E_\mu^\text{rec} < 5$ GeV. The events with reconstructed muon energy less than 5 GeV do not seem to be sensitive to NSI due to small NSI-induced matter effects: $L_\nu \epsmutau V_\text{CC} \lesssim \Delta m^2/(2 E_\nu)$ at small energies.
	
	\item As we go to higher energies, the mass-squared difference-induced neutrino oscillations die down, while the NSI-induced matter effect term $L_\nu \epsmutau V_\text{CC}$ increases. Therefore, large effects of NSI are observed at high energies. 
	
	\item The effect of NSI is also larger at higher baselines, since in this case, neutrinos pass through inner and outer cores that have very high matter densities (around 13 g/cm$^3$ and 11 g/cm$^3$, respectively), and hence higher values of $L_\nu \epsmutau V_\text{CC}$.
	
	\item In many of the $(E_\mu^\text{rec},\,\cos\theta_\mu^\text{rec})$ bins, NSI gives rise to an excess in $\mu^-$ events and a deficit in $\mu^+$ events, or vice versa. If $\mu^-$ and $\mu^+$ events are not separated, this would lead to a dilution of information.  Therefore, muon charge information is a crucial ingredient in the search for NSI. We have also seen this feature in the $\log_{10}[L_\mu^\text{rec}/E_\mu^\text{rec}]$ distributions of $\mu^-$ and $\mu^+$ events in Sec.\,\ref{sec:events-L/E}.
	
\end{itemize}

%%%%%%%%%%%%%%%%%%%%%%%%%%%%%%%%%%%%%%%%%%%%%%%%%%%%%%%%%%%%%%%%%%
%%%%%%%%%%%%%%%%%%%%%%%%%%%%%%%%%%%%%%%%%%%%%%%%%%%%%%%%%%%%%%%%%
\section{Identifying NSI through the Shift in the Location of Oscillation Dip}
\label{sec:reco-dip}
%%%%%%%%%%%%%%%%%%%%%%%%%%%%%%%%%%%%%%%%%%%%%%%%%%%%%%%%%%%%%%%%%
%%%%%%%%%%%%%%%%%%%%%%%%%%%%%%%%%%%%%%%%%%%%%%%%%%%%%%%%%%%%%%%%%

To study the effect of NSI in oscillation dip, we focus on the ratio of upward-going (U) and downward-going (D) muon events as the observable to be studied as defined in Eq.~\ref{eq:U/D_def}. One advantage of the use of the U/D ratio would be to nullify the effects of systematic uncertainties like flux normalization, cross sections, overall detection efficiency, and energy dependent tilt error, as can be seen later. Note that the up-down symmetry of the detector geometry and detector response play an important role in this. 

\begin{figure}
	\centering
	\includegraphics[width=0.45\linewidth]{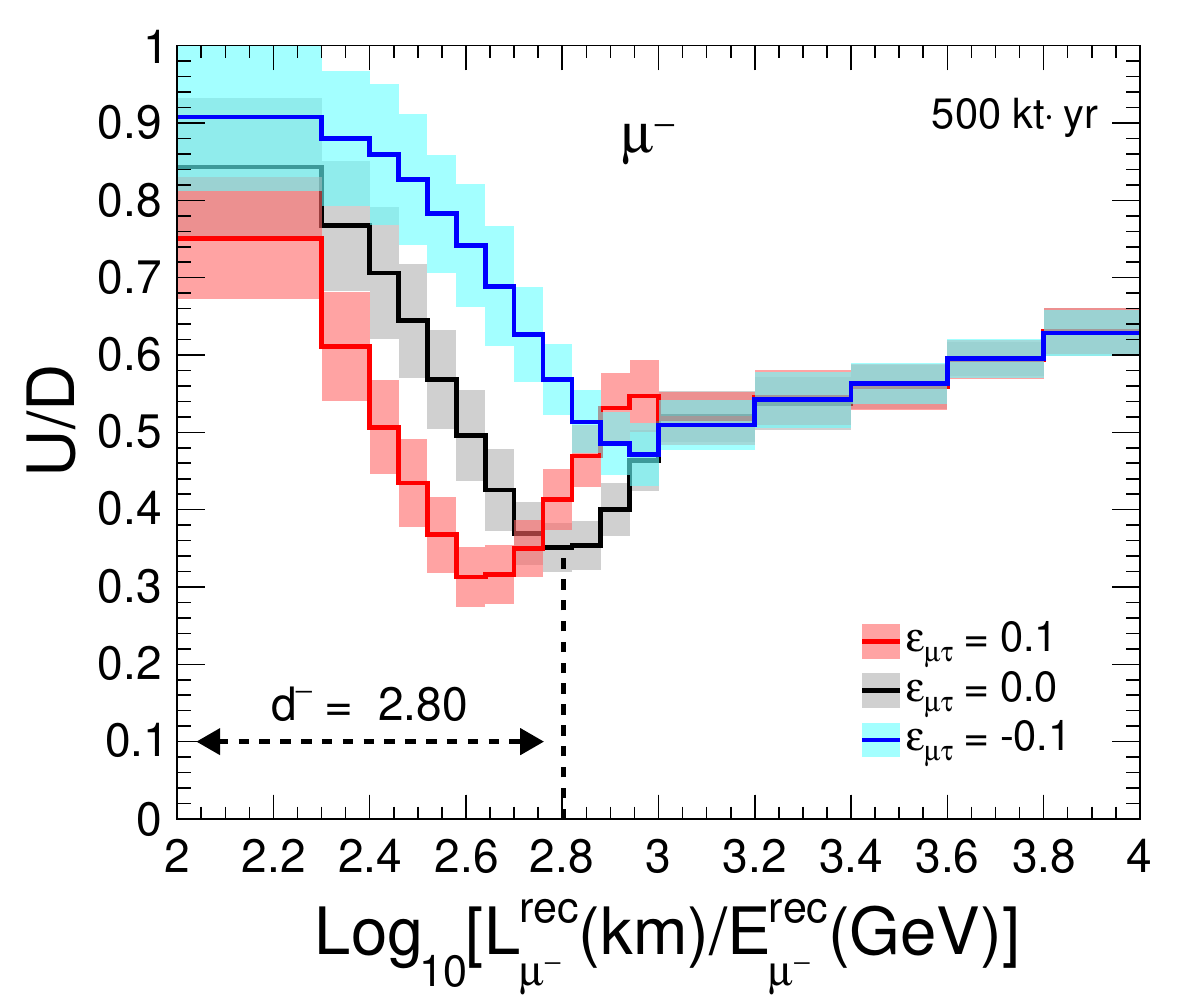}
	\includegraphics[width=0.45\linewidth]{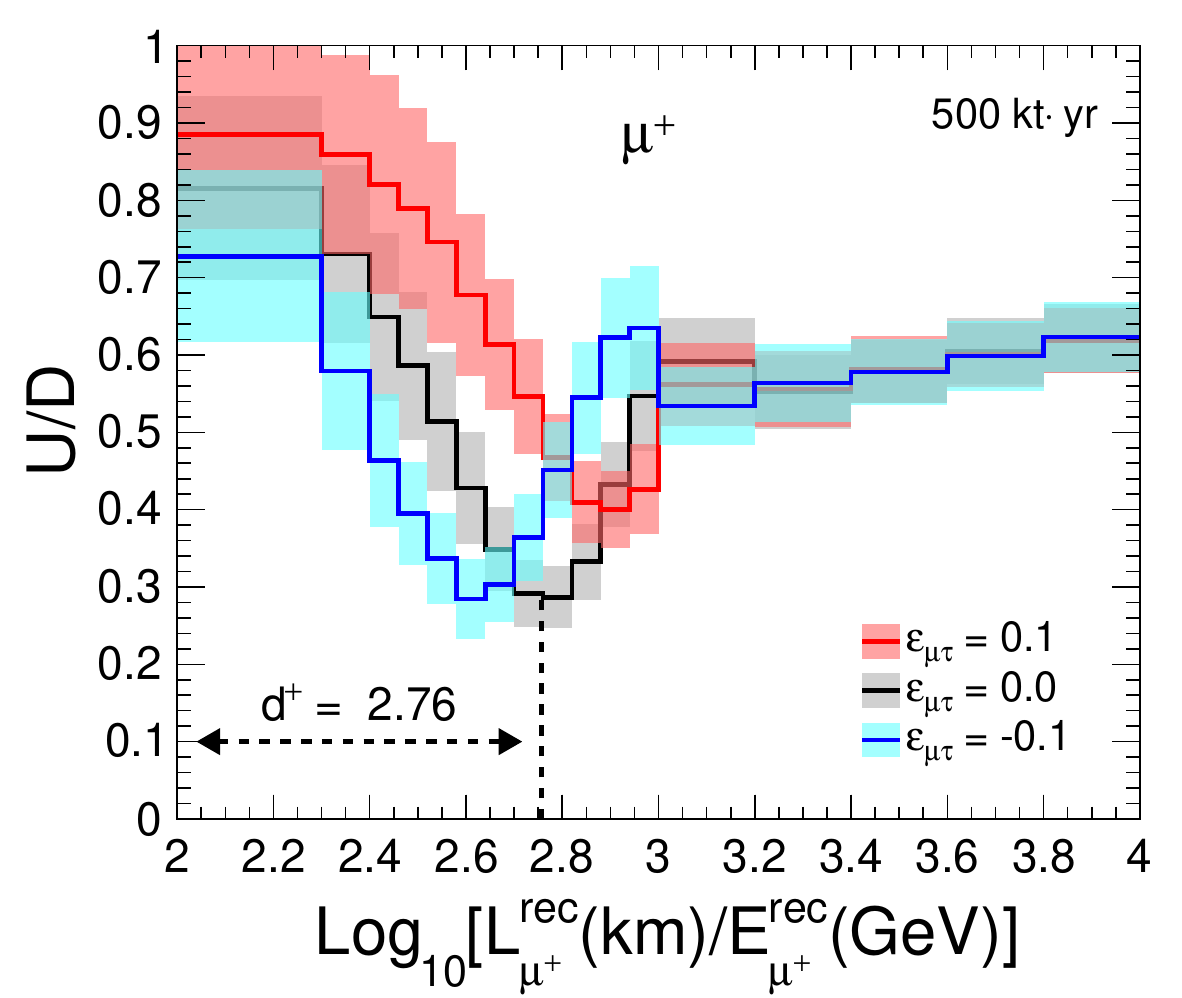}
	\caption{The $\log_{10}(L_\mu^\text{rec}/E_\mu^\text{rec})$ distributions of ratio of upward-going and downward-going $\mu^-$ (left panel) and $\mu^+$ (right panel) events using 10-year data of ICAL. The black curves correspond to SI ($\varepsilon_{\mu\tau} = 0$), whereas red and blue curves are for $\varepsilon_{\mu\tau} = 0.1$ and -0.1, respectively. The statistical fluctuations shown by shaded boxes are the root-mean-square deviation of 100 independent distributions of U/D ratio, each for 10 years, whereas the mean of these distributions are shown by solid curves. We consider normal mass ordering, and the benchmark values of oscillation parameters given in Table~\ref{tab:osc-param-value}.~\cite{Kumar:2021lrn}}
	\label{fig:osc_dip_10yr_NSI}
\end{figure}

Fig.~\ref{fig:osc_dip_10yr_NSI} presents the $L_\mu^{\text{rec}}/E_\mu^{\text{rec}}$ distributions of U/D with $\varepsilon_{\mu\tau} = 0$ (SI) and $\pm 0.1$, for an exposure of 10 years at ICAL. The mass ordering is taken as NO. The statistical fluctuations shown in the figure are the root-mean-square deviation obtained from the simulations of 100 independent sets of 10 years data. Around $\log_{10}[L_\mu^\text{rec}/E_\mu^\text{rec}] \sim 2.4 - 2.6$, we see a significant modification in the U/D ratio due to the presence of non-zero $\varepsilon_{\mu\tau}$ ($\pm 0.1$), particularly in the location of the oscillation dip. The dip shifts towards left or right ( i.e. to lower or higher values of $L_\mu^\text{rec}/E_\mu^\text{rec}$, respectively) from that of the SI case with non-zero $\varepsilon_{\mu\tau}$. This shift for $\mu^-$ and $\mu^+$ events is observed to be in the opposite directions, as is expected from the discussions in Sec.\,\ref{subsec:osc-dip}. For example, for $\varepsilon_{\mu\tau} >0$, the location of oscillation dip in $\mu^-$ events gets shifted towards smaller $L_\mu^\text{rec}/E_\mu^\text{rec}$ values. On other hand, in $\mu^+$ events, the oscillation dip shifts to higher values of $L_\mu^\text{rec}/E_\mu^\text{rec}$. For $\varepsilon_{\mu\tau} <0$, this shift is in the opposite directions. Moreover, as the oscillation dip shifts towards the higher value of $L_\mu^\text{rec}/E_\mu^\text{rec}$, the dip becomes shallower due to the effect of rapid oscillation at high $L_\nu/E_\nu$. This effect is visible in both, $\mu^-$ and $\mu^+$. It can be easily argued that the modification in oscillation dips of $\mu^-$ and $\mu^+$ due to non-zero $\varepsilon_{\mu\tau}$ is the reflection of how the survival probabilities of neutrino and antineutrino change in the presence of non-zero $\varepsilon_{\mu\tau}$, as expected from Eq.\,\ref{eq:osc-min-lbye} in Sec.~\ref{subsec:osc-dip}.

Note that for a neutrino detector that is blind to the charge of muons, the dip itself would get diluted since different average inelasticities of neutrino and antineutrino events at these energies would lead to slightly different dip locations in $\mu^-$ and $\mu^+$ distributions. Moreover, the shift of oscillation dip towards opposite directions in $\mu^-$ and $\mu^+$ events due to non-zero $\varepsilon_{\mu\tau}$ would further contribute to the dilution of NSI effects on the dip location. On the other hand, a detector like ICAL that has the charge identification capability, not only provides independent undiluted measurements of dip locations in $\mu^-$ and $\mu^+$ events, but also provides a unique novel observable that can cleanly calibrate against the value of $\epsmutau$, as we shall see in the next section.

%%%%%%%%%%%%%%%%%%%%%%%%%%%%%%%%%%%%%%%%%%%%%%%%%%%%%%%%%%%%%%%%%%%%%%
\subsection{A Novel Variable $\Delta d$ for Determining $\epsmutau$}
\label{subsec:delta-d}
%%%%%%%%%%%%%%%%%%%%%%%%%%%%%%%%%%%%%%%%%%%%%%%%%%%%%%%%%%%%%%%%%%%%%%

\begin{figure}
	\centering
	\includegraphics[width=0.45\linewidth]{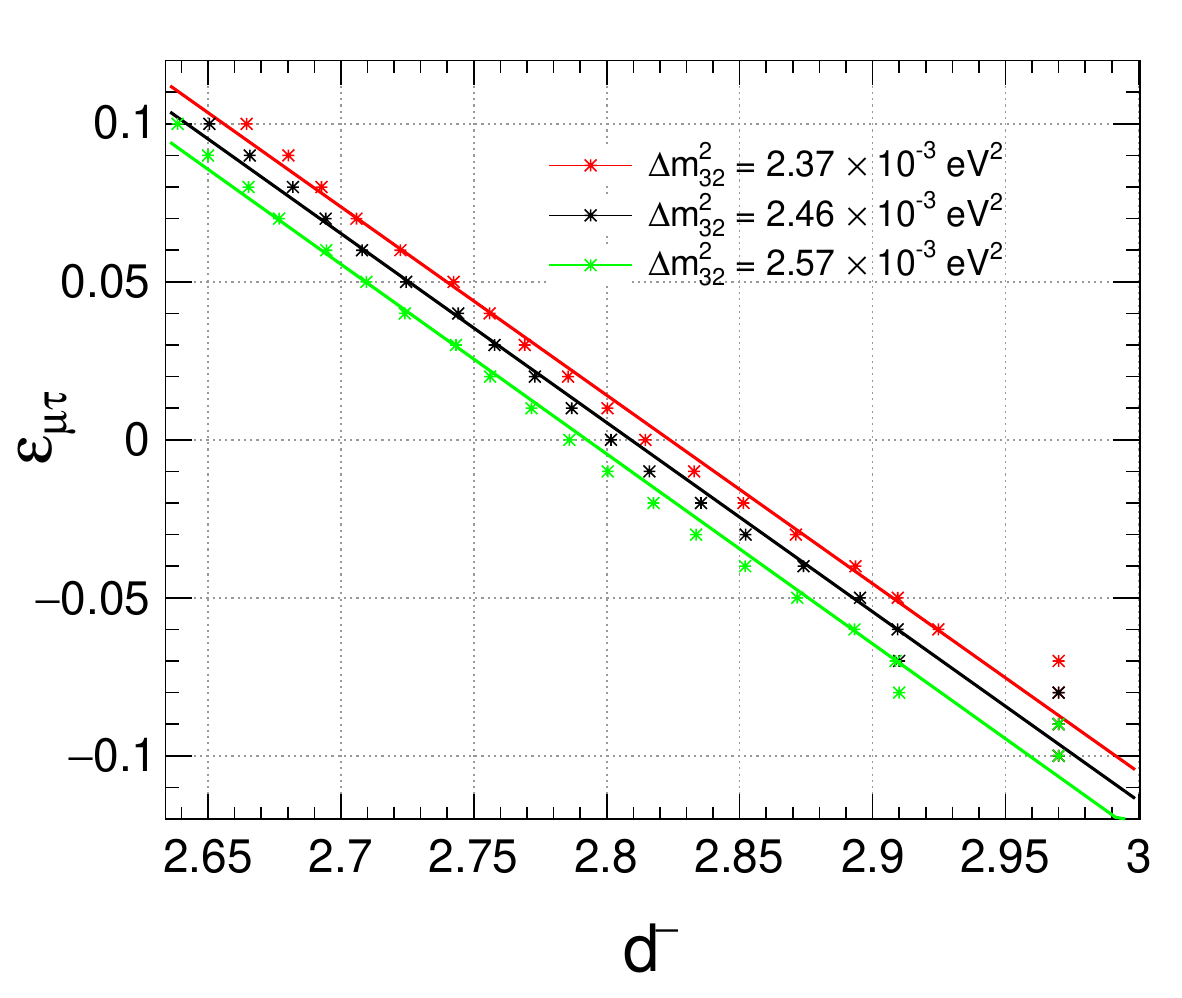}
	\includegraphics[width=0.45\linewidth]{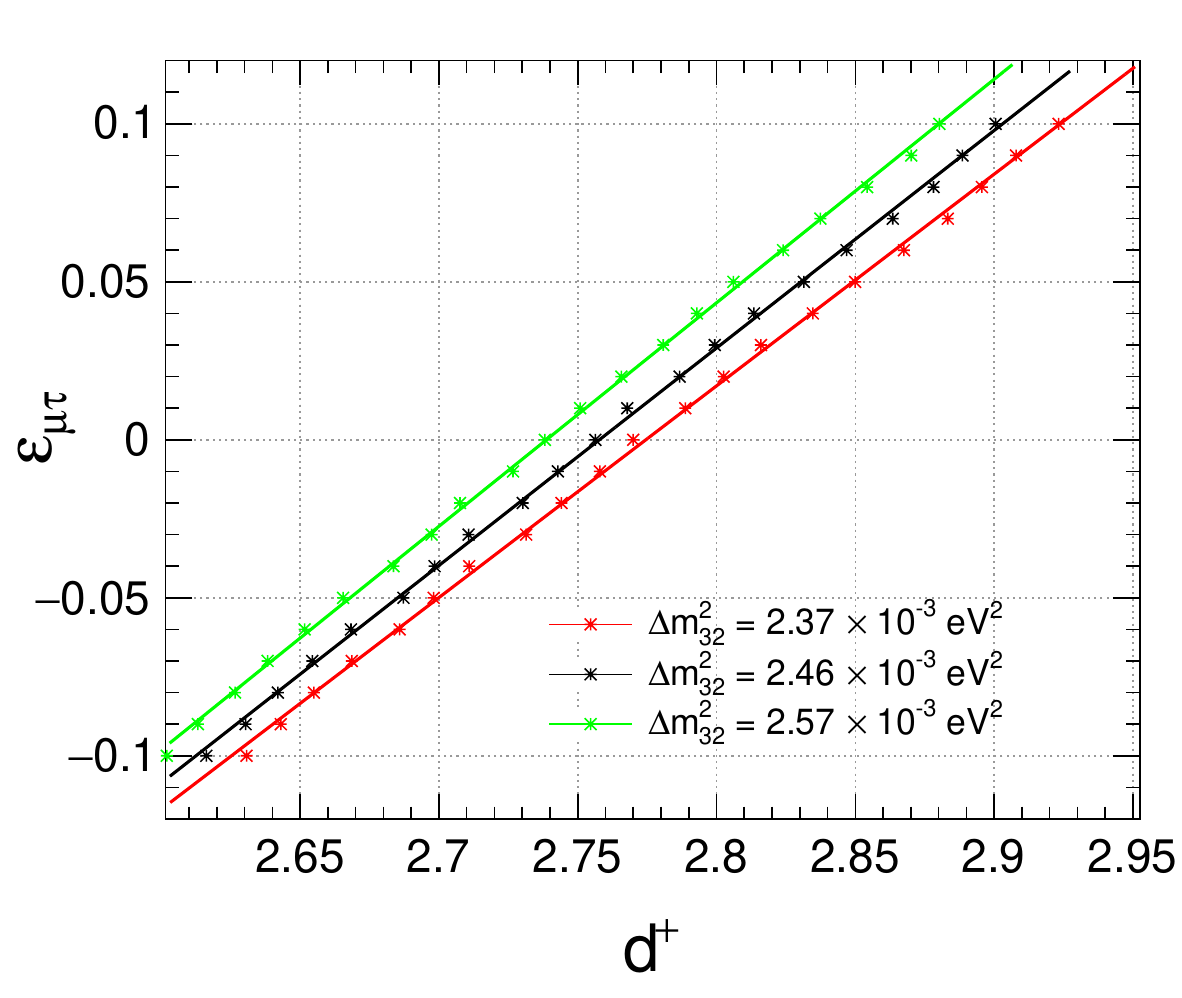}
	\includegraphics[width=0.45\linewidth]{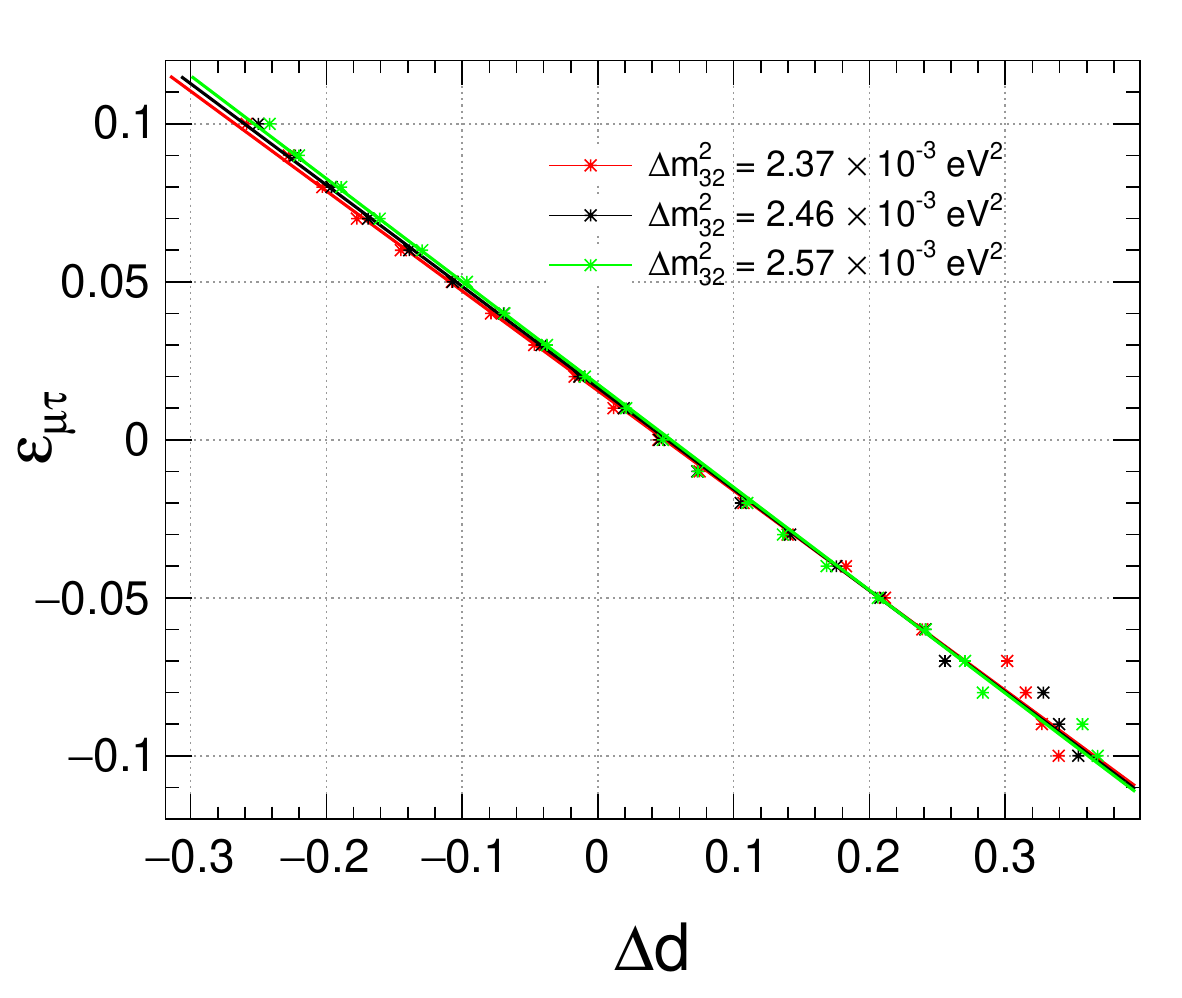}
	\caption{Upper panels: The calibrations of $\varepsilon_{\mu\tau}$ with the dip locations in U/D ratios of $\mu^-$ (left panel) and $\mu^+$ (right panel) events separately. Lower panel: The calibration of $\varepsilon_{\mu\tau}$ with difference in dip locations of $\mu^-$ and $\mu^+$. Note that for $-0.1<\varepsilon_{\mu\tau}<-0.07$, the dip location is simply the center of the $\log_{10}[L_\mu^\text{rec}/E_\mu^\text{rec}]$ bin with minimum U/D ratio since the fitting is not stable for these values. We consider normal mass ordering, and the benchmark values of oscillation parameters given in Table~\ref{tab:osc-param-value}.~\cite{Kumar:2021lrn}}
	\label{fig:effect_dmsq32_NSI}
\end{figure}

Since the oscillation dip locations in the U/D distributions of both, $\mu^-$ and $\mu^+$ events, shift to lower / higher values of $L_\mu^{\text{rec}}/E_\mu^{\text{rec}}$ depending on the value of $\epsmutau$, the dip location in either of these distributions may be used to estimate $\varepsilon_{\mu\tau}$, independently. We have already developed a dip-identification algorithm~\cite{Kumar:2020wgz} as described in Sec.~\ref{sec:dip_fitting}, where the dip location is not simply the lowest value of U/D ratio, but takes into account information from the surrounding bins that have a U/D ratio below a  threshold value. The algorithm essentially identifies the cluster of contiguous bins that have a U/D ratio lower than any surrounding bins, and fits these bins with a quadratic function whose minimum corresponds to the dip. The value of $\log_{10}[L_\mu^\text{rec}/E_\mu^\text{rec}]$ corresponding to this minimum is termed as the dip location. We denote the dip location for $\mu^-$ and $\mu^+$ events as $d^-$ and $d^+$, respectively.

We obtain the calibration of $\varepsilon_{\mu\tau}$ with the dip locations for $\mu^-$ and $\mu^+$ events separately, using 1000-year MC data. In the top panels of Fig.\,\ref{fig:effect_dmsq32_NSI}, we present the calibration curves of $\epsmutau$ with $d^-$ (left panel) and $d^+$ (right panel), for three different values of $\Delta m^2_{32}$. The straight line nature of the calibration curves can be understood qualitatively using Eq.~\ref{eq:lbye-dip-calibration}. For a majority of events ($E_\mu^\text{rec} <25$ GeV), the $\epsmutau V_\text{CC} E_\nu$ term is much smaller than $\Delta m^2_{32}$ term, and therefore the linear expansion as shown in Eq.~\ref{eq:lbye-dip-calibration} is valid.

In Eq.~\ref{eq:lbye-dip-calibration}, the dominating $\Delta m^2_{32}$ dependence of the dip location is through the first term. We can get rid of this dependence by introducing a new variable
\begin{equation}
\Delta d = d^- - d^+\,,
\label{eq:def-deltad}
\end{equation}
which is expected to be proportional to $\epsmutau$, and can be used to calibrate for $\epsmutau$. As may be seen in the bottom panel of Fig.\,\ref{fig:effect_dmsq32_NSI}, this indeed removes the dominant  $\Delta m^2_{32}$-dependence in the calibration of $\epsmutau$. Note that some $\Delta m^2_{32}$-dependence still survives, which may be seen in the slightly different slopes of the calibration curves for different $\Delta m^2_{32}$ values. However, this dependence is clearly negligible, as it appears in Eq.~\ref{eq:lbye-dip-calibration} as a multiplicative correction to $\epsmutau$. We have thus obtained a calibration for $\epsmutau$ that is almost independent of $\Delta m^2_{32}$.

Note that $\varepsilon_{\mu\tau} = 0$ does not imply $\Delta d =0$, since different matter effects and different inelasticities in neutrino and antineutrino channels cause the dips to arise at slightly different $\log_{10}[L_\mu^\text{rec}/E_\mu^\text{rec}]$ values in these two channels even in the absence of NSI.

%%%%%%%%%%%%%%%%%%%%%%%%%%%%%%%%%%%%%%%%%%%%%%%%%%%%%%%%%%%%%%%%%%%%%%%
\subsection{Constraints on $\varepsilon_{\mu\tau}$ from the Measurement of $\Delta d$}
\label{subsec:eps-limit-delta-d}
%%%%%%%%%%%%%%%%%%%%%%%%%%%%%%%%%%%%%%%%%%%%%%%%%%%%%%%%%%%%%%%%%%%%%%

\begin{figure}
	\centering
	\includegraphics[width=0.7\linewidth]{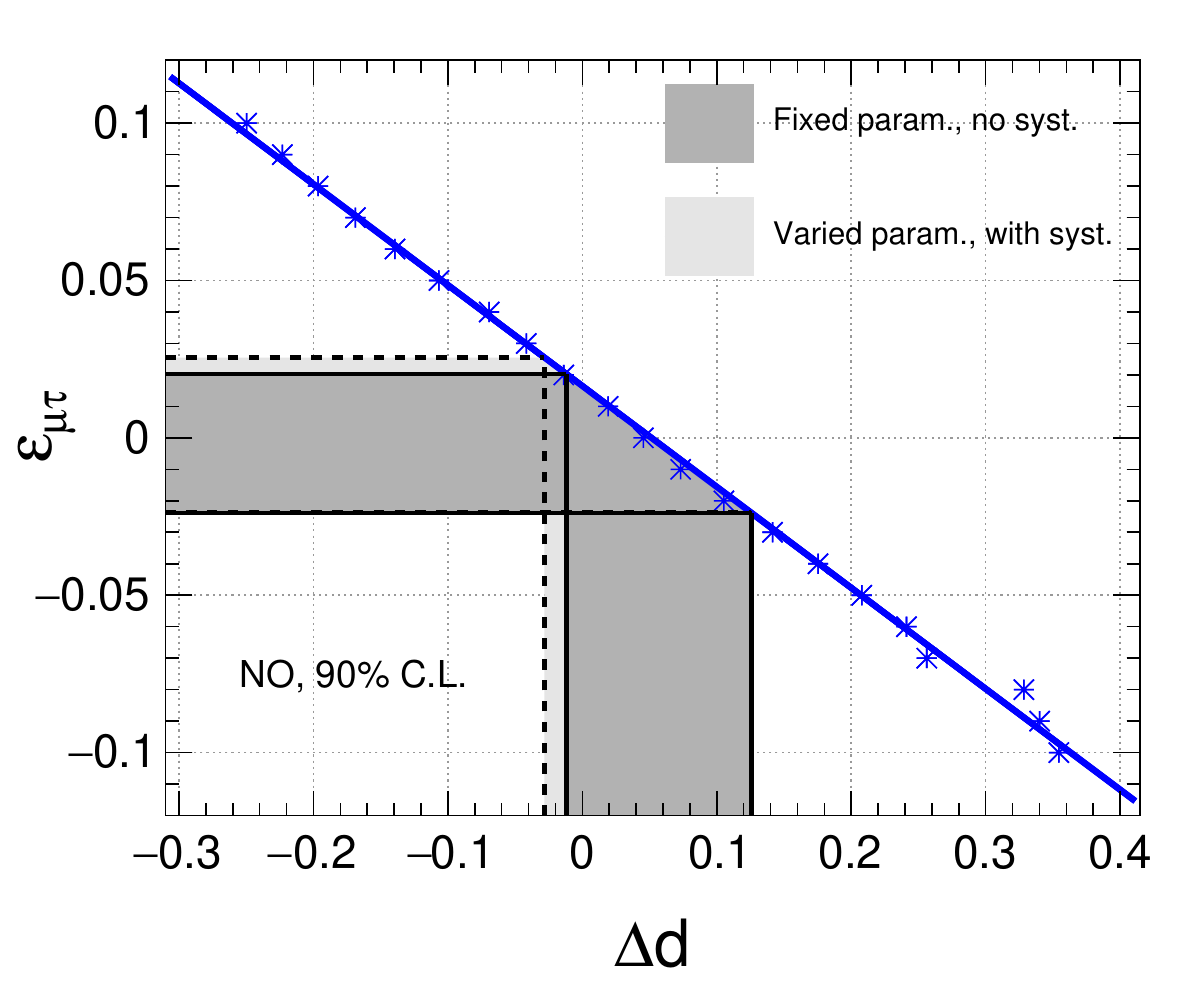}
	\caption{The blue stars and the blue solid line indicate the calibration curve of $\epsmutau$ against $\Delta d$, obtained using 1000-year MC data with normal mass ordering. The gray bands show the expected 90\% C.L. limits on $\Delta d$ (vertical bands), and hence on $\epsmutau$ (horizontal bands), with 500 kt$\cdot$yr of ICAL exposure, when the actual value of $\varepsilon_{\mu\tau}=0$. The light (dark) gray  bands show the limits when the errors on oscillation parameters, and the impact of systematic uncertainties are included (excluded), as discussed in Sec.~\ref{subsec:eps-limit-delta-d}. For fixed-parameter case, we use the benchmark values of oscillation parameters given in Table~\ref{tab:osc-param-value}.~\cite{Kumar:2021lrn}}
	\label{fig:osc_dip_results_NSI}
\end{figure}

We saw in the last section that one can infer the value of $\varepsilon_{\mu\tau}$ in the observed data from the calibration curve between $\Delta d$ and $\varepsilon_{\mu\tau}$, independent of the actual value of $\Delta m^2_{32}$. We obtain the calibration curve using the 1000-year MC-simulated sample, and  show it with the solid blue line in Fig.~\ref{fig:osc_dip_results_NSI}.

For estimating the expected constraints on $\varepsilon_{\mu\tau}$, we simulate 100 independent sets of data, each for 10-year exposure of ICAL with the benchmark values of oscillation parameters as given in Table~\ref{tab:osc-param-value}, and $\varepsilon_{\mu\tau} = 0$. In Fig.~\ref{fig:osc_dip_results_NSI}, the range indicated along the x-axis by dark gray band is the expected 90\% C.L. range of the measured value of $\Delta d$, while the range indicated along the y-axis by the dark gray band is the expected 90\% C.L. limit on $\varepsilon_{\mu\tau}$. From the figure, we find that one can expect to constrain $\epsmutau$ to $-0.024 < \varepsilon_{\mu\tau} < 0.020$ at 90\% C.L. with 10 years of data.

Since the calibration was found to be almost independent of the actual value of $\Delta m^2_{32}$, we expect that even if the actual value of $\Delta m^2_{32}$ were not exactly known, the results will not change. We have confirmed this by generating 100 independent sets of 10-year exposure, where the $\Delta m^2_{32}$ values used as inputs follow a Gaussian distribution $\Delta m^2_{32} = (2.46 \pm 0.03) \times 10^{-3}$ eV$^2$, in accordance to the range allowed by the global fit~\cite{NuFIT,Esteban:2020cvm,deSalas:2020pgw,Marrone:2021} of available neutrino data.

We also checked the impact of uncertainties in the values of the other oscillation parameters. We first simulated 100 statistically independent unoscillated data sets. Then for each of these data sets, we take 20 random choices of oscillation parameters, according to the Gaussian distributions
\begin{align}
\Delta m^2_{21} &= (7.4 \pm 0.2) \times 10^{-5} \mbox{ eV}^2 \,,\nonumber \\
\Delta m^2_{32} &= (2.46 \pm 0.03) \times 10^{-5} \mbox{ eV}^2 \,, \nonumber \\
\sin^2 2\theta_{12} &= 0.855 \pm 0.020 \,, \nonumber \\
\sin^2 2\theta_{13} &= 0.0875 \pm 0.0026 \,, \nonumber \\
\sin^2 \theta_{23} &= 0.50 \pm 0.03 \,,
\end{align}
guided by the present global fit \cite{NuFIT}. We keep $\delta_\text{CP}=0$, since its effect on $\nu_\mu$ survival probability is known to be highly suppressed in the multi-GeV energy range~\cite{Kopp:2007ne}. This procedure effectively enables us to consider the variation of our results over 2000 different combinations of oscillation parameters, to take into account the effect of their uncertainties.

We do not observe any significant dilution of results due to the variation over oscillation parameters. This is expected, since
(i) solar oscillation parameters $\Delta m^2_{21}$ and $\theta_{12}$ do not contribute significantly to the $\nu_\mu$ survival probability in the multi-GeV range,
(ii) the reactor mixing angle $\theta_{13}$ is already measured to a high precision,
(iii) the mixing angle $\theta_{23}$ does not affect the location of the oscillation minimum at the probability level, and
(iv) our procedure of using $\Delta d$ (see Eq.~\ref{eq:def-deltad}) to determine $\epsmutau$ minimizes the impact of $\Delta m^2_{32}$ uncertainty, as shown in Sec.~\ref{subsec:delta-d}.

In addition, we take into account the systematic uncertainties in the neutrino fluxes and cross sections following the procedure described in Sec.~\ref{sec:Calib-LbyE}. When the oscillation parameter uncertainties and all the five systematic uncertainties mentioned above are included, it is found that the method based on the shift in the dip locations may constrain $\epsmutau$ to $- 0.025 < \epsmutau < 0.024$ at 90\% C.L.. The results denoting the effects of these uncertainties have been shown with light gray bands in Fig.~\ref{fig:osc_dip_results_NSI}. 

%%%%%%%%%%%%%%%%%%%%%%%%%%%%%%%%%%%%%%%%%%%%%%%%%%%%%%%%%%%%%%%%%
%%%%%%%%%%%%%%%%%%%%%%%%%%%%%%%%%%%%%%%%%%%%%%%%%%%%%%%%%%%%%%%%
\section{Identifying NSI through Oscillation Valley}
\label{sec:reco-valley}
%%%%%%%%%%%%%%%%%%%%%%%%%%%%%%%%%%%%%%%%%%%%%%%%%%%%%%%%%%%%%%%%
%%%%%%%%%%%%%%%%%%%%%%%%%%%%%%%%%%%%%%%%%%%%%%%%%%%%%%%%%%%%%%%

In this section, we explore the features of the oscillation valley in the presence of NSI (non-zero $\varepsilon_{\mu\tau}$).  We have already observed in Sec.~\ref{subsec:osc-valley} that in the presence of non-zero $\varepsilon_{\mu\tau}$, the oscillation valley in the oscillogram has a curved shape that depends on the extent of $\epsmutau$, its sign, and whether one is observing the $\mu^-$ or $\mu^+$ events. Therefore, the curvature of oscillation valley is expected to be a useful parameter for the determination of $\varepsilon_{\mu\tau}$.  

%%%%%%%%%%%%%%%%%%%%%%%%%%%%%%%%%%%%%%%%%%%%%%%%%%%%%%%%%%%%%%%%%%
\subsection{Curvature of Oscillation Valley as a Signature of NSI}
\label{subsec:curv-valley}
%%%%%%%%%%%%%%%%%%%%%%%%%%%%%%%%%%%%%%%%%%%%%%%%%%%%%%%%%%%%%%%%%%%

\begin{figure}
	\centering
	\includegraphics[width=0.45\linewidth]{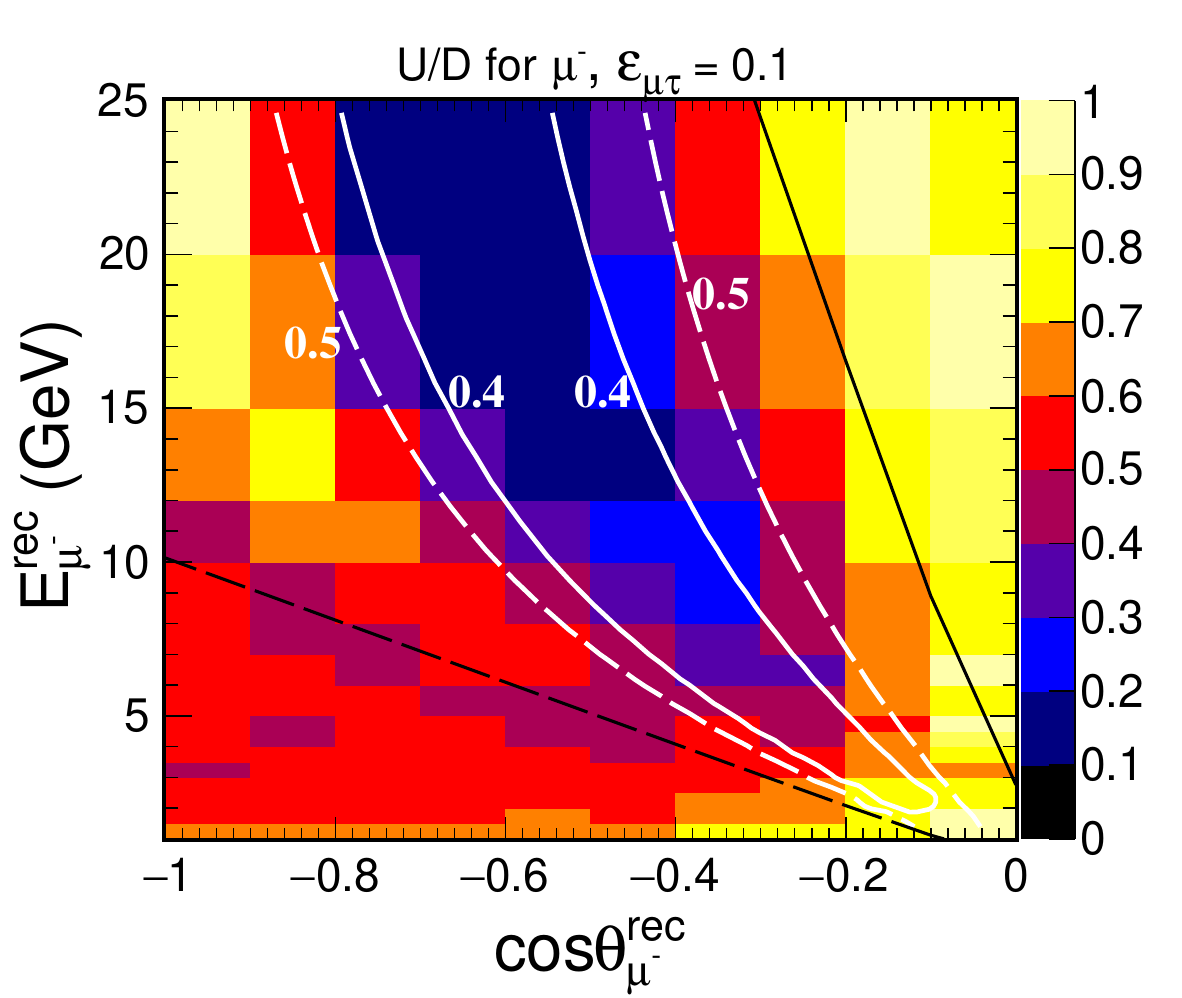}
	\includegraphics[width=0.45\linewidth]{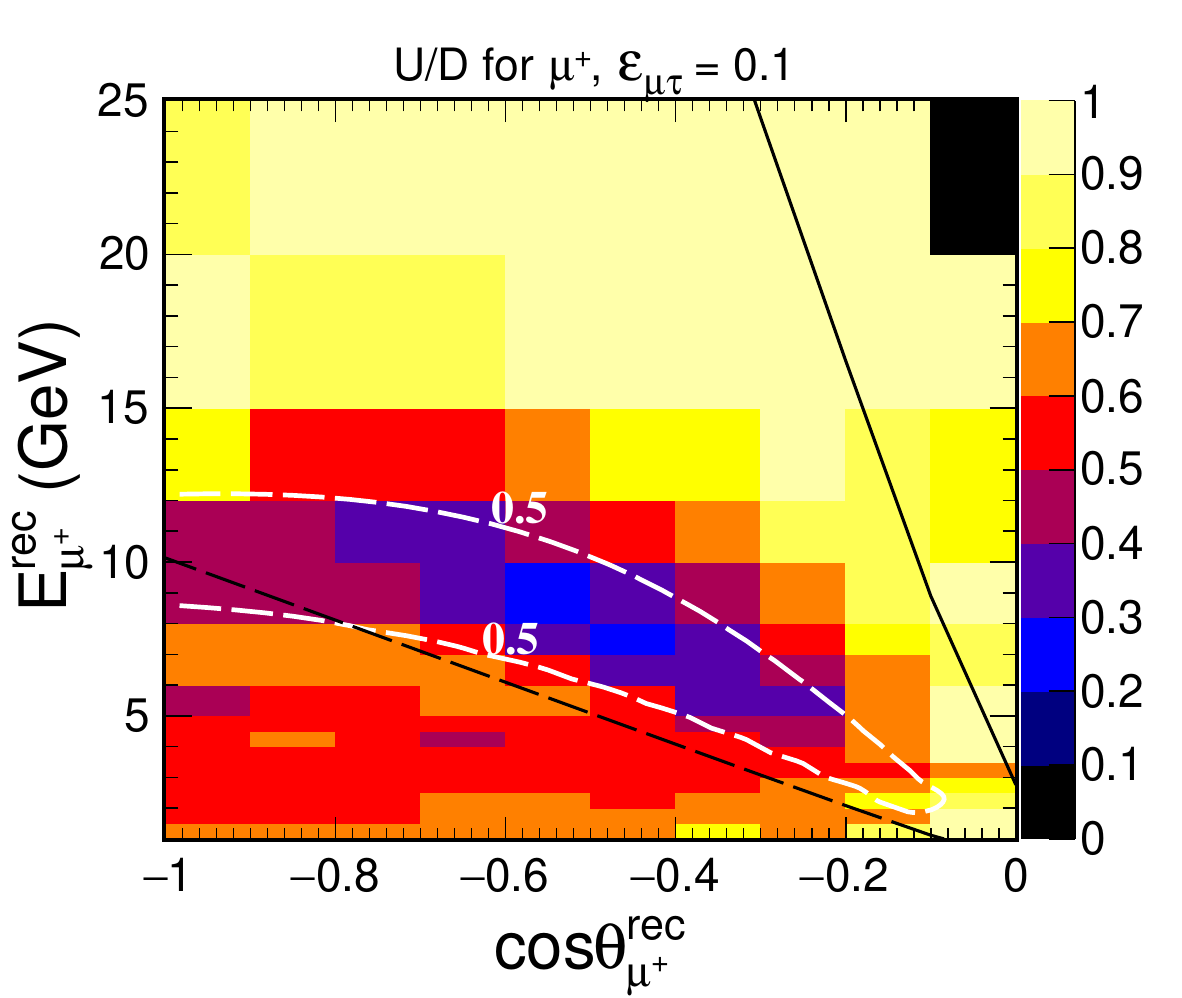}
	\includegraphics[width=0.45\linewidth]{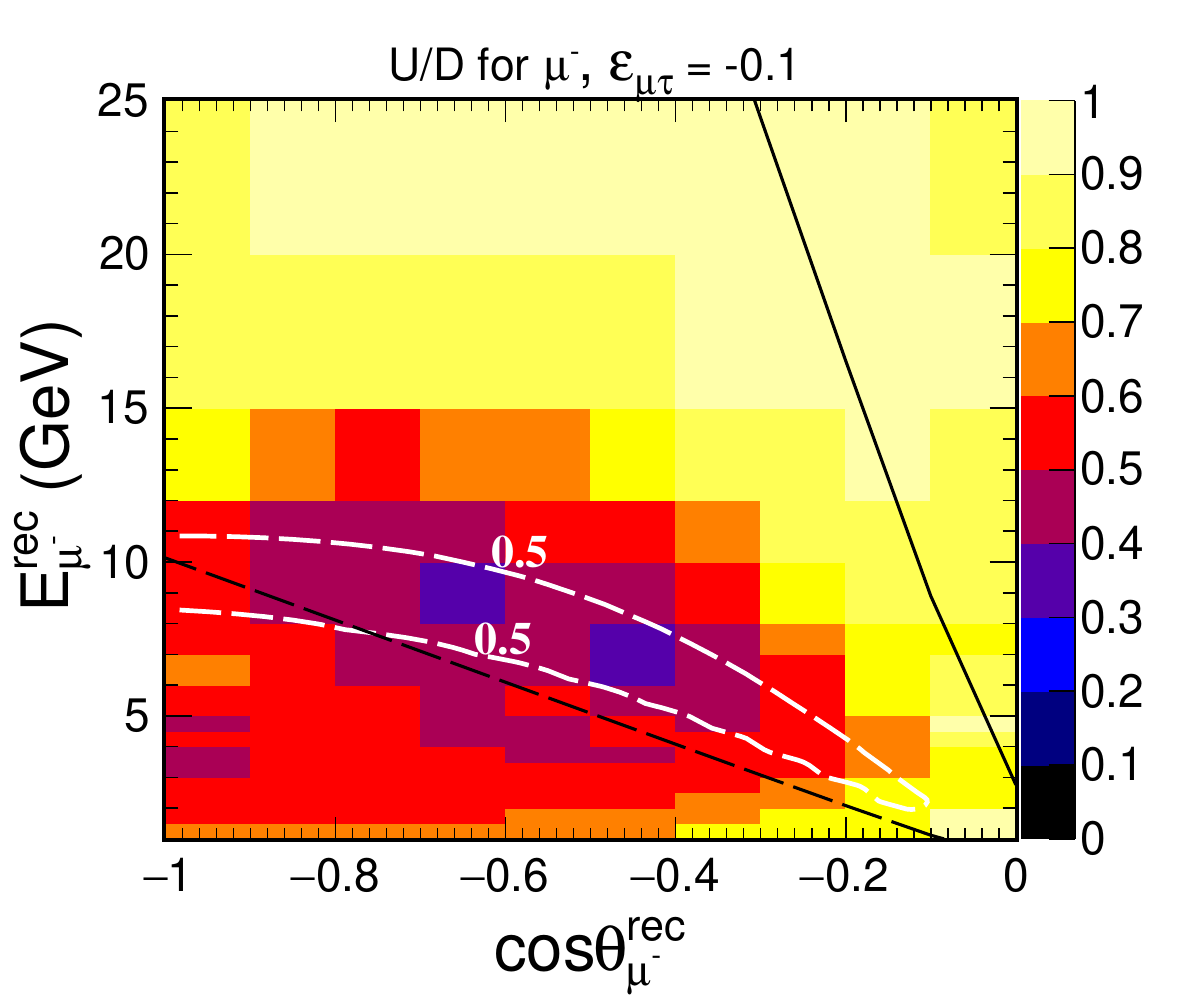}
	\includegraphics[width=0.45\linewidth]{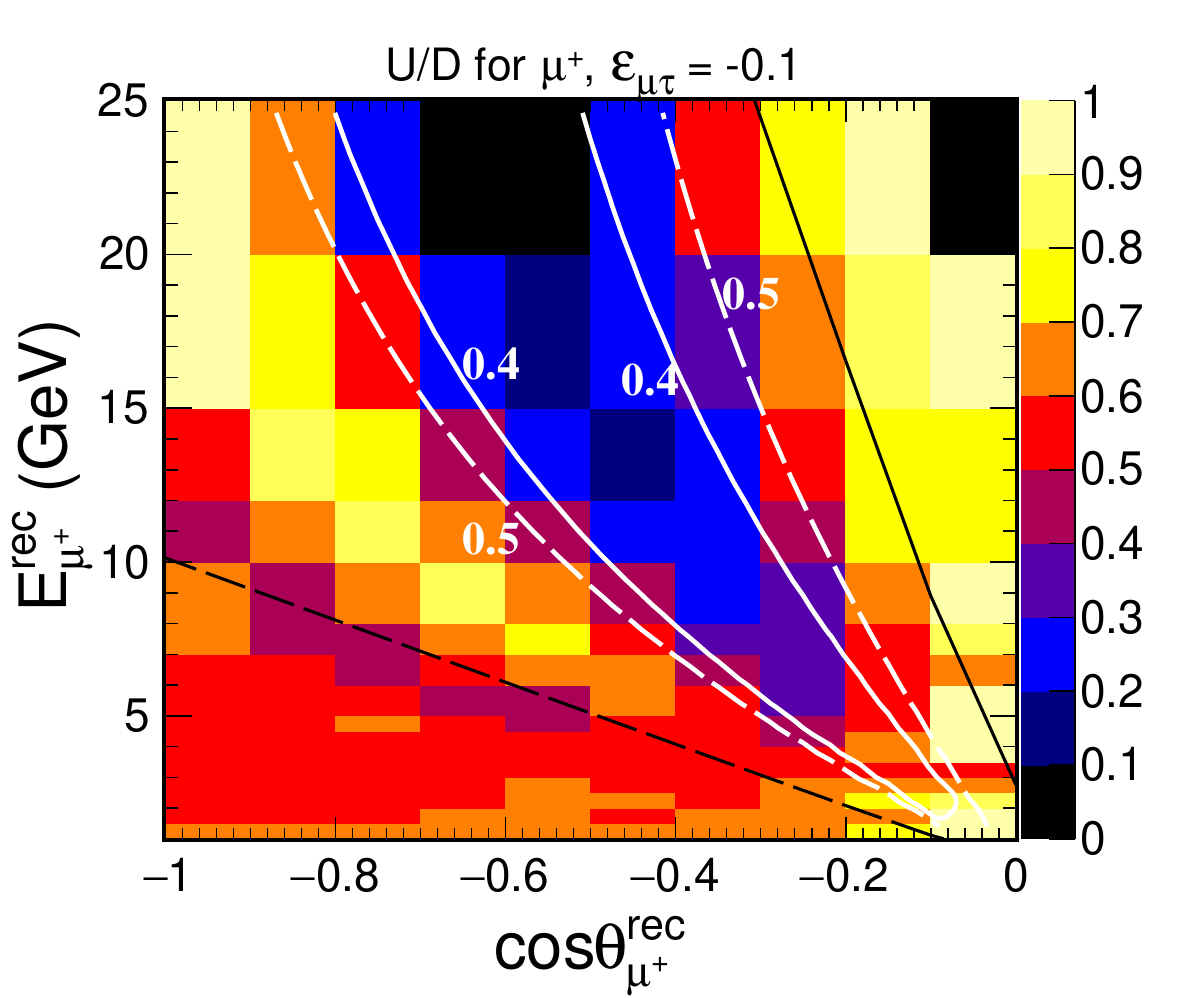}
	\caption{The distributions of U/D ratios in ($E_\mu^\text{rec}$, $\cos\theta_\mu^\text{rec}$) plane with non-zero $|\varepsilon_{\mu\tau}|$ (+0.1 in upper panels and -0.1 in lower panels) as expected at ICAL in 10 years. Left and right panels are for $\mu^-$ and $\mu^+$ events, respectively.  The white solid and dashed curves correspond to contours with fitted U/D ratio equal to 0.4 and 0.5, respectively, obtained after fitting of the oscillation valley with function $F_\alpha$ (see Eq.~\ref{eq:fitting-func-nsi}). The black solid and dashed lines correspond to $\log_{10}[L_\mu^\text{rec}/E_\mu^\text{rec}] =$ 2.2 and 3.1, respectively. We consider normal mass ordering, and the benchmark values of oscillation parameters given in Table~\ref{tab:osc-param-value}.~\cite{Kumar:2021lrn}}
	\label{fig:osc_valley_10yr_NSI}
\end{figure}

Figure~\ref{fig:osc_valley_10yr_NSI} presents the distributions of U/D ratios of $\mu^-$ and $\mu^+$ events separately in the $(E_\mu^\text{rec}\,,\cos\theta_\mu^\text{rec})$ plane, for two non-zero $\varepsilon_{\mu\tau}$ values (i.e. $\pm 0.1$). For the sake of demonstration (in order to reduce the statistical fluctuations in the figure and emphasize the physics), we show the binwise average of 100 independent data sets, each with 10-year exposure of ICAL. In contrast to the linear nature of oscillation valley with SI ($\varepsilon_{\mu\tau} =0$), the valley is curved in the presence of NSI. The direction of bending of the oscillation valley is decided by sign of $\varepsilon_{\mu\tau}$ as well as whether it is $\mu^-$ or $\mu^+$. An important point to note is that the nature of curvature of the oscillation valley, that we observed in the ($E_\nu, \cos\theta_\nu$) plane with neutrino variables (see Fig.~\ref{fig:osc_valley_neutrino_NSI}) with NSI, is preserved in the reconstructed oscillation valley in the plane of reconstructed muon observables ($E_\mu^\text{rec}, \cos\theta_\mu^\text{rec}$). This is due to the excellent energy and angular resolutions of the ICAL detector for muons in the multi-GeV energy range. 

We retrieve the information on the bending of the oscillation valley by fitting with an appropriate function, which is a generalization of Eq.\,\ref{eq:fitting-func-sm}. We introduce an additional free parameter $\alpha$ that characterizes the curvature of the oscillation valley. The function we propose for fitting the oscillation valley in the presence of $\varepsilon_{\mu\tau}$ is  
\begin{equation}
F_\alpha(E_\mu^\text{rec}, \cos\theta_\mu^\text{rec}) = Z_\alpha + N_\alpha \cos^2
\left(m_\alpha \frac{\cos\theta_\mu^\text{rec}}{E_\mu^\text{rec}}  + \alpha\,
\cos^2\theta_\mu^\text{rec}  \right),
\label{eq:fitting-func-nsi} 
\end{equation}
where $Z_\alpha$, $N_\alpha$, $m_\alpha$, and $\alpha$ are the free parameters to be determined from the fitting of the U/D ratio in the ($E_\mu^\text{rec},\,\cos\theta_\mu^\text{rec}$) plane. Equation\,\ref{eq:fitting-func-nsi} is inspired by the survival probability as given in Eq.\,\ref{eq:NSI-pmumu-nu-final-omsd}. In Eq.~\ref{eq:fitting-func-nsi}, the parameter $\alpha$ enters multiplied with  $\cos^2\theta_\mu^\text{rec}$, where one factor of $\cos\theta_\mu^\text{rec}$ includes the $L$ dependence of oscillation. The other factor of $\cos\theta_\mu^\text{rec}$ is to take into account the baseline dependence of the matter potential. While this is a rather crude approximation, it is sufficient to provide us a way of calibrating $\epsmutau$, as will be seen later in the section. In Eq.\,\ref{eq:fitting-func-nsi}, the parameters $Z_\alpha$ and $N_\alpha$ fix the depth of valley, while $m_\alpha$ and $\alpha$ determine the orientation and curvature of oscillation valley, respectively. Therefore, $m_\alpha$ and $\alpha$ together account for the combined effect of the mass-squared difference $\Delta m^2_{32}$ and the NSI parameter $\varepsilon_{\mu\tau}$ in oscillations. 

As in the analysis in Sec.~\ref{sec:2D-fit-procedure}, we restrict the range of $\log_{10}[L_\mu^\text{rec}/E_\mu^\text{rec}]$ to 2.2 -- 3.1, to remove the unoscillated part ($<2.2$) as well as the region with significant matter effects ($>3.1$). The cut on the U/D value is applied at 0.9. In Fig.\,\ref{fig:osc_valley_10yr_NSI}, the contours of the fitted U/D ratio of 0.4 and 0.5, shown with white solid and dashed curves, respectively, clearly indicate that the function $F_\alpha$ works well to reproduce the curved nature of the oscillation valley with non-zero $\varepsilon_{\mu\tau}$. We see that the oscillation valley for $\mu^-$ with positive (negative) $\varepsilon_{\mu\tau}$ closely resembles that for $\mu^+$ with negative (positive) $\varepsilon_{\mu\tau}$. This happens due to the degeneracy in the sign of $\varepsilon_{\mu\tau}$ and the sign of matter potential $V_\text{CC}$ -- these appear in the combination $\epsmutau V_\text{CC}$ in the survival probability expression in Eq.~\ref{eq:NSI-pmumu-nu-final-omsd}. One can see that for $\mu^+$ ($\mu^-$) with $\epsmutau =0.1$ ($-0.1$), the contour with the fitted U/D ratio of 0.4 is absent. This is because the valley is shallower in these cases. This feature corresponds to a similar one observed in Fig.~\ref{fig:osc_dip_10yr_NSI}, where the oscillation dip is shallower for $\mu^+$ ($\mu^-$) with $\epsmutau =0.1$ ($-0.1$). 

%%%%%%%%%%%%%%%%%%%%%%%%%%%%%%%%%%%%%%%%%%%%%%%%%%%%%%%%%%%%%%%%%%%%%%
\subsection{Constraints on $\varepsilon_{\mu\tau}$ from Oscillation Valley}
\label{subsec:eps-limit-osc-valley}
%%%%%%%%%%%%%%%%%%%%%%%%%%%%%%%%%%%%%%%%%%%%%%%%%%%%%%%%%%%%%%%%%%%%%

We saw in the previous section that the shape of the oscillation valley in the muon observables is well-approximated by the function in Eq.\,\ref{eq:fitting-func-nsi}, with the two parameters $\alpha$ and $m_\alpha$. Therefore, we propose to use these two parameters for retrieving $\varepsilon_{\mu\tau}$. At ICAL, since the data on $\mu^-$ and $\mu^+$ events are available separately, we can have two independent fits for  the U/D ratios in $\mu^-$ and $\mu^+$ events. This would give us independent determinations of the parameter pairs $(\alpha^-, m_\alpha^-)$ and $(\alpha^+, m_\alpha^+)$, for $\mu^-$ and $\mu^+$ events, respectively. One can then get the calibration of $\varepsilon_{\mu\tau}$ in the plane of $\alpha$ and $m_\alpha$, for $\mu^-$ and $\mu^+$ events, separately. For other experiments that do not have charge identification capability, there would be a single calibration curve of  $\varepsilon_{\mu\tau}$ with $\alpha$ and $m_\alpha$, based on the U/D ratio of muon events without charge information. 

\begin{figure}
	\centering
	\includegraphics[width=0.7\linewidth]{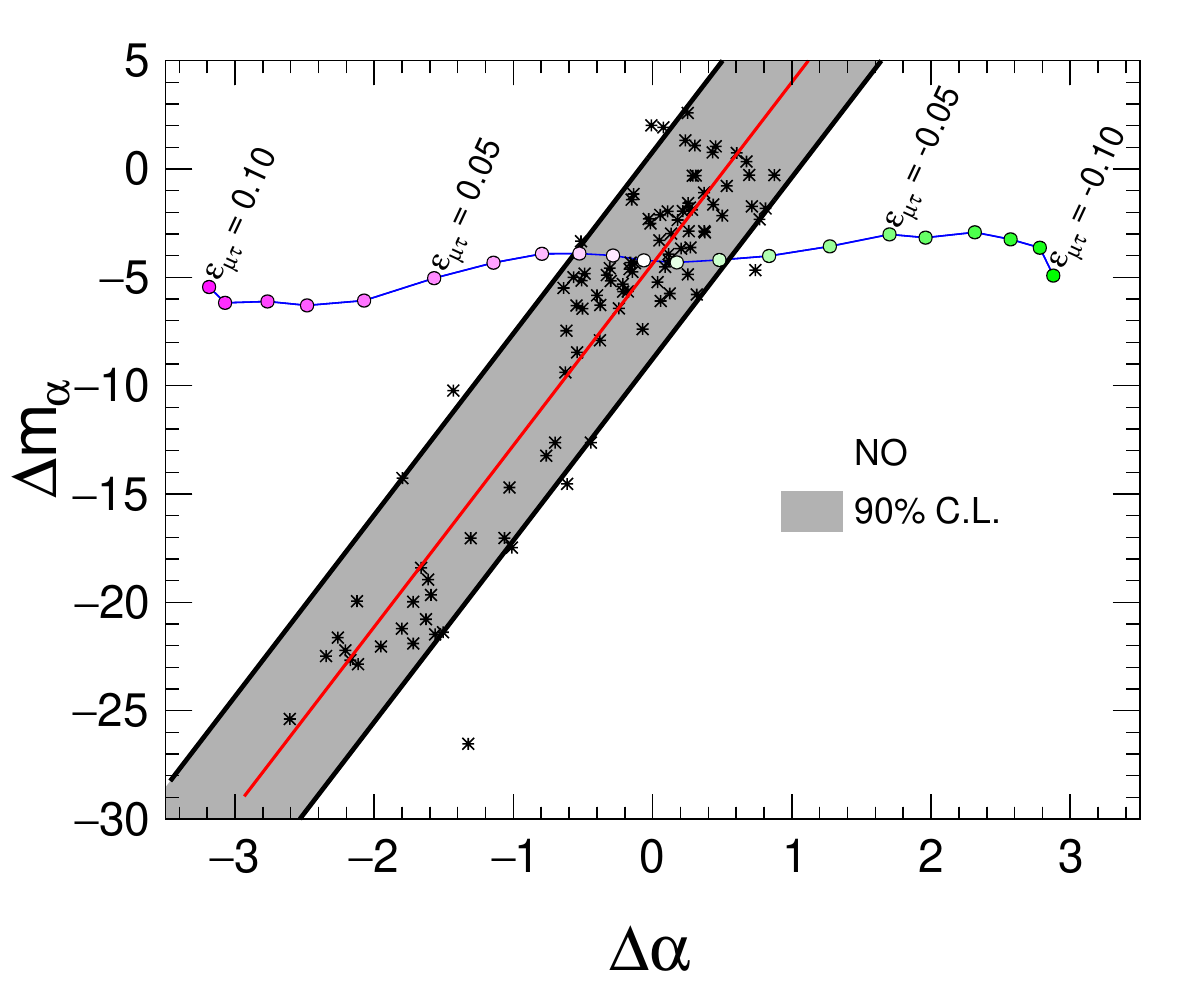}
	\caption{The blue curve with colored circles shows the calibration of $\epsmutau$ in the $(\Delta \alpha, \Delta m_\alpha)$ plane, obtained from 1000-year MC. The black stars are the values of $(\Delta \alpha,\Delta m_\alpha$) obtained from 100 independent data sets, each for 10-year data of ICAL, when $\epsmutau=0$. The gray band represents the expected 90\% C.L. region in this plane for a 10-year exposure at ICAL, when $\epsmutau=0$. The part of the blue calibration curve covered by the gray band gives the expected 90\% C.L. bounds on $\epsmutau$. We consider normal mass ordering, and the benchmark values of oscillation parameters given in Table~\ref{tab:osc-param-value}.~\cite{Kumar:2021lrn}}
	\label{fig:osc_valley_results_NSI}
\end{figure}

Building up on our experience in the analysis of the oscillation dip in the presence of NSI (see Sec.\,\ref{subsec:delta-d}), in this analysis of oscillation valley, we use the following observables that quantify the $\mu^-$ vs. $\mu^+$ contrast in the values of the parameters $\alpha$ and $m_\alpha$: 
\begin{equation}
\Delta \alpha = \alpha^- - \alpha^+\,, \hspace{1cm} 
\Delta m_\alpha = m_{\alpha^-} - m_{\alpha^+} \,.
\end{equation}
The blue curve with colored circles in Fig.\,\ref{fig:osc_valley_results_NSI} shows the calibration of $\varepsilon_{\mu\tau}$ in ($\Delta \alpha, \,\Delta m_\alpha$) plane, obtained by using the 1000-year MC sample. It may be observed that the calibration of $\varepsilon_{\mu\tau}$ is almost independent of $\Delta m_\alpha$, and $\varepsilon_{\mu\tau}$ is almost linearly  proportional to $\Delta \alpha$.

In order to  estimate the extent to which $\epsmutau$ may be constrained at ICAL with an exposure of 10 years, we simulate the statistical fluctuations by generating 100 independent data sets, each corresponding to 10-year exposure and $\epsmutau=0$. The black dots in Fig.\,\ref{fig:osc_valley_results_NSI} are the values of $\Delta \alpha$ and $\Delta m_\alpha$ determined from each of these data sets. The figure shows the results, with the benchmark value of oscillation parameters given in Table~\ref{tab:osc-param-value}.

It is observed that with a 10-year data, the statistical fluctuations lead to a strong correlation between the values of $\Delta \alpha$ and $\Delta m_\alpha$, and we have to employ a two-dimensional fitting procedure. We fit the scattered points in $(\Delta \alpha$, $\Delta m_\alpha$) plane with a straight line. The best fitted straight line is shown by the red color in Fig.\,\ref{fig:osc_valley_results_NSI}, which corresponds to the average of these points, and indeed passes through the calibration point obtained from 1000-year MC events for $\epsmutau=0$. The 90$\%$ C.L. limit on $\varepsilon_{\mu\tau}$ is then obtained based on the measure of the perpendicular distance of every point from the red line. The region with the gray band in Fig.\,\ref{fig:osc_valley_results_NSI} contains 90\% of the points that are closest to the red line. The expected 90\% C.L. bounds on $\epsmutau$ can then be determined from the part of the blue calibration curve covered by the gray band. Any value of the pair $(\Delta \alpha, \Delta m_\alpha)$ that lies outside the gray band would be a smoking gun signature for nonzero  $\epsmutau$ at 90$\%$ C.L.. Figure \,\ref{fig:osc_valley_results_NSI} indicates that if $\epsmutau$ is indeed zero, our method, as applied at the ICAL detector, would be able to constrain it to the range $-0.022<\varepsilon_{\mu\tau}<0.021$ at 90\% C.L., with an exposure of 500 kt$\cdot$yr.

We explore the possible dilution of our results due to the uncertainties in oscillation parameters and the five systematic errors, following the same procedure described in Sec.~\ref{subsec:eps-limit-delta-d}. This yields the expected 90\% C.L. bounds as $-0.024<\varepsilon_{\mu\tau}<0.020$, thus keeping them almost unchanged.

%%%%%%%%%%%%%%%%%%%%%%%%%%%%%%%%%%%%%%%%%%%%%%%%%%%%%%%%%%%%%%%%%%%%
%%%%%%%%%%%%%%%%%%%%%%%%%%%%%%%%%%%%%%%%%%%%%%%%%%%%%%%%%%%%%%%%%%%%
\section{Summary and Concluding Remarks}
\label{sec:NSI_conclusion}
%%%%%%%%%%%%%%%%%%%%%%%%%%%%%%%%%%%%%%%%%%%%%%%%%%%%%%%%%%%%%%%%%%%%%
%%%%%%%%%%%%%%%%%%%%%%%%%%%%%%%%%%%%%%%%%%%%%%%%%%%%%%%%%%%%%%%%%%%%%

Atmospheric neutrino experiments are suitable for exploring the non-standard interactions of neutrinos due to the accessibility to high energies ($E_\nu$) and long baselines ($L_\nu$) through the Earth, for which the effect of NSI-induced matter potential is large. The NSI matter potential produced in the interactions of neutrino and matter fermions may change the oscillation pattern. The multi-GeV neutrinos, in particular, offer the advantage to access the interplay of NSI and Earth matter effects. In this chapter, we have explored the impact of the NSI parameter $\varepsilon_{\mu\tau}$ on the dip and valley patterns in neutrino oscillation probabilities, and proposed a new method for extracting information on $\epsmutau$ from the atmospheric neutrino data.

Detectors like ICAL, that have good reconstruction capability for the energy, direction, and charge of muons, can reproduce the neutrino oscillation dip and valley patterns in their reconstructed muon data, from $\mu^-$ and $\mu^+$ events separately. The observables chosen to reproduce these patterns are the ratios of upward-going (U) and downward-going (D) events, which minimize the effects of systematic uncertainties. We analyze the U/D ratio in bins of reconstructed muon observables $L_\mu^\text{rec}/E_\mu^\text{rec}$, and in the two-dimensional plane $(E_\mu^\text{rec}, \cos\theta_\mu^\text{rec})$. We show that nonzero $\epsmutau$ shifts the locations of oscillation dips in the reconstructed $\mu^-$ and $\mu^+$ data in opposite directions in the $L_\mu^\text{rec}/E_\mu^\text{rec}$ distributions. At the same time, nonzero $\epsmutau$ manifests itself in the curvatures of the oscillation valleys in the $(E_\mu^\text{rec}, \cos\theta_\mu^\text{rec})$ plane, this bending being in opposite directions for $\mu^-$ and $\mu^+$ events. The direction of shift in the dip location and the direction of the bending of the valley also depend on the sign of $\epsmutau$, as well as on the mass ordering.  ICAL, therefore, will be sensitive to sgn($\epsmutau$), once we know the neutrino mass ordering. We present our results in the normal mass ordering scenario, the analysis for the inverted mass ordering scenario will be exactly analogous. 

In the oscillation dip analysis, we introduce a new observable $\Delta d$, corresponding to the difference in the locations of dips in $\mu^-$ and $\mu^+$ event distributions. We show the calibration of $\epsmutau$ against $\Delta d$ using 1000-year MC sample at ICAL, and demonstrate that it is almost independent of the actual value of $\Delta m^2_{32}$.  We incorporate the effects of statistical fluctuations corresponding to 10-year simulated data, uncertainties in the neutrino oscillation parameters, and major systematic errors, by simulating multiple data sets. Including all these features, it is possible to constrain the NSI parameter in the range $-0.025 < \epsmutau < 0.024$ at 90\% C.L. with 500 kt$\cdot$yr exposure.

In the oscillation valley analysis, we propose a function $F_\alpha$ that quantifies the curvature ($\alpha$) and orientation ($m_\alpha$) of the oscillation valley in the ($E_\mu^\text{rec}$, $\cos_\mu^\text{rec}$) plane. This analysis may be performed for the $\mu^-$ and $\mu^+$ events separately, leading to independent measurements of $\epsmutau$. However, we further find that the differences in these parameters for the $\mu^-$ and $\mu^+$ events are quantities that are sensitive to $\epsmutau$, and show the calibration for $\epsmutau$ in the $(\Delta \alpha, \Delta m_\alpha)$ plane using 1000-year MC sample at ICAL. Incorporating the effects of statistical fluctuations, uncertainties in the neutrino oscillation parameters, and major systematic errors, the NSI parameter can be constrained in the range $-0.022 < \epsmutau < 0.021$ at 90\% C.L. with 500 kt$\cdot$yr exposure.

Our analysis procedure is quite straightforward and transparent, and is able to capture the physics of the NSI effects on muon observables quite efficiently. It builds upon the major features of the impact of $\epsmutau$ on the oscillation dip, and identifies the contrast in the shift in dip locations of $\mu^-$ and $\mu^+$ as the key observable that would be sensitive to NSI. It exploits the major pattern in the complexity of the two-dimensional event distributions in the ($E_\mu^\text{rec}$, $\cos_\mu^\text{rec}$) plane by fitting to a function approximately describing the curvature of the valley. The two complementary methods of using the oscillation dips and valleys provide a check of the robustness of the results. The shift of the dip location and the bending of the oscillation valley, and survival of these effects in the reconstructed muon data, are deep physical insights obtained through the analysis in this chapter.

Note that, though the dip and the valley both arise from the first oscillation minimum, the information content in the valley analysis is more than that in the dip analysis. For example, (a) the features of the valley (like curvature) in Fig.~\ref{fig:osc_valley_10yr_NSI} cannot be predicted from the shift in the dip as seen in Fig.~\ref{fig:osc_dip_10yr_NSI},  and (b) the shift in the dip may have many possible sources, but the specific nature of bending of the valley would act as the confirming evidence for NSI. Therefore, physics unravelled from the valley analysis is much richer than that from the dip.

Note that muon charge identification plays a major role in our analysis. In a detector like ICAL, the measurement of $\epsmutau$ is possible independently in both the $\mu^-$ and $\mu^+$ channels. The oscillation dips and valleys in $\mu^-$ and $\mu^+$ event samples have different locations and shapes. As a result, the information available in the dip and valley would be severely diluted in the absence of muon charge identification. But more importantly, the quantities sensitive to $\epsmutau$ in a robust way turn out to be the ones that reflect the differences in the properties of oscillation dips and valleys in $\mu^-$ and $\mu^+$ events. For example, $\Delta d$ is the observable identified by us, whose calibration against $\epsmutau$ is almost independent of the actual value of $\Delta m^2_{32}$. For $\Delta d$ to be measured at a detector, the muon charge identification capability is necessary. The data from ICAL will thus provide the measurement of a crucial observable that is not possible at other large atmospheric neutrino experiments like Super-K or DeepCore/ IceCube. At future long-baseline experiment like DUNE, that will have access to 1 -- 10 GeV neutrino energy and separate data sets for neutrino and antineutrino, our methodology to probe NSI using oscillation-dip location can also be adopted, by replacing the U/D ratio with the ratio of observed number of events and the predicted number of events without oscillations.

We expect that further exploration of the features of the oscillation dips and valleys, observable at ICAL-like experiments, would enable us to probe neutrino properties in novel ways.

\end{refsegment}
\cleartooddpage
\chapter{Neutrino Oscillation Tomography of Earth}
\label{chap:tomography}
\begin{refsegment}
	
%==========================
\section{Introduction and motivation}
\label{sec:introduction}
%==========================

Neutrinos are elusive particles, but they are capable of reaching places inaccessible by any other means. The tiny interaction cross section enables neutrinos to even pass through solid objects like Earth because they only interact via weak interactions. The atmospheric neutrinos travel long distances through the Earth and undergo coherent elastic forward scattering with the ambient electrons inside the Earth which leads to the modification of neutrino oscillations. When neutrinos pass deep through the mantle, the Mikheyev-Smirnov-Wolfenstein (MSW) resonance~\cite{Wolfenstein:1977ue,Mikheev:1986gs,Mikheev:1986wj} starts playing an important role in neutrino oscillations around 6 to 10 GeV of energies. On the other hand, the core-passing neutrinos with energies in the range of 3 to 6 GeV experience a different kind of resonant effect which is known as neutrino oscillation length resonance (NOLR)~\cite{Petcov:1998su,Chizhov:1998ug,Petcov:1998sg,Chizhov:1999az,Chizhov:1999he} or parametric resonance~\cite{Akhmedov:1998ui,Akhmedov:1998xq}. These density-dependent matter effects can be used to reveal the distribution of matter inside the Earth. 

The neutrinos have the potential to throw some light on the internal structure of Earth via neutrino absorption, oscillations, and diffraction. The large interaction cross section of neutrinos at energies above a few TeV results in the attenuation of neutrinos during propagation through Earth~\cite{Gandhi:1995tf}. The idea of exploring Earth's interior using neutrino absorption was discussed first time by Placci and Zavattini in a preprint in 1973~\cite{Placci:1973}, and Volkova and Zatsepin in a talk in 1974~\cite{Volkova:1974xa}. There are numerous studies considering neutrinos from different sources, such as man-made neutrinos~\cite{Placci:1973,Volkova:1974xa,Nedyalkov:1981,Nedyalkov:1981pp,Nedyalkov:1981yy,Nedialkov:1983,Krastev:1983,DeRujula:1983ya,Wilson:1983an,Askarian:1984xrv,Volkova:1985zc, Tsarev:1985yub,Borisov:1986sm,Tsarev:1986xg,Borisov:1989kh,Winter:2006vg}, extraterrestrial neutrinos~\cite{Wilson:1983an,Kuo:1995,Crawford:1995,Jain:1999kp,Reynoso:2004dt} and atmospheric neutrinos~\cite{Gonzalez-Garcia:2007wfs,Borriello:2009ad,Takeuchi:2010,Romero:2011zzb}. This is often termed as ``Earth tomography''~\cite{DeRujula:1983ya}. The neutrino-based absorption tomography of Earth has been performed using the multi-TeV atmospheric neutrino data at the IceCube detector~\cite{Donini:2018tsg} where the mass of Earth is measured using neutrinos. 

In the past few decades, a significant progress has been made in the precision measurement of neutrino oscillation parameters. The discovery of non-zero reactor mixing angle $\theta_{13}$ have opened a unique avenue for tomography of the Earth using matter effects in neutrino oscillations at the multi-GeV range of energies. The neutrino oscillation tomography of Earth based on matter effects has been considered by the study of man-made beams~\cite{Ermilova:1986ph,Nicolaidis:1987fe,Ermilova:1988pw,Nicolaidis:1990jm,Ohlsson:2001ck,Ohlsson:2001fy,Winter:2005we,Minakata:2006am,Gandhi:2006gu,Tang:2011wn,Arguelles:2012nw}, atmospheric~\cite{Agarwalla:2012uj,Rott:2015kwa,Winter:2015zwx,Bourret:2017tkw,Bourret:2019wme,Bourret:2020zwg,DOlivo:2020ssf,Kumar:2021faw,Maderer:2021aeb,Kelly:2021jfs,Denton:2021rgt,Capozzi:2021hkl}, solar~\cite{Ioannisian:2002yj,Ioannisian:2004jk,Akhmedov:2005yt,Ioannisian:2015qwa,Ioannisian:2017chl,Ioannisian:2017dkx,Bakhti:2020tcj}, and supernova~\cite{Lindner:2002wm,Akhmedov:2005yt} neutrinos. The third possibility of Earth tomography using the study of diffraction pattern produced by coherent neutrino scattering in crystalline matter inside Earth is technologically not feasible~\cite{Fortes:2006}.

The current understanding of the structure of Earth is provided by the seismic studies~\cite{Robertson:1966,Brush:1980,Dziewonski:1981xy,Loper:1995,Alfe:2007} where the propagation of seismic waves inside the Earth reveals the properties of matter. The reflection and refraction of earthquake waves is used to understand the proprieties of materials inside Earth. According to the present understanding, the Earth consists of concentric shells of different densities and compositions. The outermost surface of Earth is made up of solid crust, below which we have a viscous mantle made up of silicate oxide. The mantle is followed by a high-density core of iron-alloy. The information carried by the seismic waves may get altered on its way. On the contrary, the information on the interaction of neutrinos with ambient electrons (so-called Earth matter effect) remains unaltered when neutrinos travel long distances inside Earth. But owing to its weak interaction nature, usually event rates are not very large in neutrino experiments, and we need to compensate for it by using massive detectors and large exposures. A neutrino detector with good resolution in the multi-GeV range of energy and direction of neutrino will be able to observe modified event distribution due to neutrino oscillations in the presence of matter.  

Since ICAL would be able to detect neutrinos and antineutrinos separately in the multi-GeV range of energies covering baselines over a wide range of 10 to $10^4$ km, ICAL can explore Earth matter effects separately in neutrino (by observing $\mu^-$ events) and antineutrino (by observing $\mu^+$ events) modes through their mass-induced flavor oscillations inside the Earth which plays an important role to probe the inner structure of Earth. The MSW resonance around 6 to 10 GeV of neutrino energies can provide crucial information about the mantle. On the other hand, the NOLR/parametric resonance experienced by the vertically upward-going neutrinos with large baselines having energies of around 3 to 6 GeV can help us in probing the high-density core. In this work, we will study the impact of the presence of various layers inside Earth on neutrino oscillations and perform statistical analysis to establish the presence of a high-density core inside Earth by ruling out the mantle-crust profile with respect to (w.r.t.) the core-mantle-crust profile.

In Sec.~\ref{sec:earth_model}, we discuss the internal structure of Earth known from seismic studies and describe the profiles of Earth to be probed by neutrino oscillations in this work. The oscillation probabilities in the presence of matter governed by various profiles of Earth are described in Sec.~\ref{sec:tomography_probability}. The good directional resolution at ICAL is used to identify neutrinos passing through core, mantle, and crust in Sec.~\ref{sec:events_layer} which also describes the resultant distribution of reconstructed muon events for these neutrinos passing through a particular set of layers. The method for statistical analysis is explained in Sec.~\ref{sec:statistical analysis} which is followed by the results in terms of the statistical significance for establishing core and ruling out alternative profiles of Earth in Sec.~\ref{sec:results}. Finally, we conclude in Sec.~\ref{sec:tomography_conclusion}.

%==========================
\section{A Brief Review of the Internal Structure of Earth}
\label{sec:earth_model}
%==========================

The seismic studies have revealed that Earth consists of concentric shells, which are crust, mantle, and core, each of them is further divided into subshells with different properties~\cite{Robertson:1966,Loper:1995,Alfe:2007}. The average radius of Earth ($R_E$) is about 6371 km. The radius of the core is almost half the radius of Earth, whereas the density of the core is twice that of the mantle. Along with seismic studies, the information about the Earth is also contributed by the gravitational measurements, which have revealed that the gravitationally measured mass of Earth ($M_E$) is about $(5.9722 \pm 0.0006) \times 10^{24}$ kg~\cite{Luzum:2011,astro_almanac}, and the measured moment of inertia is found to be $(8.01736 \pm 0.00097) \times 10^{37}$ kg m\textsuperscript{2}~\cite{Chen:2014}. Since the experimentally measured moment of inertia is lower than the expected one ($2/5M_ER_E^2 \sim 9.7 \times 10^{37}$ kg m\textsuperscript{2}) for a uniform distribution of matter inside the Earth, the density is expected to be higher deep inside the Earth with more concentration of material towards the center. The average density of Earth is about 5.51 g/cm\textsuperscript{3} which is more than the density of rocks found on the surface. This observation also indicates that the interior of Earth is expected to have a higher density.

Now, if we talk about the layers of Earth, the outermost layer of crust is made up of solid rocks and has the lowest density among all layers~\cite{Robertson:1966,Alfe:2007}. Under the crust, we have the mantle, which consists of extremely hot rocks that are solid in the upper mantle but highly viscous plastic in the lower mantle. The mantle is followed by the high-density core, which is mainly composed of iron and nickel. The core can further be divided into outer core and inner core. 

The seismic waves propagating through the bulk of Earth can be classified into two categories which are the pressure (P) waves and shear (S) waves~\cite{Robertson:1966}. The P-waves cause the particles of the medium to oscillate in the direction of motion, whereas the S-waves result in the vibration of particles in the perpendicular direction. Since the outer core does not allow the transmission of S-waves along with a significant drop in the velocity of P-waves, the outer core is likely to be in a liquid state with its viscosity as low as that of water~\cite{Robertson:1966}. The inner core was discovered by Miss I. Lehmann in 1936 by observing its higher P-waves velocity~\cite{Lehmann:1936}. Further, the physical state of the inner core is expected to be solid, which may be possible even at such high temperatures because of a significant increase in the melting point of iron and other heavy elements at the tremendous pressure present inside the inner core~\cite{Birch:1964}.

%========================	
\begin{figure}[htb!]
	\centering
	\includegraphics[width=0.4\textwidth]{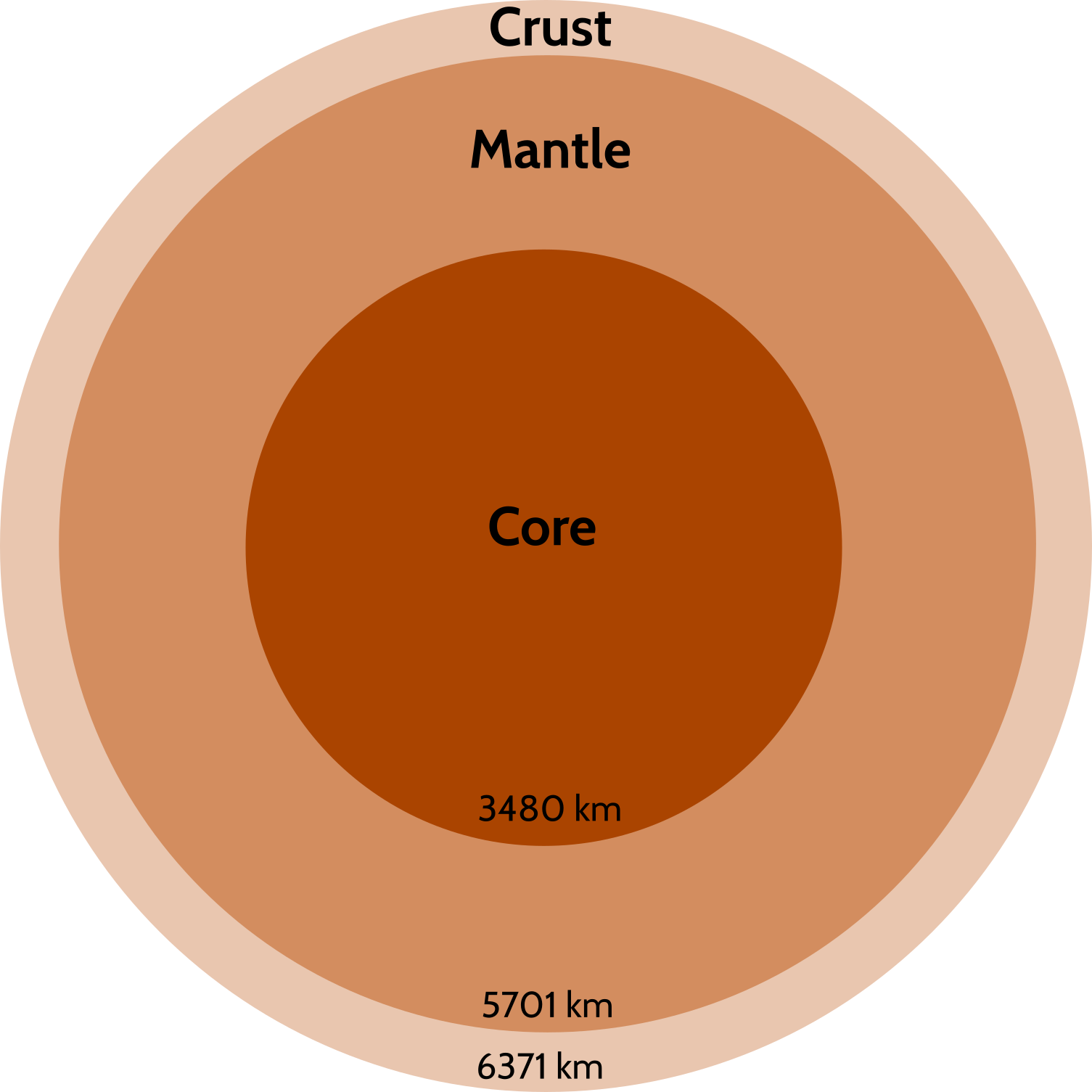}
	\includegraphics[width=0.5\textwidth]{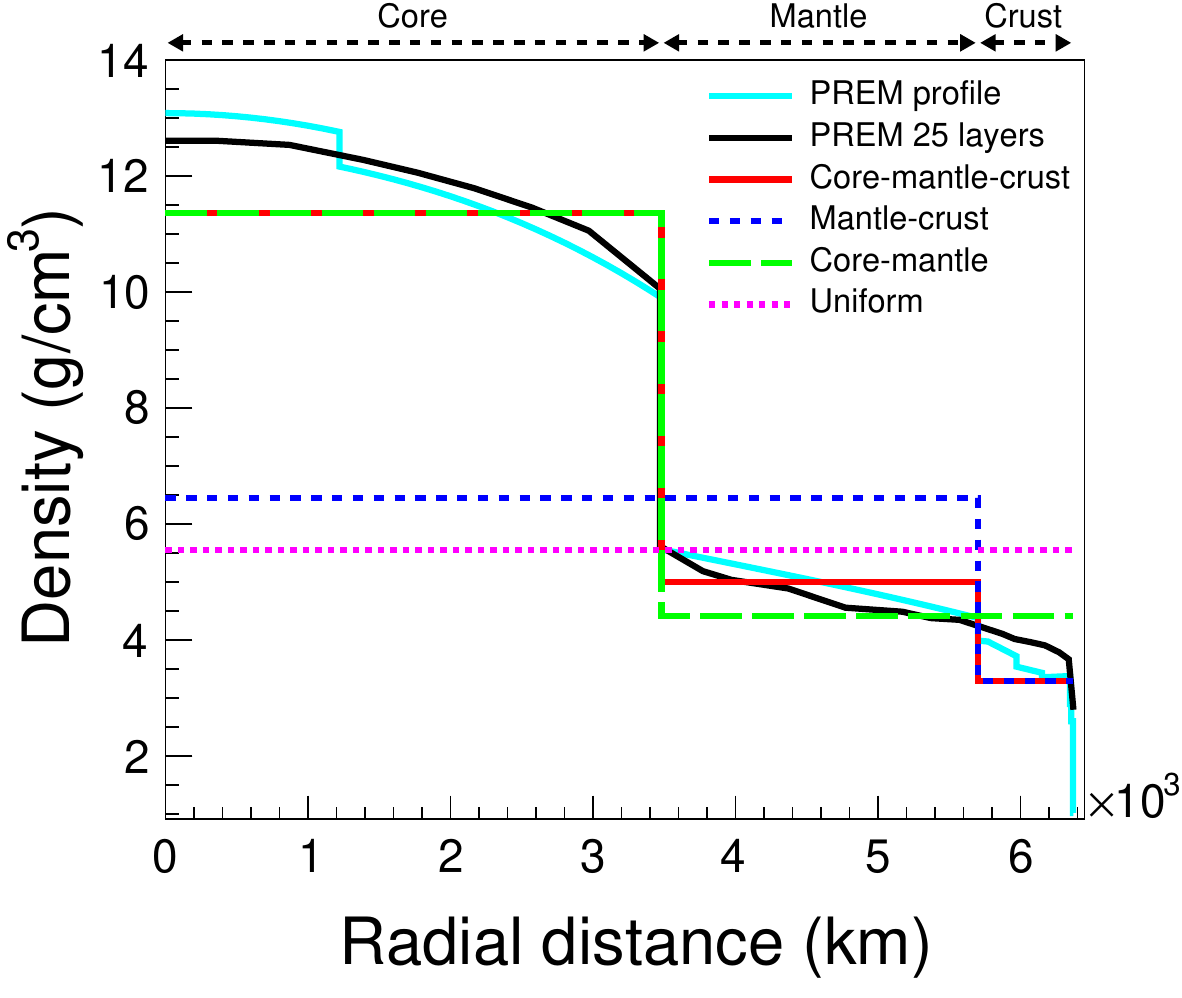}
	\caption{Left: Three-layered profile of Earth. Right: Density distributions of profiles of Earth as functions of radial distance from the center of Earth. Note that the total mass of the Earth is the same in all the profiles.~\cite{Kumar:2021faw}}
	\label{fig:three-layer-model}
\end{figure}
%========================

The measurement of velocities of seismic waves is utilized to develop the Preliminary Reference Earth Model (PREM)~\cite{Dziewonski:1981xy} as shown by the cyan curve in the right panel of Fig.~\ref{fig:three-layer-model} where the density is shown as a function of radial distance \textit{i.e.} the distance of a layer from the center of Earth. We would like to mention that in the actual PREM profile, the Earth is divided into 81 layers. But what we use here as a PREM profile~\cite{Dziewonski:1981xy} for the sake of computational ease is a 25-layered profile of Earth (black curve) that preserves all the important features of the Earth profile. We have checked that whatever conclusion, we have drawn in this work, will not alter whether we take 25 layers or 81 layers.

Guided by the PREM profile of Earth, we consider a three-layered profile of Earth as shown in the left panel of  Fig.~\ref{fig:three-layer-model}. The innermost layer is the core which is followed by the mantle, and the outermost layer is the crust. The density distribution for the three-layered structure is shown by the red curve in the right panel of Fig.~\ref{fig:three-layer-model}. The layer boundaries and their densities for the three-layered profile of Earth are mentioned in Table~\ref{tab:profiles}.

%========================	
\begin{table}[htb!]
	\centering
	\begin{tabular}{|c|c|c|}
		\hline \hline
		Profiles & Layer boundaries (km) & Layer densities (g/cm\textsuperscript{3})\\
		\hline
		PREM & 25 layers & 25 densities\\
		Core-mantle-crust & (0, 3480, 5701, 6371) & (11.37, 5, 3.3) \\
		Mantle-crust  & (0, 5701, 6371) & (6.45, 3.3)\\
		Core-mantle  & (0, 3480, 6371) & (11.37, 4.42)\\
		Uniform  & (0, 6371) & (5.55) \\
		\hline \hline
	\end{tabular}\\
	\caption{The boundaries and densities of layers for various profiles of Earth considered in this analysis. The radius and mass of Earth remain invariant for all these profiles.~\cite{Kumar:2021faw}}
	\label{tab:profiles}
\end{table}
%========================

Since neutrino oscillations occur in the vacuum also, one of the important tasks is to rule out the vacuum hypothesis and feel the presence of matter. For vacuum, we consider the density of the Earth to be zero. We further consider alternative profiles of the Earth as mentioned in Table~\ref{tab:profiles} for testing against the three-layered profile using neutrino oscillations. While considering alternative profiles of Earth, we assume the radius and the mass of Earth to be invariant\footnote{Note that the moment of inertia of Earth can also be considered as an additional invariant quantity on which the information is obtained from gravitational studies independent of seismology.}. The dashed blue curve in the right panel of Fig.~\ref{fig:three-layer-model} shows the mantle-crust profile, which has a two-layered structure with mantle and crust where the core and mantle are fused together. The core-mantle profile has a two-layered structure with core and mantle where the crust is merged into the mantle as shown by the dashed green curve in the right panel of Fig.~\ref{fig:three-layer-model}. The uniform density profile is shown by the dotted pink curve in the right panel of Fig.~\ref{fig:three-layer-model}. 

Since the distributions of densities in these profiles of the Earth are different from each other, we expect the neutrino oscillation probability to modify differently in the presence of matter governed by these profiles. In Sec.~\ref{sec:tomography_probability}, we discuss the effect of these profiles of the Earth on neutrino oscillation probabilities. 

%==========================
\section{Effect of Various Density Profiles of Earth on Probability Oscillograms}
\label{sec:tomography_probability}
%==========================
 
ICAL is sensitive to atmospheric muon neutrinos and antineutrinos in the multi-GeV range of energy. After traveling long distance inside the Earth, the initial muon neutrino $\nu_\mu$ at production may survive as muon neutrino $\nu_\mu$ at detection with survival probability $P(\nu_\mu \rightarrow \nu_\mu)$ whereas an electron neutrino $\nu_e$ may oscillate to muon neutrino $\nu_\mu$ with appearance probability $P(\nu_e \rightarrow \nu_\mu)$. The muon neutrino events detected at ICAL are contributed by both survival $(\nu_\mu \rightarrow \nu_\mu)$ as well as appearance $(\nu_e \rightarrow \nu_\mu)$ channels. 

\begin{figure}[htb!]
	\centering
	\includegraphics[width=0.48\linewidth]{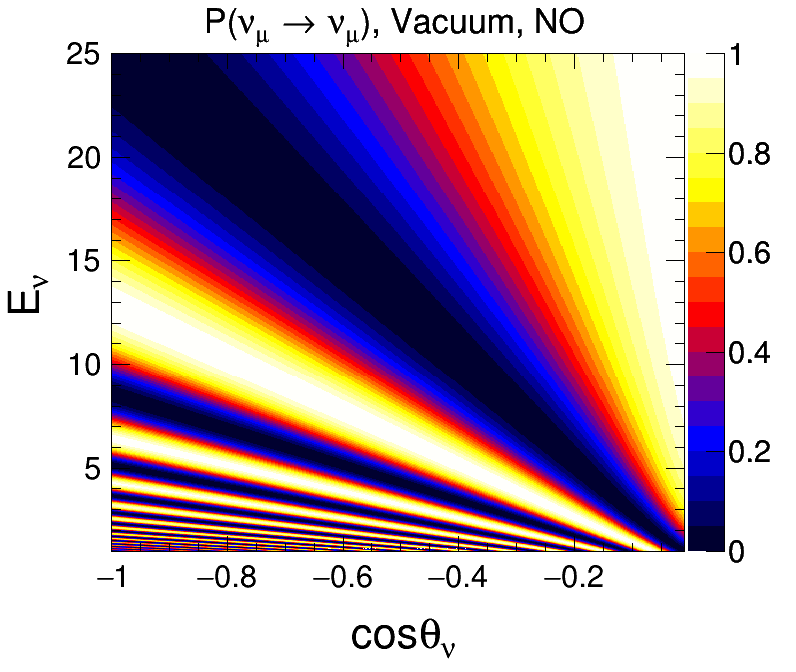}
	\includegraphics[width=0.48\linewidth]{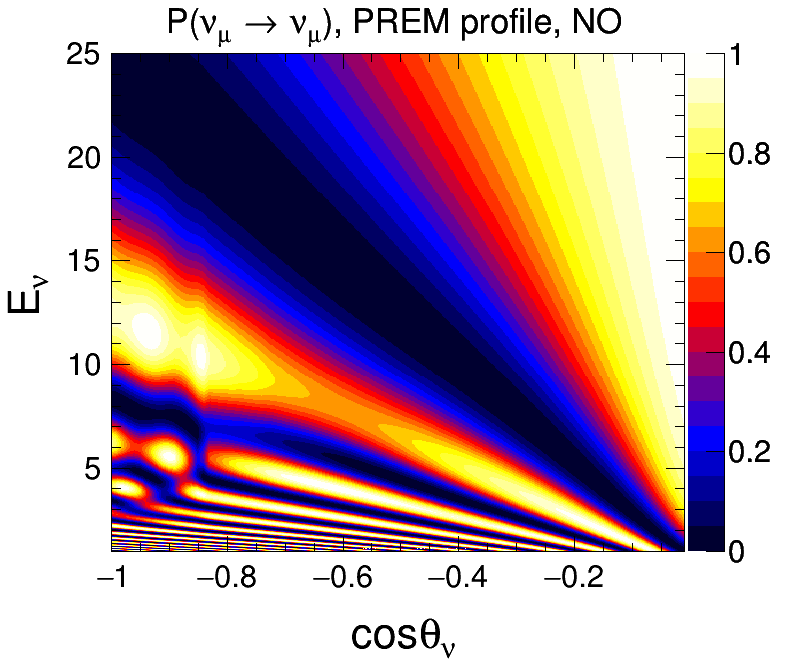}
	\includegraphics[width=0.48\linewidth]{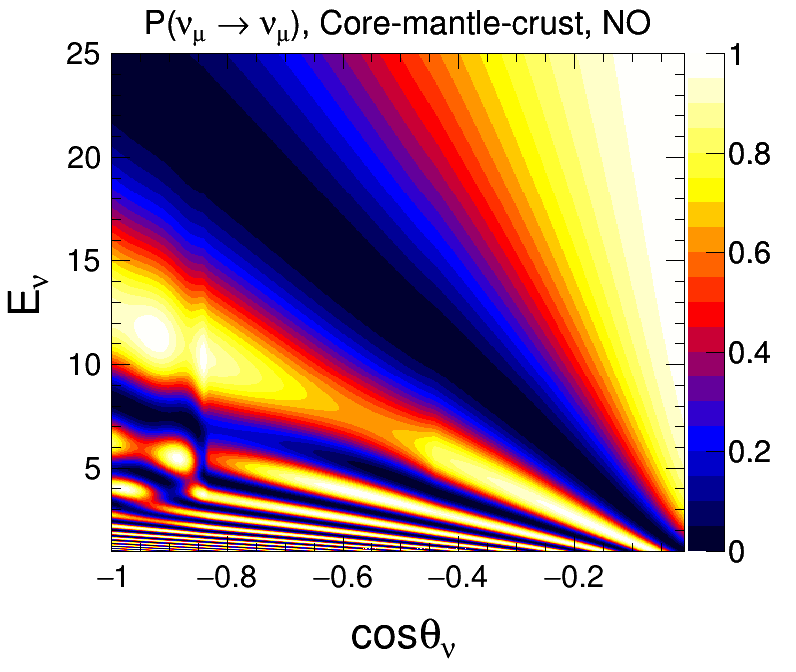}
	\includegraphics[width=0.48\linewidth]{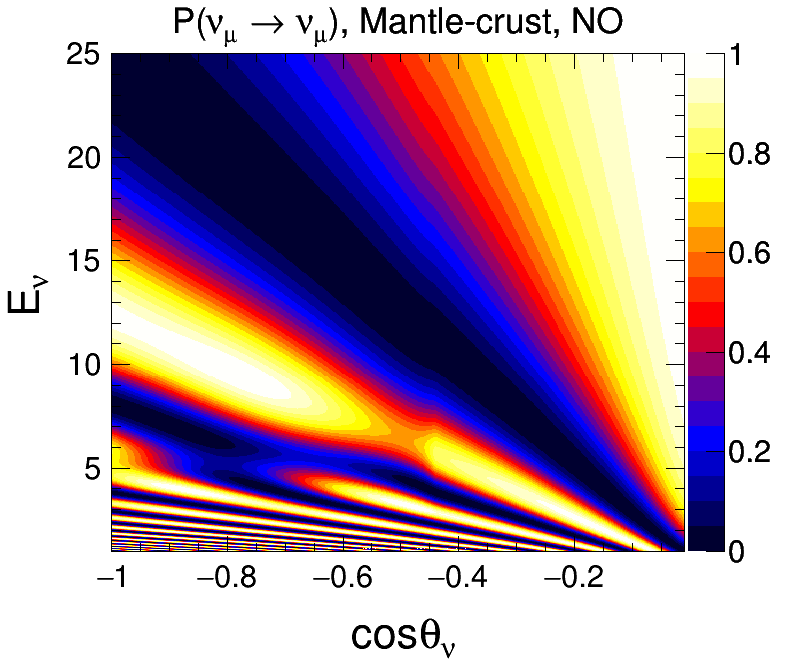}
	\includegraphics[width=0.48\linewidth]{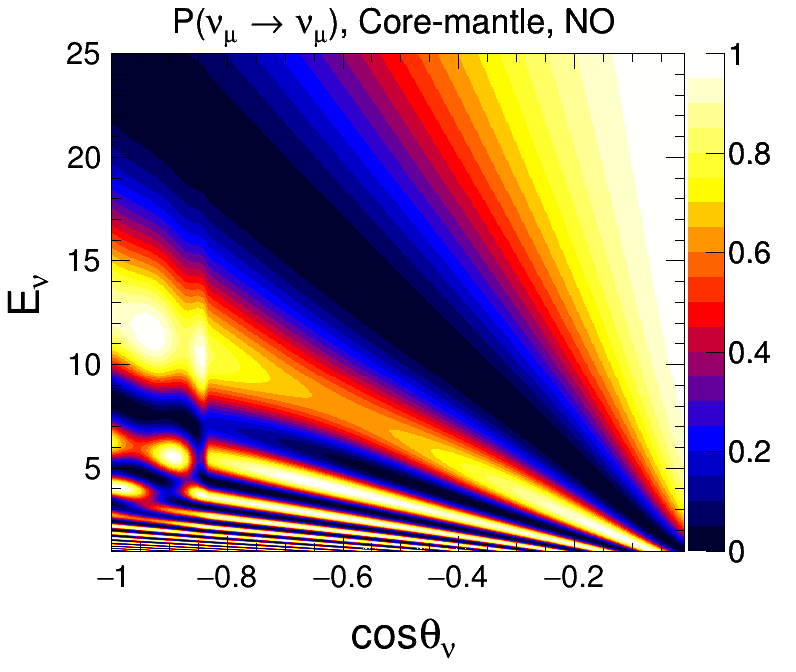}
	\includegraphics[width=0.48\linewidth]{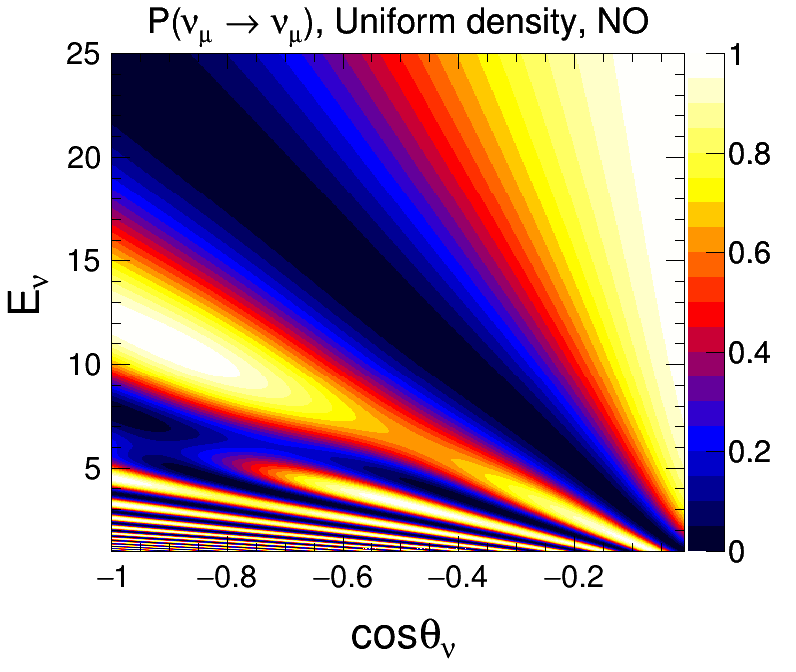}
	\caption{$P(\nu_{\mu} \rightarrow \nu_{\mu})$ oscillograms considering various density profiles of Earth.
		We take the three-flavor oscillation parameters from Table~\ref{tab:osc-param-value}. We assume NO and $\sin^2\theta_{23} = 0.5$.~\cite{Kumar:2021faw}}
	\label{fig:tomography_Puu_oscillogram_model}
\end{figure}

\begin{figure}[htb!]
	\centering
	\includegraphics[width=0.48\linewidth]{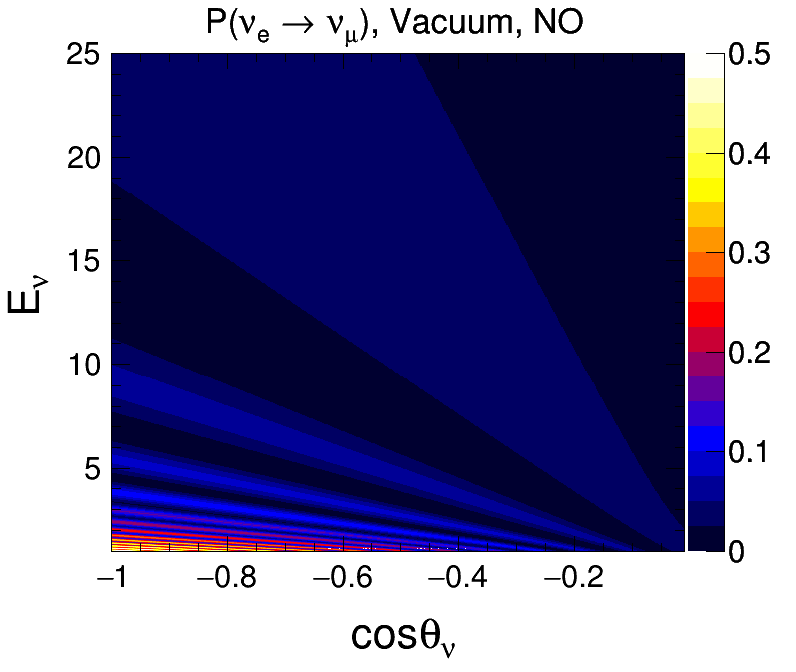}
	\includegraphics[width=0.48\linewidth]{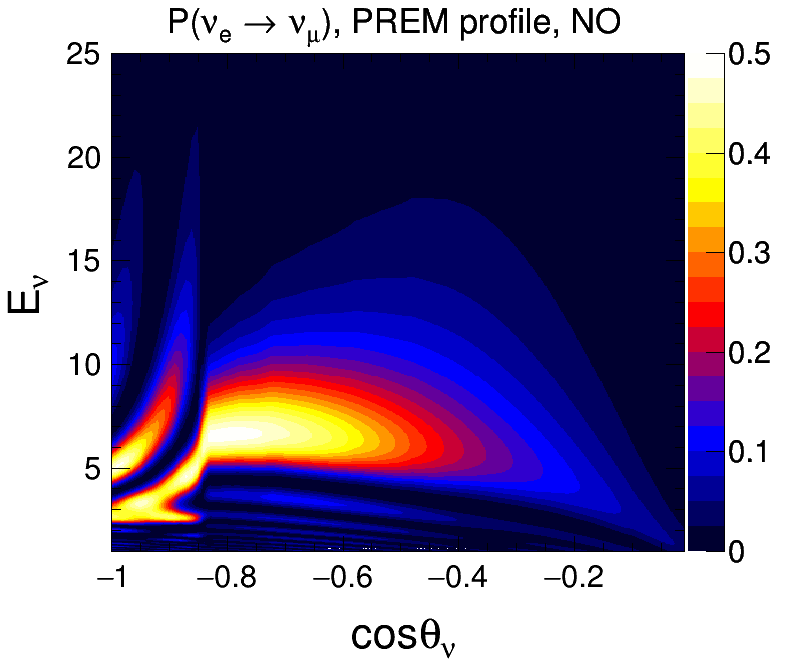}
	\includegraphics[width=0.48\linewidth]{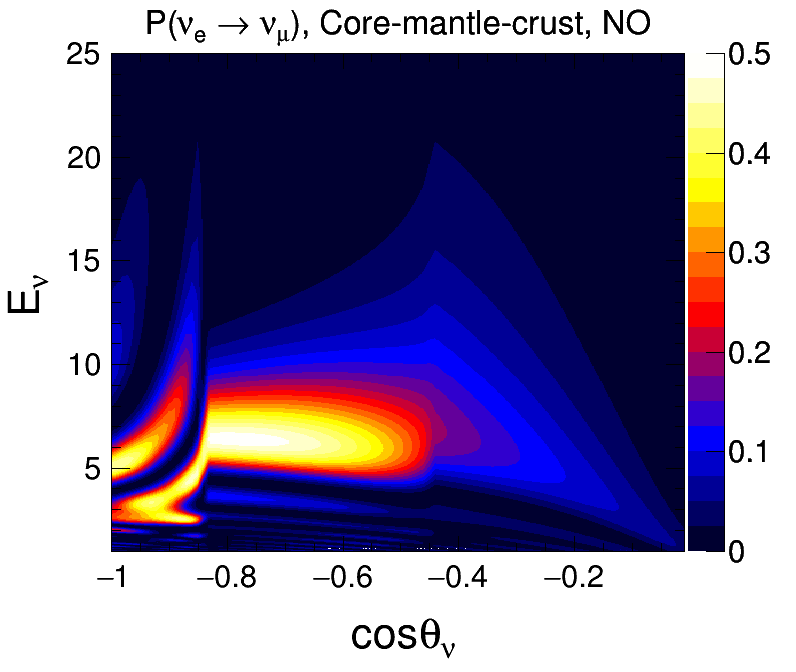}
	\includegraphics[width=0.48\linewidth]{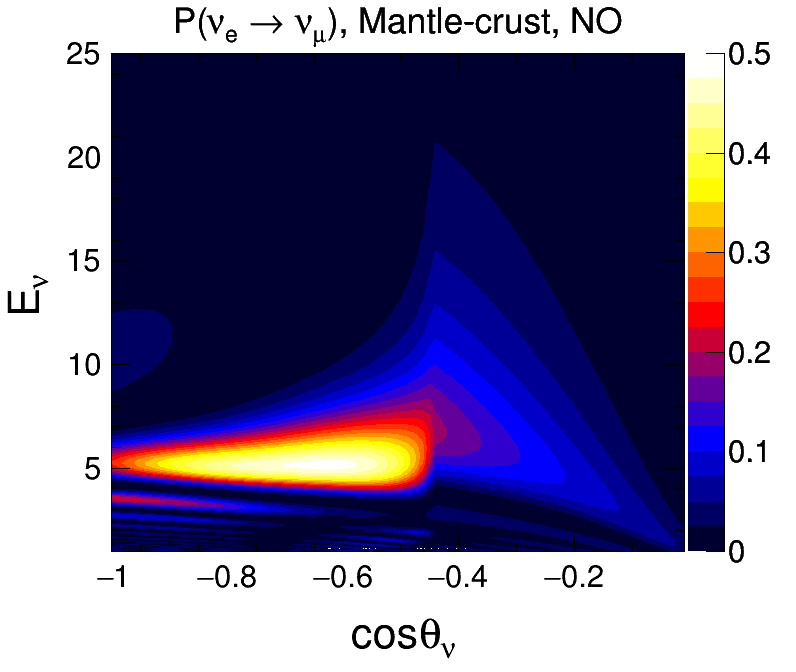}
	\includegraphics[width=0.48\linewidth]{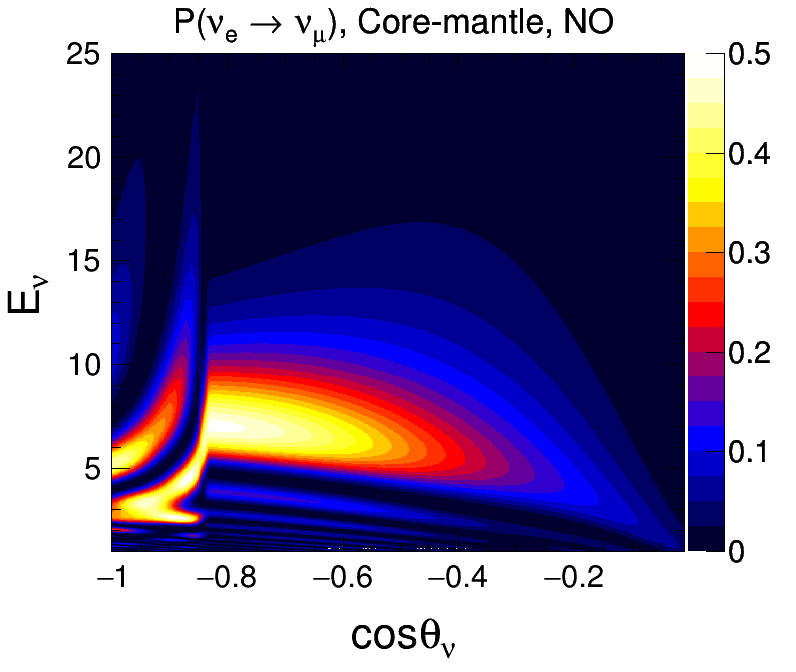}
	\includegraphics[width=0.48\linewidth]{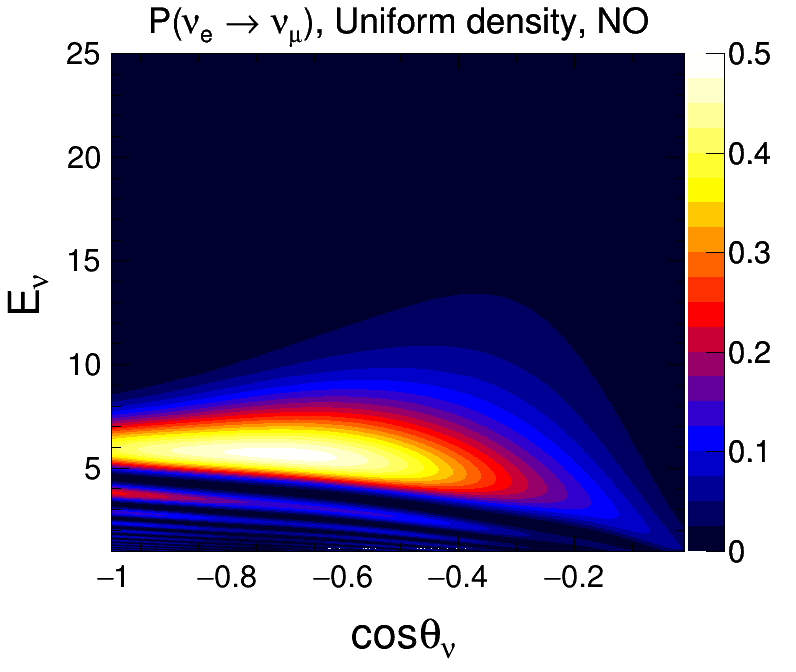}
	\caption{$P(\nu_{e} \rightarrow \nu_{\mu})$ oscillograms considering various density profiles of Earth.
		We take the three-flavor oscillation parameters from Table~\ref{tab:osc-param-value}. We assume NO and $\sin^2\theta_{23} = 0.5$.~\cite{Kumar:2021faw}}
	\label{fig:tomography_Peu_oscillogram_model}
\end{figure}

In this analysis, we use the values of benchmark oscillation parameters mentioned in Table~\ref{tab:osc-param-value}. In Fig.~\ref{fig:tomography_Puu_oscillogram_model}, we present the oscillograms for $\nu_\mu$ survival channel $(\nu_\mu \rightarrow \nu_\mu)$ in the plane of $\cos\theta_\nu$ vs. $E_\nu$ considering NO in the three-flavor neutrino oscillation framework using various profiles of Earth. In Fig.~\ref{fig:tomography_Peu_oscillogram_model}, we show the same for $\nu_\mu$ appearance channel $(\nu_e \rightarrow \nu_\mu)$. In Figs.~\ref{fig:tomography_Puu_oscillogram_model} and~\ref{fig:tomography_Peu_oscillogram_model}, $\cos\theta_\nu = 1$ corresponds to the downward-going neutrino and -1 to the upward-going neutrino. In both these figures, we study six different profiles of the Earth which are i) vacuum, ii) PREM, iii) core-mantle-crust, iv) mantle-crust, v) core-mantle, and vi) uniform density.  

%========================
%========================

%========================
%========================

\begin{itemize}
	\item \textbf{Vacuum:} The left panel of the first row in Fig.~\ref{fig:tomography_Puu_oscillogram_model} shows the survival probability $P(\nu_{\mu} \rightarrow \nu_{\mu})$ in vacuum where we can identify the first oscillation minimum as a dark blue diagonal band which starts from ($E_\nu = 1$ GeV, $\cos\theta_\nu = 0$) and ends at ($E_\nu = 25$ GeV, $\cos\theta_\nu = -1$). This diagonal band is named as ``oscillation valley'' in chapters ~\ref{chap:dip_valley} and \ref{chap:NSI}. The higher-order oscillation minima and maxima in vacuum are shown with thinner bands of blue and yellow colors, respectively, in the lower-left triangle. The left panel in the first row in Fig.~\ref{fig:tomography_Peu_oscillogram_model} shows $P(\nu_e \rightarrow \nu_\mu)$ for the case of vacuum oscillation where we do not see any matter effect. 	
	
	\item \textbf{PREM profile:} The right panel in the first row in Fig.~\ref{fig:tomography_Puu_oscillogram_model} shows the survival probability $P(\nu_\mu \rightarrow \nu_\mu)$ in the presence of matter with the PREM profile. The oscillation valley can be observed along with matter effect. The red patch around $-0.8 < \cos\theta_{\nu} < -0.5$ and $6  \text{ GeV} < E_{\nu} < 10  \text{ GeV}$ corresponds to MSW resonance whereas yellow patches around $\cos\theta_{\nu} < -0.8$ and $3 \text{ GeV} < E_{\nu} < 6 \text{ GeV}$ is due to the NOLR/parametric resonance. The right panel in the first row in Fig.~\ref{fig:tomography_Peu_oscillogram_model} shows $P(\nu_e \rightarrow \nu_\mu)$ in the presence of matter with the PREM profile where we can identify the MSW resonance as a single yellow patch around $-0.8 < \cos\theta_\nu < -0.5$ whereas the NOLR/parametric resonance can be seen as two yellow patches around $ \cos\theta_{\nu} < -0.8$. A sharp transition is observed around the boundary of core and mantle at $\cos\theta_\nu = -0.84$.
	
	\item \textbf{Core-mantle-crust profile:} The survival probability $P(\nu_\mu \rightarrow \nu_\mu)$ with the three-layered profile of core-mantle-crust is shown in the left panel of the second row in Fig.~\ref{fig:tomography_Puu_oscillogram_model} where we can identify the MSW resonance as well as the NOLR/parametric resonance  similar to the case of PREM profile. The left panel in the second row in Fig.~\ref{fig:tomography_Peu_oscillogram_model} shows $P(\nu_e \rightarrow \nu_\mu)$ for core-mantle-crust profile where we can identify the MSW resonance as well as the NOLR/parametric resonance.  Here, we can observe two sharp transitions at core-mantle boundary ($\cos\theta_\nu = -0.84$) and mantle-crust boundary ($\cos\theta_\nu = -0.45$). 
	
	\item \textbf{Mantle-crust profile:} The right panel of the second row in Fig~\ref{fig:tomography_Puu_oscillogram_model} shows the survival probability $P(\nu_\mu \rightarrow \nu_\mu)$ for the case of the two-layered profile of mantle-crust where we can observe that the MSW resonance is modified significantly and the NOLR/parametric resonance  is not visible. This indicates that the absence of core modifies the matter effects significantly. Due to the absence of core, the NOLR/parametric resonance  as well as the sharp transition around $\cos\theta_\nu = -0.84$ is absent in $P(\nu_e \rightarrow \nu_\mu)$ for mantle-crust profile as shown in the right panel of the second row in Fig.~\ref{fig:tomography_Peu_oscillogram_model}.
	
	\item \textbf{Core-mantle profile:} For the case of the two-layered profile of core-mantle shown in the left panel of the third row in Fig.~\ref{fig:tomography_Puu_oscillogram_model}, the MSW resonance, as well as the NOLR/parametric resonance, are observed clearly for the survival probability $P(\nu_\mu \rightarrow \nu_\mu)$ which indicate that the absence of crust does not affect the matter effects by a large amount. For the core-mantle profile shown in the left panel of the third row in Fig.~\ref{fig:tomography_Peu_oscillogram_model}, the matter effects for $P(\nu_e \rightarrow \nu_\mu)$  are the same as observed in the case of the three-layered profile, but the sharp transition around $\cos\theta_\nu = -0.45$ is absent because we do not have the mantle-crust boundary in this profile.
	
	\item \textbf{Uniform density:} The right panel of the third row in Fig.~\ref{fig:tomography_Puu_oscillogram_model} shows the survival probability $P(\nu_\mu \rightarrow \nu_\mu)$ for the case of uniform density inside Earth where we can identify the MSW resonance, which is disturbed by a small amount. The NOLR/parametric resonance  is absent, which is a sign of the absence of core. In the right panel of the third row of  Fig.~\ref{fig:tomography_Peu_oscillogram_model}, $P(\nu_e \rightarrow \nu_\mu)$ is shown for uniform density inside Earth where we can find that the NOLR/parametric resonance, as well as two sharp transitions, are absent. This is because we do not have the core and any boundaries between layers.
\end{itemize}

Thus, we may infer from these plots that the presence of mantle and core results in the MSW resonance and the NOLR/parametric resonance, respectively, whereas boundaries between layers result in sharp transitions. We would like to mention that we have used NO for these plots where significant matter effects is observed in the neutrino channel, and antineutrinos feel negligible matter effect. If we consider the case of IO, antineutrinos will feel the significant matter effects rather than neutrinos.  Our aim is to observe these features in the reconstructed muon observables at ICAL in 10 years.

%==========================
\section{Reconstructed Events at ICAL}
\label{sec:tomography_events}
%==========================

We generate the reconstructed $\mu^-$ and $\mu^+$ events at ICAL in 10 years following the procedure described in Sec.~\ref{sec:event_simulation}. For the case of NO, the 50 kt ICAL detector would detect about 4614 reconstructed $\mu^-$ and 2053 reconstructed $\mu^+$ events in 10 years with a total exposure of 500 kt$\cdot$yr using three-flavor neutrinos oscillation with matter effects considering 25-layered PREM profile of Earth as shown in Table~\ref{tab:tomography_events}. The ns timing resolution of RPCs~\cite{Dash:2014ifa,Bhatt:2016rek,Gaur:2017uaf} enables ICAL to distinguish between upward-going and downward-going muon events, which are also shown in Table~\ref{tab:tomography_events}. We also show events considering the three-layered profile of core-mantle-crust as well as the vacuum where we can observe some difference in upward-going events only which have experienced matter effect. It is important to note that there is not much difference in total event rate for these profiles, but the final result receives a contribution from binning of these events, which is possible because of good resolution of energy and direction of reconstructed muons at ICAL. The directions of these reconstructed muons can be used to get information about the regions in the Earth through which neutrinos have traversed, as we discuss in Sec.~\ref{sec:events_layer}.

%--------------------------------------------------
\begin{table}[htb!]
	\begin{center}
		\begin{tabular}{|l|c|c|c|c|c|c|}
			\hline \hline
			\multirow{2}{*}{Profiles} & \multicolumn{3}{c|}{Reconstructed $\mu^-$ events}  & \multicolumn{3}{c|}{Reconstructed $\mu^+$ events}\\ \cline{2-7}
			& Upward & Downward & Total & Upward & Downward & Total \\
			\hline
			PREM & 1654 & 2960 & 4614 & 741 & 1313 & 2053 \\
			Core-Mantle-Crust & 1659 & 2960 & 4619 & 739 & 1313 & 2052 \\
			Vacuum & 1692 & 2960 & 4652 & 745 & 1313 & 2057 \\
			\hline \hline
		\end{tabular}
		\caption{The total number of reconstructed $\mu^-$ and $\mu^+$ events expected in the upward and downward direction at the 50 kt ICAL detector in 10 years which is scaled from 1000-year MC data. We take the three-flavor oscillation parameters from Table~\ref{tab:osc-param-value}. We assume NO and $\sin^2\theta_{23} = 0.5$.~\cite{Kumar:2021faw}}
		\label{tab:tomography_events}
	\end{center}
\end{table}
%------------------------------------------------------------------

%==========================
\section{Identifying Neutrino Events Passing through Different Layers of Earth}
\label{sec:events_layer}
%==========================

\begin{figure}[htb!]
	\centering
	\includegraphics[width=0.6\linewidth]{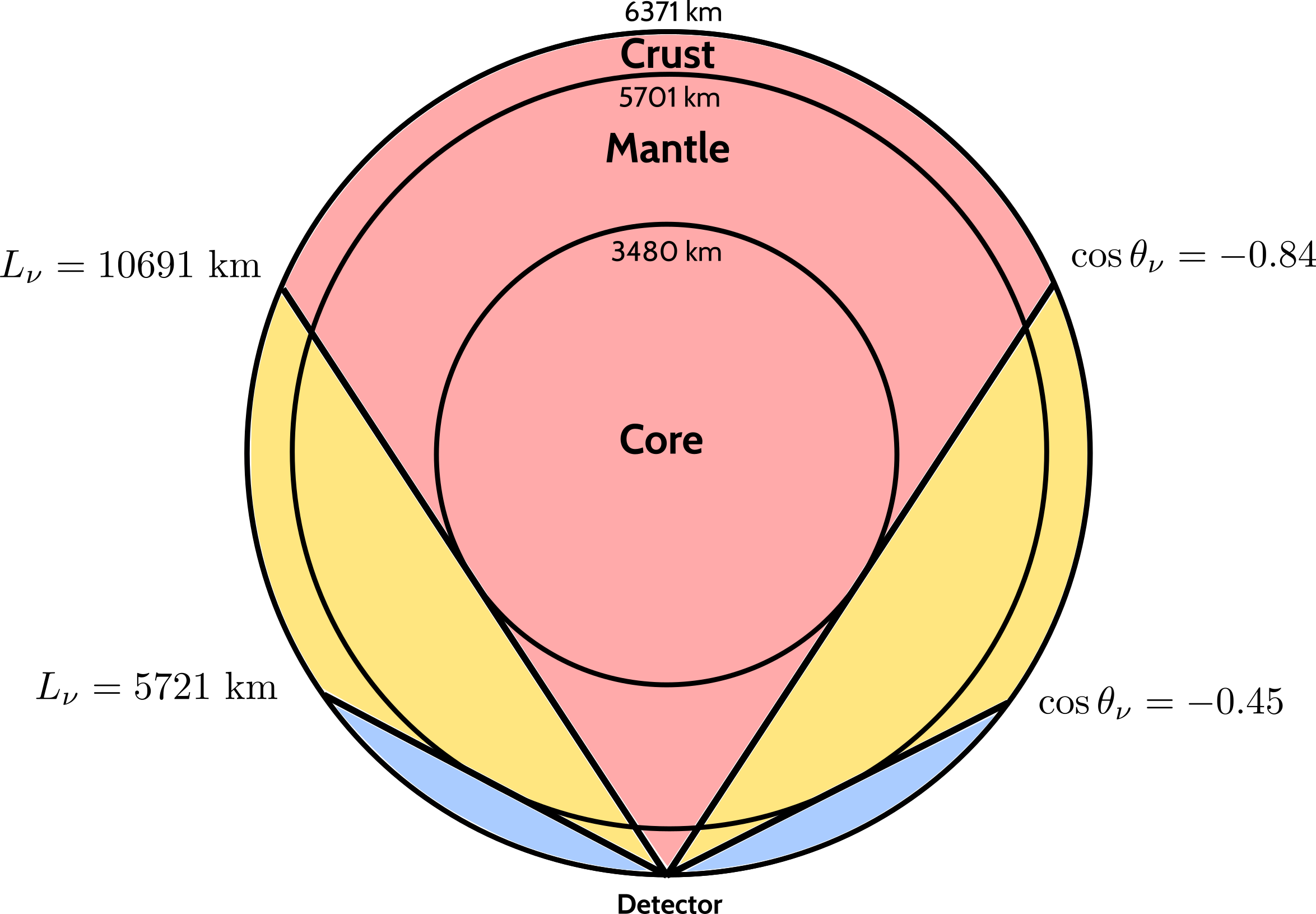}
	\caption{Neutrinos passing through regions consisting of a particular set of layers of Earth depending upon the zenith angle.~\cite{Kumar:2021faw}
	}
	\label{fig:neutrino-path}
\end{figure}

The atmospheric neutrinos cover a wide range of baselines ($L_\nu$) from 15 km to 12757 km that correspond to downward and upward directions, respectively. The vertically upward-going neutrinos pass through a set of layers of Earth depending upon their directions, as shown in Fig.~\ref{fig:neutrino-path}. The vertically upward-going neutrinos with $\cos\theta_\nu < -0.84$ pass through crust-mantle-core region as shown by pink color in Fig.~\ref{fig:neutrino-path}. The yellow region in Fig.~\ref{fig:neutrino-path} with $-0.84 <\cos\theta_\nu < -0.45$ shows the neutrino events passing through crust-mantle region. The neutrinos which pass through only crust are shown by the blue color region in Fig.~\ref{fig:neutrino-path}.

%==========================
%==========================

%========================
\begin{table}[htb!]
	\centering
	\begin{tabular}{|c|c|c|c|c|}
		\hline\hline
		Regions & $\cos\theta_{\nu}$ & $L_\nu$ (km) & $\mu^{-}$ Events & $\mu^{+}$ Events\\
		\hline
		Crust-mantle-core&	(-1.00, -0.84) & (10691, 12757) & 331 & 146 \\
		Crust-mantle & (-0.84, -0.45) & (5721, 10691) & 739 & 339\\
		Crust & (-0.45, 0.00) & (437, 5721) & 550 & 244 \\ 
		Downward & (0.00, 1.00) & (15, 437) & 2994 & 1324 \\ 
		Total & (-1.00, 1.00) & (15, 12757) & 4614 & 2053 \\ 
		\hline\hline
	\end{tabular}
	\caption{Reconstructed $\mu^-$ and $\mu^+$ events expected at ICAL for 500 kt$\cdot$yr exposure for neutrinos passing through various regions depending upon zenith angle of neutrino. These reconstructed muon events for 10 years are scaled from 1000-year MC data. We consider three-flavor neutrino oscillations in the presence of matter with the PREM profile. We take the three-flavor oscillation parameters from Table~\ref{tab:osc-param-value}. We assume NO and $\sin^2\theta_{23} = 0.5$.~\cite{Kumar:2021faw}}
	\label{tab:layer-passing-event}
\end{table}
%========================

Table~\ref{tab:layer-passing-event} shows the expected number of events at ICAL for 500 kt$\cdot$yr exposure for neutrinos passing through different regions shown in Fig.~\ref{fig:neutrino-path}. Here, we consider three-flavor neutrino oscillations in the presence of matter with the PREM profile of Earth. ICAL would detect about 331 (146) $\mu^-$ ($\mu^+$) events corresponding to the crust-mantle-core passing neutrinos (antineutrino). About 739 $\mu^-$ and 339 $\mu^+$ events would be detected for crust-mantle passing neutrinos and antineutrinos, respectively. The events passing through only crust would be about 550 and 244 for $\mu^-$ and $\mu^+$, respectively. 

Note that the total number of events for reconstructed $\mu^-$ (4614) and $\mu^+$ (2053) are the same in Table~\ref{tab:layer-passing-event} and Table~\ref{tab:tomography_events} for the PREM profile case. Also, it is worthwhile to mention that the reconstructed downward-going $\mu^-$ and $\mu^+$ events mentioned in Table~\ref{tab:layer-passing-event} are a bit different as compared to the downward-going events as mentioned in Table~\ref{tab:tomography_events} for the PREM profile case. It happens because of the angular smearing caused by the kinematics and the finite angular resolution of the detector. Because of this angular smearing, we may have differences in the directions of neutrinos and reconstructed muons, which may force a downward-going neutrino (near horizon) to appear as an upward-going reconstructed muon event.

%========================
\begin{figure}[htb!]
	\centering
	\includegraphics[width=0.42\linewidth]{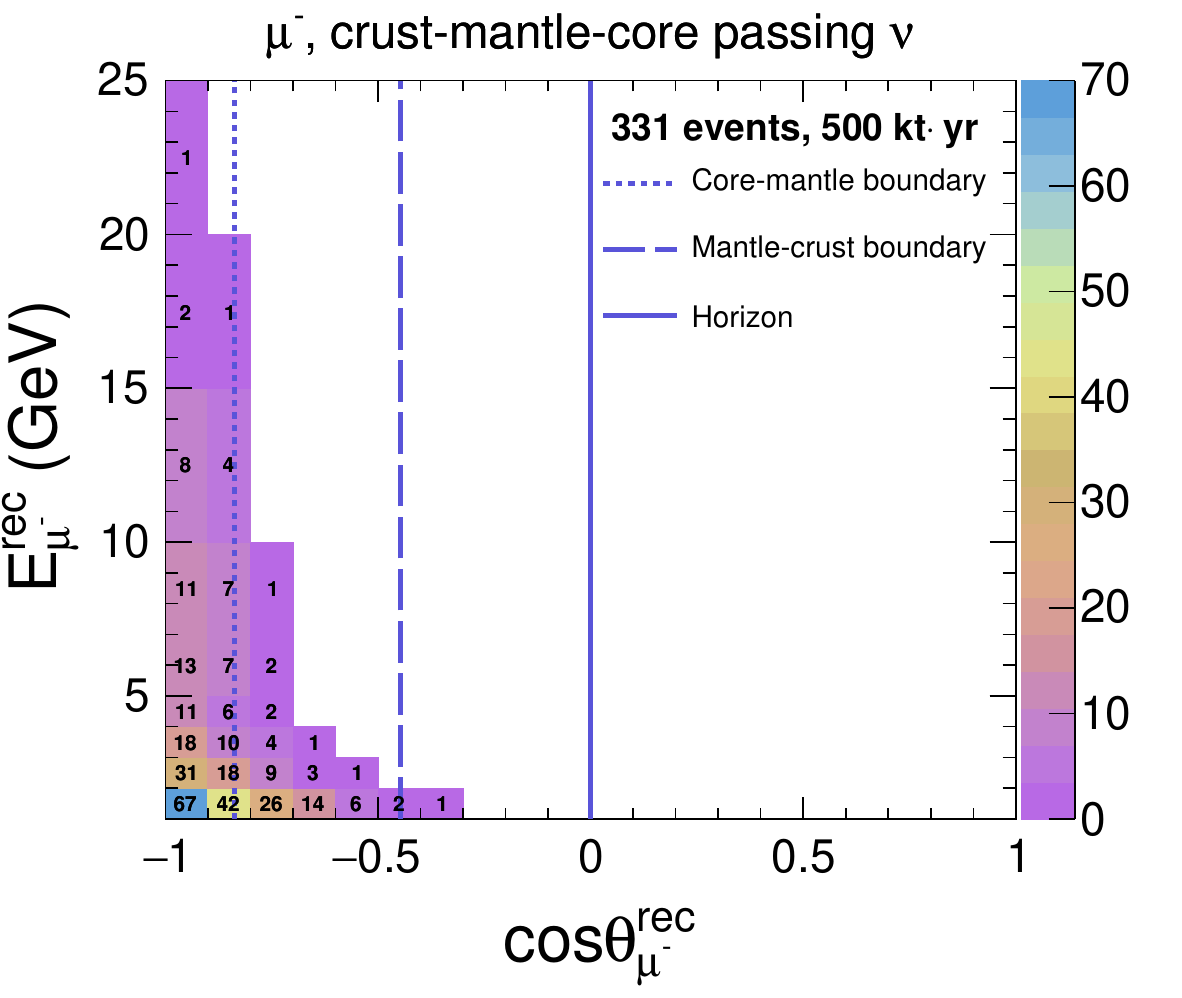}
	\includegraphics[width=0.42\linewidth]{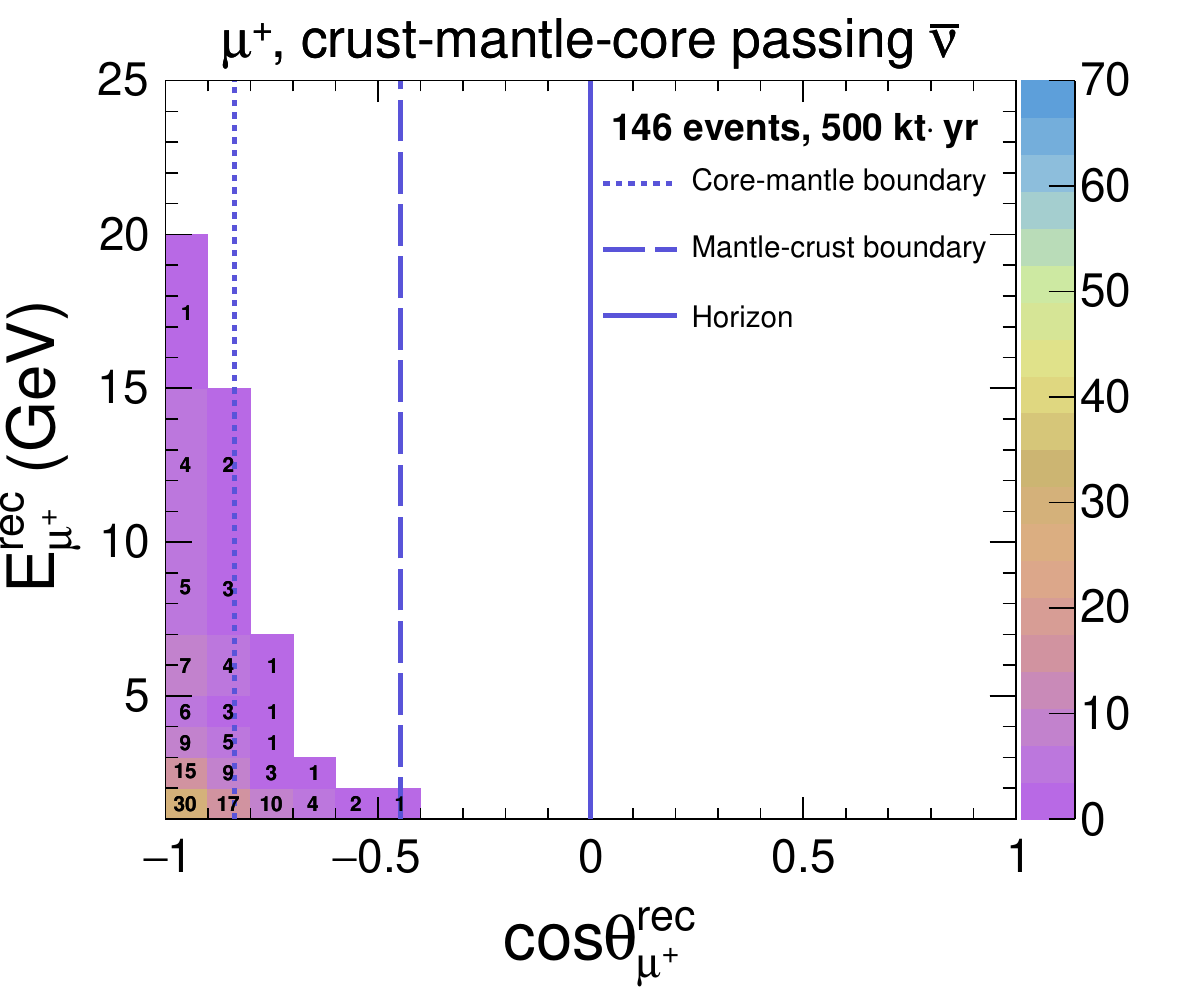}
	\includegraphics[width=0.42\linewidth]{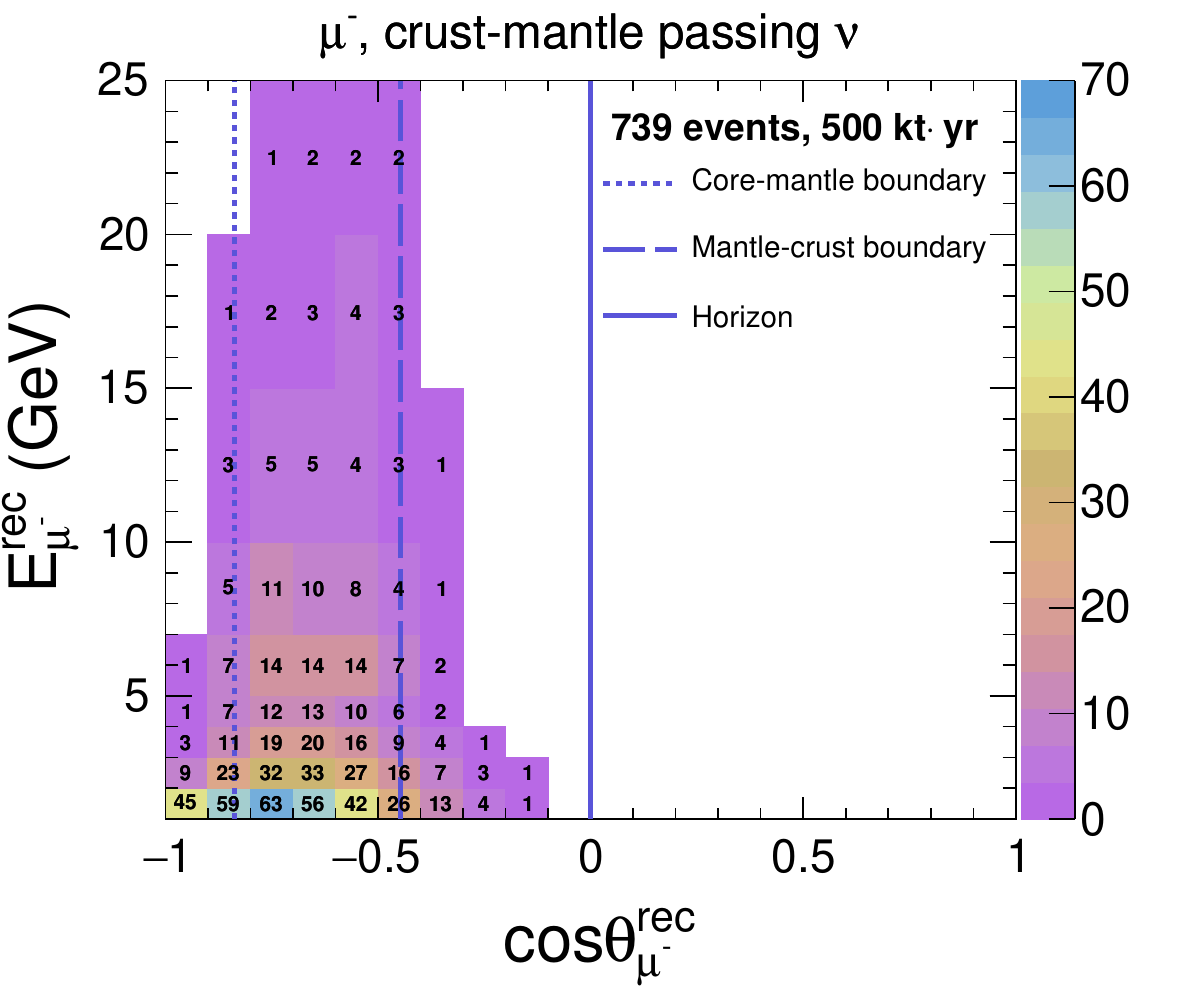}
	\includegraphics[width=0.42\linewidth]{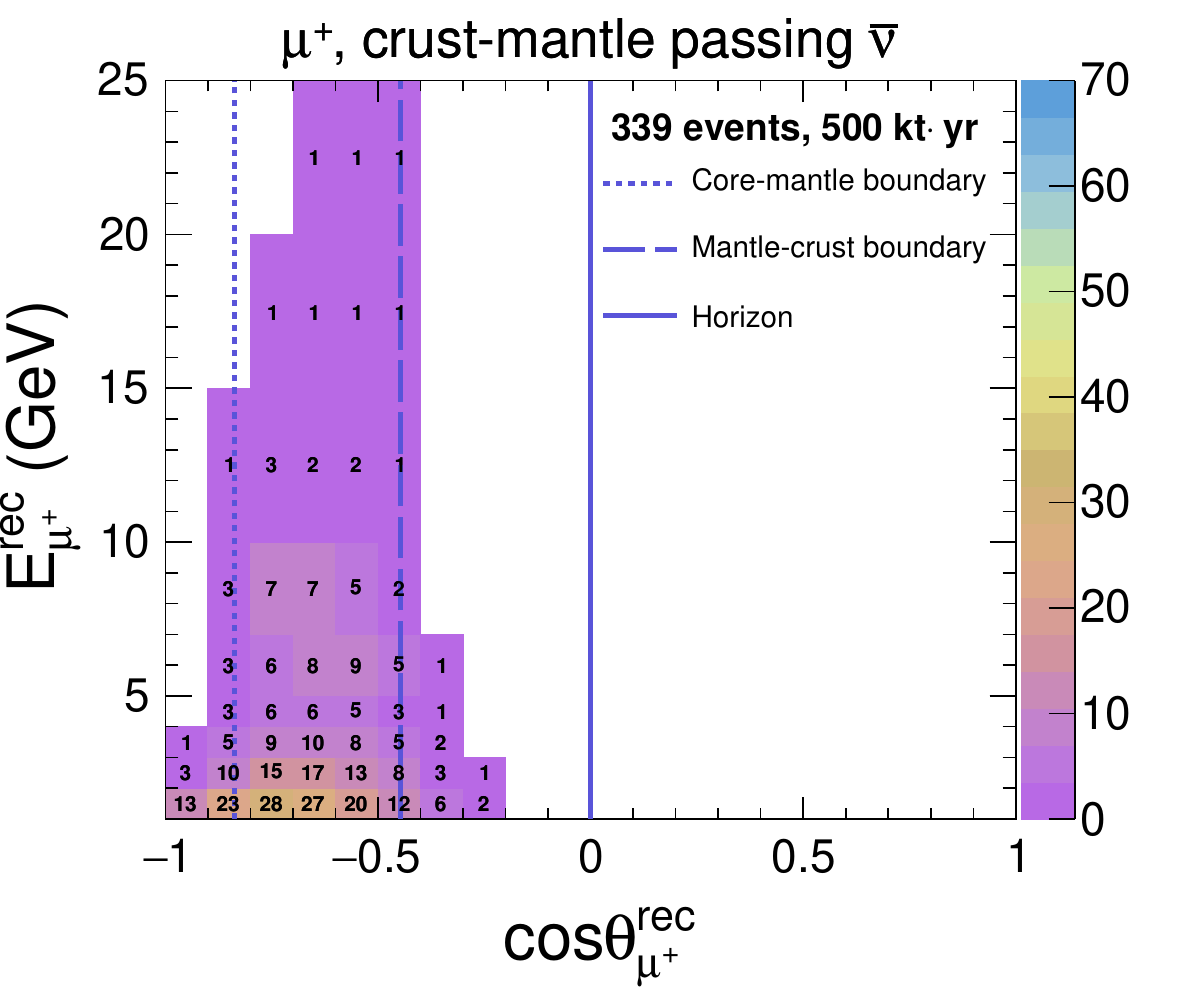}
	\includegraphics[width=0.42\linewidth]{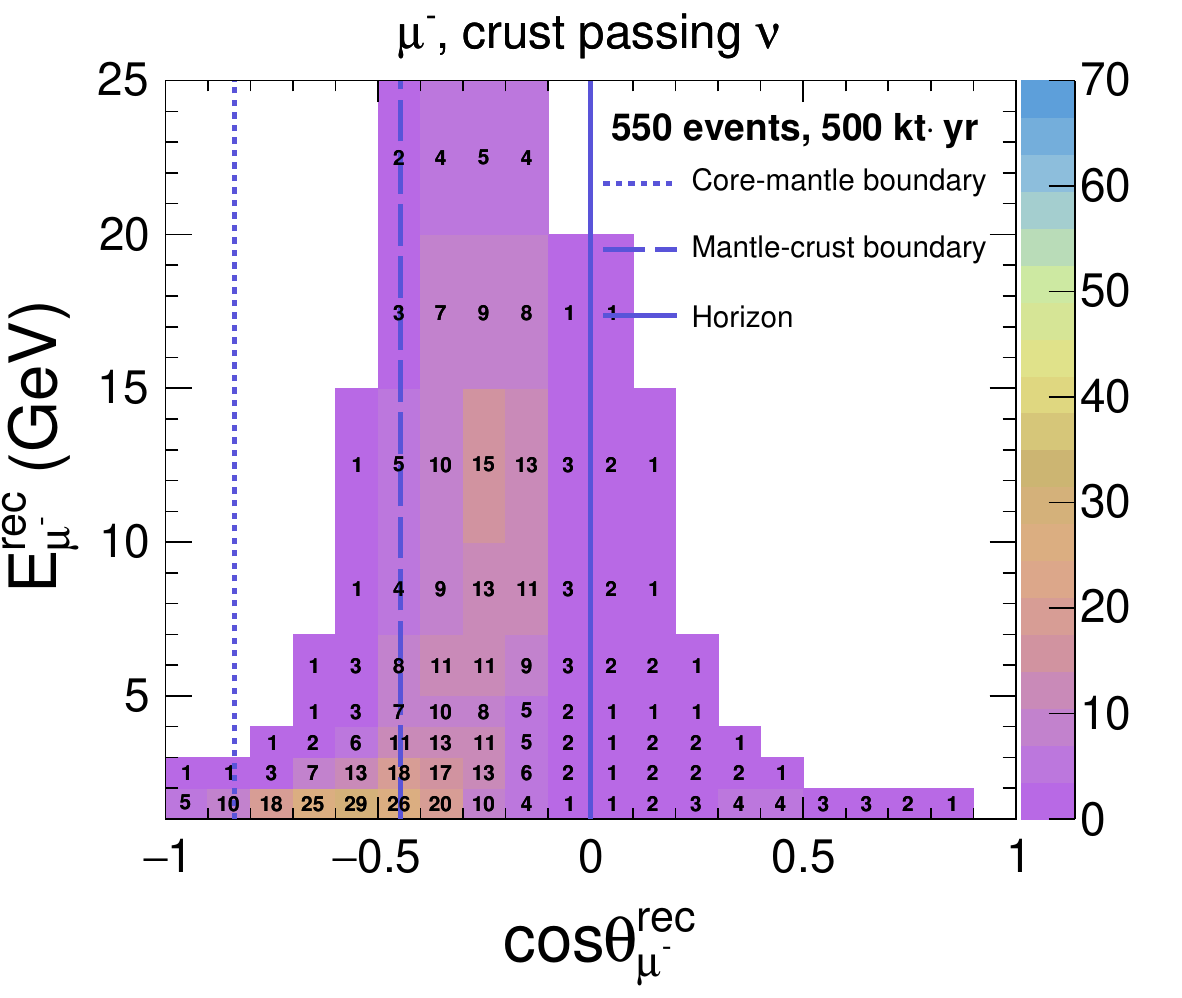}
	\includegraphics[width=0.42\linewidth]{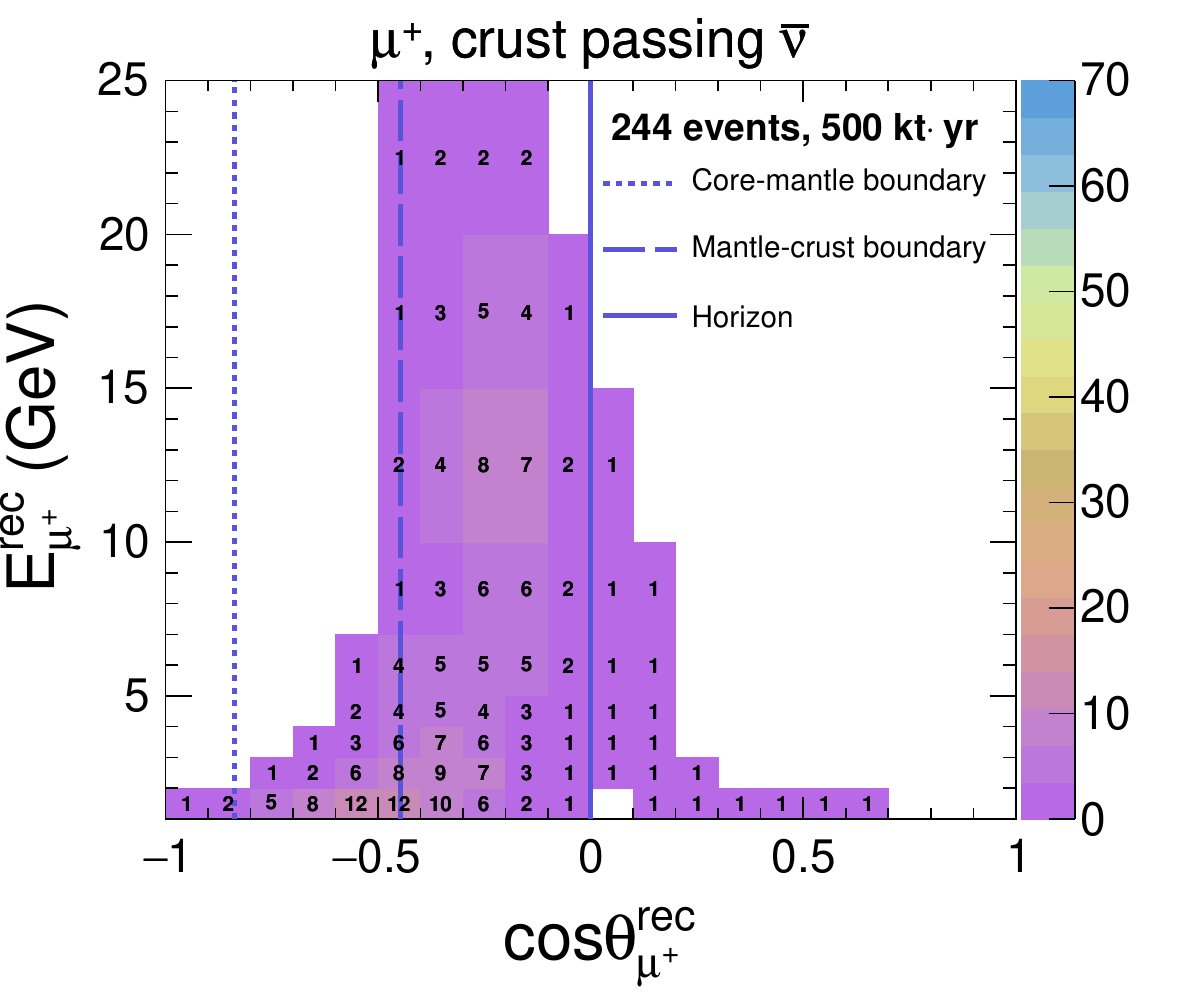}
	\caption{Reconstructed muon event distribution at ICAL for 500 kt$\cdot$yr exposure for neutrinos traversed through various regions in the Earth. These reconstructed muon events for 10 years are scaled from 1000-year MC data. We consider three-flavor neutrino oscillations in the presence of matter with the PREM profile. We take the three-flavor oscillation parameters from Table~\ref{tab:osc-param-value}. We assume NO and $\sin^2\theta_{23} = 0.5$. The top, middle, and bottom panels show the distribution of reconstructed muon events for the parent neutrinos passing through the crust-mantle-core, crust-mantle, and crust, respectively. The dotted, dashed, and solid vertical blue lines correspond to the core-mantle boundary, mantle-crust boundary, and horizontal direction, respectively. The Left (right) panels show the reconstructed $\mu^-$ ($\mu^+$) events.~\cite{Kumar:2021faw}}
	\label{fig:event_dist_nu_density_zones}
\end{figure}
%========================

We would like to point out that while using reconstructed muon observables, the difference in the directions of muon and neutrino due to angular smearing may cause a deterioration in the capability of ICAL to identify the region traversed by the neutrino. Figure~\ref{fig:event_dist_nu_density_zones} shows event distributions of reconstructed muons in ($E_\mu^\text{rec}$, $\cos\theta_\mu^\text{rec}$) plane for neutrinos passing through different regions. For demonstrating reconstructed event distributions for 500 kt$\cdot$yr exposure, we have chosen a binning scheme such that we have total 9 bins in $E_\mu^\text{rec}$ and 20 bins in $\cos\theta_\mu^\text{rec}$. For $E_\mu^\text{rec}$, we have 4 bins of 1 GeV in the range 1 -- 5 GeV, 1 bin of 2 GeV in the range 5 -- 7 GeV, 1 bin of 3 GeV in the range 7 -- 10 GeV, and 3 bins of 5 GeV in the range 10 -- 25 GeV, whereas uniform bins of 0.1 are used for $\cos\theta_\mu^\text{rec}$ in the range of -1 to 1. In Fig.~\ref{fig:event_dist_nu_density_zones}, the vertical dotted blue line shows the core-mantle boundary, whereas the vertical dashed blue line shows the mantle-crust boundary. The horizontal direction is shown with a solid blue line.

The left panel in the first row in Fig.~\ref{fig:event_dist_nu_density_zones} shows the distribution of reconstructed $\mu^-$ events for neutrinos passing through crust-mantle-core region. Here, we can observe that although the actual neutrinos are present only on the left side of the dotted blue line, a few reconstructed muons get smeared into other regions also. The left panel in the second row in Fig.~\ref{fig:event_dist_nu_density_zones} shows the event distribution of reconstructed $\mu^-$ events for neutrinos passing through crust-mantle region \textit{i.e.} between dotted and dashed blue lines. Although most of the events remain between dotted and dashed blue lines, some events smear into other regions also. The reconstructed $\mu^-$ event distribution for crust passing neutrinos is shown in the left panel of the third row in Fig.~\ref{fig:event_dist_nu_density_zones}. A similar kind of smearing is observed for the distributions of reconstructed $\mu^+$ events as shown in the right panels in Fig.~\ref{fig:event_dist_nu_density_zones}. Thus, we can say that the good directional resolution at ICAL enables the reconstructed muon events to preserve the information about the regions traversed by neutrinos.

%==========================
\section{Simulation Method}
\label{sec:statistical analysis}
%==========================

%==========================
\subsection{Binning Scheme}
%==========================

%------------------------------------------
\begin{table}[htb!]
	\centering
	\begin{tabular}{|c|c|c|c c|}
		\hline \hline
		Observable & Range & Bin width & \multicolumn{2}{c|}{Number of bins} \\
		\hline
		\multirow{4}{*}{ $E_\mu^\text{rec}$ (GeV)} & [1, 4] &  0.5 & 6 & \rdelim\}{4}{7mm}[12] \\ 
		& [4, 7] & 1 & 3  & \\
		& [7, 11] & 4 & 1  & \\
		& [11, 21] & 5 & 2  & \\
		\hline 
		\multirow{3}{*}{$\cos\theta_\mu^\text{rec}$} & [-1.0, -0.4] & 0.05 & 12  & \rdelim\}{3}{7mm}[21]\\
		& [-0.4, 0.0] & 0.1 & 4  & \\
		& [0.0, 1.0] & 0.2 & 5  & \\
		\hline 
		\multirow{3}{*}{ ${E'}_\text{had}^\text{rec}$ (GeV)} & [0, 2] &  1 & 2 & \rdelim\}{3}{7mm}[4] \\ 
		& [2, 4] & 2 & 1  & \\
		& [4, 25] & 21 & 1  & \\
		\hline \hline
	\end{tabular}
	\caption{The binning scheme considered for reconstructed observables $E_\mu^\text{rec}$, $\cos\theta_\mu^\text{rec}$, and ${E'}_\text{had}^\text{rec}$ for $\mu^-$ as well as $\mu^+$ events.~\cite{Kumar:2021faw} }
	\label{tab:binning-2D-10years}
\end{table}
%-----------------------------------------------

In this work, we are harnessing the matter effects to understand the distribution of matter inside the Earth. The binning scheme used in Ref.~\cite{Devi:2014yaa} is optimized to probe the Earth's matter effects considering $E_\mu^\text{rec}$, $\cos\theta_\mu^\text{rec}$, and ${E'}_\text{had}^\text{rec}$ as observables. In this binning scheme, $E_\mu^\text{rec}$ is considered in the range of 1 -- 11 GeV whereas ${E'}_\text{had}^\text{rec}$ is having a range of 0 -- 15 GeV. We have modified this binning scheme by adding two bins of 5 GeV for $E_\mu^\text{rec}$ in the range  11 -- 21 GeV whereas last bin of ${E'}_\text{had}^\text{rec}$ is increased up to 25 GeV. The resulting binning scheme is shown in Table~\ref{tab:binning-2D-10years} where we have total 12 bins in $E_\mu^\text{rec}$, 21 bins in $\cos\theta_\mu^\text{rec}$ and 4 bins in ${E'}_\text{had}^\text{rec}$. We would like to mention that the bin sizes are chosen following the detector resolutions such that there is a sufficient number of events in each bin. Although the matter effects are experienced by upward-going neutrinos only, we have considered $\cos\theta_\mu^\text{rec}$ in the range of -1 to 1 because downward-going events help in increasing overall statistics as well as minimizing normalization uncertainties in atmospheric neutrino events. This also incorporates those upward-going (near horizon) neutrino events that result in downward-going reconstructed muon events due to angular smearing during neutrino interaction as well as reconstruction. We have considered the same binning scheme for $\mu^-$ as well as $\mu^+$.

%==========================
\subsection{Numerical Analysis}
%==========================

In this analysis, the $\chi^2$ statistics is expected to give median sensitivity of the experiment in the frequentist approach~\cite{Blennow:2013oma}. We define the following Poissonian $\chi^2_{-}$ for $\mu^-$ in $E_\mu^\text{rec}$, $\cos\theta_\mu^\text{rec}$, and ${E'}_\text{had}^\text{rec}$ observables as considered in Ref.~\cite{Devi:2014yaa}:

\begin{equation}\label{eq:chisq_mu-}
\chi^2_- = \mathop{\text{min}}_{\xi_l} \sum_{i=1}^{N_{{E'}_\text{had}^\text{rec}}} \sum_{j=1}^{N_{E_{\mu}^\text{rec}}} \sum_{k=1}^{N_{\cos\theta_\mu^\text{rec}}} \left[2(N_{ijk}^\text{theory} - N_{ijk}^\text{data}) -2 N_{ijk}^\text{data} \ln\left(\frac{N_{ijk}^\text{theory} }{N_{ijk}^\text{data}}\right)\right] + \sum_{l = 1}^5 \xi_l^2\,,
\end{equation}

where, 
\begin{equation}
N_{ijk}^\text{theory} = N_{ijk}^0\left(1 + \sum_{l=1}^5 \pi^l_{ijk}\xi_l\right)\,,
\end{equation}
$N_{ijk}^\text{theory}$ and $N_{ijk}^\text{data}$ represent the expected and observed number of $\mu^-$ events, respectively, in a given $(E_\mu^\text{rec}, \cos\theta_\mu^\text{rec}, {E'}_\text{had}^\text{rec})$ bin, whereas $N_{ijk}^0$ are the number of events without considering systematic uncertainties. In this analysis, we use the method of pulls~\cite{Gonzalez-Garcia:2004pka,Huber:2002mx,Fogli:2002pt} to consider five systematic uncertainties following Refs.~\cite{Ghosh:2012px,Thakore:2013xqa}: flux normalization error (20\%), cross section error (10\%), energy dependent tilt error in flux (5\%), error in zenith angle dependence of flux (5\%), and overall systematics (5\%).

Following the same procedure, we define $\chi^2_{+}$ for $\mu^+$, which will be calculated separately along with $\chi^2_{-}$. The total $\chi^2_\text{ICAL}$ for ICAL is calculated by adding $\chi^2_{-}$ and $\chi^2_{+}$
\begin{equation}\label{eq:chisq_total}
\chi^2_\text{ICAL} = \chi^2_{-} + \chi^2_+.
\end{equation}
We use the benchmark choice of oscillation parameters given in Table~\ref{tab:osc-param-value} as true parameters for simulating data. In theory, first of all, the $\chi^2_\text{ICAL}$ is minimized with respect to pull variables $\xi_l$ and then, marginalization is done for oscillation parameters  $\sin^2\theta_{23}$ in the range (0.36, 0.66), $\Delta m^2_\text{eff}$ in the range (2.1, 2.6) $\times10^{-3}$ eV\textsuperscript{2}, and mass ordering over NO and IO. The solar oscillation parameters $\sin^2 2\theta_{12}$ and $\Delta m^2_{21}$ are kept fixed at their true values given in Table~\ref{tab:osc-param-value} while performing the fit. As far as the reactor mixing angle is concerned, we consider a fixed value of $\sin^2 2\theta_{13} = 0.0875$ both in data and theory since this parameter is already very well measured~\cite{Marrone:2021,NuFIT,Esteban:2020cvm,deSalas:2020pgw}.
Throughout this analysis, we consider $\delta_\text{CP} = 0$ both in data and theory.

%==========================
\section{Results}
\label{sec:results}
%==========================

For statistical analysis, we simulate the prospective data assuming the three-layered core-mantle-crust profile as the true profile of the Earth. The statistical significance of the analysis for ruling out the mantle-crust profile with respect to the core-mantle-crust profile is quantified in the following way
\begin{equation}\label{eq:chisq_diff}
\Delta \chi^2_\text{ICAL-profile} = \chi^2_\text{ICAL}~ (\text{mantle-crust}) - \chi^2_\text{ICAL}~ (\text{core-mantle-crust})
\end{equation}
where, $\chi^2_\text{ICAL}$ (mantle-crust) and  $\chi^2_\text{ICAL}$  (core-mantle-crust) is calculated by fitting prospective data with mantle-crust profile and core-mantle-crust profile, respectively. Since the statistical fluctuations are suppressed to calculate Asimov (or median) sensitivity,  we have $\chi^2_\text{ICAL}~ (\text{core-mantle-crust}) \sim 0$. 

%==========================
\subsection{Effective Regions in $(E_\mu^\text{rec}, \cos\theta_\mu^\text{rec})$ Plane to Validate Earth's Core}
\label{sec:results_core}
%==========================

The sensitivity of ICAL towards various density profiles of Earth mainly stems from the Earth's matter effects experienced by neutrinos and antineutrinos while they travel long distances inside the Earth. For a given mass ordering, the Earth's matter effects felt by neutrinos and antineutrinos are different, which in turn alter the neutrino and antineutrino oscillation probabilities in a different fashion. In this work, while distinguishing between various density profiles of the Earth, the major part of the sensitivity comes from neutrino (antineutrino) mode if the true mass ordering is assumed to be NO (IO). We have elaborated on this issue in the next paragraph. On the contrary, while determining the sensitivity of ICAL towards neutrino mass ordering, both neutrino and antineutrino events contribute irrespective of the choice of true mass ordering~\cite{Ghosh:2012px,Devi:2014yaa}. 

The sensitivity of ICAL to rule out the simple two-layered mantle-crust profile of the Earth in theory while generating the prospective data with the three-layered core-mantle-crust profile mostly comes from $\mu^-$ ($\mu^+$) events if the true mass ordering is NO (IO). In the fixed-parameter scenario, we obtain the median $\Delta \chi^2_\text{ICAL-profile}$ (Data: core-mantle-crust, theory: mantle-crust) of 6.90 (4.10) if the true mass ordering is NO (IO). Note that when NO is our true choice, the contribution towards the fixed-parameter $\Delta \chi^2$ from $\mu^-$ ($\mu^+$) events is 6.85 (0.05). We see a completely opposite trend when IO is our true choice. For the true IO scenario, the contribution towards the fixed-parameter $\Delta \chi^2$ from $\mu^-$ ($\mu^+$) events is 0.02 (4.08).

\begin{figure}
	\centering
	\includegraphics[width=0.48\linewidth]{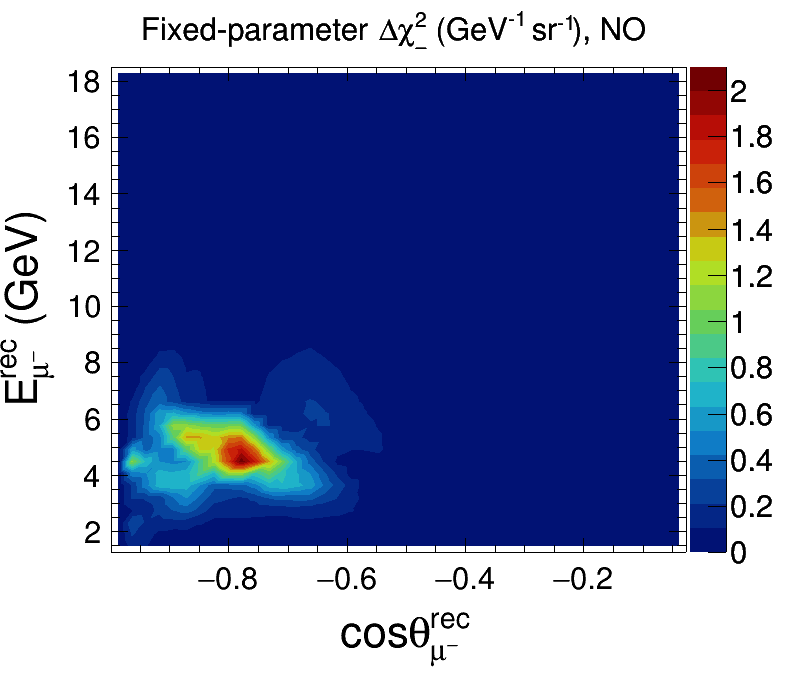}
	\includegraphics[width=0.48\linewidth]{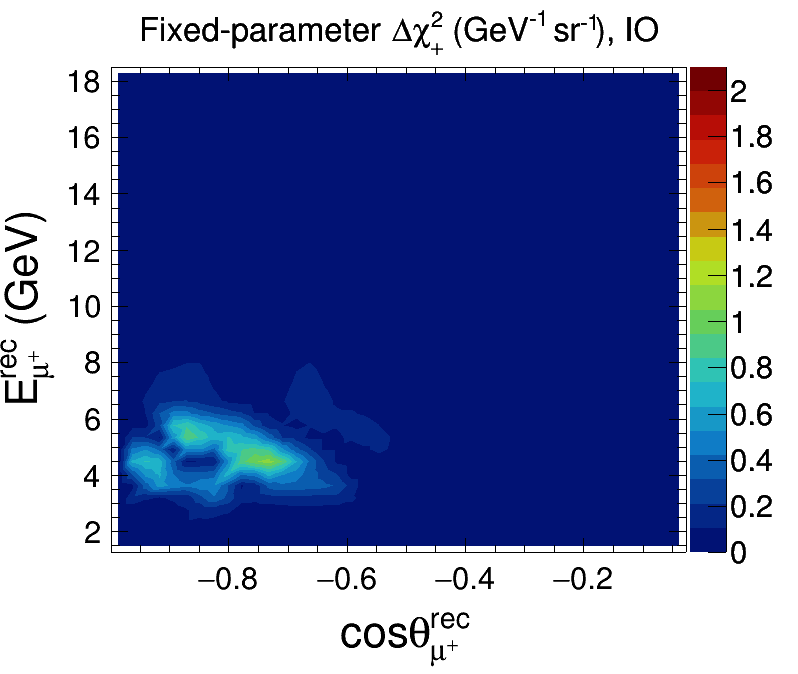}
	\caption{The distribution of fixed-parameter $\Delta \chi^2_{-}$ ($\Delta \chi^2_+$) with NO (IO) without pull penalty term for ruling out the mantle-crust profile in theory w.r.t. the core-mantle-crust profile in data in ($E_\mu^\text{rec},\; \cos\theta_\mu^\text{rec}$) plane as shown in the left (right) panel. Note that $\Delta \chi^2_{-}$ and $\Delta \chi^2_{+}$ is plotted in the unit of GeV\textsuperscript{-1} sr\textsuperscript{-1} where we have divided them by $2\pi\times \text{bin area}$. The $\Delta \chi^2_-$ ($\Delta \chi^2_+$) for IO (NO) is not significant, and hence, not shown here. We take the three-flavor oscillation parameters from Table~\ref{tab:osc-param-value}.~\cite{Kumar:2021faw}}
	\label{fig:chisq_contour}
\end{figure}

To identify the ranges of energy and direction which are contributing significantly to $\Delta \chi^2_\text{ICAL-profile}$ for ruling out the two-layered mantle-crust profile in theory against the three-layered core-mantle-crust profile in data, we have plotted the distribution of fixed-parameter $\Delta \chi^2_{-}$ and $\Delta \chi^2_+$ without pull penalty term\footnote{$\Delta \chi^2_{-}$ and $\Delta \chi^2_+$ are calculated without pull penalty $\sum_{l = 1}^5 \xi_l^2$ (see Eq.~\ref{eq:chisq_mu-}) to explore contributions from each bin in  ($E_\mu^\text{rec},\; \cos\theta_\mu^\text{rec}$) plane for $\mu^-$ and $\mu^+$ events, respectively.} as the contribution  towards $\Delta \chi^2$ from $\mu^-$ and $\mu^+$ events, respectively in ($E_\mu^\text{rec},\;\cos\theta_\mu^\text{rec}$) plane as shown in Fig.~\ref{fig:chisq_contour}. The left panel of Fig.~\ref{fig:chisq_contour} shows the distribution of $\Delta \chi^2_{-}$ (GeV\textsuperscript{-1} sr\textsuperscript{-1}) for NO in the plane of ($E_\mu^\text{rec},\;\cos\theta_\mu^\text{rec}$) where we can observe that the sensitivity to rule out Earth's core is contributed significantly by bins of higher baselines and multi-GeV energies in the range of 3 to 7 GeV of the reconstructed muons. The baselines with significant contribution correspond to the region around the boundary of core and mantle, where the matter density gets modified significantly during the merger of core and mantle to form the two-layered profile being probed here. 

We would like to mention that the detector response is already optimized by the ICAL collaboration for these core-passing events in the above-mentioned multi-GeV energy range as described in Sec.~\ref{sec:event_reco}. Since the reconstructed muon energy threshold of 1 GeV is much lower than the energies contributing to the sensitivity of ICAL toward the Earth's matter effect, the sensitivity of ICAL towards validating Earth's core is not going to be affected by the possible fluctuations around the energy threshold of 1 GeV in the ICAL detector. The contribution of $\Delta \chi^2_{+}$ for NO is negligible and hence not shown here. In the same fashion, the right panel of Fig.~\ref{fig:chisq_contour} shows the distribution of $\Delta \chi^2_{+}$ (GeV\textsuperscript{-1} sr\textsuperscript{-1}) for IO where also, the contribution appears from the lower energies and higher baselines. The contribution of $\Delta \chi^2_{+}$ for IO is smaller than that for $\Delta \chi^2_{-}$ for NO because the lower cross-section for antineutrino results in the lesser statistics of $\mu^+$ events compared to $\mu^-$ events. For the case of IO, the contribution of $\Delta \chi^2_{-}$ is not significant.  

%==========================
\subsection{Sensitivity to Validate Earth's Core with and without CID}
\label{sec:results_core_margin}
%==========================

Till now, we have shown the fixed-parameter results, but now for final results, we marginalize over oscillation parameters $\sin^2\theta_{23}$, $\Delta m^2_\text{eff}$ and mass ordering while incorporating systematic errors as explained in Sec.~\ref{sec:statistical analysis}. The total statistical significance includes contributions from both $\mu^-$ as well as $\mu^+$ as shown in Eq.~\ref{eq:chisq_total}. Here, we calculate the statistical significance to rule out the alternative profiles of Earth in theory with respect to the three-layered profile of core-mantle-crust in data as shown in Table~\ref{tab:chisq_analysis_results}. We have also compared alternative profiles of Earth in theory with respect to the PREM profile~\cite{Dziewonski:1981xy} in MC data. We would like to remind you that the PREM profile is with 25 layers as described in Sec.~\ref{sec:earth_model} by the solid black line in the right panel of Fig.~\ref{fig:three-layer-model}.

We can observe in Table~\ref{tab:chisq_analysis_results} that the $\Delta \chi^2_\text{ICAL-profile}$ for ruling out the vacuum in theory with respect to the three-layered profile of core-mantle-crust in data is 4.65 for NO (true) with CID, which shows that ICAL has good sensitivity towards the presence of matter effects. In the absence of CID, this $\Delta \chi^2_\text{ICAL-profile}$ drops to 2.96, which shows that the capability of ICAL to distinguish $\mu^-$ and $\mu^+$ is crucial to observe the matter effects. For the case of IO (true), these numbers decrease further because, in this case, most of the contribution comes from $\mu^+$ that has lesser statistics due to a lower cross-section of antineutrinos compared to neutrinos.

%========================
\begin{table}[hbt!]
	\begin{center}
		\begin{tabular}{|c|c|c|c|c|c|}
			\hline \hline
			\multirow{3}{*}{MC Data} & \multirow{3}{*}{Theory} & \multicolumn{4}{c|}{$\Delta \chi^2_\text{ICAL-profile}$}\\ \cline{3-6}
			& & \multicolumn{2}{c|}{NO(true)} & \multicolumn{2}{c|}{IO(true)}\\ \cline{3-6}
			& & with CID & w/o CID & with CID & w/o CID\\
			\hline
			Core-mantle-crust & Vacuum    & 4.65 & 2.96 & 3.53 & 1.43 \\
			Core-mantle-crust & Mantle-crust& 6.31 & 3.19 & 3.92 & 1.29 \\
			Core-mantle-crust &	Core-mantle & 0.73 & 0.47 & 0.59 & 0.21 \\
			Core-mantle-crust & Uniform   & 4.81 & 2.38 & 3.12 & 0.91 \\
			& & &  & & \\
			PREM profile & Core-mantle-crust   & 0.36 & 0.24 & 0.30 & 0.11 \\
			PREM profile & Vacuum & 5.52 & 3.52 & 4.09 & 1.67 \\
			PREM profile & Mantle-crust & 7.45 & 3.76 & 4.83 & 1.59 \\
			PREM profile & Core-mantle & 0.27 & 0.18 & 0.21 & 0.07 \\
			PREM profile & Uniform  & 6.10 & 3.08 & 3.92 & 1.18 \\
			\hline \hline
		\end{tabular}
	\end{center}
	\caption{Ruling out the alternative profiles of Earth at the median $\Delta \chi^2$ level. We marginalize over oscillation parameters $\sin^2\theta_{23}$, $\Delta m^2_\text{eff}$, and mass ordering in theory, whereas the remaining oscillation parameters are kept fixed at their benchmark values as mentioned in Table~\ref{tab:osc-param-value}. The third and fourth (fifth and sixth) columns show results considering NO (IO) as true mass ordering in data. The results in the third and fifth (fourth and sixth) columns are with (without) the charge identification capability of ICAL.~\cite{Kumar:2021faw}}
	\label{tab:chisq_analysis_results}
\end{table}
%========================

Since we have found that ICAL can sense the presence of matter effects, now we can calculate the statistical significance to identify the profile that satisfies the distribution of matter inside Earth. The $\Delta \chi^2_\text{ICAL-profile}$ for ruling out the two-layered coreless profile of mantle-crust in theory with respect to the three-layered core-mantle-crust profile in the prospective data is about 6.31 for NO (true) with CID, and this is the sensitivity with which ICAL can validate the presence of core inside Earth in the context to PREM model. For the case of IO (true), this result drops to 3.92. 

We find that the trend in the final results with marginalization is the same as observed for the fixed-parameter case. After marginalization, the contributions from $\mu^-$ ($\mu^+$) events towards the $\Delta \chi^2_\text{ICAL-profile}$ for validating Earth's core is 6.09 (0.21) for NO as the true choice of mass ordering. If IO is the true mass ordering, then we see an opposite trend where the contribution towards the $\Delta \chi^2_\text{ICAL-profile}$ from $\mu^-$ ($\mu^+$) is 0.09 (3.82) after marginalization.

We would like to mention that if we do not incorporate hadron energy information and just use ($E_\mu^\text{rec}$, $\cos\theta_\mu^\text{rec}$) binning scheme from Table~\ref{tab:binning-2D-10years} then the $\Delta \chi^2_\text{ICAL-profile}$ for validating Earth's core after marginalization over oscillation parameters is about 3.20 for NO (true) with CID. Thus, we can say that the incorporation of hadron energy information improves the sensitivity of ICAL towards validating Earth's core.

The $\Delta \chi^2_\text{ICAL-profile}$ for ruling out the core-mantle profile in theory with respect to the core-mantle-crust profile in data is smaller than 1, which shows that the matter effects caused by crust is not significant. For ruling out the uniform distribution of matter in theory, we get  $\Delta \chi^2_\text{ICAL-profile}$ as 4.81, which indicates the capability of ICAL to feel the non-uniformity in density distribution inside Earth. 

We would like to mention that it does not make much difference if we perform analysis using the simple three-layered profile instead of the PREM profile and save computational time. The $\Delta \chi^2_\text{ICAL-profile}$ for the three-layered profile of core-mantle-crust in theory with respect to 25-layered PREM profile (as shown by the black line in the right panel of Fig.~\ref{fig:three-layer-model}) in data is 0.36 (0.30) for NO (IO) which shows that irrespective of the choice of the ordering of neutrino masses in nature, the analysis of atmospheric neutrino data with the simplified three-layered profile is a legitimate choice. Note that if we generate our prospective data with the more refined PREM profile having 25 layers and try to distinguish it from our hypothetical mantle-crust profile in theory, then we get a slightly increased $\Delta \chi^2_\text{ICAL-profile}$ of 7.45 for NO and 4.83 for IO.

%==========================
\subsection{Impact of Marginalization over Various Oscillation Parameters}
\label{sec:results_margin_impact}
%==========================

The sensitivity of ICAL to differentiate various density profiles of Earth may get deteriorated due to the uncertainties in neutrino oscillation parameters. To understand the impact of uncertainties of individual oscillation parameters on the sensitivity of ICAL to rule out an alternative profile of Earth while generating prospective data with the three-layered profile of core-mantle-crust, we marginalize over one oscillation parameter at a time in theory as shown in Table~\ref{tab:margin_impact_chisq}. In data, we take NO as true mass ordering and use the benchmark values of oscillation parameter given in Table~\ref{tab:osc-param-value}. 

%========================
\begin{table}[hbt!]
	\begin{center}
		\begin{tabular}{|l|l|c|c|c|c|c|}
			\hline \hline
			\multirow{3}{*}{MC Data} & \multirow{3}{*}{Theory} & \multicolumn{5}{c|}{$\Delta \chi^2_\text{ICAL-profile}$} \\ \cline{3-7}
			&  & Fixed & \multicolumn{4}{c|}{Marginalization over}\\ \cline{4-7}
			&  & parameter & $\sin^2\theta_{23}$ & $|\Delta m^2_\text{eff}|$ & $\pm|\Delta m^2_\text{eff}|$ & All \\ \hline
			Core-mantle-crust &	Mantle-crust & 6.90 & 6.36 & 6.84 & 6.84 & 6.31 \\
			Core-mantle-crust &	Vacuum & 6.80 & 6.44 & 5.16 & 4.94 & 4.65 \\
			PREM &	Mantle-crust & 7.88 & 7.47 & 7.81 & 7.81 & 7.45 \\
			PREM &	Vacuum & 7.71 & 7.28 & 6.10 & 5.89 & 5.52 \\
			\hline \hline
		\end{tabular}
	\end{center} 
	
	\caption{The impact of marginalization over oscillation parameters $\sin^2\theta_{23}$, $|\Delta m^2_\text{eff}|$, and mass ordering on the sensitivity of ICAL to rule out the alternative profile of Earth at the median $\Delta \chi^2$ level. We assume true mass ordering as NO in data. The $\Delta \chi^2_\text{ICAL-profile}$ for the fixed-parameter case is given in the third column. The marginalized $\Delta \chi^2_\text{ICAL-profile}$ obtained after performing minimization separately over $\sin^2\theta_{23}$, $|\Delta m^2_\text{eff}|$, and $\Delta m^2_\text{eff}$ (with both mass orderings) in theory are given in fourth, fifth and sixth columns, respectively. The marginalized $\Delta \chi^2_\text{ICAL-profile}$ after performing combined minimization over $\sin^2\theta_{23}$, $\Delta m^2_\text{eff}$, and both mass orderings in theory is given in the last column. The remaining oscillation parameters are kept fixed at their benchmark values as mentioned in Table~\ref{tab:osc-param-value}.~\cite{Kumar:2021faw}}
	\label{tab:margin_impact_chisq}
\end{table}
%========================

The third column of Table~\ref{tab:margin_impact_chisq} shows the fixed-parameter $\Delta \chi^2_\text{ICAL-profile}$ where we have not marginalized over any oscillation parameters in theory. In the fourth column, we marginalize over $\sin^2\theta_{23}$ in the range (0.36, 0.66) in theory and keep the other oscillation parameter fixed at their benchmark values as mentioned in Table~\ref{tab:osc-param-value}. Similarly, we marginalize over $|\Delta m^2_\text{eff}|$ in the range (2.1, 2.6) $\times \,10^{-3} ~\text{eV}^2$ with same mass ordering (NO) in theory and data as shown in the fifth column. In the sixth column, we marginalize over $\Delta m^2_\text{eff}$ while considering both mass orderings in theory which effectively varies $\Delta m^2_\text{eff}$ in the range (-2.6, -2.1) $\times 10^{-3}~\text{eV}^2$ and (2.1, 2.6) $\times 10^{-3}~\text{eV}^2$. Finally, in last column, we shows $\Delta \chi^2_\text{ICAL-profile}$ with marginalization over $\sin^2\theta_{23}$, $\Delta m^2_\text{eff}$, and both mass orderings in theory.

The median $\Delta \chi^2_\text{ICAL-profile}$ which is the sensitivity of ICAL to rule out the two-layered profile of mantle-crust while generating prospective data with the three-layered profile of core-mantle-crust, is 6.90 when no marginalization is performed over any oscillation parameter as shown in the first row of Table~\ref{tab:margin_impact_chisq}. After marginalization over $\sin^2\theta_{23}$, $\Delta m^2_\text{eff}$, and both mass orderings in theory, the above-mentioned $\Delta \chi^2_\text{ICAL-profile}$ drops to 6.31. Here, marginalization over $\sin^2\theta_{23}$, in theory, affects the sensitivity most.	

Similarly, when we rule out vacuum scenario in theory by generating data with the core-mantle-crust profile, we obtain $\Delta \chi^2_\text{ICAL-profile}$ of 6.80 if we do not marginalize over any oscillation parameter in theory as shown in the second row of Table~\ref{tab:margin_impact_chisq}. This $\Delta \chi^2_\text{ICAL-profile}$ reduces to 4.65 if we marginalize over $\sin^2\theta_{23}$, $\Delta m^2_\text{eff}$, and both mass orderings in theory. We observe that in this case, the marginalization over $\Delta m^2_\text{eff}$, and both mass orderings substantially reduces the $\Delta \chi^2_\text{ICAL-profile}$. 

From the above-mentioned observations, we can conclude that the marginalization over oscillation parameters has a large impact when we attempt to distinguish between various density profiles at ICAL. In the future, the more precise determination of oscillation parameters will help us to rule out the alternative profiles with better sensitivity at ICAL. The above findings hold if we generate the prospective data with the 25-layered PREM profile instead of the three-layered profile of core-mantle-crust and differentiate it against the mantle-crust or vacuum profile in theory. 

%==========================
\subsection{Impact of Different true Choices of $\sin^2\theta_{23}$}
\label{sec:results_th23}
%==========================

\begin{figure}
	\centering
	\includegraphics[width=0.49\linewidth]{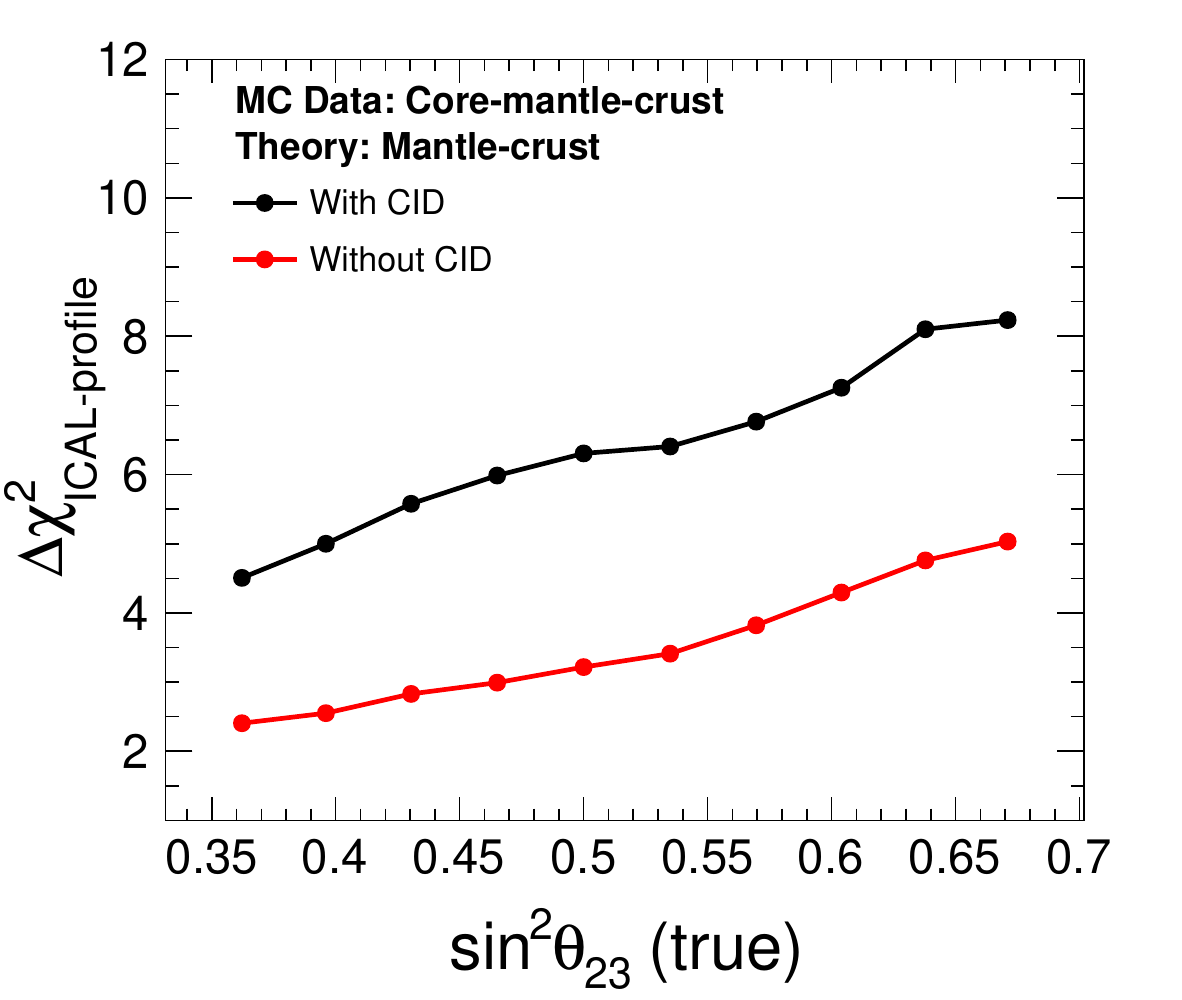}	\includegraphics[width=0.49\linewidth]{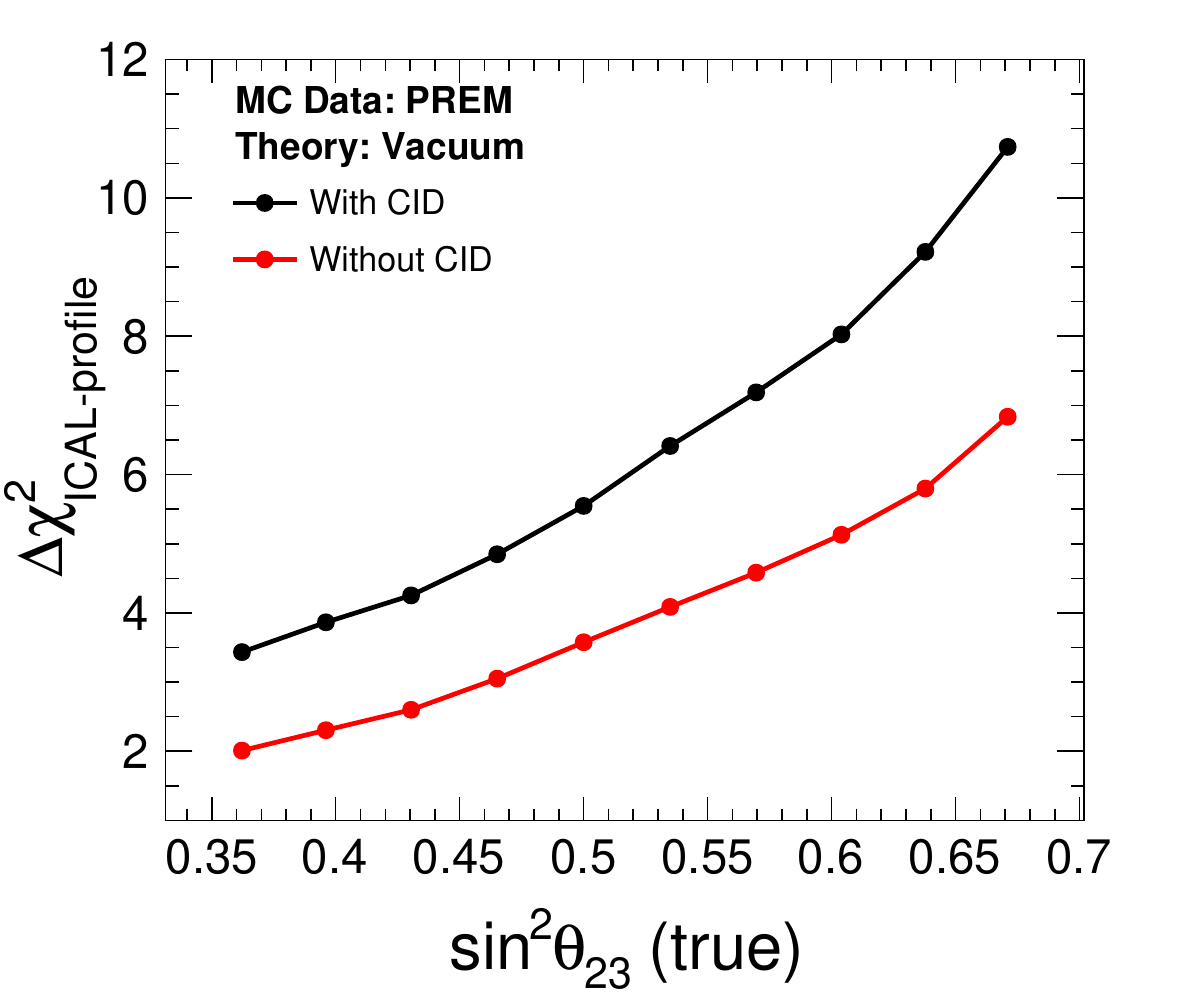}
	\caption{The median $\Delta \chi^2_\text{ICAL-profile}$ as a function of the choice of $\sin^2\theta_{23}$ in data. The median $\Delta \chi^2_\text{ICAL-profile}$ is the sensitivity with which we can validate Earth's core by ruling out the mantle-crust profile in theory w.r.t. the core-mantle-crust profile in data as shown in the left panel. The right panel shows the sensitivity at median $\Delta \chi^2_\text{ICAL-profile}$ level to rule out vacuum scenario in theory w.r.t. the PREM profile in data. In both the panels, the black (red) curves show the sensitivities with (without) charged identification capability of ICAL. Note that we marginalize over oscillation parameters $\sin^2\theta_{23}$, $\Delta m^2_\text{eff}$, and mass ordering, whereas the remaining oscillation parameters are kept fixed at their benchmark values as mentioned in Table~\ref{tab:osc-param-value}. We assume mass ordering as NO in data.~\cite{Kumar:2021faw}}
	\label{fig:chisq_th23_var}
\end{figure}

So far, we have taken in our analysis, $\sin^2\theta_{23}(\text{true}) = 0.5$ as our benchmark choice but the recent global fit data also indicates that $\theta_{23}$ may not be maximal, it can either lie in the lower octant where $\sin^2\theta_{23} < 0.5$ or the higher octant where $\sin^2\theta_{23} > 0.5$. Needless to mention that $\theta_{23}$ is the most uncertain oscillation parameter at present apart from $\delta_\text{CP}$. So, now, it is legitimate to see how the sensitivity of ICAL towards validating the Earth's core may change if, in nature, $\theta_{23}$ (true) turns out to be non-maximal. To analyze this, we are presenting Fig.~\ref{fig:chisq_th23_var} where, in the x-axis, we have varied the choice of $\sin^2 \theta_{23}$ in data in the range 0.36 to 0.66, and in the y-axis, we are evaluating the median $\Delta \chi^2_\text{ICAL-profile}$, the sensitivity with which we can validate Earth' core (left panel) and rule out vacuum scenario in theory with respect to PREM profile in data (right panel). Here, we marginalize over oscillation parameters $\sin^2\theta_{23}$ in the range of 0.25 to 0.75, $\Delta m^2_\text{eff}$ in the range of (2.1, 2.6) $\times 10^{-3}~\text{eV}^2$ and both the mass orderings NO as well as IO, whereas the remaining oscillation parameters are kept fixed at their benchmark values as mentioned in Table~\ref{tab:osc-param-value}. 

%%========================
%%========================

The dominant contribution of matter effect appears in term of $\sin^2\theta_{23}$ for survival probability $P(\nu_\mu\rightarrow \nu_\mu)$ as well as appearance probability $P(\nu_e\rightarrow \nu_\mu)$ as shown by series expansion in Ref.~\cite{Akhmedov:2004ny}. $P(\nu_\mu\rightarrow \nu_\mu)$ decreases almost linearly with $\sin^2\theta_{23}$ whereas $P(\nu_e\rightarrow \nu_\mu)$ increases linearly. Since the contribution of appearance $(\nu_e\rightarrow \nu_\mu)$ channel is smaller than that of survival $(\nu_\mu\rightarrow \nu_\mu)$ channel, the net matter effect does not cancel out completely and shows almost linear dependence on $\sin^2\theta_{23}$.

This linear dependence of matter effect on $\sin^2\theta_{23}$ results in an increasing $\Delta \chi^2_\text{ICAL-profile}$ with $\sin^2\theta_{23}$(true) as shown in both panels in Fig.~\ref{fig:chisq_th23_var} because $\Delta \chi^2_\text{ICAL-profile}$ in both the cases are driven by matter effect. Thus, we can say that the Earth's core can be validated with a higher confidence level if, in nature, $\theta_{23}$ is found to be lying in the higher octant. We can also observe in both cases that $\Delta \chi^2_\text{ICAL-profile}$ is higher if the charge identification capability is present. Thus, the presence of charge identification capability is crucial in validating Earth's core (left panel) as well as ruling out vacuum scenario (right panel).

%==========================
\section{Summary and Concluding Remarks}
\label{sec:tomography_conclusion}
%==========================

Atmospheric neutrinos travel long distances inside Earth and feel the presence of matter effects that depend upon the density distribution inside Earth. Neutrino oscillation tomography utilizes the matter effects experienced by neutrinos to unravel the internal structure of Earth. Guided by the PREM profile, we use a three-layered density profile of Earth where we have core, mantle, and crust. For comparison, we consider alternative profiles of Earth -- mantle-crust, core-mantle, and uniform density. 

In Sec.~\ref{sec:tomography_probability}, we show the effect for various profiles of Earth on the neutrino oscillations in survival $(\nu_\mu\rightarrow\nu_\mu)$ and appearance $(\nu_e\rightarrow\nu_\mu)$ channels. We observe that the presence of mantle and core result in the MSW resonance and NOLR/parametric resonance, respectively. On the other hand, the presence of a boundary between layers results in a sharp transition in oscillation probabilities in $P(\nu_e\rightarrow\nu_\mu)$ channel.

Table~\ref{tab:tomography_events} shows that about 4614 $\mu^-$ and 2053 $\mu^+$ events are expected at ICAL for 500 kt$\cdot$yr exposure considering three-flavor neutrino oscillations in the presence of matter with the PREM profile. Utilizing the neutrino directions, we estimate that about 331 $\mu^-$ and 146 $\mu^+$ core-passing events would be detected at ICAL in 10 years as shown in Table~\ref{tab:layer-passing-event}. The events passing through the crust-mantle region and only crust are also shown in Table~\ref{tab:layer-passing-event}. In Fig.~\ref{fig:event_dist_nu_density_zones}, we can observe that the information about the region traversed by neutrinos is preserved even after reconstruction as muon events, but some of the reconstructed muons may get smeared into other regions due to reaction kinematics and finite detector resolution.

After identifying the events passing through various regions inside Earth, we perform statistical analysis to differentiate between two profiles of Earth using atmospheric neutrino events at ICAL. We would like to mention that $\Delta \chi^2$ for the determination of mass ordering is contributed by both neutrino and antineutrino irrespective of the choice of true mass ordering. On the other hand, in our study, where we are contrasting between different profiles of Earth for a given mass ordering, $\Delta \chi^2$ is mostly contributed by neutrino for NO (true) and antineutrino for IO (true). We estimate statistical significance at $\Delta \chi^2$ level to rule out the coreless profile of mantle-crust with respect to the core-mantle-crust profile as given by Eq.~\ref{eq:chisq_diff}. Figure~\ref{fig:chisq_contour} shows that the significant contribution to $\Delta \chi^2_{-}$ (NO) and $\Delta \chi^2_+$ (IO) is received from higher baselines and lower energies which is the region around the boundary between core and mantle. The density in this region gets significantly modified in the absence of a core.  

We show the final results in Table~\ref{tab:chisq_analysis_results} in terms of $\Delta \chi^2_\text{ICAL-profile}$ to rule out the alternative profiles in theory with respect to the core-mantle-crust profile in data. For final results, $\Delta \chi^2_\text{ICAL-profile}$ is marginalized over oscillation parameters $\sin^2\theta_{23}$, $\Delta m^2_\text{eff}$ and mass ordering. The results for the coreless profile of mantle-crust in theory with respect to the core-mantle-crust profile in the prospective data show that the presence of Earth's core in the context of PREM model can be validated at $\Delta \chi^2_\text{ICAL-profile}$ of 6.31 for NO (true) and 3.92 for IO (true) using 500 kt$\cdot$yr exposure at ICAL with charge identification capability. On the other hand, if we generate our prospective data with a more refined PREM profile of the Earth having 25 layers and contrast it with our hypothetical profile of the Earth consisting of only mantle and crust in theory, then we get a slightly enhanced $\Delta \chi^2_\text{ICAL-profile}$ of 7.45 for NO (true) and 4.83 for IO (true). Important to note that in the absence of charge identification capability of ICAL, these sensitivities deteriorate significantly to 3.76 for NO (true) and 1.59 for IO (true).

We demonstrate that the sensitivity to rule out the alternative profiles of Earth deteriorates with marginalization. This indicates that with the improvement in the precision of oscillation parameters in the future, the alternate profiles of Earth can be ruled out with better sensitivity. In Fig.~\ref{fig:chisq_th23_var}, we show that the sensitivity to validate Earth's core increases as we increase the true value of $\sin^2\theta_{23}$. Thus, the presence of Earth's core can be validated at higher sensitivity if $\theta_{23}$ is found to be lying in the higher octant. It is important to note that the presence of charge identification capability is an important feature of ICAL, which significantly improves the results for studies involving matter effects. We hope that the analysis performed in this chapter will open a new vista for the ICAL detector at the upcoming INO facility.

\end{refsegment}

\cleartooddpage
\chapter{Summary and Concluding Remarks}
\label{chap:summary}
\begin{refsegment}
Neutrinos were proposed to explain the phenomenon of beta decay. The discovery of neutrinos happened after 30 years using the intense source of nuclear reactor. Later, neutrinos were detected from various sources like atmosphere, Sun, accelerators, reactors, supernova, and radioactivity in Earth. The anomalies in solar and atmospheric neutrino fluxes were resolved with the discovery of neutrino oscillations by Super-Kamiokande experiment using the data of atmospheric neutrinos. Over the past two decades, most of the neutrino oscillation parameters have been measured with good precision. The CP-phase $\delta_{\rm CP}$ and atmospheric mixing angle $\theta_{23}$ still possess large uncertainties. Another important aim of future neutrino experiments are the determination of neutrino mass ordering and the octant of $\theta_{23}$. These details are described in chapters~\ref{chap:intro} and \ref{chap:neutrino_oscillations}.

Atmospheric neutrinos have played an important role in the precision measurement of oscillation parameters $\Delta m^2_{32}$ and $\theta_{23}$. Atmospheric neutrinos have access to a multi-GeV range of energies over a wide range of baselines. The $W$-mediated charged-current interactions of upward-going neutrinos with the ambient electrons inside Earth result into the matter effects that modify the oscillation patterns for neutrinos and antineutrinos differently. The separate measurements of matter effects in neutrinos and antineutrinos can help us determine the neutrino mass ordering. 

The upcoming 50 kt Iron Calorimeter (ICAL) detector at the India-based Neutrino Observatory (INO) aims to detect atmospheric muon neutrinos and antineutrinos separately in the energy range of about 1 to 25 GeV covering the baselines from about 15 to 12750 km. ICAL consists of stacks of iron layers as passive detector elements and Resistive Plate Chambers (RPCs) sandwiched between them as active detector elements. The charged-current interactions of neutrinos in the iron layers result into the production of muons and hadrons. The magnetic field of 1.5 T enables ICAL to distinguish between $\mu^-$ and $\mu^+$ events and hence, $\nu_\mu$ and $\bar{\nu}_\mu$. The muon results into a track-like event, whereas hadron leads to a shower-like event. The ICAL detector is described in detail in chapter~\ref{chap:ICAL}. Now, we described the work included in this thesis in chapters~\ref{chap:RPC_response} to \ref{chap:tomography}. 

In chapter~\ref{chap:RPC_response}, we describe the studies on response uniformity of RPC. We study the effect of non-uniform resistivity of graphite layer on detector response. A ROOT-based mathematical framework has been developed to simulate the charge transport in graphite layer. An experimental setup is developed using linear stage, and picoammeter to measure the non-uniform resistivity of graphite layer which is given as an input to the simulation. The effect of non-uniform resistivity is studied by simulating the potential buildup on the application of high voltage on one edge of the graphite layer. The simulations predicted that the potential distribution is uniform at the saturation but the time-constant gets affected by the non-uniform surface resistivity. To verify these predictions experimentally, a high-impedance probe is developed using an operational amplifier in voltage follower circuit. The experimentally measured distributions of potential and time-constant are found to be in good agreement with simulated distributions. 

In chapter~\ref{chap:dip_valley}, we show that an atmospheric neutrino experiment like ICAL can observe oscillation dip and oscillation valley features in reconstructed muon observables. First, we show the oscillation dip feature in $\nu_\mu$ survival probability as a function of $L_\nu/E_\nu$. The oscillation dip corresponds to the situation when oscillation is maximum, i.e., $P(\nu_\mu\rightarrow \nu_\mu) =  0$. The oscillation dip manifests as the oscillation valley in the plane of ($E_\nu$, $\cos\theta_\nu$) in the form of a diagonal band. Next, we demonstrate that the oscillation dip can be observed at ICAL using the ratio of upward-going ($U$) and downward-going ($D$) reconstructed muon events as a function of $L_\mu^\text{rec}/E_\mu^\text{rec}$. We propose a dip-identification algorithm to find the location of the dip that is used to measure the value of $\Delta m^2_{32}$. We also show that the value of atmospheric mixing angle $\theta_{23}$ can be measured using the ratio of total $U/D$ ratio. We further demonstrated that the oscillation valley feature can be observed in the plan of $(E_\mu^\text{rec},\cos\theta_\mu^\text{rec}$) using $U/D$ ratio at ICAL. We show that the oscillation valley can be fitted to obtain the alignment that is used to measure the value of $\Delta m^2_{32}$. Note that in both these approaches, we measure the value of $\Delta m^2_{32}$ independently for $\mu^-$ and $\mu^+$ channels. In our methods of oscillation dip and valley, we also incorporate the statistical fluctuations, systematic errors, and uncertainties in neutrino oscillation parameters using multiple sets of simulated data. These approaches provide a new way of measuring atmospheric oscillation parameters, which is complementary to the $\Delta\chi^2$ approach. Since long-baseline experiments have access to only a fixed baseline, they can perform only oscillation dip analysis. On the other hand, the atmospheric neutrino experiments like Hyper-K, ORCA, IceCube-Upgrade, and DUNE-atmospheric can perform oscillation dip as well as oscillation valley analyses. The observation of oscillation dip and valley would provide an orthogonal approach to establish the nature of neutrino oscillations and hence make the neutrino oscillation picture more robust.

In chapter~\ref{chap:NSI}, we propose a new approach to probe neutral-current non-standard interactions (NSIs) of neutrinos using oscillation dip and valley in reconstructed muon observables at ICAL. We focus on the flavor-changing NSI parameter $\epsmutau$, which has a maximum impact on the survival probability of muon neutrinos. We observe that the non-zero value of $\epsmutau$ results in the shifts in the dip locations, which are opposite for $\mu^-$ and $\mu^+$. The direction of shift of dip location also depends upon the sign of $\epsmutau$ as well as mass ordering. We introduce a novel variable $\Delta d$ representing the difference in dip locations for $\mu^-$ and $\mu^+$, which has sensitivity towards the magnitude as well as the sign of $\epsmutau$ and is independent of $\Delta m^2_{32}$. We use the difference in dip locations for $\mu^-$ and $\mu^+$ to place constraint on $\epsmutau$ using 500 kt$\cdot$yr exposure at the ICAL detector. As far as the oscillation valley is concerned, it bends in the presence of NSI parameter $\epsmutau$. These bendings of oscillations valleys are opposite for $\mu^-$ and $\mu^+$. The direction of bending of oscillation valley also depends on the sign of $\epsmutau$ and mass ordering. We demonstrate that the contrast in the curvatures of oscillations valleys for $\mu^-$ and $\mu^+$ can also be used to estimate the bounds on $\epsmutau$. The estimated precision on $|\epsmutau|$ using oscillation dip and valley measurements is about 2\% at 90\% C.L. using 500 kt$\cdot$yr exposure at ICAL. The effects of statistical fluctuations, systematics errors, and uncertainties in neutrino oscillation parameters are incorporated using multiple sets of simulated data. It is important to note that the charge identification capability of ICAL plays an important role in observing the opposite shifts in dip locations and the contrasts in curvatures of valleys for $\mu^-$ and $\mu^+$. We expect that the further exploration of features of oscillation dip and valley at experiments like ICAL would help us understand the properties of neutrinos in novel ways. 

The interior of Earth consists of unknown regions and extreme environments which are inaccessible directly. The information about the internal structure of Earth has been obtained using gravitational measurements and seismic studies, which depend upon the gravitational interaction and electromagnetic interaction, respectively. On the other hand, atmospheric neutrinos can use the weak interactions with the ambient electrons inside Earth to probe the internal structure of Earth. The idea of tomography of Earth using the absorption of neutrinos has been explored in many studies. Significant progress in the precision measurements of neutrino oscillation parameters has opened a unique avenue for using neutrino oscillations as a tool to probe the interior of Earth. The neutrino oscillation tomography exploits the matter effects experienced by neutrinos to reveal the internal structure of Earth. In chapter~\ref{chap:tomography}, we show that the atmospheric neutrinos at 50 kt ICAL detector can be used to explore the core of the Earth. We show that the presence of mantle results in the MSW resonance, whereas the presence of core gives rise to the neutrino oscillation length resonance or parametric resonance. Using the excellent direction resolution of ICAL, we can probe the regions through which neutrino has traversed. In 10 years, ICAL would detect about 331 $\mu^-$ and 146 $\mu^+$ events corresponding to the core-passing neutrinos. We demonstrate that the presence of the core of Earth can be validated with a sensitivity of 7.45 (4.83) for normal (inverted) mass ordering by ruling the mantle-crust profile with respect to the PREM profile of Earth. Note that in the absence of charge identification capability of ICAL, this sensitivity would deteriorate to 3.76 (1.59) for normal (inverted) ordering. The huge amount of atmospheric neutrino data going to be collected at the next-generation neutrino experiments would significantly improve the prospects of neutrino oscillation tomography of Earth. The combined approaches of gravitational measurements, seismic studies, geoneutrino detection, neutrino absorption, and neutrino oscillation would start an era of multi-messenger tomography of Earth. 

We believe that the physics analyses performed in this thesis would open new ways of studying neutrino oscillations, probing BSM scenarios, and pave the way for neutrino as a tool for tomography of Earth. In the experiment part of this thesis, the framework developed to simulate charge transport can be a stepping stone for a more detailed simulation of devices like RPCs. 

\section*{Future Scope}
	The framework developed to simulate charge transport in the graphite layer can be extended to simulate more complex geometries like a three-dimensional device of RPC. In the present work, our focus was on developing and bench-marking the simulation framework; hence, the input conditions were kept constant. For future analysis, we can also extend the simulation framework to accept the time-dependent inputs, which can help us simulate the propagation of signals. This can open the possibility of studying attenuation, reflection, and collection of signals. Therefore, we believe that the simulation and experimental approach presented in this thesis could be useful for a more detailed understanding of the working of devices like RPC.
	
	The approach of using oscillation dip and valley to measure the neutrino oscillation parameters and probe the non-standard interactions of neutrinos is novel and complementary to traditional analyses. The rich features of oscillation dip and valley can be used to explore other BSM physics scenarios. A more refined analysis can extract more information from the oscillation valley, which can, not only probe new physics models, but also help us in removing certain degeneracies, just like the measurement of $\epsmutau$ was independent of $\Delta m^2_{32}$. Access to $\mu^-$ and $\mu^+$ channels also provide an opportunity to look for features that are different for neutrinos and antineutrinos. Therefore, we believe that the oscillation dip and valley have the potential to help us explore unique features of neutrino oscillations. 
	
	The atmospheric neutrinos have access to wide ranges of energies, and baselines, which enable them to reach the unexplored regions of Earth. The precisely measured atmospheric neutrino data in next-generation neutrino experiments can reveal the more detailed structure of Earth. Most importantly, the neutrino oscillation tomography of Earth would exploit the weak interactions with electrons inside the Earth which is complementary to gravitational and seismic measurements. Neutrino tomography can emerge as a practical application with the potential to shed light on pressing issues like the location of the core-mantle boundary, the chemical composition of the core, and the state of matter inside the core, etc. Therefore, the era of multi-messenger tomography would enable us to improve our understanding of our planet.
	
\end{refsegment}

\cleartooddpage
\appendix
%%===============================
%\bibliographystyle{JHEP}
%\bibliography{references.bib}
%%================================
\printbibliography %(Print bibliography: You have to run 'bibtex')

\end{document}